\documentclass[a4paper,12pt,openany]{book}
\usepackage[applemac]{inputenc} % Spécifie que tu utilise des encodages de caractères 'à la macintosh'
\usepackage[T1]{fontenc} % Cette ligne est utile. L'intérêt est surtout avec les caractères accentués et la production de pdf. C'est surtout utile en français mais je conseille de la garder
\usepackage{geometry} % Pour une bonne définition des marges (ici valeurs par défaut)
\usepackage{amssymb,amsmath} % Pour charger les extensions de l'AMS avec les extensions de symboles correspondantes
\usepackage{graphicx} % Pour intégrer des graphiques
\usepackage[english]{babel} % Quelques changements cosmétiques dépendant  de la langue.
\usepackage{mathrsfs} % Pour des jolies lettres rondes majuscules10
\usepackage{hyperref}
\newcommand{\kB}{\ensuremath{k_{\mathrm{B}}}} % constante de Boltzmann avec un k italique et un B droit en indice
\newcommand{\ave}[1]{\ensuremath{\left \langle {#1} \right \rangle}} % moyenne avec des <> (en fait c'est un caractère spécifique et plus joli pour cet usage) de taille ajustable en fonction de l'argument
\newcommand{\abs}[1]{\ensuremath{\left\lvert{#1}\right\rvert}} % valeur absolue de taille ajustable en fonction de l'argument
\newcommand{\eps}{\ensuremath{\varepsilon}} % le epsilon qu'aime bien fabrice, plutôt une notation pour une variable ce qui permet de différentier de la permittivité
\newcommand{\ket}[1]{\ensuremath{\left \lvert {#1} \right \rangle}} % le ket de la MQ de taille ajustable en fonction de l'argument (|x>)
\newcommand{\var}{{\ensuremath{\mathrm{var}}}} % Notation de la variance (en lettre droite comme les autres fonctions sous LaTeX)
\newcommand{\fano}{\ensuremath{\mathscr{F}}} % Notation du facteur de Fano en lettre cursive (cf notre PRA)
\newcommand{\spectral}{\ensuremath{\mathscr{S}}} % Notation de la densité spectrale en lettre cursive (cf notre PRA)
\newcommand{\ii}{\ensuremath{\mathfrak{i}}} 
\newcommand{\jj}{\ensuremath{\mathfrak{j}}}
\newcommand{\gf}{\ensuremath{\mathfrak{g}}}
\newcommand{\al}{\ensuremath{\alpha}}

\newcommand{\de}{\ensuremath{\delta}}
\newcommand{\De}{\ensuremath{\Delta}}
\newcommand{\om}{\ensuremath{\omega}}
\newcommand{\Om}{\ensuremath{\Omega}}
\newcommand{\J}{\ensuremath{\mathcal{J}}}
\newcommand{\C}{\ensuremath{\mathcal{C}}}
\newcommand{\E}{\ensuremath{\mathcal{E}}}
\newcommand{\N}{\ensuremath{\mathcal{N}}}
\newcommand{\Q}{\ensuremath{\mathcal{Q}}}
\newcommand{\D}{\ensuremath{\mathcal{D}}}
\newcommand{\G}{\ensuremath{\mathcal{G}}}
\newcommand{\R}{\ensuremath{\mathcal{R}}}
\newcommand{\T}{\ensuremath{\mathcal{T}}}
\newcommand{\V}{\ensuremath{\mathcal{V}}}

\newcommand{  \p   }   { \ensuremath{      \big{(}          }               }
\newcommand{  \q   }   {  \ensuremath{      \big{)}        }                }

\newlength{\figwidth} % dimension utilisée dans les montages de figure sur une page

\begin{document}

%%%%%%%%%%%%%%%%%%%%%%%%%%%%%%%%%%%%%%%%%%%%%%%%%%%

\begin{titlepage} 

\title{\Huge \textbf{Quiet Lasers}}

\author{%
Jacques \textsc{Arnaud}
\thanks{Mas Liron, F30440 Saint Martial, France},
Laurent \textsc{Chusseau}
\thanks{Institut d'\'Electronique du Sud, UMR n°5214 au CNRS, Universit\'e Montpellier II, F34095 Montpellier, France},
Fabrice \textsc{Philippe}
\thanks{LIRMM, UMR n°5506 au CNRS, 161 rue Ada, F34392 Montpellier, France}
}
\maketitle
\thispagestyle{empty}

\end{titlepage}

%%%%%%%%%%%%%%%%%%%%%%%%%%%%%%%%%%%%%%%%%%%%%%%%%%%

\frontmatter
\tableofcontents

%%%%%%%%%%%%%%%%%%%%%%%%%%%%%%%%%%%%%%%%%%%%%%%%%%%

\chapter{Foreword} 

This book provides simple ways of evaluating the amplitude and linewidth of stationary laser oscillators operating near some angular frequency $\omega_o$. Fluctuations about mean values are of paramount importance in communication systems and sensors. We are particularly concerned with "quiet" lasers, that is lasers whose output power does not fluctuate much in the course of time. The devices considered may be pictured as boxes containing conservative elements such as capacitances, inductances, and (non-reciprocal) circulators, supplied in energy by a constant electrical current $J$. The optical field is treated as an ordinary function of time that does not possess independent degrees of freedom and is not directly measurable. Static energy is converted to optical energy with the help of negative conductances, while the device output is a positive conductance $G$ (or many conductances) representing absorption of light by a photo-detector (or photo-detectors). These are collections of two-level atoms resonant with the field, which brings some of them from the ground state to the excited state. From there, the atomic electrons acquire a large energy with the help of a static field, and thus, transitions occur at definite times. The complete system under consideration is set up once for all, and we are only concerned with the spectral densities (or cross-spectral densities) of the stationary  time series. We associate with (positive or negative) conductances independent random current sources $c(t)$ whose spectral densities are $\hbar\omega_o\abs{G}$, and enforce the law of conservation of energy. Since (positive or negative) conductances do not store energy,  the electrical power fed into a conductance is equal to the optical power that it delivers, the random current sources being taken into account in the power balance. The conductances may depend on parameters, for example on the number of electrons in the conduction band of semi-conductors. But because the conductance variations are small one may suppose that the random-current spectral densities are unaffected. Quite generally, when the input power (often called the "pump") does not fluctuate in the course of time, the same is true for the oscillator output, at least at small Fourier frequencies ($\Omega\to 0$), irrespectively of the value of the various (conservative) parameters that enter. The linewidth is obtained by first evaluating the instantaneous frequency deviation. The results obtained on the basis of this semi-classical method coincide in every details with the corresponding Quantum Optics results. 

General considerations concerning Physics, Mathematics, Circuits, Statistical Mechanical and Interaction between atoms and classical fields are outlined. The nature of random-current sources is clarified by way of examples. We next consider the linear regime applicable to below-threshold oscillators and the linearized regime. We treat the effect of gain compression, electrical feedback, inhomogeneous broadening, ring-type lasers and four-level-atoms laser oscillators. It is shown that light beams may be (non-linearly) amplified without noise increment. The above concepts were presented by one of us in Cargèse in 1989\,\cite{Arnaud1989}. Because there exist now-a-days ample means of locating research works, only few references to previous publications are given. Experimental results are omitted.
   
%%%%%%%%%%%%%%%%%%%%%%%%%%%%%%%%%%%

\mainmatter
\chapter{Introduction}\label{introduction}

%%%%%%%%%%%%%%%%%%%%%%%%%%%%%%%%%%%%

The following quotation defines the nature of the semi-classical method employed in this book:

\begin{quote}
I am not seeking the meaning of the light quanta in the vacuum but rather in places where emission and absorption occur, and I assume that what happens in the vacuum is rigorously described by Maxwell's equations (Max Planck, letter to Einstein, 1907, see\,\cite{Seelig1954}).
\end{quote}

Laser noise impairs the operation of optical communication systems and the measurement of small displacements or small rotation rates with the help of optical interferometry. Even though laser light is far superior to thermal light, minute fluctuations restrict the ultimate performances. Signal-to-noise ratios, displacement sensitivities, and so on, depend mainly of the spectral densities, or correlations, of the photo-currents. It is therefore important to have at our disposal formulas enabling us to evaluate these quantities for configurations of practical interest in a form as accurate and as simple as possible. We are concerned with basic concepts leaving out practical considerations and experimental results. Non-essential noise sources such as mechanical vibrations are ignored. Real lasers involve many secondary effects that are presently neglected for the sake of clarity. For example, because of the large size of the cavity in comparison with wavelength, lasers tend to oscillate on more than one mode. Even if the side-mode powers are much reduced with the help of distributed feed-backs or secondary cavities, small-power side modes may significantly influence laser-noise properties, particularly near the shot-noise level. Side-mode powers should probably be less than 40 dB below the main mode power to be insignificant. In the case of gas lasers, multiple levels, atomic collisions, thermal motions, and so on, may strongly influence noise properties, but these effects are neglected here.

\paragraph{A quiet classical oscillator}

The purpose of this paragraph is to show that the clock invented by Huygens in the 18~th-century is a "quiet oscillator" in the the sense that the dissipated power does not vary much in spite of the random environment. The events, however, are not quantized, that is they not identical. Accordingly, they need not occur regularly in time. In similar conditions, photo-detection events are quantized and the event statistics is sub-Poisson.

An oscillator position $x(t)$ varies essentially as a sinusoidal function of time with a period $\mathcal{T}$, but the amplitude and phase of that oscillation may undergo small fluctuations. We say that an oscillator is "quiet" when the dissipated power, viewed as the oscillator output\footnote{Instead of being dissipated locally, the oscillator power could be carried away and absorbed at a distant location. This is what usually occurs. However, from a theoretical stand-point, whether the power is dissipated locally or far away is immaterial under ideal conditions.}, does not vary much in the course of time. Our main physical argument is that this is the case when the input power is steady in time. As a means of introducing the subject we describe below the so-called "grand-mother clock" that consists primarily of a pendulum. Damping due to air molecules is compensated for, on the average, by an escapement mechanism driven by a falling weight.  The molecules in our model are being picked up at random, and thus the absolute temperature T does not enter. It is acknowledged that according to the Bernoulli and Maxwell theory the molecules are in fact moving randomly and the pendulum motion is damped by collision with these molecules. Our unusual model has been selected to simplify the calculations, but it should be equivalent to the realistic one. The purpose of this book is to explain why lasers driven by a steady pump give at their output sub-Poisson photo-electrons. The principle of power conservation at small Fourier frequencies is general, and is exemplified by the pendulum. In the case of lasers, on the other hand, the photo-electrons carrying all the same energy, as recalled above, power regulation entails sub-Poisson event statistics. 

The basic element of a grand-mother clock is a weight $W$ suspended at the end of a weightless bar of length $L$ in the earth gravitational field $g$. As was first shown by Galileo the oscillation period $\mathcal{T}=2\pi \sqrt{L/g}$ does not depend on the oscillation amplitude as long as this amplitude remains small, a condition that we assume fulfilled. The period $\mathcal{T}$ does not depend either on the weight value according to the equivalence principle: inertial mass equals gravitational mass. For simplicity we suppose that the pendulum period is unity, that is $\mathcal{T}=1$s. This amounts to selecting some appropriate $L$ value, considering that $g\approx 9.81$ m/s$^2$. We also suppose that $W=1$ so that  the highest weight altitude $E$ represents the pendulum energy since the kinetic energy then vanishes. In the following we denote by $E_k, ~k=1,2,...$ the pendulum energies at successive periods of oscillation. This energy gets decremented by a random damping mechanism to be specified below, and incremented by a regular escapement mechanism.
\begin{figure}
\includegraphics[scale=0.8]{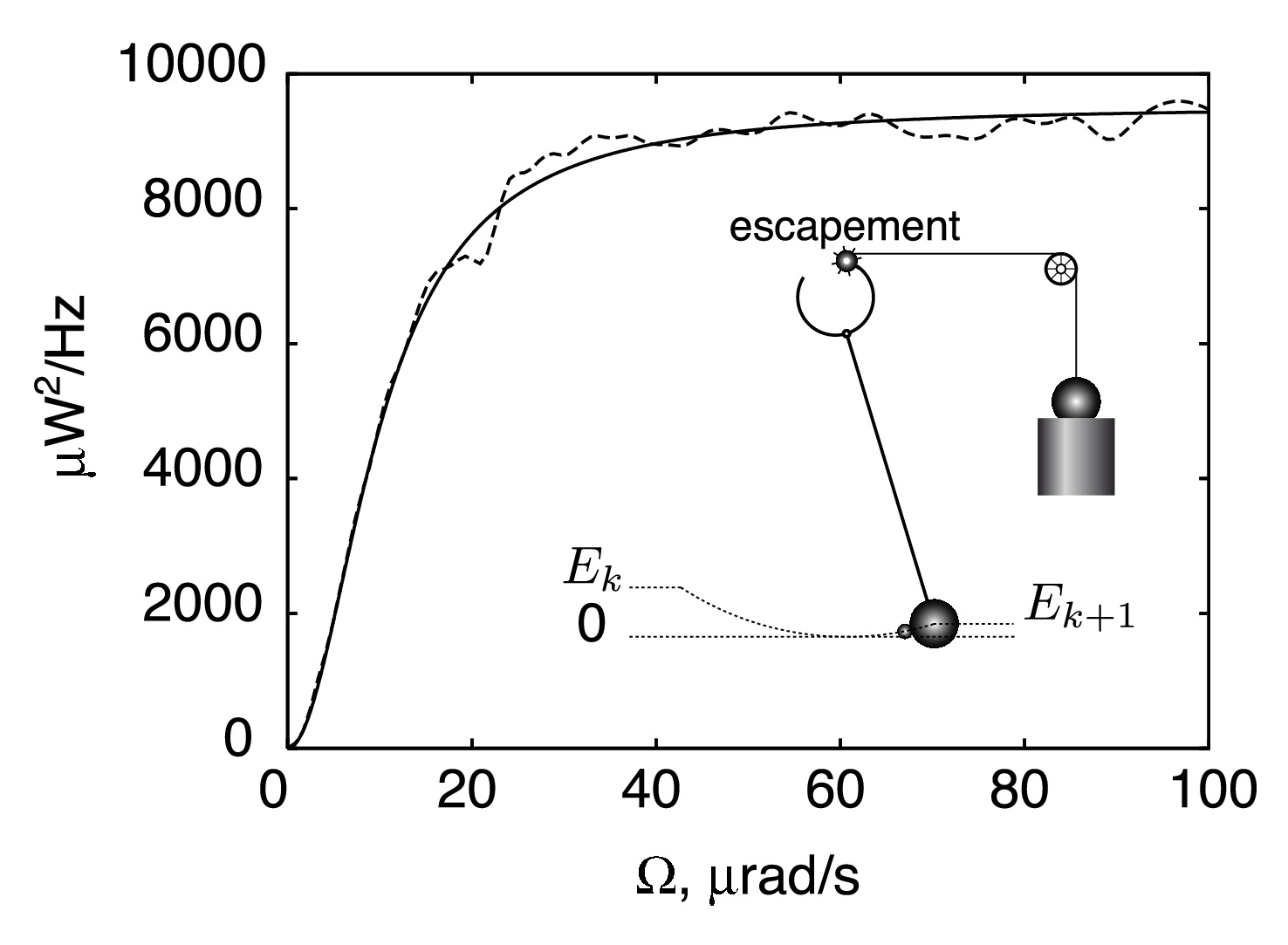}
\caption{ The figure represents the "grand-mother" clock, discovered by Huygens in the 18 th century. The pendulum consists of a weight at the end of a weightless bar. In our model, damping is caused by molecules of weight $w$ being raised by the pendulum from the lowest to the highest weight level, with probability $p\ll1$. Damping is compensated for, on the average, by an escapement mechanism driven by a falling weight delivering a constant energy $\delta$ per period. The curves show the spectral density of the dissipated power $P(t)$ as a function of the Fourier frequency $\Omega/2\pi$, numerically evaluated (irregular curve) and obtained analytically (smooth curve). This clock, in its idealized form, is a quiet oscillator in the sense that in spite of the randomness introduced by damping the dissipated power does not fluctuate at small Fourier frequencies. This figure illustrates the behavior of high-power laser diodes driven by a constant current.}
\label{figgm}
\end{figure}
Let us first describe the damping mechanism. The pendulum, with energy $E_k$, is supposed to pick up with probability $p\ll1$ at each period a molecule of weight $w$ at rest at the lowest level, and to release it at the highest level $E_{k+1}$ (see Fig. \ref{figgm}). Because the probability $p\ll1$, the molecule-picking events form a Poisson process. The average inter-event time for a Poisson process is known to be $1/p$. Note that we are considering only time intervals much larger that the pendulum period $\mathcal{T}=1$. Raising a molecule of weight $w$ from altitude $0$ to altitude $E_{k+1}$ amounts to reducing the pendulum energy from $E_k$ to $E_{k+1}=E_k-wE_{k+1}$ according to the law of energy conservation. It follows that, if a molecule-picking event occurs (a rare event), we have: $E_{k+1}=E_k/(1+w)\approx (1-w)E_k,~w\ll1$. 

In order to maintain a constant oscillation amplitude, at least on the average, a power supply is required.  Power is delivered by a weight suspended at the end of a cord. An escapement mechanism (crudely represented in the figure) allows the suspended weight to drop by a fixed height at each swing of the pendulum, thereby delivering to it a constant energy, or power since $\mathcal{T}=1$, that we denote $\delta$. The pendulum average energy $\ave{E}$ is obtained by equating the input power $\delta$ and the average absorbed power $pw\ave{E}$. Thus, $\ave{E}=\delta/pw$. It is appropriate in numerical calculations to begin with a pendulum energy equal to $\ave{E}$. At every period ($k=1,2,...$) we add to the pendulum energy the energy delivered by the escapement mechanism, that is $E_{k+1}=E_{k}+\delta$. We also select a random number $x$ uniformly distributed between 0 and 1. If $x<p$ (a rare event), we subtract from the pendulum energy the molecule-raising energy: $E_{k+1}=(1-w)E_{k}$. While the power supply is constant in time according to the above discussion, the damping mechanism has a random character. Our purpose is to evaluate the power released by the molecules as a function of time. As said above, this energy is generated at times (called "events") corresponding to a Poisson process. But the energy released by the molecules varies from event to event. If a molecule-picking event occurs when the pendulum energy is higher than usual, that event absorbs a larger-than-usual energy. This is how one can explain in a qualitative manner the mechanism behind dissipation regulation. As Fig. \ref{figgm} shows, the spectral density of this so-called "marked" Poisson process,  generated numerically, is in excellent agreement with the analytical formula given below, obtained essentially according to the principles employed for high-power laser diodes later in this book
\begin{align}
\label{intro}
\spectral(\Om)=\frac{\delta^2/p}{1+(pw/\Om)^2}.
\end{align}
This formula shows that the output power spectral density vanishes as the Fourier angular frequency $\Om\to 0$. This must be the case because, as we discussed earlier, for slow variations, conservation of energy implies conservation of power, the stored energy being then negligible. In our numerical application, we have selected $w=1$mN, $\de=10\mu$J, $p=0.01$ and thus the average inter-event time is one hundred times the oscillation period. To conclude, the fact that oscillator outputs do not fluctuate much when the power supply is steady is a general result related to energy conservation. When the events are quantized, the event statistics must therefore be sub-Poisson.

\paragraph{The laser model}

As is the case for mechanical oscillators, laser oscillators consist primarily of conservative devices such as interconnected capacitances, inductances, (non-reciprocal) circulators, resonating at angular frequency $\omega\equiv2\pi/\mathcal{T}$. Gain is provided by a negative conductance $-G_e$ (subscript "$e$" for emitting). Loss is modeled by a positive conductance $G$. Here again we observe that whether the absorbing conductance is connected directly to the laser, or instead at the end of a matched transmission line, is immaterial as far as the theory is concerned. The quantity of interest is the power $P(t)$ dissipated in the latter conductance, which models an ideal photo-detector generating at discrete times $t_k,~k=...-1,0,1,2...$, photo-electrons, each delivering the same energy. The quantity of interest is a set of event times described by a probability law. For the case of a perfectly quiet laser oscillator the time intervals between successive events are almost constant. Power may be supplied to a laser diode, for example, by a very large inductor delivering a nearly constant electrical current to the negative conductance $-G_e(n,R)$. If a stable oscillation is to be sustained at some level, that conductance must be a decreasing function of the power-emission rate $R$. Or it may be an increasing function of an electronic population $n$, which itself decreases when the oscillation intensity increases, with a time delay. This delay causes relaxation oscillations to occur unless the laser power is large. Note that $n$ is an integer and that the electrons have all the same electric charge. However, when  $n$ is large, one may define its derivative with respect to time $dn/dt$, according to the weak-noise approximation. Other important parameter in laser-diode noise theory are the $\alpha$-factors, which tell us how capacitances vary as a result of electronic population variations, and the $K\equiv 1+h^2$ factor, usually associated with gain guidance. Precise definitions will be provided in due time.

\paragraph{What is being measured and predicted statistically to second order?} 

In our model fluctuations are caused exclusively by the random current sources associated with absorbing and emitting elements. These are related to the laws of Statistical Mechanics, as we shall see. The system reacts to these random sources in different ways depending on the particular configuration considered in order that the law of average-energy conservation be fulfilled. Considerations of causality are irrelevant for the stationary systems presently considered, the set up being defined once for all. All we have to do is to observe the times $t_k, k=...,-1,0,1,2...$ at which photo-electron events occur. These event energies having reached a classical level they occur at well-defined times. It is understood that a large number of macroscopically identical set-ups have been fabricated, so that a large number of time series are recorded. The probabilities predicted by the theory may be verified by comparison with this ensemble of records. We restrict ourselves to the second-order probability $p(t_1,t_2)$ that an event occurs at $t_1$ and another (not necessarily the next one) at $t_2$, to within $dt_1,dt_2$. Because of stationarity, the spectral density mentioned earlier, $\spectral(\Omega)$, is the Fourier transform of $p(0,\tau)$ with respect to $\tau$, according to the Wiener-Khintchin theorem. One may also record the number of photo-electrons arriving within some time interval. The relationships between these quantities follow from the theory of point processes. One may also ask the following question: Given some measurements made on a system, can we predict what measurements would be performed on a modified system? In the case where a linear, cold and reflection-less attenuator is introduced in front of a detector, the time series follow from the original time series by the process of "thinning" (also called "decimation"). In general an answer to the question just asked cannot be given.

\paragraph{The Quantum Theory}

In the conservative devices mentioned earlier (capacitances, inductances...) the current consists of a very large number of moving electrons that act collectively. They may be treated according to the Classical Circuit theory, rather than from the many-electrons Schrödinger equation. The Schrödinger equation enters, however, in the treatment of the atoms that compose negative or positive conductances. From a known classical preparation, the Schrödinger equation predicts from the Born rule the probability that the electron be found in the lower or upper state if a measurement is being performed. In either case the electron delivers a classical signal and is left in the state it has been found in, according to Bohr\footnote{Bohr does not specify precisely how one may go from a quantum system such as an atom to a classical signal. The suggestion has been made that this is the result of the interaction of the atom with the environment, followed by a "Quantum Darwinism" process that selects a particular classical outcome (Zurek, 2009)\,\cite{Zurek2009}.}. In general, we suppose that the conductances are known before-hand and little attention is brought to the atom Quantum Theory. Most of our conclusions rest instead on Statistical Mechanics.

\paragraph{Comparison with Quantum Optics results, and with the semi-classical "phasor" theory:}

The final expressions obtained are in exact agreement with those derived from Quantum Optics methods whenever the latter are available. This is the case in particular for sub-Poisson lasers. The statement often heard that sub-Poissonian photon statistics always require quantization of the optical field is unfounded. On the other hand, theories found in most of the Optical Engineering literature rest on the concept that the classical oscillating field is supplemented by a random field due to "spontaneous emission in the mode". Such theories give reasonably accurate results only at high noise levels. They involve parameters whose values are difficult to establish beforehand.

\paragraph{Main approximations}

Only two limiting cases will be considered, namely the linear regime and the linearized regime. In the linear regime optical potentials and currents are proportional to the noise sources. The response of linear systems to specified sources is straighforward, but dispersion effects need investigation. This regime is applicable to lasers below the so-called "threshold" driving current and, usually, to attenuators and amplifiers. The \emph{linearized} regime is applicable to well-above-threshold lasers. In that regime one evaluates average optical potentials and currents ignoring the noise sources. Next, one supposes that the \emph{deviations} of the optical potentials and currents from their average values, denoted by $\De$, are proportional to the noise sources. The latter enter again when powers are being evaluated, that is, current noise sources are not given for free, so to speak, but they do enter in the power balance. This is because we take this effect into consideration that our theory differs from previous semi-classical "phasor" theories. The intermediate situation in which the system is neither linear nor can be linearized that may occur for closed-to-threshold lasers is not considered. As said above, we treat only the stationary regime found when a laser is driven by a constant current and no element is prescribed as being time-dependent, in which case photo-detection events form a stationary point process. 

Spontaneous atomic decay is neglected for the sake of simplicity. In laser diodes employing semi-conducting materials the bottom of the conduction band is filled up with $n$ electrons, according to the Fermi-Dirac distribution. Likewise, there are $n$ holes at the top of the valence band. Because the electrons fill up the available states, the static potential $U$ across the diode slightly exceeds $E_g/e$, where $E_g\approx \hbar\omega_g$ denotes the semiconductor energy gap. The conduction electrons pile up, so to speak, so that there are at most two electrons per level according to the Pauli principle. The same observation applies to the holes. The rate equations that we shall introduce later on entail random fluctuations of $n$, and thus fluctuations $\De U$ of the potential $U$. This fluctuation is very small, yet measurable. One may also measure the correlation between $\De U$ and the detected rate fluctuation $\De D$. This correlation may be defined in such a way that it is independent of any linear optical loss that may occur between the laser and the detector. From our view-point, the fluctuation $\De U$ is a small effect that may initially be neglected. 

\paragraph{Laser spectral width}

The light spectrum is a well defined quantity. To observe it, is suffices to insert between the laser and the photo-detector a narrow-band, reflexion-less, cold and linear filter whose response is centered at some frequency $\om_m\approx \om$. The average photo-current rate $\ave{D(t)}$ is proportional to the light spectral density $\spectral(\om_m)$. In the \emph{linearized regime}, the light spectrum may be evaluated by first neglecting amplitude fluctuations and considering the frequency fluctuation $\De\om(t)$. The latter is obtained by considering that the random current sources are of the form $C'(t)+\ii C''(t)$, where $C'(t),~ C''(t)$ are uncorrelated random currents whose spectral densities are $\hbar\om \abs{G}$. The "instantaneous" frequency fluctuations follow from elementary circuit considerations. Experimentally, frequency noise may be converted to photo-current noise through a dual-detector arrangement or through heterodyning.

\paragraph{Summary of the concepts}

We begin with a simple assertion, namely that configurations having the same energy have the same probability to occur (Statistical Mechanics). We end up with the relative noise (to be defined) for various laser oscillators configurations. Time reversibility holds not only for the fundamental laws of Physics such as the equations of Classical and Quantum Mechanics, but also in Statistical Mechanics if one restricts ourselves to reversible engines\footnote{The transfer of energy from the power supply, for example a very large inductance in which some current flows, to the potential sink in which photo-electrons end up, is not, however, reversible in usual circumstances. }. Three principles are employed: 1) The law of conservation of energy. 2) The fact that configurations of equal energy are equally likely to occur. 3) The law of conservation of the electric charge. From these three principles one derives that to any conductance $G$ one must associate a random current source $c(t)$ whose spectral density is $\hbar\omega_o \abs{G}$, where $\hbar\omega_o$ denotes the level energy difference for atoms resonating with the electrical field. 

Let us clarify the different approaches presented. First of all, we would like to answer the following question: Since in our model the input power is a constant and spontaneous emission is neglected, why, fundamentally, is the detected rate $D(t)$ fluctuating? We have explained why, in the case of a single detector and at small Fourier frequencies, $D(t)$ in fact does not fluctuate. But the question remains in the case of two (or more) detectors, or at non-zero Fourier frequencies. The answer comes from Statistical Mechanics or from Quantum Mechanics with a phenomenological parameter. The two are closely related in concept as there were historically (Carnot $\to$ Boltzmann $\to$ Einstein $\to $ Schrödinger). Indeed any cycle treated by the methods of Quantum mechanics \emph{must} obey the laws of Statistical Mechanics. The prescription we end up with consists, as described above, of associating to any conductance a complex random current source proportional to the absolute value of that conductance. Equivalently, one may supplement any absorbed or generated rate (rate $\equiv$ power/$\hbar\omega$) with a complex fluctuation $r(t)$ whose real and imaginary parts are at the shot-noise level, that is, have spectral densities equal to the average rate. The question is thus: What justifies the existence of such random current or rate sources?

What is properly "Quantum" in both disciplines (Statistical Mechanics and Quantum Mechanics) is the concept of identical objects. In the chapter on Statistical Mechanics, we describe a heat engine having reservoirs at two different altitudes. In each of these reservoirs there are weight-one balls and a greater number of possible locations. The exchange of two balls, picked up at random, between the two reservoirs, enables one, not only to recover the Carnot expressions for the work performed by an ideal heat engine and its efficiency, but also the correct expressions for the fluctuations. Here the identical objects could be macroscopic. But in the field of lasers, the identical objects are electrons, whose electrical charge and mass are given to us by nature. 

Going further along the same lines, we consider a cavity containing $n$ two-level atoms initially in the excited state. The basic principle of Statistical Mechanics says that configurations corresponding to the same energy are equally likely to occur. This principle tells us for example how many atoms are in the excited state on the average in a state of equilibrium at some temperature. The situation becomes slightly more complicated if we assume that the cavity may contain a resonant field with energy $E$. Calling $m$ the integral part of $E/\hbar \omega$, we notice that $n$ may be decremented by one provided that $m$ be incremented by one, the principle of conservation of energy being then fulfilled, or the converse. This kind of reasoning provides us with the probability distribution of $m$. But, starting from $n$ atoms in any state, one may wonder how the atom-field interaction may lead to the Statistical Mechanical result. This is achieved by introducing the Einstein rates of stimulated emission and absorption. But this is done in this book for a single resonator instead of a "black-body" (large multi-mode cavity, as was done early in the 19 th century. A reversibility principle is enforced. It is found that, to obtain the established equilibrium result, these rates must be supplemented by the random rates $r(t)$ mentioned earlier.

Let us now turn to the Quantum Mechanical treatment, based on the Schröd\-inger equation. We consider an atom in the ground state at time $t=0$. The atom is in that state because a downward transition just occurred. The atom is then submitted to a classical resonator field whose frequency is equal to the atomic levels energy difference, divided by $\hbar$. A standard treatment shows that the probability that the atom be in the excited state at time $t$ is $\sin^2(\frac{\Omega_R}{2}t)$, where $\Omega_R^2$ is proportional to the field energy. It is plausible that the probability that the next downward transition occurs at time $t$ (to within $dt$) is given by $\sin^2(\frac{\Omega_R}{2}t)$ multiplied by a phenomenological constant $2\gamma$. This assertion may be proved to be valid in the limit of small $\gamma$. We then obtain the statistics of successive events, which is not Poisson. However, when a large number of point processes are superposed the statistics of the resulting process is Poisson. This means that even for small $\gamma$-values, but a large number of independent electrons, the event statistics is Poisson. For large values of $\gamma$, on the other hand, the statistics may be shown to be Poisson even for a single electron. We then show that such conclusions require the introduction of random sources of the kind mentioned above. To conclude, there are (at least) two main arguments supporting the view that random sources of a special kind must be introduced whenever a circuit involves absorbing or emitting elements, one based on Classical Statistical Mechanics, and one based on Quantum Mechanics. The relation existing between absorption and fluctuations is of course well-known since the Nyquist contribution. We attempt here, however, to recover the needed results from first principles through partly heuristic considerations.

If the existence of these random sources is accepted, it becomes, ironically, more difficult to understand why, in spite of them, the  photo-current \emph{does not} fluctuate at small Fourier frequencies, as asserted earlier. Indeed, the random sources associated with different conductances being \emph{independent}, they cannot "conspire", so to speak, to lead to this simple result. It is essential to appreciate that the conductances react to the random sources (both their own and the sources associated with the other conductances, transformed by the circuit) locally, in such a way that the law of conservation of energy be fulfilled. That is, the output rate from a conductance must be equal at any instant to the input rate, since a conductance does not store energy. This occurs because the conductances are allowed to vary through small changes of the parameters. Since the law of energy conservation is enforced at every conductances, and the rest of the circuit consists exclusively of conservative elements, it is clear that the sum of the power rates entering into the various ports of the complete conservative system vanishes at low Fourier frequency. Finally, let us emphasize that the random sources that we introduce are not measurable by themselves. Measurements refer exclusively to photo-detection point processes.

\paragraph{Chapters content}

In the first chapter, we first give an account of the most relevant results in Physics, lists mathematical results relating to deterministic or random functions of time, discusses the classical circuit theory, the laws of Statistical mechanics and offer methods of establishing that the spectral density of noise sources associated with a conductance is proportional to the absolute value of that conductance. We present the theory of electron-field interaction. The basic assumptions made is that the system considered (including the sources and absorbers of energy) is stationary and that the number of particles is large. 

Next we consider the linear and linearized regime, and the relative noise of idealized laser diodes at high power levels. At such high powers, the time derivative of the number $n$ of electrons in the conduction band may be neglected and no relaxation-oscillation occurs. Gain compression (explicit dependence of the gain on the emitted rate) is neglected at that point.  The more general theory is given later on. A schematic simulating lasers Fourier-frequency response is described. 

\paragraph{Summary of the book}

To summarize, this book shows that results such as the one just cited may be derived from a semi-classical theory in which the optical field is treated as a classical function of time and does not possess independent degrees of freedom. Any medium may be described by a circuit consisting of conservative elements such as capacitances, inductances, and (non-reciprocal) circulators, and positive or negative conductances that represent the field-to-atom coupling. Noise sources are associated with such conductances. The laws of conservation of energy and electrical charge are enforced and consideration is given to the second law of Thermodynamic. Only stationary linear or linearized regimes are considered. A number of particular configurations are treated for the sake of illustration. In some chapters only the final result is given, after some introductory material. Specifically:

\begin{itemize}
\item The linear regime. In that regime the conductances $G$ are constant. We first consider a given incident beam and describe how it can be characterized with the help of photo-detectors, introducing the concept of relative noise $\N$. Next we consider the noise properties of attenuators and amplifiers, and evaluate the linewidth of linear oscillators. 

\item The linearized regime. We now suppose that the conductances depend on the electron number $n$, and employ a first-order expansion, namely $G(n)=G(\ave{n})+\frac{dG}{dn}\Delta n$, which is accurate as long as $\Delta n\ll \ave{n}$. One first solve for the steady-state equations, ignoring the noises sources. Next, we express the conservation of the energy and of the electrical charge in the form $dn(t)/dt =J-R(t)$, where $J$ denotes the constant rate of charge injection, and $R(t)$ the emitted electromagnetic rate (power divided by $\hbar\omega$). Here we assume that $n$ is a large number, so that its derivative with respect to time makes sense. Calculations may be performed in terms of currents and potentials, or in terms of forward and backward rates. One finds that, at high power, the relative noise $\N$ increases smoothly as a function of the Fourier frequency $\Omega$. In contradistinction, at low power, $\N(\Omega)$ exhibits a large peak referred to as a relaxation oscillation.

\item We evaluate laser linewidths within the above linearized regime. One method consists of calculating first the spectral density of the instantaneous optical frequency fluctuation $\Delta \omega(t)$. For a simple configuration, this is done by writing that the total current, sum of a deterministic current which depends on $\omega(t)$ and a complex random current $C(t)$, vanishes. The laser linewidth is proportional to that spectral density. Alternatively, we may evaluate optical phase fluctuations.

\item  We discuss the effect of electrical feedback. Part of the detected rate is amplified, ideally without added noise, and applied to an amplitude or phase modulator on the input or output light beam. Finally we show that an optical amplifier of a special kind ($C$-amplifier) employing electrical feedback (and possibly gain compression) has the remarkable property of preserving fluctuations. 

\item We provide some information concerning the electrical properties of semiconductors, such as the optical gain, the $\alpha$-factor, etc.... If the absolute value of a negative conductance (expressing optical gain) decreases when the emitted light rate $R$ increases, we speak of "gain compression". This effect may possibly be due to spectral-hole burning. The laser amplitude may get stabilized by this effect alone. Conductances are then of the form $G(n,R)$. The small explicit dependence of $G$ on $R$ tends to damp the relaxation oscillations. A numerical simulation exhibits the phenomenon of spectral-hole burning in a semi-conductor and suggests some value for the dependence of $G$ on $R$.

\end{itemize}

References to previous works are not given because we found it too difficult to interpret them adequately. Fortunately, it is presently easy to track down through electronic means previous contributions.

\paragraph{Appendix to the introduction: Resonator energy statistics}

We will be mostly concerned in this book with high power lasers, so that the optical field energy is much larger than $\hbar\om$, where $\hbar$ denotes the Planck constant and $\om$ the oscillation frequency. (we set for convenience $\hbar\om=1$). It is nevertheless interesting to speculate on what may happen from a semi-classical view-point at small field energies, $E\approx 1$. If we allow atoms to stay for a sufficiently long time in a cold environment, they almost certainly are in the ground state. Initially, we have no knowledge concerning the optical resonator field energy. But if cold atoms are interacting with the resonator repeatedly, the integer part of the field energy eventually vanishes: $m=0$. If another cold atom interacts with the field, one must presume that the resonator field phase changes during the interaction process in such a way that the atom exits in the ground state, since otherwise the field energy would be negative. If, next, an atom in the excited state interacts with the resonator, this atom may end up in the ground state, in which case $m=1$, and so on. If it ends up in the excited state, then $m$ remains equal to zero. Because in general we only know the \emph{probability} that the atom is initially in the excited state, we end up with a probability law $p(m)$ for $m$. The only knowledge we may have concerning the amplitude and phase of the field itself is through observations such as the ones we have just described: measurements are made on atoms only. We end up with the conclusion that the field energy distribution is proportional to $\exp(-m/T)$, where $m≥0$, the number of light quanta, is the integer part of the field energy, and $T$ the absolute temperature. The average energy thereby obtained is the same as the one deduced from resonator quantization, but the energy variance tends to the classical limit at large temperatures, while this is not the case for quantized resonators. Given the resonator energy, we can plausibly ascribe a random phase to the field.

\newpage

\chapter{Physics}\label{history}

According to the latin poet Lucretius, a follower of Democritus, there are no forbidden territories to knowledge: "...we must not only give a correct account of celestial matter, explaining in what way the wandering of the sun and moon occur and by what power things happen on earth. We must also take special care and employ keen reasoning to see where the soul and the nature of mind come from". And indeed, the three most fundamental questions: what is the origin of the world? what is life? what is mind? remain subjects of scientific examination. Needless to say, the present book addresses much more restricted questions.

We will first recall how Physics evolved from the early times to present, no attempt being made to follow strictly the course of history. The theory of light or particle motion and the theory of heat followed independent paths for a long time. The Einstein contributions proved crucial to re-unite these two fields early in the 20th century. We may distinguish "pictures" based on our in-born or acquired concepts of space and time that may not answer all legitimate questions nor be accurate in every circumstances, and complete theories. Quantum theory is considered by most physicist as being accurate and complete, although questions of interpretation remain open. We will consider in some detail the theory of waves and trajectories that are essential to understand the mechanisms behind vacuum-tube and laser operation. 

\section{Early times}\label{early}

From the time of emergence of the amphibians, earth, a highly heterogeneous stuff, is our living
place. On it, we experience a variety of feelings. We feel the pull of gravity, breath air, get heat from the fire and the sun, and feed on plants growing on earth and water. Our experience, both as
human beings and as physicists, is based on these living conditions. One may presume that natural selection led humans to an intuitive understanding of geometrical-physical-chemical quantities such as space, time, weight, warmth, flavor, and so on. At some point in the evolutionary process a degree of abstraction, made possible by an enlarged brain, facilitated our fight for survival. An example of abstract thinking is the association with space of the number 3, corresponding to the number of perceived dimensions. People "in the street" may however wish to distinguish the two horizontal-plane dimensions and the vertical dimension, considering that, for the latter, up and down are non-equivalent directions. It was not appreciated in the ancient times that the distinction between "up" and "down" is caused by the earth gravitational field, and that people living on the other-side of the earth have the same feelings as we do in their every-day life, even though, with respect to our own reference frame, they are "up-side-down". As we shall see, analogous considerations may apply to time, according to Boltzmann.

 Another naturally evolving concept is indeed the distinction between past and future and physical causality: matter acts on matter only at a later time. The so-called "arrow of time" is a much debated subject. According to Boltzmann, in an infinite universe, there may be large-scale spontaneous fluctuations of the entropy (that one may crudely describe as expressing disorder). Past $\to$ future would correspond to the direction of increasing entropy. There may be times where the entropy decreased instead of increasing. But the distinction is purely a matter of convention (in analogy with the "up and down" distinction mentioned above). This view point is consistent with the fact that the fundamental equations of Physics are (with the exception of the rarely occurring neutral-kaon decay) invariant under a change from $t$ to $-t$. There are objections to the Boltzmann view-point, however: the world is not large enough and is not old enough. Most recent authors would rather ascribe the time arrow to cosmic evolution, with the universe starting at the "big-bang" time in a state of very low entropy. A good easy-reading book on the subject is by H.C. Von Baeyer\,\cite{Von-Baeyer1999}. In the present book mainly reversible processes are considered and the problem of the time "arrow" does not arise.
 
In contrast with the rational view concerning causality, the magic way of thinking presupposes the existence of causal relationships between our desires, fears, or incantations, and facts. Now-a-days, magic thinking co-exists with rational thinking probably because it gives people sharing similar beliefs a sense of togetherness and helps a few individuals acquire authority and power. The consequences of irrationality are often too remote to be of concern to most. 

The control of fire by man some 500 000 years ago and drastic climatic changes that occurred, mainly in Europe, some 23 000 years ago, trigerred evolutionary events. Likewise, the practice of growing crops made possible a population explosion some 10 000 years ago, particularly in Egypt, and gave an incentive for measuring geometrical figures, precisely accounting for elapsed times, and measuring weights. Let us now consider more precisely what is meant by matter, space and heat. 

Empedocle ($\sim$500 BC) viewed the world as being made up of four elements, namely earth, water, air and fire. These elements remain a source of inspiration for poets and scientists alike, but they are not considered anymore as having a fundamental nature. Democritus ($\sim$400 BC) pictured reality as a collection of interacting particles that cannot be split ("a-toms"). Aristotle wrote in his Metaphysics VIII: "Democritus apparently assumes three differences in substances; for he says that the underlying body is one and the same in material, but differ in shape, position, and inter-contact". This picture remains accurate. There are many books devoted to the discoverers of the ancient time. We consulted the Ref.\,\cite{Pleiade1957}. 

The present work is not concerned with the cosmos per se. Yet, one cannot ignore that observations of the sky have been a source of inspiration in the past and remain very much so at present. Early observers distinguished stars from planets, the latter moving apparently with respect to the former. The ancient Greeks (Ptolemeus) conceived a complicated system of rotating spheres aimed at explaining the apparent motion of these celestial objects. Aristarque (310-230 BC), however, realized that the earth was rotating about itself and about the sun, the latter being considered to be located at the center of the universe. This \emph{heliocentric} system was rediscovered by Copernic (1473-1543) and popularized by G. Bruno (burned at stake in Rome in 1600 for heresy). Next came the establishment of the three laws of planetary motion by Kepler, the dynamical explanation of these laws by Newton, which involves a single universal constant, namely $G$.

When two bodies are in thermal contact they tend to reach the same temperature. Thus, two differently constructed thermometers may be calibrated one against the other by placing them in the same bath and comparing their readings. In the case of thermal contact the hotter body loses an amount of heat gained by the colder one but the converse never occurs. It may well be that the condition of heat-engine reversibility, discovered by Carnot in 1824\,\cite{EJP3}, could have been made at a much earlier time and could have served as a basis for subsequent developments in Physics. A more detailed history of Statistical Mechanics is given below and in Section \ref{Chistory}.

\section{How physicists see the world}\label{how}

Beyond a qualitative understanding of the nature of heat, early observers were able to perform measurements of temperature and gas pressure with fair accuracy. Temperatures were measured through the expansion of gases at atmospheric pressure, linear interpolation being made between the freezing (0°C) and boiling (100°C) water temperatures. The concept of absolute zero of temperature emerged through the observation that extrapolated gas volumes would vanish at a negative temperature, now known to be -273.15°C($\equiv$ zero kelvin). The Classical Theory of Heat was established in the 18th and 19th centuries mainly by Black, Carnot and Boltzmann. The major contribution is due to Carnot who introduced the concept of heat-engine \emph{reversibility}. The fact that hot bodies radiate power was known very early (some reptiles possess highly-sensitive thermal-radiation detectors). It is however only in the 19th century that the proportionality of the total radiated power to the fourth power of the absolute temperature was established. Difficulties relating to the theory of blackbody radiation led Planck and Einstein around 1900 to the conclusion that Classical Physics ought to be replaced by a more fundamental theory, namely the Quantum Theory, even though important conclusions may be reached without it. Another motivation for studying in some detail the theory of heat is that lasers are in some sense heat engines. They may be ``pumped'' by radiations originating from a hot body such as the sun. But, just as is the case for heat engines, a cold body is also required to absorb the radiation resulting from the de-excitation of the lower atomic levels. Lasers are able to convert heat into work in the form of radiation, but their efficiency is limited by the second law of thermodynamics. Output-power average values and fluctuations may be similar for lasers and heat engines. 

The grand picture we now have is that of a world 13 billions years old and 13 billions light-years across containing about 10$^{11}$ galaxies. Apparently, 80 \% of matter is in a dark form, of unknown nature, that helped galaxy formation. Our own galaxy (milky way) contains about $10^{11}$ stars and possesses at its center a spinning black hole with a mass of 4 millions solar masses. Eight planets (mercury, venus, earth, mars, jupiter, saturne, uranus, neptune) are revolving around our star (sun).  Penzias and Wilson discovered in 1965 the cosmic background microwave radiation, which accurately follows the Planck law for a temperature of 2.73 kelvins. This cosmic black-body radiation is almost isotropic. Yet, minute changes of intensity according to the direction of observation have been measured, which provide precious information concerning the state of the universe some 300 000 years after the "big-bang". Numerous observations relating to ordinary stars such as the sun, neutron stars, quasars, black holes are particularly relevant to high-energy physics. It is expected that gravitational waves emitted for example by binary stars or collapsing stars will be discovered within the next decades. Their detection may require sophisticated laser interferometers operating in space. In such interferometers, laser noise plays a crucial role. Reactors aim at creating on earth conditions similar to those occurring in the sun interior, i.e., temperatures of millions of kelvins, and to deliver energy, perhaps by the year 2050. An alternative technique employs powerful lasers shooting at a deuterium-tritium target. A reduction of the laser-beam wave-front fluctuations are essential in that application.

\section{Epistemology}\label{epis}

Epistemology is the study of the origin, nature, methods and limits of knowledge. Undoubtedly, Physics is an experimental science. Its purpose is to predict the outcome of observations, or at least average values of such observations over a large number of similar systems, from few principles using Mathematics as a language. Observations are required to set aside as much as possible human subjectivity. This is done by performing a large number of "blind" experiments, the same procedure being repeated again and again in independent laboratories. A physical theory should be "falsifiable", that is, one should be able to realize, or at least conceive, an experiment capable of disproving it. 

The average value $\ave{a}$ of a quantity $a$ is calculated by summing $ap(a)$, where $p(a)$ is the probability of $a$. It is apparently difficult to provide an unambiguous definition of the word probability. According to mathematicians, "there is a fundamental mistrust in probability theory among physicists. The need to extract as comprehensive information as possible from a given set of data is in many cases not as pressing as in other fields since active experiments can be repeated in principle until the obtained results  satisfy preset precision requirements. In other fields, the available data should be exploited with every conceivable care and effort". As data comes in our estimate of $p(a)$ improves, and eventually approaches an objective value, defined according to the frequentists view-point. 

In practice, most scientific progresses were accomplished with the help of intuitively-appealing \emph{pictures}, describing how things happen in our familiar three-dimensional space and evolve in the course of time. These pictures are supposed to tell us how things are behind the scene, or to suggest calculations whose outcome may be compared to experimental results. Let us quote Kelvin: ``I am never content until I have constructed a mechanical model of the subject I am studying. If I succeed in making one, I understand; otherwise I do not''. But many models, helpful at a time, need often be discarded in favor of more abstract view-points. The Democritus picture of reality has been worked out in modern time by Bernoulli, Laplace and a few others. Given perfectly accurate observations made at some time, called "initial conditions", the theory is supposed to predict the outcome of future observations if the system observed is not perturbed meanwhile. Poincaré, however, pointed out that for some systems, e.g., three or more interacting bodies in Celestial Mechanics, the error grows quickly in the course of time when the initial conditions are not known with perfect accuracy. In some cases the system evolves into a so-called "deterministic chaos". The equations that describe ideal motions are time reversible, so that when the system is known with perfect accuracy at a time its state in the past as well as in the future is predictable. Postdiction makes sense if measurements have been made in the past but were not revealed to the physicist. What we have just described is sometimes referred to as the Classical Paradigm. 

Reality is surely a concept of practical value. Anyone wishes to distinguish reality, as something having a degree of permanency, from illusions or dreams that are transitory in nature. On some matters, the opinions of a large number of people are sought, supposing that their agreement would prevent individual failures. In that sense, reality may exist independently of observers and be \emph{revealed} by observations. We adopt the Bohr view-point that observations relate only to complete set ups, including the preparation and measurement devices, the latter being considered classical. The object to be measured should be able to switch another object involving a large number of degrees of freedom from one metastable state to another. As said before, measurements are made on atoms (or electrical charges) only, not on fields. If we introduce a device such as an absorber (whose macroscopic properties have been separately established) between a laser and a detector, one must in principle consider the properties of the new device as a whole, including the laser, the absorber and the detector. It is only in special circumstances that the result of measurements performed on the modified set-up may be predicted on the basis of measurements made on the previous one.

We are not concerned in the present paper with Physics in general but only with stationary configurations. The system is allowed to run in an autonomous manner, that is without any external action impressed upon it, and there is a continuous record of the times at which photo-electrons are emitted. Systems on which we may act from the outside are not considered, and accordingly the law of causality is not relevant. Photo-electrons may be accelerated to such high energies by static fields that no ambiguity occurs concerning their occurrence time. The question asked to the physicist then resembles the one asked to people attempting to recover missing letters from impaired manuscripts: can you determine the missing letters from the known part of the text? In the present situation one would like to be able to tell whether an event occurred during some small time interval, given the rest of the record. Or at least give the \emph{probability} that such an event occurs in the specified time interval. In other words, given a large collection of similar systems, on what fraction of them does an event occur? Instead of being given impaired records, we may be given information concerning the various components that constitute the system, such as lenses, semi-conductors, and so on, characterized by earlier, independent measurements. These measurements are deterministic in nature because they are performed in the classical high-field regime. In view of the observed uncertainty, random noise sources must be introduced somewhere in the theory. We consider that the random noise sources are located at emitters and absorbers, are time-symmetric with respect to one another, and are unobservable. Whether they originate from a sub-quantum process (that would fix the value of the Planck constant $\hbar$), is unknown.

\section{Waves and trajectories}\label{waves}

Physics courses usually first describe how the motion of masses may be obtained from the Newtonian equations. But it might be preferable to let students get first  familiarity with classical waves, for example by observing capillary waves on the surface of a mercury bath. Such waves are described by a real function of space and time that one may denote $\psi(x,t)$ in one space dimension. One reason (to be explained in more detail subsequently) to consider waves as being of primary interest is that the law of refraction follows in a logical manner from the wave concept, but does not from the ray concept. Once wave concepts have been sufficiently clarified, the many-fold connections existing between waves on the one hand, and particles or light rays on the other hand, may be pointed out. Note that, historically, the motions of macroscopic bodies and light rays were established first (around 1600) and the properties of waves later on (around 1800 for light and 1900 for particles). Few precise results concerning waves seem to have been reported at the time of the ancient Greece. Yet, casual observation of the sea under gently blowing winds suffices to reveal important features. Had such observations been made, the course of discoveries in Science would perhaps have been quite different from the one that actually occurred. 

Waves at the surface of constant-depth seas propagate at constant speed $u$. In realistic conditions there is some dissipation and the wave amplitude may decrease but the wave speed remains essentially unchanged. This is a striking example of a physical object whose speed does not vary, no force being impressed upon it. The only condition required is that the medium parameters (the sea depth in the present situation) do not vary from one location to another. 
 
In the 1630s Galileo observed that macroscopic objects move at a constant speed when no force is exerted upon them, in contradiction with the then-prevailing Aristotle teaching. A related finding by Galileo is the principle of special relativity: The laws of Physics established in some inertial laboratory are the same in another laboratory moving at a constant speed with respect to the first. In the year 1637 Descartes proposed the following interpretation for the refraction of light rays at the interface between two transparent media such as air and water. Descartes associates with a light ray a momentum that he calls "determination" having the direction of the ray and a modulus depending on the medium considered but not on direction. He observes that the $x$-component of the momentum should not vary at the interface as a consequence of the uniformity of the system in that direction, justifying this assertion by a mechanical analogy, namely a ball traversing a thin paper sheet. The law of refraction asserting that $\cos (\theta_1)/\cos (\theta_2)$, where the angles are defined with respect to the $x$-axis and the subscripts 1,2 refer to the two media, does not depend on the ray direction, follows from the above concepts. Note that Descartes was only concerned with trajectories in space, i.e., he was not interested in the motion of light pulses in time, so that questions sometimes raised as to whether light pulses propagate faster or slower in air or in water are not relevant to his discussion. 

No one at the time suggested that there may be a connection between particles or light rays on the one hand, and waves on the other hand. The wave properties of light were discovered by Grimaldi, reported in 1665, and explained by Huygens in 1678. The wave properties of particles were discovered much later by Davisson and Germer in 1927. In modern terms the Galileo, Descartes (and later Newton) concepts imply that particles and light rays obey ordinary differential equations. But without the wave concept the law of refraction for light or for particles relies on observation and intuition rather than logic. 

A wave packet has finite duration but includes many wave crests. A key concept is that of group velocity defined as the velocity of the peak of a wave packet, or short pulse. In particular, what is usually called the "velocity" of a (non-relativistic) body is the group velocity of its associated wave. But usually wave packets spread out in the course of time. In the non-linear regime though, wave packets, called \emph{solitons}, may exhibit particle-like behavior in the sense they do not disperse. Bore-like solitary waves created by horse-drawn barges were first reported in 1844 by Russell.

Let us be more precise about waves. As said above, waves are very familiar to us, particularly gravity waves (not to be confused with the Einstein gravitational waves) on the sea generated by wind, or capillary waves generated on the surface of a lake by a falling stone. Simple reasoning and observations lead among other results to the law of refraction. Waves are defined by a real function $\psi(x,t)$ for one space coordinate $x$, and time $t$, obeying a partial differential equation. If the wave equation is unaffected by space and time translations we may set $\psi(x,t)=f(x-ut)$ for arbitrary speeds $u$. This results into an ordinary differential equation for the function $f(x)$ which in general admits solutions. Let us begin our discussion with monochromatic (single-frequency) waves propagating in the $x$ direction in a conservative linear and space-time invariant medium. The wavelength $\lambda$ is the distance between adjacent crests at a given time. We define the wave number $k=2\pi/\lambda$. The wave-period $\mathcal{T}$ is the time it takes a crest to come back, at a given location. We define the frequency $\om=2\pi/\mathcal{T}$. It follows from the above definitions that the velocity of a crest, called the phase velocity, is $u=\om/k$. Such waves propagate at constant speed without any action being exerted on them. For linear waves there is a definite relationship between $\om$ and $k$ independent of the wave amplitude, called the dispersion equation. For gravity waves in deep non-viscous waters we have, for example, $\om=\sqrt{gk}$, where $g\approx 9.81 $m/s$^2$ is the earth acceleration. When the water depth $h$ is not large compared with wavelength (shallow water), the dispersion relation involves $h$ as a parameter. 

The above considerations may be related to mechanical effects. Indeed, if a wave carrying a power $P$ is fully absorbed, the absorber is submitted to a force $F$ satisfying the relation $P/\om=F/k$. This ratio, called "wave action", depends on the nature of the wave but does not vary if some parameter is changed smoothly, either in space or in time. For a wave of finite duration $\tau$, the energy collected by the absorber is $E=P\tau$ and the momentum received (product of its mass and velocity) is $p=F\tau$. 

If the water depth $h$ is changed at time $t=0$ from, say, 1m to 2m, it is observed that $k$ is unchanged as a consequence of the wave continuity. But invariance of $k$ implies a frequency change since the  dispersion equation depends on $h$. In that case, the wave speed changes at time $t$. Conversely,  If the water depth $h$ changes at some location $x=0$ from, say, 1m to 2m, it is observed that $\om$ is unchanged as a consequence of the wave continuity. But invariance of $\om$ implies a wave number change since the  dispersion equation depends on $h$. In that case the wave speed changes at $x=0$.

Consider now a monochromatic wave (fixed frequency $\om$) propagating in two dimensions with coordinates $x$, $y$. The direction of propagation is defined as being perpendicular to the crests and the wavelength $\lambda=2\pi/k$ is defined as the distance between adjacent crests at a given time. But one may also define a wavelength $\lambda_x$ in the direction $x$ as the distance between adjacent crests \emph{in the $x$-direction} at a given time. Let the wave be incident obliquely on the interface between two media, the $x$-axis. For gravity waves the two media may correspond for example to $h(y)=1$m$, y>0$ and $h(y)=2$m$, y<0$.  Because of the continuity of the wave, $\lambda_x$ is the same in the two media. If we further assume that the propagation is \emph{isotropic}, that is, that $k$ does not depend on the direction of propagation of the wave in the $x,y$ plane, the law of refraction follows, namely that $k_x=k_1\cos (\theta_1)=k_2\cos (\theta_2)$, where the subscripts 1,2 refer to $y>0$ and $y<0$ respectively, and the angles $\theta$ are defined with respect to the interface, that is, to the $x$-axis. The law of refraction therefore follows from wave continuity and isotropy alone. 

Questions relating to the velocity of light pulses are important for the transmission of information. A wave-packet containing many wave crests moves at the so-called "group velocity" $v=d\om/dk$, which often differs much from the phase velocity $u$ defined above. Considering only two waves at frequency $\om$ and $\om+d\om$, the relation $v=d\om/dk$ may be visualized as a kind of Moiré effect. Wave crests move inside the packet, being generated at one end of the packet and dying off at the other end. For waveguides we have $uv=c^2, v<c,u>c$. For matter waves associated with a particle the group velocity $v$ coincides with the particle velocity. Since the energy $E=p^2/\p 2m\q$ and $p=mv$, a previous relation reads $p^2/\p2m\om\q=p/k$. It follows that $u=\om/k=p/2m=v/2$. For gravity waves the dispersion relation gives instead $u=2v$. A general result applicable waves propagating in loss-less media is that the group velocity $v$ is the ratio of the transmitted power $P$ and the energy stored per unit length. The group velocity never exceeds the speed of light $c$ in free space. In the presence of losses, the situation is more complicated.

Wave solutions of the form $\psi(x,t)=\psi(x-ut)$, where $\psi(x)$ is some given function and $u$ a constant, exist also for non-linear wave equations. When the $\psi(x)$ function is localized in $x$, the invariant wave-form is called a solitary wave. In some cases, solitary waves exhibit transformations akin to those of particles when two waves collide and are called "solitons" in the sense that the soliton integrity is being preserved.

As said before, most continuous media may be modeled by discrete circuits. For example, a transmission line may be modeled by series inductances and parallel capacitances. Free space may be modeled by electrical rings in which electrical charges move freely and magnetic rings in which (hypothetical) magnetic charges would move freely. If each electrical ring is interlaced with four magnetic rings and conversely, the Maxwell equations in free space obtain in the small-period limit.

Under confinement along the $x$-direction, waves at some fixed frequency $\om$ may be viewed as superpositions of "transverse modes". For a transverse mode the wave-function factorizes into the product of a transverse function $\psi(x;\om)$ and a function of the form $\exp(ik(\om)z-i\om t)$. Another connection between waves and rays rests on the representation of transverse modes by ray \emph{manifolds}. These are not however independent rays. A phase condition is imposed on them that leads to approximate expressions of $\psi(x;\om)$ and $k(\om)$. Note an analogy with Quantum-Mechanics stationary states, $z$ and $t$ being interchanged.

Thus the wave-particle connection is many fold. First the medium in which the wave propagates may be approximated by a discrete sequence of elements, for example a periodic sequence of springs and masses for acoustical waves and electrical inductance-capacitance circuits for electromagnetic waves, with a period allowed to tend to zero at the end of the calculations. One motivation for introducing this discreteness is that computer simulations require it anyway. A more subtle one is that some divergences may be removed in that way. We have mentioned above capillary waves on a mercury bath. They may be treated by considering the forces binding together the mercury molecules and their inertia, ending up with equations of fluid mechanics. Like-wise, acoustical waves in air may be described through the collision of molecules in some limit (isothermal or adiabatic). Second, wave modes may be described approximately (WKB approximation) by ray \emph{manifolds}. Third, one may consider the behavior of wave packets in the high-frequency limit and liken the wave packets trajectories to those of macroscopic bodies.

We have described the motion of light and particles in terms of waves. Semi-classical theories such as the one employed in the present paper rest indeed on wave concepts, namely Quantum Mechanics for describing electrons, and Circuit Theory for describing the relationship between potentials and currents. The speed of light in free-space does not enter into the Quantum Theory. In circuit theory, one may discretize space, and then ignore the speed of light. When particles such as electrons are electrically charged they may be accelerated to arbitrarily large energies by static electrical potentials. Being then in the classical domain there is no ambiguity concerning their arrival time. Uncharged point particles such as neutrons could conceivably be accelerated similarly by gravitational fields, even though this may turn out to be difficult in practice.

\section{Atoms and elements}\label{el}

Around 1927 it was proposed by de Broglie, and subsequently verified experimentally, that a wave of wave-number $k=mv/\hbar$ should be associated with electrons of mass $m$ and velocity $v$. According to the semi-classical theory, the Planck constant $\hbar$ enters in Physics through the relation $k=mv/\hbar$, relating wavelength and momentum. ($\hbar$ was introduced earlier in the theory of black-body radiation, in 1900). An approximate solution for the motion of an electron following a closed classical path in the neighborhood of a positively charged nucleus thus amounts to prescribe that an integral number $n$ of wavelengths $2\pi/k$ fits along the closed classical path.  These discrete solutions are called "stationary states" and $n$ is essentially the principal quantum number. According to the Pauli principle, at most two electrons (with spin $±\hbar/2$) may be ascribed to each of these states. At zero absolute temperature and without excitation by other particles, only the lowest-energy states are filled with electrons. Different elements (H, He, Li...) differ by the number $Z=1,2,3...$ of protons in their nuclei. Hydrogen and helium nuclei, and their isotopes, came up early after the "big-bang". The other elements were formed in the interior of stars, and subsequently dispersed into space.

In the next paragraph we recall how the chemical and electronic properties of the various elements found in nature follow from the above principle, and describe what happens when atoms get closer and closer to one another to form crystals. Then we recall the basic properties of semi-conductors.  

\section{Electron states}\label{elements}

We summarize below the most basic concepts concerning elements found in nature and their electron states. The simplest element is the hydrogen atom consisting of a proton with an electrical charge $e$ and a mass much larger than the electron mass $m$, so that for most purposes the proton may be considered as being fixed in space. This proton attracts one electron of charge $-e$ so that the assembly is neutral. According to Classical Mechanics the electron may circle around the proton at a distance $r$ with a velocity $v$ such that the centrifugal force is balanced by the attraction from the proton, namely $mv^2/r=e^2/\p 4\pi\epsilon_o r^2\q$. From this view-point any distance $r$ may occur, the velocity $v$ being appropriately chosen. According to Quantum Theory a wave-length $2\pi\hbar/mv$ is associated with electrons moving at velocity $v$. The resonance condition is that an integral number $n$ of wavelengths fits within the electron path perimeter $2\pi r$. According to this model, due to Bohr, there is only a discrete sequence of allowed electron energies, corresponding to $n=1,2...$. The more exact theory due to Schrödinger leads to symmetrical ground states, called $s$-states, and anti-symmetrical 3-times degenerate first-excited states, called $p$-states. 

The elements found in nature (roughly 100) were classified by Mendeleïev in 1869 on empirical grounds. As said above, helium nuclei consist of two protons, lithium nuclei of three protons, and so on, with an equal number of electrons, so that atoms are electrically neutral. There may be various numbers of neutrons bound to the protons that depart from the number of protons by a few units, corresponding to different isotopes, some of them being unstable. Neutrons are considered at one point as an example of multilevel system (bismuth nuclei immersed in a magnetic field). Most elements have an outer layer consisting of a number of electrons going from 1 (e.g., sodium) to 8 (e.g., neon). Particularly important are 3-5 crystals, such as gallium-arsenide.

\section{Semi-conductors}\label{semi}

Our purpose here is to give readers unfamiliar with solid-state physics an overview of the most important phenomena. For silicon, the number of outer electrons is 4. Two silicon atoms (or more) get bound to one another by exchanging electrons of opposite spins (covalent binding). When two atoms are approaching one another, their electronic states get perturbed. As it happens, the isolated-atom electron $s$-state acquires an energy greater than the isolated-atom electron p-states. For a large number $N$ of atoms, the atomic separation $a$ sets up at a value that minimizes the total energy. The original s-states then split into $N$ states that are so-closely spaced in energy that they form an almost continuous band of states called the conduction band. The original 3-fold degenerate $p$-states split into $N$ states that are so-closely spaced in energy that they form three almost continuous band of states called the valence bands. Because the degeneracy is lifted these three bands should be distinguished. They are called respectively the heavy-hole band, the light-hole band and the split-off band. For our purposes, only the heavy-hole band needs be considered. 

The separation in energy between the bottom of the conduction band and the top of the valence band is called the band gap $E_g$, often expressed in electron-volts. At $T$=0K, the lower-energy valence band is filled with electrons while the higher-energy conduction band is empty. At that temperature the electrons are unable to respond to an external field because no state is available to them (except perhaps at extremely-high fields). If an electron is introduced in the conduction band by some means it moves in response to an electrical field with an apparent mass $m_c$ smaller than the free-space mass $m$. If, on the other hand, an electron is removed from the valence band one says that a "hole" has been introduced. This hole is ascribed a positive charge $e$ and a mass usually larger than $m$. 

When two materials having different band gaps are contacted the band gap centers align approximately, and potential steps occur both in the conduction and valence bands. In the case of a double-hetero-junction the lower-band-gap material is sandwich between two higher-band-gap materials. The potential steps tend to confine both free electrons and free holes in the central low-band-gap material (e.g., GaAs). Being confined in the same volume electrons and holes easily interact.

As the band-gap decreases electrons may undergo virtual transfers from one band to the other more easily. As a consequence the material is more easily polarized by external (static or optical) electrical fields. In other words, the material permittivity $\epsilon(\om)$ increases as the band gap decreases. This is why the permittivity (or refractive index) of the low-band-gap gallium-arsenide is significantly larger than the permittivity (or refractive index) of the large-band-gap aluminum arsenide. When a small-band-gap semi-conductor (GaAs) is sandwiched between two higher-band-gap semiconductors (AlAs), the higher-index material may guide optical waves. This fact is important for the guidance of optical waves in laser diodes employing double-hetero-junctions. An happy circumstance is therefore that electrons, holes, and light, may all get confined in the central part of the double-hetero-junctions considered. 

Gallium possesses 3 electrons in the outer shell and arsenide possesses 5 electrons. Equal numbers of these atoms may associate to form a crystal of gallium-arsenide (Ga-As), a material particularly important in Opto-Electronics. The reason for this importance is that, unlike silicon, this is a "direct band-gap" material. In direct band-gap materials the minimum of the conduction-band energy and the valence-band maximum energy correspond to the same electron momentum. Accordingly, electrons lying at the bottom of the conduction band may get easily transferred to the top of the valence band, and conversely, the law of momentum conservation being then fulfilled, the optical field momentum being negligible. In such a process, an energy $E_g$ is absorbed by light through stimulated or spontaneous emission processes. Unfortunately, this energy may also be absorbed by another electron (Auger effect) that subsequently cascades down, its energy being converted into heat. 

Finally, one should say a word about doping, considering as an example a silicon crystal. When a small number of silicon atoms are replaced by arsenic atoms, these atoms, referred to as "impurities", easily deliver an electron (n-doping). Conversely, when a small number of silicon atoms are replaced by gallium atoms these atoms easily capture electrons ($p$-doping). A p-n diode consists of two contacting semi-conductors, one with $p$-doping and one with n-doping. Electron currents may be injected into p-n diodes, and in particular into double-hetero-junctions. This is the current referred to in this paper as the laser-diode driving current $J$.

The above discussion hopefully provides the essential concepts that one needs to get some understanding of the electrical behavior of laser diodes. Note that we denote by $z$ the coordinate along which the optical wave propagates (junction plane) and by $x$ the direction perpendicular to the semiconductor layers. Guidance along the transverse $y$ direction is also considered.

\section{Detectors and sources}\label{detectors-sources}

In subsequent sections we discuss sources of electromagnetic radiations and ways of detecting them. It is appropriate to consider first detectors because there exist natural sources of radiation such as the sun, and the difficulty was initially to detect such radiations rather than to generate them. Detectors convert high-frequency radiation into slowly varying currents. The mode of operation of some detectors, called "classical detectors", may be explained on the basis of the Classical Equations of Electron Motion. For others, called "quantum detectors", the Quantum Theory is required. Conversely, sources convert slowly varying currents into high-frequency radiation. The mode of operation of some sources, called "classical sources", may be explained on the basis of the Classical Equations of Electron Motion. For others, called "quantum sources", the Quantum Theory of Electron Motion is required.

The first known light detector is the eye. Modern detectors were first vacuum tubes operating with a low-work function cathode and accelerating potentials. There exist now quantum detectors whose mode of operation is based on the phenomenon of stimulated absorption. An early man-made generator of high-frequency radiation is a vacuum-tube called the "reflex klystron". The main light sources are today hot bodies and lasers.

\section{Classical detectors}\label{cdetectors}

Classical detectors are diodes that exhibit non-linear current-potential characteristics. If a sinusoidal potential is applied to the diode the current then exhibits a non-zero average value, which is a measure of the applied sinusoidal-potential amplitude.

\begin{figure}
\setlength{\figwidth}{0.45\textwidth}
\centering
\begin{tabular}{cc}
\includegraphics[width=\figwidth]{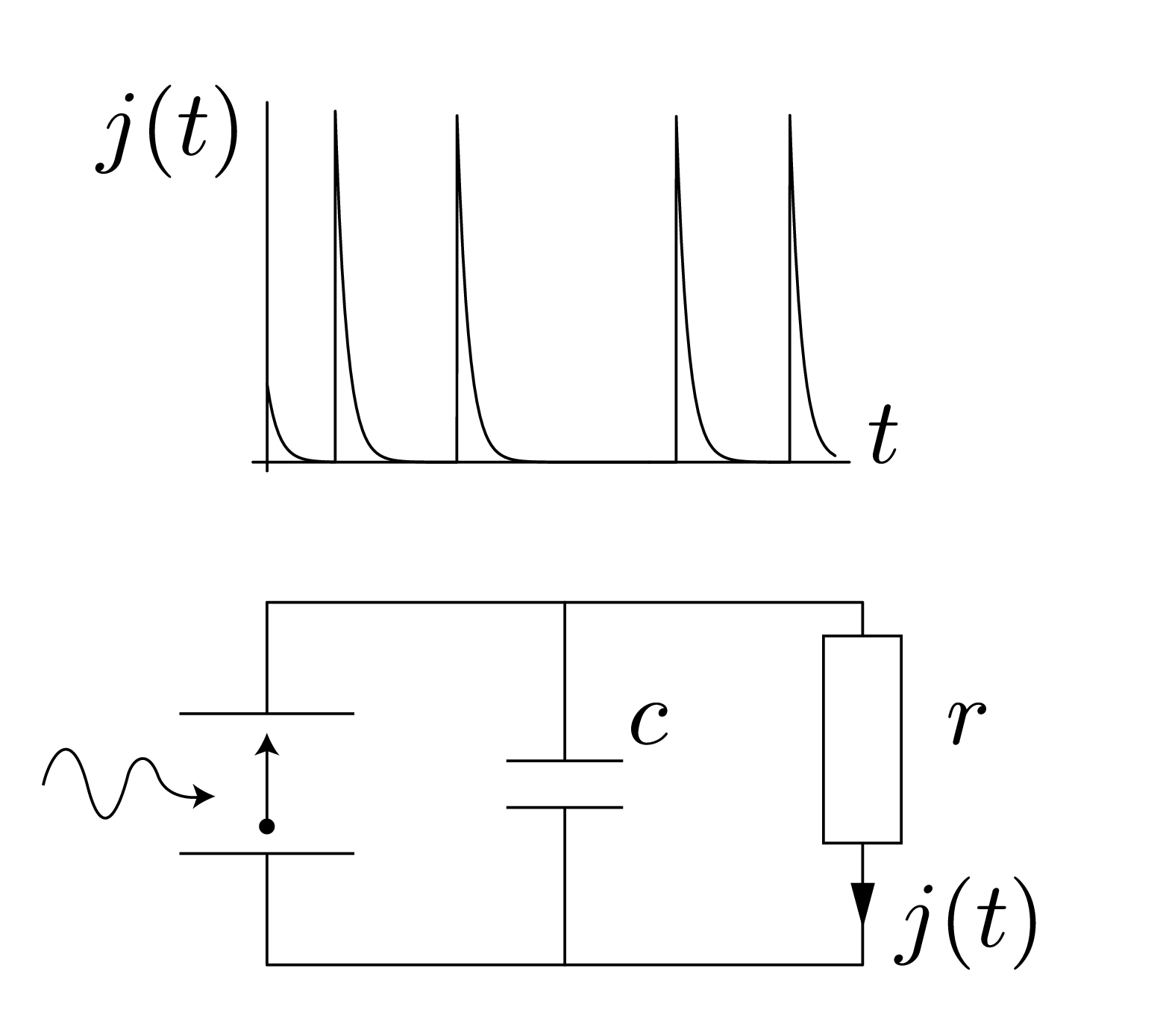} & \includegraphics[width=\figwidth]{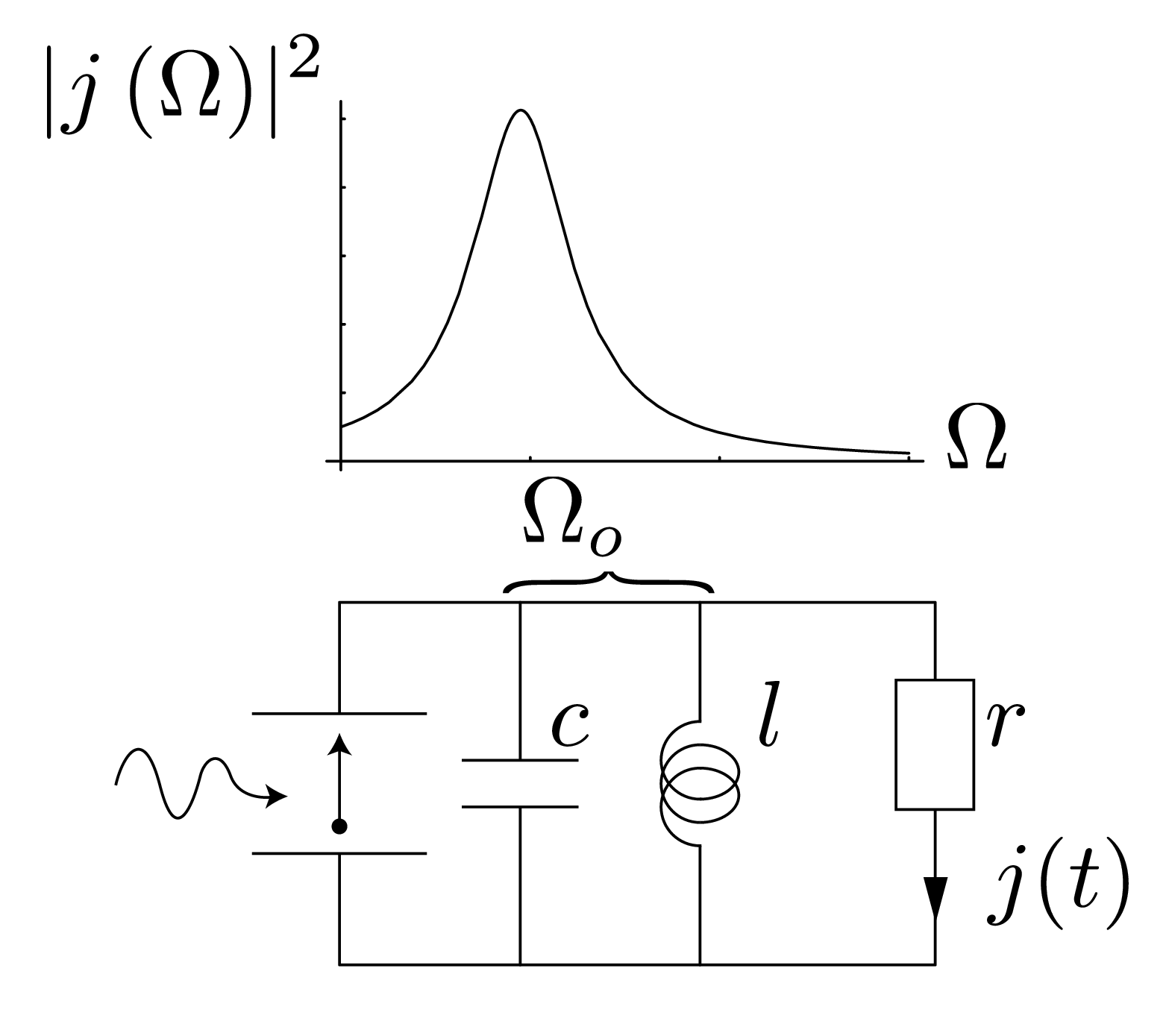} \\
(a) & (b) \\
\end{tabular}
\caption{In a), on top the photo-current is represented as a function of time, the output circuit capacitance (lower part) being taken into account. In b) the photo-current spectrum (on top) for the case where the output circuit (lower part) is a resonating circuit tuned at some Fourier frequency $\Om_o$.}
\label{photodetection}
\end{figure}

Let us recall the basic mode of operation of conventional electronic diodes, photo-detectors and photo-multipliers. Conventional electronic diodes are made up of two parallel plates (labeled in what follows the lower and upper plates) separated by a distance $d$ in vacuum. The lower plate, called "anode" is at zero potential by convention, and the upper plate, called "cathode", is raised at the potential $-U$ with $ U>0$. Suppose that at time $t=0$ an electron is freed from the upper plate and attracted by the anode\footnote{To achieve this, the cathode "work function" energy must be overcome by heat (thermo-ionic emission), high electric fields (field emission), electrons (secondary emission), or light (photo-electric emission). Electrons may be freed by thermal motion provided $\kB T$ be of the order of the metal work-function. If nickel is coated with barium oxide, a temperature of 1000 K may suffice. Field emission occurs with kilo-volt potentials if the cathode has the shape of a needle. Electrons may be freed by light provided $\hbar\om$ exceeds the metal work-function, where $\hbar$ denotes the Planck constant and $\om$ the light frequency. Visible light for example is adequate when the cathode is coated with cesium. The non-zero initial electron velocities are presently neglected, that is, the initial electron momentum $p(0)=0$. Electrons in a metal are bound to it because they are attracted by their image charge. They may escape, though, because of a tunneling effect whose understanding requires Quantum Mechanics. But once the electron is sufficiently far away from the cathode, the Classical Equations of Electron Motion are appropriate.}. Considering only absolute values, the electron momentum increases linearly with time $t$ according to the law $p(t)=eUt/d$, where $-e$ denotes the electron charge, until it reaches the anode at time $\tau=d\sqrt{2m/eU}$, where $m$ denotes the electron mass. The electron kinetic energy is then converted into heat. In the following $\tau$ is set equal to zero. Fig.~\ref{photodetection} illustrates in a), the photo-current, represented as a function of time. Because the output circuit capacitance is taken into account each electron arrival corresponds to an exponentially-decaying pulse of the form $\exp(-t/rc)$. In b), photo-current spectrum for the case where the output circuit is a resonating circuit tuned at some Fourier frequency $\Om_o=1/\sqrt{\ell c}$.

In the case of photo-multipliers, the electron kinetic energy, instead of being dissipated into heat, is employed, at least in part, to free two or more electrons from the anode (secondary-emission effect). The latter are accelerated by a third plate, and so on, so that each electron freed from the cathode by light gets converted into an electron bunch containing $n$ electrons, for example, $n=10^6$. The situation is the same as if the absolute value $e$ of the electron charge had been multiplied by $n$. The purpose of photo-multipliers is not to improve the signal-to-noise ratio, which may only degrade. It is to raise the signal to such high levels that the thermal noise of subsequent electronic amplifiers is rendered negligible. 

In temperature-limited thermo-ionic diodes the electronic density is so low that the Coulomb interaction between electrons may be neglected. The electrons are emitted independently of one-another and their emission times are Poisson distributed (see the mathematical section). In that case the diode current fluctuation $\De j(t)$ obeys the so-called shot-noise formula, with a (double-sided) spectral density equal to $e\ave{j(t)}$. But when the electron flow is space-charge limited the current is sub-Poisson, i.e., the spectral density is much smaller than the one just given.  This effect was discovered in 1940 by Thompson, North and Harris.
% Corresponding reference not found in Phys Rev as expected, some works in these years appeared in J. Chem. Phys but not with all these authors. 

The detectors considered above have been idealized for the sake of simplicity neglecting, e.g., dark currents and thermal noise. As said before, the current flowing out of photo-detectors may be viewed as a sum over the relative integer $k$ of delta-functions of the form $e\delta(t-t_k)$, where the $t_k$ are the occurrence times. If this current is transmitted through a low-pass filter such as the one shown in Fig.~\ref{photodetection}, individual pulses may overlap, however, and not be distinguishable any more from one another. The current fluctuation then resembles gaussian white noise (normal process) irrespectively of the event-times statistics. This process is (statistically speaking) time reversible, a property preserved by linear systems.

When radiation is absorbed by a small body the latter temperature increases, resulting in some change of the body properties (bolometer). According to recent reports the sensitivity of some bolometers may be on the order of a light-quantum energy.

\section{Quantum detectors}\label{qdetectors}

When light impinges on cold atoms, the atoms may move from their ground state to some excited state and, from then, get ionized with the help of an independent source of energy. In that way photo-electrons are generated, which occur at specific (but random) times. This mechanism could also take place at low frequencies, but then extremely cold atoms would be needed. Otherwise the atoms could get excited thermally, instead of being excited by the impinging radiation. If the radiation power is large, the beam may be spread out to a mosaic of detectors, so that the photo-electrons are well separated in time from one another in each detector. The corresponding event processes should then be superposed (without any time shift of course. Otherwise we would always end up with a Poisson process).

Quantum photo-detectors (sometimes referred to as "narrow-band" photo-detectors), involve two electron energy levels, coupled to continua, and operate through the process of stimulated absorption. Ideally, the device is reversible in the sense that the electrical energy may be converted back into light energy through the process of stimulated emission. In contradistinction, conventional photo-diodes necessarily dissipate energy in the form of heat. Of particular interest is the visible-light-photon-counter which has a high quantum efficiency.

\section{Classical sources}\label{csources}

The first high-frequency oscillator was probably a triode, with a feed-back mechanism from the anode to the grid controlling the current flow. The reflex klystron, discovered by the Varian brothers in 1937, may deliver electromagnetic radiation up to a frequency of about 10 GHz. It employs space-charge-limited cathodic emission. As recalled above, Thompson and others discovered in 1940 that the current emitted by space-charge-limited cathodes is strongly sub-Poisson. The emitted radiation is therefore expected to be sub-Poisson. This however has apparently not been observed. Reflex klystrons were mostly employed as low-noise local oscillators in radar heterodyne receivers until they were superseded by solid-state devices. 

Reflex klystrons involve two anodes made up of grids presumed to be transparent to electrons and separated by a distance $d$. Two cathodes are located just outside the anodes. The lower one emits electrons, while the upper one plays the role of a reflector\footnote{Usually the reflector is raised at a potential slightly lower than that of the emitting cathode to prevent electrons from being captured. It also helps finely tune the klystron oscillation frequency.} so that the electron motion as a function of time is a zig-zag path. The two anodes are part of a resonator. When the device oscillates an alternating potential $v(t)$ appears between the two anodes. If the electron emission time is appropriate, the electron looses its energy giving it up to the oscillating potential through an induced current. But since the electron emission times are uniformly distributed along the time axis, the net interaction with the field vanishes. Accordingly, initially, the electrons do not deliver any energy to the oscillating potential. It is as a result of the field action on the electron trajectories that non-zero energy exchanges between the field and the electrons may occur. This effect is called "bunching". Once an electron has lost most of its energy it gets captured by the anodes, and instantaneously jumps from the anode to the cathode through the static potential\footnote{In conductors the number of electrons is close to the number of atoms, a huge number (precisely, in copper there are two electrons per atom). It follows that electrons are moving at very low speeds, on the order of 1$\mu$m/s, even for large currents. Accordingly, the above statement that jumps are instantaneous may seem surprising. As a matter of fact, electrons appearing on one plate are not the same as the electrons hitting the other plate. In conductors the electrical charge should be best viewed as a continuous incompressible fluid undergoing collective motion.}. The static-potential energy then gets reduced by  $eU$. In some sense, the electron plays an intermediate role. Indeed, the net effect of a complete electronic cycle is that, for each electronic event, the static potential source delivers an energy $eU$ to the oscillating potential source $v(t)$. 

A phenomenon akin to stimulated absorption may be understood similarly. This time, we suppose that the electron is emitted by one of the anodes. Without an alternating field this electron would remain permanently in the neighborhood of the anodes. However, a resonance with the alternating potential may force the electron to oscillate along the $x$ axis with increasing amplitude until its energy reaches the value $eU$, in which case it gets captured by one of the cathode. The net effect of this electronic process is that the alternating source gives energy to the static source, the opposite of what was discussed in the previous paragraph.

\section{Quantum sources}\label{qsources}

The first man-made quantum oscillators involving discrete matter levels were masers, operating at microwave frequencies. Subsequently maser action was discovered to occur naturally near some stars. The first laser, generating visible light, was discovered by Maiman in 1959. The fact that space-charge-limited cathodes generate light with sub-Poisson statistics was first demonstrated by Teich and Saleh in 1983. 

The best-known light source is thermal radiation. A hot body like the sun radiates energy. The energy inside a closed cavity at absolute temperature $T$ contains an energy given by the law discovered in 1900 by Planck. An important feature of this law is that it involves a previously unknown universal constant $\hbar$ with the dimension of action or angular momentum (energy$\times$time). From the Quantum Mechanical view point, heat excites electrons to atomic levels higher in energy than the ground state energy.  These electrons then may decay spontaneously to the ground state by emitting light. A similar mechanism is at work in the so-called "light-emitting diodes" but the spectrum then, whose center is essentially defined by the semiconductor band-gap, is narrow.

A conventional neon tube generates light because the electric discharge excites neon atoms that subsequently decay to the ground state, thereby emitting ultraviolet light (subsequently converted into visible light) by the process of spontaneous emission, similar to what happens in thermal sources, but with a narrower spectrum. The so-called Helium-Neon laser\footnote{Helium plays the role of a "buffer gas", allowing the lower neon level to get depopulated.}, radiating light at a free-space wavelength of 0.63 $\mu$m, differs from conventional neon tubes in that two mirrors located at both ends, and facing each others, force the emitted light to move back and forth in the tube. Light gets amplified by the process of stimulated emission, and damped by the process of stimulated absorption. The former exceeds the latter when there are more atoms in the higher state than in the lower state (population inversion). To achieve this condition the lower-level population must be reduced through spontaneous decay to even-lower levels (3-levels lasers). Eventually a steady state of oscillation is reached. The emitted light spreads out in free space as little as is allowed by the laws of diffraction, and the laser light is nearly monochromatic (single frequency). The laser linewidth, though small, is of major importance in some applications. Laser diodes (also called injection lasers) employ  a semi-conductor with a doping that delivers electrons (n-type) and a doping that absorbs electrons (p-type).

To summarize, lasers essentially consist of single-mode resonators containing
three-level atoms or other forms of matter with a supply of energy called the pump and a sink of
energy, perhaps an optical detector. Or else the generated light is dissipated and converted into heat. As said before, detectors  convert the light
energy into a sequence of electrical pulses corresponding to
photo-detection events. When the pump is non-fluctuating the emitted light does not fluctuate
much. Precisely, this means that the variance of the number of
photo-detection events observed over a sufficiently long period of
time is much smaller than the average number of events.  Light having
that property is said to be sub-Poisson.

\section{Quantum Electrodynamics}\label{quantum-light}

Historically, the concept that light should be quantized appeared around 1905 on the basis of thoughts expressed by Einstein, but was formalized later on by Dirac. Measurements on black-body spectra were performed around 1900 with the help of gratings of appropriate periods, and described by a formula that involves the universal constant $\hbar$. On the other hand, the wave properties of electrons were discovered only decades later because the concept that electrons might possess wave-like behavior ought to wait for the observation that atoms emit light at well-defined frequencies, and because of the technical difficulty of sending electrons emitted from a small-area source on crystals (playing the role of gratings) in a very good vacuum. The interpretation of the observed diffraction patterns involves the constant $\hbar$. It is perhaps not preposterous to suggest that these two key discoveries could have occurred in the reversed order. Had this be the case, the Planck constant would have been considered as being fundamentally related to atomic behavior, and the subsequent appearance of the same constant in black-body radiation would have been viewed as a consequence of the atomic theory. More precisely, the classical theory tells us that the average energy $\ave{E}\equiv E$ of a resonator equals the absolute temperature $T$. The resonator action $f\equiv E/\om$ therefore obeys the equation $df/dx+f^2=0$, where $x\equiv \om/T$. Apparently, it did not occur to the physicists at the time that the divergence was removed merely by adding a constant on the right-hand-side of that equation, that is solving $df/dx+f^2=(C/2)^2$. Setting $C=\hbar$, the Planck law of black-body radiation is obtained. In that (admittedly heuristic) approach, the discontinuity concept and the entropy does not enter.

The majority view point is that light possesses independent degrees of freedom corresponding to some number of "photons". However, a recent author\,\cite{Smilga2009} says the following: "The apparent degrees-of-freedom of the photon field reflect
the kinematical degrees-of-freedom of the two-particle state space of
massive fermions, rather than independent degrees-of-freedom of the
photon field. The excellent agreement of [his] model with the experimental value of the fine-structure constant $\alpha\approx 1/137.036$ can
be considered as an experimental proof that, in principle, all two-particle
states, contributing to the electromagnetic interaction, have been correctly
accounted for. At the same time this result shows evidence that the photon
field does not possess degrees-of-freedom of its own. It rather relays the
kinematical degrees-of-freedom of the charged particles". It is fair to say that this view-point is not shared by most physicists.

\section{Non-relativistic approximation.}\label{nonrelat}

Some ancient philosophers thought that the speed of light is infinite. If it were the case it would be immaterial to say that light propagates from the sun to the eye (say) or the converse. The non-relativistic approximation is applicable to an hypothetical world in which the speed of light in free space would be arbitrarily large, the other constants ($\hbar,\epsilon_o,e,m$, defined in Section \ref{notation}) remaining as they are. Out of the latter constants we may define the electron spin along some quantization axis $±\hbar /2$, the Bohr radius $ a_o=4\pi\epsilon_o\hbar^2/me^2\approx0.53~ 10^{-10}$ meters, a speed unit $v_o=e^2/4\pi\epsilon_o\hbar=\hbar/a_om\approx 2.19~10^6 $ meters/second, a nominal metal plasma frequency $\om_p\equiv v_o/a_o\approx 4.13 ~10^{16}$rad/s, and the electron magnetic moment, equal to the Bohr magneton $e\hbar/2m$. A large part of Physics may be obtained on the basis of such a non-relativistic approximation. However, in order to define an inductance from its geometrical dimensions we need the free-space permeability $\mu_o\equiv \frac{1}{\epsilon_o c^2}$. We therefore keep such terms in the Circuit Theory.

\section{Finite speed of light.}\label{finite}

Römer discovered in 1676 through a kind of Doppler effect, using the motion of a Jupiter satellite as a clock, that light propagates at a finite speed $c\approx 300 ~000$ km/s. The law of causality, as it is presently understood, then implies that light propagates from the sun to the eye, for example. The Maxwell theory of electromagnetic waves suggests that radiated heat, as well as light, consists of electromagnetic waves of some sort. In 1862 Maxwell wrote "[electromagnetic waves travel] at a speed so nearly that of light that it seems we have strong reason to conclude that light itself (including radiant heat and other radiations) is an electromagnetic disturbance in the form of  waves propagated through the electromagnetic field according to electromagnetic laws.”

Note that if $v_o$ is written as $\al c$, where $\al\approx 1/137.036$ denotes the fine-structure constant, the non-relativistic approximation amounts to setting $\al$ as equal to zero. Because of the finite speed of light, a number of small corrections to the non-relativistic theory, on the order of $\al$, were observed. In particular, the spin-orbit splitting of electron states in atoms, the electron anomalous magnetic moment, the Casimir effect. $\gamma$-rays originating from collapsing super-massive stars located at distances from earth ranging from 7 to 13 billions light years have been observed with an energy $\hbar \om\approx 1$ micro-joule. It is not clear whether such $\gamma$-ray bursts may be viewed on earth as point particles.

\section {Quantum Optics.}\label{quantumoptics}\label{qopt}

Following proposals by Einstein in 1905 and Dirac later on, the concept that the optical field should be quantized, that is treated as an operator rather than a classical function of time, is almost universally accepted. A comprehensive reference is\,\cite{Mandel1995}. There are however a number of effects, initially ascribed to light quantization, that may receive a classical explanation. Such explanations are simpler mathematically than the Quantum Optics methods, and provide an intuitive feeling for what is going on. 

Quantum Optics provide us in particular with the probability of an atom, initially in the lower state and interacting for some time with a resonator, be found after that time in the upper state. This probability depends on the initial state of the resonator, formally akin to the state of harmonic mechanical oscillators. This is a (coherent or incoherent) superposition of eigen-functions labeled by $m$. For example, the thermal state, prepared by letting the resonator interact with a bath at some temperature $T$, consists of an incoherent superposition of $m$ states obeying the Boltzmann law of distribution. The number state $<m>$ (labeled: "non-classical state") means that the cavity contains exactly $m$ "photons". These states could be defined by their preparation procedures. Quantum Optics tells us for example that if an atom in the lower state enters into a resonator in the vacuum state ($<m>$=0), it necessarily exits in the lower state. This conclusion may be difficult to explain semi-classically. Indeed, the resonator contains some field, even at $T=0$. If we employ the classical Rabi equations of interaction between a quantized electron and a classical field, we end up with the conclusion that there is a non-zero probability that the atom exits in the upper state, thereby contradicting the law of conservation of energy. However, in the semi-classical theory, one must take into account the back action of the atom on the amplitude and phase of the resonator field.  

Let us recall some of the arguments given in favor of the photon concept, which evolved into the modern second-quantization procedures. Light-quanta (later on called "photons") were introduced by Einstein on the basis of the following argument.  Consider a collection of two-level atoms in a state of thermal equilibrium with the black-body radiation field. When an atom in the upper state decays to the lower state by emitting light spontaneously it recoils if the light  emission is \emph{directed} but would not if light were radiated (almost) isotropically. Einstein calculations indicate that a directed emission is required if the Maxwellian atomic velocity distribution is to be recovered. The Einstein picture fits well with the view that light is emitted only if it is directed toward some absorber. This is sometimes phrased as follows: every photon is virtual, being sooner or later absorbed.

Another argument in favor of the concept that light consists of lumps of energy $\hbar \om$ is the observation that when a light beam of constant small intensity is incident on an ideal photo-detector (i.e., free of dark current and thermal noise) photo-current events sometimes occur long before the required optical energy $\hbar \om$ has been collected, in apparent violation of the law of conservation of energy. However, one should require that only the law of \emph{average} energy conservation be enforced. For a single system, there exists no independent way of measuring the "light intensity" as a function of time. The only information one may obtain concerning the intensity of a light beam is through the output of photo-detectors, and this brings us back simply to the observation made. A number of authors have shown that many effects that were at a time supposed to prove the reality of the photon concept may be interpreted in a semi-classical manner.

Other Quantum Optics treatments consider atoms in either their upper (pumping atoms) or lower (detecting atoms) states introduced at specific times into the optical cavity and spending there a fixed time $\tau$. Whether the atoms leaving the cavity are in their lower or upper state may be measured, and the corresponding probabilities may be evaluated. However, the flying-atoms configurations just described is quite different from those discussed in this book, which are \emph{stationary}. The purpose of this book is not to challenge the validity of Quantum Optics methods, but to see how far one can go with a semi-classical approach, as far as stationary lasers are concerned. 

\section{Physical paradigms.}\label{paradigm} 

A physical "paradigm" rests on a number of universal constants, on particles parameters, and on recipes to relate the theory to observations. For example Newtonian Celestial Mechanics employs a single universal constant, namely $G$, point particles have as sole parameter their mass (referred to that of a particular piece of platinum, called the kilogram). Given the position and speed of the particles at a given time, the theory provides their positions and speeds at all times.

The present theory employs as universal constants $\hbar$, $\frac{1}{4\pi\epsilon_o}$, the Boltzmann constant $\kB$ being set as unity by an appropriate choice of temperature unit. $G$ is used only in simulations given for the sake of illustration. $\mu_o=1/\epsilon_o c^2$ is used only in the circuit theory. Particles (electrons) are characterized by their mass, electrical charge, and, in some circumstances, their spin and magnetic moment. Given the elements constitutive of a particular device, set up once for all, we determine the photo-electrons statistics which may, in principle, be measured with unlimited accuracy.

\section{Units and notations}\label{notation}

Our notations and conventions may differ from those employed by engineers, physicists, or experimentalists, which are not fully consistent. We attempted to follow the majority rule unless this leads to confusion. To simplify formulas we sometimes set as unity quantities such as the characteristic conductance of transmission lines. Otherwise, SI units (see below) are employed throughout. The term "light quantum" means to us that the field energy divided by $\hbar\om$ may be written as the sum of an integer $m$ and a non-integer part comprised between 0 an 1. The integer part is said to consist of $m$ "light quanta", the latter having no physical connotation. We call "oscillator" a mechanical oscillating system, and "resonator" an optical cavity.

\section{Numerical values.}\label{sec_num}

Useful numerical values in the realm of Non-Relativistic Physics are
\begin{align}\label{eps}
G~\textrm{(Newton gravitational constant)}&\approx 6.67~10^{-11}~\textrm{SI}\nonumber\\
g~\textrm{(earth gravitational acceleration at see level)}&\approx 9.81~\textrm{meters per second squared}\nonumber\\
e~\textrm{(absolute electron charge)}&\approx1.60~10^{-19}~\textrm{coulombs}\nonumber\\
m~\textrm{(electron mass)}&\approx9.10~10^{-31}~\textrm{kilograms}\nonumber\\
\hbar ~\textrm{(Planck constant divided by $2\pi$)}&\approx1.05~10^{-34}~\textrm{joules}\times \textrm{second}\nonumber\\
\kB~\textrm{(Boltzmann constant)}~&\approx1.38~10^{-23}~\textrm{joules/kelvin}\nonumber\\
\frac{1}{4\pi\epsilon_\circ}=10^{-7}\p 2.99792458~10^{8}\q^2~\textrm{farads/meter}
\end{align}
The latter value is \emph{exact}, i.e., not subjected to revision as a consequence of later measurements, and involves a finite number of digits. The constants $e$, $m$, $\hbar$, $4\pi\epsilon_\circ$, $\kB$ are the main ones that enter into the present theory. In the circuit theory we set the free-space permeability $\mu_\circ=4\pi 10^{-7}$. We sometimes employ as energy unit the electron-volt $\approx1.60~10^{-19}$ joules.

\section{Notation.} 

We list here only some of the notation employed in this paper. Different functions are distinguished by explicitly writing out their arguments. For example the Fourier transform of a function $\psi(x)$ is denoted by $\psi(k)$, i.e., with the same symbol, even though these are different functions. When the arguments are similar the same function, however, is intended. For example $w(t)$ and $w(\tau)$ represent the same function. One should not confuse a constant $U$ (no argument) with a function $U(x)$, for example. As usual, $\cos^2(x)$ means $\p\cos(x)\q^2$, and likewise for other trigonometric functions. $\psi^\star(x)$ is equivalent to $\left(\psi(x)\right)^\star$.

For a two-state electron the lower and upper levels ("working levels") are denoted 1 and 2, or "a" (absorbing) and "e" ("emitting") levels, respectively. Usually, spontaneous decay from level 2 to level 1 is neglected. The decay time, if not infinite, is denoted by $\tau_s$. For 4-level electrons, levels of increasing energy are denoted 0,1,2,3. Pumping occurs from 0 to 3. The decay time of a resonator with loss is denoted by $\tau_p$.

The probability that $a$ be larger than $b$ is denoted by $p(a>b)$. We employ double-side spectral densities, so that the usual shot-noise formula $2e\abs{J}$ is written as $e\abs{J}$, i.e., without a factor of 2 ($-e$ denotes the electron charge and $J$ the average current). We may perform two kinds of averaging. Quantum Mechanical averaging relate to an ensemble of similarly prepared systems (that is, macroscopic differences are supposed to be too small to affect the averages. Quantum and Statistical averages may be functions of time, and they may possibly be averaged further over time.

We recall here a notation commonly employed in Electrical Engineering. Usually, an amplifier is loaded with a nominal conductance such as 20 milli-siemens (resistance of 50 $\Om$) and the input impedance is equal to that of the load. If the input power of an amplifier is $P_{in}$ and the output power is $P_{out}$, the amplifier gain in decibel (abbreviation "dB") is defined as $10\log_{10}(P_{out}/P_{in})$. If the amplifier is linear the gain does not depend on the input power. In terms of the potentials $V_{in},V_{out}$ at the input and output ports, the gain reads $20\log_{10}(\abs{V_{out}/V_{in}})$ dB, because powers are proportional to the modulus-squares of the potentials in the situation considered. Likewise, an attenuation is defined as $10\log_{10}(P_{in}/P_{out})=-10\log_{10}(P_{out}/P_{in})$. A gain of 3dB means that the input power is multiplied by a factor close to 2. Note that dBm means decibels above a power of 1 mW. For example, 30dBm represents approximately a power of 1 watt.

In schematics, current sources are represented by circles with an arrow in them, while potential sources are represented by a circle with + and - signs, to define what is meant by positive current or positive potential, as shown later in Fig.~\ref{circuitfig}. By potential (or current) \emph{sources} we mean potentials (or currents) that do not depend on the current (potential) delivered. These are sometimes referred to as: "prescribed classical sources".

We will be mainly concerned of rates (number of events per unit time) of the form $Q+\De Q(t)$, where $Q$ denotes the average rate and $\De Q(t)$ a small fluctuation. We call $\spectral_{\De Q}$ the double-sided spectral density of $\De Q(t)$. This is in general a function of the (angular) Fourier frequency $\Om$. The relative noise $\N$ is defined by
\begin{align}\label{defN}
Q\N(\Om)\equiv  \frac{\spectral_{\De Q}(\Om)}{Q}-1
\end{align}
The relative noise $\N$ is expressed in seconds. It is unaffected by cold linear reflexion-less attenuations. $\N=0$ for Poisson processes, and is negative for sub-Poisson processes. When the laser output is split into two beams directed to two detectors with detection rates $\mathcal{D}_1(t)$ and $\mathcal{D}_2(t)$ respectively, the laser output rate is \emph{defined} as $\mathcal{Q}(t)\equiv \mathcal{D}_1(t)+\mathcal{D}_1(t)$.

\paragraph{Main parameters}

The injected electron rate $J$ is supposed to be a known constant. Because losses (other than those due to photo-detectors) and spontaneous emission are neglected, the total average output photo-electron rate $Q$ is equal to $J$. The expression of  the relative noise $\N$ as a function of the Fourier frequency $\Om$ depends on a number of parameters. We assume that the detector is represented by a positive conductance $G(\om)$. The active conductance is in general an admittance $Y(T,\om,n,R)\equiv G(T,\om,n,R)+\ii B(T,\om,n,R)  $, where $T$ denotes the absolute temperature, $\om$ the optical frequency, $n$ the number of electrons in the conduction band for a semi-conductor, and $R$ the output light-quanta rate. The conductance (real part of $Y$) is often denoted by $-G_e, G_e>0$. Below, we omit partial derivation signs when they are unnecessary.

\begin{itemize} 

\item $\beta\equiv 1/T$ is the absolute temperature reciprocal. Usually, $T=300$K. We set $k_B$=1.

\item $\tau_p$ denotes the resonator field lifetime, that is the average time that a light pulse would spent in the resonator assuming that the gain medium has been suppressed. More precisely, for an $L-C$ resonating circuit and a conductance $G$, we have $\tau_p=C/G$. In laser diode $\tau_p$ is on the order of one ns.

\item If $G(n)$ denotes the gain of a semiconductor as a function of the number $n$ of electrons in the conduction band, we define the differential gain as $g\equiv (n/G)(dG/dn)$. Usually for semiconductors $g\approx 2$.

\item If $Y(n)\equiv G(n)+\ii B(n)$ denotes the admittance of a piece of semiconductor, we define $\alpha\equiv (dB/dn)/(dG/dn)$. $\alpha$ vanishes for a symmetrical $G(\om)$ curve at peak gain, and may in general be on the order of 2.

\item We define the dispersion factor $h\equiv (dG/d\om)/(dB/d\om)$ and $K\equiv 1+h^2$. $K$ is often called the linewidth enhancement factor.

\item If $G(n,R)$ depends explicitly of the emitted rate, we define the gain compression factor $\kappa\equiv-(R/G)(dG/dR)$. The effect is often being referred to as the "non-linear gain" effect.

\end{itemize} 

\newpage

\chapter{Mathematics}\label{math}

Ideally, the present section should derive all the mathematical results subsequently employed from axioms. This goal is not accomplished for lack of space, time (and ability). Often, physical intuition or numerical calculations help. For example, the celebrated Wiener-Khintchin theorem that relates spectrum and correlation was first obtained by Einstein in an intuitive manner. We recall in the present chapter elementary mathematical formulas and less known results. It may be helpful to indicate which use will be made of the results given in the following sections. The descriptions below are only indicative.
\begin{enumerate}
\item Bi-complex numbers are convenient to treat sinusoidal modulations of optical frequencies.
\item The solution of third-order equations (that few people bother remembering) is given.
\item Little-known integrals are needed to obtain the linewidth of inhomogeneously broadened lasers.
\item Matrices are mostly employed to treat the response of linear circuits, either in the electrical potential-electrical current form, or in the scattering-matrix form. They are also needed to treat the so-called "density matrices" that characterize quantized electrons. The latter matrices are Hermitian and of trace $≤1$. Alternatively, one may employ generalized Rabi equations, which are first-order differential equations. 
\item Fourier transform enables us to express fluctuations as functions of the Fourier 
frequency $\Omega$.
\item The Laplace transform relates correlations and waiting times. The concept is that, given that an event occurred at $t$=0, the probability that an event occurs at time $t$ is the sum of the probabilities that this happens through one jump, two jumps, ...and so on.
\item Partitions. 1+3 is a partition of the number 4. We are interested in the number $p(n)$ of partitions of some integer $n$. This quantity is most useful to treat the Statistical Mechanics of electrons in the micro-canonical ensemble in the case of evenly-spaced levels. $n$ (denoted $r$) then represents the energy added to the system, and $p$ (denoted $W$) is the number of distinct configurations.
\item Random variables. The outcome of trowing a dice is a random variable of probability 1/6. Moments and cumulants are defined. For independent randdom variables, the cumulant of a sum is the sum of the cumulants of the individual random variables.
\item Stationary (ergodic) stochastic processes are also referred to as "noise". To each outcome of trowing a dice we associate a function of time.
\item Stationary (ordered) point processes. These are points (or events) on the time axis. If they occur independently, the process is called Poisson. If successive points are more evenly spaced than that, the process is "sub-Poisson".
\item Dirac pulses. The photo-detection current is a sum of Dirac distributions $\delta(t)$ occurring at the points just defined, to within a constant (the electron charge). We relate the photo-current spectrum $\spectral(\Omega)$ to the normalized correlation $g(t)$.
\item Photo-count variance. We relate the number of points occurring during some time interval $\mathcal{T}$ to the quantities defined above.
\item We offer a picturesque "dark room model". If people enter into a room regularly, their exit rate is  regular as well at small Fourier frequencies, because not many people may stay in the room.
\item A result from signal theory is recalled, namely the expression of the spectral density of some phase-modulated oscillation.
\end{enumerate}

\section{Complex numbers}\label{complex}

A complex number is denoted either as $z=\Re\{z\}+\ii \Im\{z\}$ or as $z=a+\ii b$, and $z^\star=a-\ii b$ denotes the complex conjugate of $z$. We denote $\abs{z}\equiv \sqrt{z z^\star}=\sqrt{a^2+b^2}$ the modulus of $z$. 

A complex notation is often employed for describing quantities that vary sinusoidally in time thatconsiderably simplifies calculations for real, causal, linear and time-invariant circuits. According to that notation, the function $i(t)=\sqrt2\abs{I}\cos(\om t+\phi)$, where the frequency $\om$ and the phase $\phi$ are real constants, is written as $i(t)=\sqrt2 \Re\{I \exp(-\ii\om t)\}$, where the complex number $I$ is defined as $I=\abs{I}\exp(-\ii \phi)$, and $\omega\equiv 2\pi \nu $ denotes the \emph{carrier} (angular) frequency. Similar definitions apply to potentials $v(t)$ varying sinusoidally in time, that is $v(t)=\sqrt2 \Re\{V \exp(-\ii\om t)\}$. We have chosen to introduce the factor $\sqrt2$ so that the average optical power, defined as the time average of the current-potential product $v(t)i(t)$ be simply equal to the real part of the product $V I^\star$, i.e., without the factor 1/2 that would otherwise occur. $V$ and $I$ are called rms (root-mean-square) complex potentials and currents, respectively, or more briefly, optical potentials and currents. The minus sign in $ \exp(-\ii\om t) $ is employed in optics because waves propagating forward in space then involve a term of the form  $ \exp(\ii kx) $, where $k$ denotes the wave-number, that is, with a plus sign. For slow variations the function $j(t)=\sqrt2\abs{J}\cos(\Om t+\phi)$ is denoted as $j(t)=\sqrt2 \Re\{J \exp(\jj\Om t)\}$, where the complex number $J$ is defined as $J=\abs{J}\exp(\jj \phi)$, and $\Omega\equiv 2\pi f$ is called the (angular) "Fourier" frequency. Even though the squares of $\ii$ and $\jj$ are both equal to -1 these two numbers should be distinguished: $\ii\jj=\jj\ii$ should not be set equal to -1.

When a source at frequency $\omega$ is modulated at frequency 
$\Omega$, the bi-complex representation described below proves useful. 
To avoid bothering with minus signs, it is convenient to set 
$i_{1}\equiv$-\ii~ and $i_{2}\equiv$\jj. Further, we set 
 $p_{1}\equiv i_{1} \omega$ and $p_{2}\equiv i_{2} 
\Omega $. The algebra of bi-complex numbers discovered by Segre in 1892, is associative and commutative. A
bi-complex number is written as
   \begin{align}\label{bi}
\mathcal {V} = a+b i_{1}+ci_{2}+d~i_{1}i_{2},
  \end{align} 
where $i_{1}^2=i_{2}^2=-1$, $i_{1}i_{2}=i_{2}i_{1}$, and $a,b,c,d$ are real numbers. Bi-complex numbers are not invertible when $a^2+b^2+c^2+d^2 = ± 2\p ad-bc\q$. To prove that, it suffices to expand the product of a bi-complex number such as $a+b~ i_{1}+c~ i_{2}+d~i_{1}i_{2}$ and a bi-complex number such as $\alpha+\beta~ i_{1}+\gamma~i_{2}+\delta~i_{1}i_{2}$, identify terms of the same nature and solve for $\alpha, \beta,\gamma,\delta$. A solution exists unless the determinant vanishes. It is useful to notice first that $(a+b~ i_{1}+c~i_{2}+d~i_{1}i_{2})(a-b~ i_{1}-c~i_{2}+d~i_{1}i_{2})=A+B~i_{1}i_{2}$, where $A=a^2+b^2+c^2+d^2$ and $B=2\p ad-bc\q$ are real numbers. Next, note that $\p A+B~i_{1}i_{2}\q \p A-B~i_{1}i_{2}\q=A^2-B^2=0$ if $A=±B$. 

 Let the real signal $v(t)$ be written as $\mathcal {V}\exp \left(\p p_{1}+ p_{2}\q t\right)$ +ccc, where "ccc" 
 means that one must add 3 terms to the one written out, one with $p_{1}$ changed to 
 $-p_{1}$, the second with $p_{2}$ changed to $-p_{2}$, and the third 
 with both $p_{1}$ and $p_{2}$ changed to $-p_{1}$ and $-p_{2}$. If $Y(p)$ denotes the usual complex circuit admittance (i.e., the 
ratio of two real polynomials in $p$),  the real electrical current flowing through the circuit is
 \begin{align}\label{theorem}
i(t)=Y(p_{1}+ p_{2})\mathcal {V}(p_{1}, p_{2}) \textrm{exp}[(p_{1}+ 
p_{2})t]+\textrm{ccc}. 
  \end{align} 
Indeed, a sinusoidally modulated sinusoidal signal is the sum of four frequencies. The response of a circuit to each of these frequencies being known, and the system being linear, the response is obtained by adding the responses to each of these four frequencies.

\section{Second and third-degree equation}\label{elem}

We give below the solution of second and third-degree equation in the form appropriate to our intended application. 

Consider the polynomial $ap^2+bp+c=0$. The solutions are
\begin{align}
\label{fou5}
p_{\pm}=\frac{-b\pm\sqrt{b^2-4ac}}{2a}.
\end{align}
For the third-degree polynomial $p^3+a_2 p^2+a_1p+a_0=0$ we evaluate sequentially 
\begin{align}
\label{fou}
q=\frac{a_1}{3}-\frac{a_2^2}{9}\qquad r=\frac{a_1a_2}{6}-\frac{a_0}{2}-\frac{a_2^3}{27}\qquad s=\sqrt{q^3+r^2}\nonumber\\
s_1=(s+r)^{1/3}\exp(\ii \pi/3)\qquad s_2=(s-r)^{1/3}\exp(\ii 2 \pi/3)\qquad s±r>0.
\end{align}
The three roots are
\begin{align}
\label{three}
p_1&=s_1+s_2-\frac{a_2}{3}\nonumber\\ 
p_2&=s_1\exp(\ii 2 \pi/3)-s_2\exp(\ii \pi/3)-\frac{a_2}{3}\nonumber\\
p_3&=-s_1\exp(\ii \pi/3)+s_2\exp(\ii2 \pi/3)-\frac{a_2}{3}.
\end{align}

\section{Useful integrals}\label{integrals}

A number of integrals from $x=-\infty$ to $x=\infty$ will be needed 
Part II. They may be obtained by contour 
integration. The method is as follows.

Recall that complex numbers are denoted by $z\equiv z'+\ii z''$, where 
$\ii^2=-1$. The complex conjugate of $z$ is denoted $z^\star\equiv 
z'-\ii z''$. Let $f(z)$ be a function of $z$ whose only 
singularities are simple poles at $z_{1},~z_{2}\ldots$. One calls 
\emph{residue} at $z_{k}$ the coefficient of $(z-z_{k})^{-1}$ in the 
(Laurent) series expansion of $f(z)$ near $z_{k}$. The integral of 
$f(z)$ along a closed counterclockwise contour is equal to $2\pi $\ii  ~
times the sum of the enclosed pole residues.

For example, closing the real axis by an upper half-circle of infinite 
radius we obtain%\textrm{d}
\begin{align}\label{residue}
\int _{-\infty}^{\infty}\frac{\textrm{d}x}{1+x^2}=\int 
_{-\infty}^{\infty}\frac{\textrm{d}x}{(x-\ii)(x+\ii)}=2\pi 
\ii~ \frac{1}{\ii+\ii}=\pi.
  \end{align}
   Here we have a single enclosed pole at $x=\ii$. The 
  coefficient of $1/(x-\ii)$ in the integrand is 
  $1/(2\ii)$ when $x=\ii$. We obtain similarly
 \begin{align}\label{first int}
\frac{1}{\pi }\int _{-\infty}^{\infty}\frac{\textrm{d}x}{(1-ax^2)^2 +x^2}=
\frac{1}{\pi }\int _{-\infty}^{\infty}\frac{\textrm{d}x~ ax^2}{(1-ax^2)^2 +x^2}=1,
  \end{align}
where $a$ denotes a non-zero constant.

Further, for application to inhomogeneously broaden lasers, let us define a weight function
 \begin{align}\label{weight}
w(x)\equiv \frac{(g-1)/\pi }{(g-1)^2 +x^2}
  \end{align}
  that reduces to the Dirac $\delta$-distribution when $g$ tends to 
  1, and 
   \begin{align}\label{Imn}
I_{mn}\equiv 8(g^2+y^2)^m \int 
_{-\infty}^{\infty}\frac{\textrm{d}x~ w(x-y)x^n}{(1+x^2)^m}. 
  \end{align}
  
  We obtain
 \begin{align}\label{Imn bis}
I_{10}&=8g\\
I_{12}&=8y^2+8g(g-1)\\
I_{20}&=4y^2(g-1)+4g^2(g+1)\\
I_{21}&=8gy\\
I_{30}&=3(g-1)y^4+6g(g^2-1)y^2+g^3(3g^2+3g+2)\\
I_{31}&=2y[y^2(g-1)+g^2(g+3)]\\
I_{32}&=(g-1)y^4+2g(g^2+3)y^2+g^3(g-1)(g+2)\\
I_{33}&=2y[(3g+1)y^2+3g^2(g-1)]\\
I_{34}&=(3g+5)y^4+6g(g^2-1)y^2+g^3(3g-2)(g-1). 
\end{align}
We also need for semiconductors
   \begin{align}\label{Imnb}
\frac{1}{\pi} \int 
_{0}^{\infty}\frac{\textrm{d}x~ \sqrt 
{x}}{(x+1)(x-a)}=\frac{1}{1+\sqrt{-a}} 
  \end{align}
if $a$ is negative, and $1/(1+a)$ if $a$ is positive. In the latter 
case, the integral is understood in principal value.

\section{Vectors and matrices}\label{vector}

A vector (bold-face letter) is a collection of complex numbers. For example, in dimension 2 
\begin{align}\label{vec}
\boldsymbol{a}=
\left( 
\begin{array}{c}
a_{1}\\
a_{2}
\end{array}
\right).
\end{align}
Transposition, indicated by a "t" in upperscript, interchanges lines and columns, 
\begin{align}\label{v1}
\boldsymbol{a}^t=
\left( 
\begin{array}{cc}
a_{1}& a_{2}
\end{array}
\right).
\end{align}
The scalar product of two vectors $\boldsymbol{a}$ and $\boldsymbol{b}$ is defined as
\begin{align}\label{v2}
\boldsymbol{a}.\boldsymbol{b}=a_1~b_1+a_2~b_2
\end{align}
The modulus square of the length of a complex vector is 
\begin{align}\label{v3}
\abs{\boldsymbol{a}}^2\equiv\boldsymbol{a}.\boldsymbol{a}^\star =a_1~a_1^\star+a_2~a_2^\star≥0.
\end{align}
A matrix is denoted for example
\begin{align}\label{vec4}
\boldsymbol{M}=
\left( 
\begin{array}{ccc}
M_{11} & M_{12}\\
M_{21} & M_{22}.
\end{array}
\right).
\end{align}
The trace of a square matrix is the sum of the diagonal elements
\begin{align}\label{vec4a}
\mathrm{tr}\,\boldsymbol{M}=M_{11} + M_{22},
\end{align}
and the determinant
\begin{align}\label{vec5}
\det\boldsymbol{M}=M_{11} M_{22}-M_{12} M_{21}.
\end{align}
The sum of two matrices is obtained by summing their elements. The product $\boldsymbol{L}$ of two matrices $\boldsymbol{M}$ and $\boldsymbol{N}$ is given by 
\begin{align}\label{vec6}
L_{ij}= M_{i1} N_{1j}+M_{i2} N_{2j}.
\end{align}
We have$\left(\boldsymbol{M}\boldsymbol{L}    \right)^t= \boldsymbol{L}^t\boldsymbol{M}^t$. When $\det\boldsymbol{M}=0$ the matrix is singular, and cannot be inverted. Otherwise, $\boldsymbol{M}^{-1}$ denotes the matrix such that $\boldsymbol{M}^{-1} \boldsymbol{M}=\boldsymbol{1}\equiv \left(\begin{array}{ccc}1&0\\0&1\end{array}\right)$.   A matrix is said to be symmetrical when $\boldsymbol{M}^t=\boldsymbol{M}$, Hermitian when $\boldsymbol{M}^{t\star}=\boldsymbol{M}$, unitary when $\boldsymbol{M}\boldsymbol{M}^{t\star}=\boldsymbol{1}$. We have $\mathrm{tr}\,\boldsymbol{A}\boldsymbol{B} =\mathrm{tr}\,\boldsymbol{B}\boldsymbol{A} $. The trace of the product of two Hermitian matrices is real.

\paragraph{Cauchy-Schwartz inequality.}

Consider two unit vectors $\boldsymbol{a},\boldsymbol{b}$, that is, such that $\abs{\boldsymbol{a}}=\abs{\boldsymbol{b}}=1$, The Cauchy-Schwartz inequality reads 
\begin{align}\label{vec7}
\abs{\boldsymbol{a}.\boldsymbol{b}^\star}^2\equiv (\boldsymbol{a}.\boldsymbol{b}^\star)(\boldsymbol{b}.\boldsymbol{a}^\star)≤1.
\end{align}
This relation is obtained by replacing $\boldsymbol{a}$ in \eqref{v3} by $\boldsymbol{a}-(\boldsymbol{a}.\boldsymbol{b}^\star)\boldsymbol{b}$. 

\paragraph{Density matrices.}

Let $\boldsymbol{\rho}$ denote an Hermitian matrix of trace 1. 

The pure-state density matrix is constructed from the vector $\boldsymbol{a}$ with $\abs{\boldsymbol{a}} ^2=1$ as
\begin{align}\label{dens}
\boldsymbol{\rho}=
\left( \begin{array}{ccc}
a_{1}\\
a_{2}
\end{array}\right)
\left(   a_1^\star~~a_2^\star  \right) 
= \left( \begin{array}{ccc}
a_{1}a_{1}^\star&a_{1}a_{2}^\star\\
a_{2}a_{1}^\star&a_{2}a_{2}^\star
\end{array}   \right)     .
\end{align}
We readily find that $\boldsymbol{\rho}^2=\boldsymbol{\rho}$, and thus $\mathrm{tr}\,\boldsymbol{\rho}^2 =\mathrm{tr}\,\boldsymbol{\rho} =1$. Note that $\boldsymbol{\rho}$ is unaffected by a change of the phase of $\boldsymbol{a}$. As examples, we may have
\begin{align}\label{ds1}
\boldsymbol{\rho}=\left(\begin{array}{ccc}   1&0\\0&0     \end{array}\right),\qquad \boldsymbol{\rho}=\frac{1}{2}\left(\begin{array}{ccc}   1&\ii\\-\ii&1     \end{array}\right).
\end{align}

 Let $\boldsymbol{\rho}_a,\boldsymbol{\rho}_b$ be two such matrices. By explicit calculation we find that $\mathrm{tr}\, \boldsymbol{\rho}_a\boldsymbol{\rho}_b =(\boldsymbol{a}.\boldsymbol{b}^\star)(\boldsymbol{b}.\boldsymbol{a}^\star)$. Thus, from \eqref{vec7} we have $\mathrm{tr}\, \boldsymbol{\rho}_a\boldsymbol{\rho}_b\leq 1$.

The mixed-state density matrix is defined as 
\begin{align}\label{dens1}
\boldsymbol{\rho}\equiv  \sum_{k}p_k\boldsymbol{\rho}_k\qquad p_k\geq 0\qquad \sum_k p_k=1,
\end{align}
where the $p_k$ may be called weights. Since $\mathrm{tr}\, \boldsymbol{\rho}_k\boldsymbol{\rho}_l\leq 1$ we obtain
\begin{align}\label{dens2}
\mathrm{tr}\, \boldsymbol{\rho}^2  =\mathrm{tr}\,\{ \sum_{k}p_k\sum_{l}p_l\boldsymbol{\rho}_k\boldsymbol{\rho}_l  \}\leq\sum_{k}p_k=1.
\end{align}
If $\mathrm{tr}\, \boldsymbol{\rho}^2  =1$ we have $(\boldsymbol{a}.\boldsymbol{b}^\star)(\boldsymbol{b}.\boldsymbol{a}^\star)=1$. The density matrix $\boldsymbol{\rho}$ is then of the pure-state form in \eqref{dens}.

Let us consider a pure-state density matrix $\boldsymbol{\rho}_k$, and suppose that the QM (Quantum Mechanical) average of some quantity, such as the power $P_k$, may be obtained from the formula
\begin{align}\label{den10}
\ave{P_k}=\mathrm{tr}\,\boldsymbol{\rho}_k \boldsymbol{P}
\end{align}
where $\boldsymbol{P}$ denotes some known Hermitian $2\times 2$ matrix. Next, suppose that the pure-state density matrix $\boldsymbol{\rho}_k$ occurs with probability $p_k$, $k=1,2...$. The statistical and QM-average of the power, denoted by a double bracket is, using the properties of the trace
\begin{align}\label{den0}
\ave{\ave{P}}_{statistical}=\sum_k p_k ~\mathrm{tr}\,\boldsymbol{\rho}_k \boldsymbol{P}=\mathrm{tr}\, \boldsymbol{\rho} \boldsymbol{P} \qquad \boldsymbol{\rho}\equiv \sum_k p_k \boldsymbol{\rho}_k.
\end{align}
From now on, the double bracket is replaced by a simple bracket.

Setting $x\equiv 2\rho_{12}',  y\equiv 2\rho_{12}'', z \equiv \rho_{22}-\rho_{11} $ (unrelated to coordinates in space), the density matrix may be written as
\begin{align}\label{dns}
     \boldsymbol{\rho}=\frac{1}{2}
  \left( 
  \begin{array}{ccc}
      1-z    &   x+\ii y     \\   
      x-\ii y&   1+z             
\end{array}  
 \right)     .
\end{align}
It follows that 
\begin{align}\label{ens1}
\mathrm{tr}\,\boldsymbol{\rho}^2 =\frac{1}{2}(1+x^2+y^2+z^2),
\end{align}
and $\mathrm{tr}\,\boldsymbol{\rho}^2\leq 1\Longleftrightarrow x^2+y^2+z^2≤1$.

\section{Fourier transforms}\label{fouriertitle}

The Fourier transform $\psi(k)$ of the  function $\psi(x)$ and the reciprocal relation are
\begin{align}
\label{fourier}
\psi(k)&=\int_{-\infty}^{\infty}{dx\exp(-\ii kx)\psi(x)}\\
\label{fourier'}
\psi(x)&=\int _{-\infty}^{\infty}{\frac{dk}{2\pi}\exp(\ii kx)\psi(k)},
\end{align}
where $\ii^2=-1$, and $k$ is called the wave-number. Note that the position of the $2\pi$ factor varies from one author and another, without of course affecting the end results\footnote{Relations similar to \eqref{fourier} and \eqref{fourier'} hold with $x$ changed to $t$, $k$ to $\om$ and (to be consistent with our conventions for optical signals) $\ii$ changed to $-\ii$. Then the element of integration in \eqref{fourier'} is $d\nu\equiv d\om/2\pi$, where $\nu$ denotes as usual the optical frequency
\begin{align}
\label{fourier1}
\psi(\om)&=\int_{-\infty}^{\infty}{dt\exp(\ii \om t)\psi(t)}\\
\label{fourier1'}
\psi(t)&=\int _{-\infty}^{\infty}{\frac{d\om}{2\pi}\exp(-\ii \om t)\psi(\om)}.
\end{align} 
In the Fourier-frequency domain, changing $\om\to\Om,\ii\to-\jj$, we write
\begin{align}
\label{fourier2}
\psi(\Om)&=\int_{-\infty}^{\infty}{dt\exp(-\jj\Om t)\psi(t)}\\
\label{fourier2'}
\psi(t)&=\int _{-\infty}^{\infty}{\frac{d\Om}{2\pi}\exp(\jj \Om t)\psi(\Om)},
\end{align}}. 
Obviously, the Fourier transform of $\p\ii k\q^n \psi(k)$ is equal to the $n$th derivative of $\psi(x)$ with respect to $x$. If $\psi(x)$ is real, we have $\psi^\star(k)=\psi(-k)$.

Note the following physical application. For particles moving in time-indepen\-dent potentials $V(x)$, stationary states $\psi(x)$ are real functions of $x$ (to within an arbitrary over-all phase factor that we set equal to 1). If furthermore $V(x)$ is an even function of $x$, $\psi(x)$ is either an even or odd function of $x$. It follows from the above considerations that the $\psi(k)$-functions are, respectively, real even or imaginary odd. In the present mathematical section we set $\hbar=1$ and do not distinguish the electron momentum $p$ from the wave-number $k$.

We will need the following expression of the Dirac $\de$-distribution\footnote{The $\de(t)$-distribution may be viewed alternatively as a function equal to $1/h$ for $-h/2<t<h/2$ and 0 otherwise, so that the area under the function is unity, letting $h$ go to zero at the end of the calculations. Many other forms of the $\de$-function may be used, with less-singular derivatives than for the one just given.}
\begin{align}
\label{delt}
\de(x)=\int_{-\infty}^{\infty}\frac{dk}{2\pi}\exp(\ii k x),
\end{align}
implying that its Fourier transform is unity.
Using this expression one may prove that 
\begin{align}
\label{planche}
\int_{-\infty}^{+\infty}\frac{dk }{2\pi}\abs{\psi(k)}^2=\int_{-\infty}^{+\infty}dx \abs{\psi(x)}^2.
\end{align}

\section{Convolution and Laplace transforms }\label{convolution}

Let us consider a real, causal, linear and time-invariant system. These conditions imply that for a potential source $v(t)$ the current $i(t)$ (or more generally the response to a source of any kind) is given by
\begin{align}
\label{conv}
i(t)=\int_{-\infty}^{+\infty}du~ h(u) v(t-u)\equiv h*v=h*y,
\end{align}
where the kernel $h(u)$ is real, equal to 0 for $u<0$, and middle stars denote convolution products. Convolutions are associative, so that parentheses in convolution products are unnecessary, and commutative. For example, for a conductance $G$, we have $i(t)=Gv(t)$, and thus $h(t)=G\de(t)$, where $\de(.)$ denotes the Dirac distribution. If $v(t)=V(p)\exp(pt)$ where $p$ denotes a complex number (not to be confused with particle momenta), $i(t)=I(p)\exp(pt)$, where $I(p)=H(p)V(p)$ and
\begin{align}
\label{convbis}
H(p)=\int_{0}^{+\infty}dt~\exp(-pt)h(t),
\end{align}
a Laplace transform, defines $H(p)$ for complex $p$. In most of this work we set $p=-\ii\om$, and $H(-\ii\om)$ is denoted $Y(\om)$ and called the admittance.  
 
In particular the Laplace transform of $\exp(\lambda t)$ is $1/(p-\lambda)$. It follows that if the reciprocal of a polynomial in $p$ may be written as a sum of terms of the form $1/(p-p_k)$ the inverse Laplace transform is easily obtained. More generally, the Heaviside theorem says that if $f(p)$ is a polynomial with distinct roots (not to be confused with probabilities) $p_k$, $k=1,2...n$ ($f(p_k)=0$), the inverse Laplace transform of $1/f(p)$ is
\begin{align}
\label{conv4}
L^{-1}\{\frac{1}{f(p)}\}=\sum_{k=1}^n \frac{\exp(p_k t)}{\bigl(  df(p)/dp  \bigr)_{p=p_k}}.
\end{align}

The Laplace transform of the convolution of any number of functions is the product of their Laplace transforms. For example, the Laplace transform of $G_k(t)\equiv w(t)*w(t)...*w(t)$ ($k$-times) is the $k$th power  $w(p)^k$ of $w(p)$, where $w(p)$ denotes the Laplace transform of $w(t)$. Thus, the Laplace transform of $G(t)\equiv w(t)+w(t)*w(t)+....$ is the sum of an infinite geometric series; 
\begin{align}
\label{convter}
G(p)=\frac{w(p)}{1-w(p)}.
\end{align}
Let us define an average waiting time
\begin{align}
\label{tmean}
\ave{t}\equiv\int_0^\infty dt~ t~w(t)=-\bigl( \frac{dw(p)}{dp}     \bigr)_{p=0}.
\end{align}
If \eqref{convter} holds, $G(t\to\infty)$ is finite and the other terms are decaying exponentials, so that $G(p\to 0)\approx G(t= \infty)/p$, we obtain
\begin{align}
\label{conv9}
\frac{1}{\ave{t}}=G(t=\infty).
\end{align}

\section{Partitions}\label{part}

A partition of $r$ is a non-increasing sequence of positive integers summing up to $r$ (not to be confused with sets of positive integers summing up to $r$, which are called compositions of $r$). The number of partitions of $r$ is denoted $p(r)$, with for example $p(6)=11$. Its generating series is 
\begin{align}\label{partition}
\sum_{r\geq 0} p(r) x^r=\prod_{n\geq1}(1-x^n)^{-1}.
\end{align}
An expression valid in the large $r$-limit is 
\begin{align}
\label{asymptotic}
p(r)\approx \frac{\exp(\pi\sqrt{2r/3})}{4r\sqrt 3}.
\end{align}

Further, the number $p(a,b,r)$ of partitions of $r$ into at most $a$ parts, none of which exceeds $b$ has for generating series the gaussian polynomials 
\begin{align}
\sum_{r\geq 0} p(a,b,r) x^r=\genfrac{[}{]}{0pt}{}{a+b}{a}\equiv\prod_{n=1}^{b}\frac{1-x^{a+n}}{1-x^n}.
\end{align}
For later use note that
\begin{align}
\label{partB}
h(x)&\equiv\ln\big(\sum_{r\geq 0} p(N,B-N,r)x^r\big)=\sum_{n=1}^{B-N}\ln(\frac{1-x^{N+n}}{1-x^n})\\\nonumber
&=\sum_{n=1}^{B-N}\big[\ln(1+\frac{N}{n})+\frac{N}{2}(x-1)+\frac{N(N+2n-6)}{24}(x-1)^2\big]+O(x-1)^3.
\end{align}
Accordingly, 
\begin{align}\label{hfun}
h'(1)&=\frac{1}{2}N(B-N),\nonumber\\
h''(1)+h'(1)&=\frac{1}{12}N(B-N)(B+1),
\end{align}
where primes (double primes) denote the first (second) derivatives with respect to the argument.

\section{Random variables}\label{random}

We first define probabilities and random variables. Next, we define stochastic processes, also called "random functions", or "noise" in Electrical Engineering. We shall mainly consider stationary processes, that is, processes unaffected by time-shifts. Ergodicity, asserting equivalence between stochastic and time averages, is assumed. We give the (Wiener-Khintchin) relation between correlations and spectral densities. Next we consider ordered point-processes, and establish relations applicable to stochastic processes of the form $\sum_k \delta(t-t_k), ~k=-1,0,1...$, where $\delta(.)$ denotes the $\delta$ distribution and the $t_k$ form a point process.

The probability $p(A)$ associated with an event $A$ is a real number comprised between 0 (impossible event) and 1 (sure event) with the following property: If event $A$ precludes event $B$ and conversely, the basic axiom is that the probability that either $A$ or $B$ occurs is  $p(A)+p(B)$. The product $AB$ of two sets of events $A$ and $B$ is the set of the events that are common to $A$ and $B$ (also called intersection of $A$ and $B$). Two events are called independent if $p(AB)=p(A)p(B)$.

To every outcome $\zeta$ of an experiment (such as throwing a dice) we associate a number $X(\zeta)$, called a random variable. $\{X≤x\}$ denotes the set of all outcomes $\zeta$ such that $X(\zeta)≤x$. The distribution function $F(x)$ is the probability $p\{X≤x\}$ that $X(\zeta)$ be less than or equal to $x$. The probability density is then defined as $P(x)=dF(x)/dx$. Since, in Physics, $x$ has usually a dimension (e.g., time) the dimension of $P(x)$ is the reciprocal of that of $x$.

If $X$ is a random variable with probability law $p(x)$ with $p(x)=0, x<0$ and the integral of $p(x)$ from $x=0$ to $\infty$ is unity, the $n$th moment $\ave{X^n}$ of $X$ is defined as the integral from 0 to $\infty$ of $p(x)x^n$. We now consider a discrete probability law $p(m), m=0,1...$ and $\sum_{m\geq 0} p(m)=1$. The $n^\mathrm{th}$ moments are defined as
\begin{align}\label{moment}
\ave{m^n}=\sum_{m\geq 0} p(m)m^n.
\end{align}
Setting
\begin{align}\label{momentbis}
h(z)=\log\sum_{m\geq 0} p(m)z^m.
\end{align}
it is straightforward to show that
\begin{align}\label{momentter}
\ave{m}&=h'(1),\\
\var(m)&=h''(1)+h'(1),
\end{align}
where a prime (double prime) denotes the first (second) derivative with respect to the argument.

Centered moments $\mu_r$ are defined by first subtracting from $X$ its average value $\ave{X}$. Cumulants $\kappa_r$ are convenient because the cumulant of the sum of two independent random variables is the sum of their respective cumulants. They are defined as 
\begin{align}\label{cumulant}
\sum_{r=1}^{\infty} \frac{\kappa_r t^r}{r!}=\log\ave{\exp(Xt)}.
\end{align}
We have: $\kappa_1=\ave{X}, \kappa_2=\mu_2=\var(X), \kappa_3=\mu_3, \kappa_4=\mu_4-3{\mu_2}^2$,\dots

\section{Stationary stochastic processes}

We are given an experiment such as throwing a dice, specified by its outcome $\zeta$. To every outcome we assign a time function $x(t; \zeta)$. This is called a stochastic process. Because the process is stationary the correlation 
\begin{align}\label{R}
R(\tau,t)\equiv \ave{x(t)x(t+\tau)}= \ave{x(0)x(\tau)}=R(-\tau),
\end{align}
where the sign $\ave{.}$ denotes a stochastic average.

The spectrum $\spectral_t(\Om)$ of a stationary process is a real non-negative even function of $\Om$. It may be obtained by considering a finite duration $\T$, evaluating of average of the modulus square of the Fourier transform of $x(t)$, dividing by $\T$, and letting $T$ go to infinity.  

Alternatively, the spectrum may be expressed as the Fourier transform of $R(\tau)$ (Wiener-Khintchine theorem)
\begin{align}\label{def}
\spectral_t(\Om)&=\lim_{\T\to \infty}\frac{1}{\T}\ave{\abs{\int_{0}^{\T}{dt~ x(t)\exp(\jj\Om t)}}^2}\nonumber\\
&=\int_{-\infty}^{+\infty}{d\tau~ R(\tau)\exp(\jj\Om \tau)}=\int_{-\infty}^{+\infty}{d\tau~ R(\tau)\cos( \Om \tau)}.
\end{align}
The two above expressions agree in the mean if the integral from 0 to $\infty$ of $\tau R(\tau)$ is finite, which we assume. For two independent processes $x(t)$ and $y(t)$ of spectral densities $\spectral_x$ and $\spectral_y$, respectively, the spectral density of  $z(t)=ax(t)+by(t)$ is $\spectral_z=\abs{a}^2\spectral_x+\abs{b}^2\spectral_y$.\footnote{We have employed above the electrical-engineering $ \exp(\jj\Om t)$ notation. To compare with the previous notation change $\Om$ to $k$, $\tau$ to $x$, and $\jj$ to $-\ii$.} 

Evaluating $\ave{\p x(\tau)±x(0)\q^2}$ we notice that $-R(0)≤R(\tau)≤R(0)$. In the special case where $x(t)$ does not depend on time $R(\tau)=\ave{x^2}$= constant. Substituting in \eqref{def} we find that $\spectral_t (\Om)=2\pi \ave{x}^2\de(\Om)$, where $\de(.)$ denotes the Dirac $\de$-distribution. We are thus led to define a reduced spectrum $\spectral(\Om)\equiv\spectral_t(\Om)-2\pi \ave{x}^2\de(\Om)$.

Conversely, the correlation may be expressed in terms of the spectrum through the inverse Fourier transform according to
\begin{align}\label{convterbis}
R(\tau)=\int_{-\infty}^{+\infty}\frac{d\Om}{2\pi}    \spectral(\Om)   \cos( \Om \tau)+\ave{x^2}.
\end{align}
If we define $y(t)\equiv x(t)-\ave{x(t)}$ we have $\ave{y(t)}=0$. The function $C(\tau)=\ave{y(0)y(\tau)}$ is called the (auto) covariance of the process $x(t)$.

The sources of noise in our theory are narrow-band current sources written as $c(t)=\sqrt 2 \p C'(t) \cos(\om_o t)+C''(t) \sin(\om_o t) \q$, where $\om_o$ denotes the average laser frequency, and $C'(t),~C''(t)$ are jointly-stationary slowly-varying real functions of time. It can be shown that $c(t)$ is wide-sense stationary if and only if $\ave{C'(t)}=\ave{C''(t)}=0$ and the auto and cross correlations fulfill the conditions $R_{C'C'}(\tau)=R_{C''C''}(\tau),R_{C'C''}(\tau)=-R_{C''C'}(\tau)$. Furthermore, we assume that the statistics is independent of a phase change, and this entails that $R_{C'C''}(\tau)=0$. Let us recall the following result. If $C'(t),~C''(t)$ are uncorrelated and their spectra $\spectral_{C'}(\om)=\spectral_{C''}(\om)$ vanish for $\abs{\om}>\om_c$, then $\spectral_{C}(\om)=\spectral_{C'}(\om-\om_o)+\spectral_{C'}(\om+\om_o)$.

\section{Stationary point processes}\label{stationary}

Point processes are sequences of increasing positive real numbers $t_{k}$, where $k=\cdots,-1,0,1,2\cdots$, see\,\cite{Cox1980}. Each $k$ value corresponds to a point (or event) occurring at time $t_k$. The average point rate is denoted by $D$. Point processes are defined by the probability $G(t)$ that, given that there is a point at $t=0$, there is a point between $t$ and $t+dt$, divided by $dt$. The normalized correlation $g(t)$ is obtained by dividing $G(t)$ by the average rate $D$. The most important point process is the Poisson process with $g(t)=1$. The "waiting time" density $w(t)$ is the probability that, given that there is a point at $t=0$, the \emph{next} point occurs between $t$ and $t+dt$, divided by $dt$. Let $d(\T)$ denote the random number of points occurring within a duration $\T$. We define $\V(\T)=var(d(\T))/D-1$. Consider as an example a Poisson process with $g(t)=1$. The probability density that the first point occurs at time $t>0$ is $\exp(-t)$, whether or not there is a point at $t=0$. If there is a point at $t=0$, $w(t)=\exp(-t)$ is the waiting-time density.

Given a point process, new point processes may be obtained through the following operations:\begin{itemize}
\item A change of time scale. In that way an inhomogeneous Poisson process may be converted into a Poisson process.
\item Thinning, in which some of the points in the original process are deleted with a constant probability, independently of all the other points. 
\item Translation of individual points.
\item Superposition, in which a number of separate processes are merged. Superposition of an arbitrarily large number of processes is a Poisson process.
\end{itemize}

\paragraph{Inhomogeneous Poisson process}

The point process considered in the present section is not stationary. It will help us to construct a stationary point process. Consider an inhomogeneous Poisson process of density $\lambda(t)$ with a point at $t=0$. This process may be reduced to a Poisson process through a transformation of the time scale $d\tau=\lambda(t)dt$. Thus the waiting-time density, that is, the probability that the next point occurs in the interval $(\tau,\tau+d\tau)$, divided by $d\tau$, reads, 
 \begin{align}
\label{ju}
w(\tau)=\lambda(\tau)\exp(-\int_0^\tau dt \lambda(t))=-\frac{d}{d\tau}\exp(-\int_0^\tau dt \lambda(t)).
\end{align}
It follows from the second form above that the integral of $w(\tau)$ from 0 to $\infty$ is unity, provided $\lambda(t)$ does not tend to 0 as $t\to\infty$. This means that the point eventually occurs. The average duration between adjacent points is, after an integration by parts
 \begin{align}
\label{juk}
\ave{\tau}=\int_0^\infty d\tau\exp(-\int_0^\tau dt \lambda(t)).
\end{align}
If, for example, $\lambda(t)=1$, we obtain $\ave{\tau}=1$ as expected. Conversely,
 \begin{align}
\label{ju5}
\lambda(t)=\frac{w(t)}{\int_t^\infty d\tau w(\tau)}=-\frac{d}{dt}\log \left( \int_t^\infty d\tau w(\tau)    \right).
\end{align}

\paragraph{Ordinary renewal process}

With $t_0=0, t_1=\tau_1, t_2=\tau_1+\tau_2,...$ and the $\tau_i, i=1,2...$ independent and distributed according to the same density $w(\tau)$, one generates an ordinary renewal process $t_k, k=1,2...$. Such a process is non-stationary, but it tends to be stationary for large times.

Let us denote
 \begin{align}
\label{juy}
w(p)=\int_0^{\infty}dt \exp(-pt)w(t)
\end{align}
the Laplace transform of the waiting-time density. Given that there is a point at $t=0$, the probability $G(t)dt$ that there is a point between $t$ and $t+dt$, is the sum of the probabilities that this occurs through one jump, two jumps,...Because the jumps are independent and have the same densities, we obtain the Laplace transform of $G(t)$ 
 \begin{align}
\label{juyi}
G(p)=\int_0^{\infty}dt \exp(-pt)G(t)
\end{align}
in the form, see Section\eqref{stationary},
 \begin{align}
\label{jyi}
G(p)=\frac{w(p)}{1-w(p)}.
\end{align}

It follows that, given the density $\lambda(t)$ of an inhomogeneous Poisson process, viewed as a renewal process, we may in principle obtain the Laplace transform of the auto-correlation function $G(t)$. As an example, suppose that $\lambda(t)=1\Longleftrightarrow w(t)=\exp(-t)$ whose Laplace transform is $w(p)=1/(1+p) $. Thus, from \eqref{jyi}, $G(p)=1/p$ and $G(t)=\lambda(t)=1$ as expected.

\section {Dirac pulses}\label{dirac}

Consider the stochastic process $\D(t)=\sum_{k}\de(t-t_k)$, where the $t_k$ form a stationary point process of density $\ave{\D(t)}\equiv D$ and $\delta(.)$ denotes the Dirac distribution. The spectrum of $\D(t)$ exhibits a singularity $2\pi D^2\de(\Om)$ at $\Om=0$ which is subtracted. The first expression in \eqref{def} give the reduced spectrum
\begin{align}\label{d'}
\spectral(\Om)=\lim_{\T\to \infty}\frac{1}{\T}\ave{\abs{\sum_{\text{allowed}~k}^{}{\exp(\jj\Om t_k)}}^2},
\end{align}
where $\Om=2\pi n/\T$, $n=1,2...$. Note that $n=0$ is not allowed, but $\Om$ can be made as small as one wishes by setting $n=1$ and letting $\T$ go to infinity. This expression is useful to evaluate spectra through numerical calculations that generate runs, each of them with a different $t_k$ sequence.

The relative noise is defined as follows
\begin{align}\label{relative noise}
\N(\Om)\equiv\frac{\spectral(\Om)}{D^2}-\frac{1}{D}.
\end{align}
This quantity vanishes if the underlying point process is Poisson, and may be negative.

The normalized correlation $g(\tau)$ is the probability that an event occurs between $\tau$ and $\tau+d\tau$, given that an event occurred at $t=\tau$, divided by $d\tau$, and normalized. We have the integral relations
\begin{align}
\label{g'}
\N(\Om)&= \int _{-\infty }^{\infty }  d\tau\bigl(g(\tau)-1\bigr)\exp(-\jj\Om \tau)\\
\label{n'}
g(\tau)-1&=\int _{-\infty }^{\infty }  \frac{d\Om}{2\pi} \N(\Om)\exp(\jj\Om \tau).
\end{align}

The motivation for introducing $g(\tau)-1$ in \eqref{g'} is that this quantity tends to 0 as $\tau$ tends to infinity because widely separated events are in that limit independent for stationary processes. The above relations are closely related to the Wiener-Khintchine relations. They can be established directly for point processes. Note that our definition of "sub-Poissonian" photo-currents is that $\N(0)<0$. This does not necessarily imply that $g(0)<1$.

\paragraph{Random deletion.}

Random deletion of events (also called "thinning" or decimation) means that each event is ascribed a probability $1-p$ of being deleted. For example, considering the first event of a given run, we flip a coin. If head, that event is preserved (probability 1/2). If tail, it is deleted. The same procedure is applied to the other events of the run and to the events of other runs, each time with a new coin flipping. Obviously the average rate $D$ of the process is multiplied by $p$. An important result is that the function $g(\tau)$ and thus the other two functions defined above, and in particular the relative noise $\N(\Om)$, are not affected. Indeed consider the case where there is one event in the time slot $[0,dt]$ and one event in the time slot $[\tau, \tau+d\tau]$, corresponding to a product of 1. In any other circumstances the product is 0. After thinning the probability of having again (1,1) is multiplied by $p^2$. But the denominator in the normalized correlation $g(\tau)$ is also multiplied by $p^2$, so that the result is unchanged. The average rate may be restored by an appropriate scaling of the time axis. But since in general $g(\tau/p)≠g(\tau)$, rescaled thinning affects the statistics with the sole exception of Poissonian processes, in which case $g(\tau)=1$.

\paragraph{Example of the relation between spectrum and correlation.}

As an example consider the relative noise of a high-power laser driven by a non-fluctuating current
\begin{align}
\label{qq}
\N(\Om)=-\frac{1}{D\p 1+\p \Om\tau_p \q^2\q },
\end{align}
where $\tau_p$ is the "life time" of the resonator. From this expression we obtain, setting $D=1$ for simplicity, that 
\begin{align}
\label{gt}
g(\tau)&=1- \int _{-\infty }^{+\infty }  \frac{d\Om}{2\pi} \frac{1}{1+\p  \Om\tau_p \q^2}\exp(\jj\Om \tau)=1-\frac{1}{2\tau_p}\exp(-\frac{\tau}{\tau_p})\\ 
\label{g0}
g(0)&=1-\frac{1}{2\tau_p}.
\end{align}
Of course $g(\tau)\to 1$ if $ \tau\to\infty$. In the present situation $g(0)<1$.

\section{Photo-count variance}

The normalized variance $\V(\T)$ of the number of events occuring during some time $\T$ and $g(\tau)$ are related as 
\begin{align}
\label{v}
\V (\T)\equiv \frac{\ave{d(\T)^{2}}-\ave{d(\T)}^2}{\ave{d(\T)}}-1&=D \int _{-\T}^{\T} d\tau(1-\frac{\abs{\tau}}{\T})\bigl(g(\tau)-1\bigr)\\
\label{v'} 2D\p g(\T)-1\q&=\frac{d^2\p \T\V (\T)\q}{d\T^2}. 
\end{align}
In the special case of a Poisson process we have $g(\tau)=1$, $\V (\T)=0$ and $\N(\Om)=0$, that is, $\spectral_{\De D}=D$.

\paragraph{Proof of the above relation}

Let $d(t)$ be the number of events occurring up to time $t$, that is the number of $k$ values such that $t_k<t$. Obviously $d(0)=0$ since the $t_k$ are positive numbers. Let us prove that for some measurement time $T$ $\ave{d(\T)}=D\T$, where $D$ is a constant called the intensity of the process.
We introduce the (positive) number $D_h(t):=d(t+h)-d(t)$ of events occurring between $t$ and $t+h$. Because the process considered is stationary $\ave{D_h(t)}$ does not depend on $t$. It is convenient to split the measurement time $\T$ into time slots of duration $h=\T/n$, labeled by $i=1,2,...n$. Eventually, we let $n$ go to infinity, so that it is unlikely that more than one event occur within any time slot. Thus, if $D_{i}\equiv D_{h}(\p i-1\q h)$ denotes the number of events occurring during slot $i=1,2...n$, we have either $D_{i}=1$ or $D_{i}=0$ and $\ave{D_i}$ does not depend on $i$. For later use note that ${D_i}^2=D_i$. The number $d(\T)$ of events occuring during the measurement time $\T$ is the sum of the $D_i$ with $i$ running from 1 to $n$, so that its average reads 
\begin{align}
\label{93}
\ave{d(\T)}=\ave{\sum_{i=1}^{n} D_{i}}=n\ave{D_i}=\frac{\T}{h}\ave{D_i}\equiv \T D,
\end{align}
where we have set $D\equiv \ave{D_i}/h$.

Because the process is stationary its auto-correlation  $\ave{D_h(t+\tau)D_h(t)}$ does not depend on $t$ for every $h>0$ and every $\tau >0$. The degree of second order coherence $g(\tau)$ is the limit of $\ave{D_h(t+\tau)D_{h}(t)}/\ave{D_h(t)}^2$ as $h$ goes to 0. Let us set for $j>i$ 
\begin{align}
\label{94}
\ave{D_{i}D_{j}}\equiv\ave{D_{i}}^2g_{n}\bigl((j-i)\frac{\T}{n}\bigr),
\end{align} 
and evaluate
\begin{align}
\label{95}
\ave{d(\T)^{2}}&=\ave{\sum_{i=1}^{n}D_{i}\sum_{j=1}^{n}D_{j}}\nonumber\\
&=n\ave{D_{i}}+2\ave{D_{i}}^2\sum_{i=1}^{n} \sum_{j=i+1}^{n}g_{n}\bigl((j-i)\frac{\T}{n}\bigr)\nonumber\\
&=\ave{d(\T)}+2\ave{D_{i}}^2\sum_{i=1}^{n} (n-i)g_{n}(\frac{i\T}{n})\nonumber\\
&=\ave{d(\T)}+2D^2\frac{\T}{n}\sum_{i=1}^{n} (\T-\frac{i\T}{n})g_{n}(\frac{i\T}{n}). 
\end{align}
In the limit $n\to \infty$ the sum may be replaced by an integral and $g_{n}$ by $g$, thus 
\begin{align}
\label{99}
\ave{d(\T)^{2}}=\ave{d(\T)}+2D^2 \int _{0}^{\T} d\tau(\T-\tau)g(\tau).
\end{align}
After slight rearranging the variance of $d(T)$ may be written in the form
\begin{align}
\label{100}
\V (\T)&\equiv\frac{\var(d(\T))}{\ave{d(\T)}}-1= \frac{\ave{d(\T)^{2}}-\ave{d(\T)}^2}{\ave{d(\T)}}-1\nonumber\\
&=2D \int _{0}^{\T} d\tau(1-\frac{\tau}{\T})\bigl(g(\tau)-1\bigr), 
\end{align}
since $\int_{0}^{\T} d\tau (1-\tau /\T)=\T/2$.
The motivation for introducing $g(\tau)-1$ in the integral is that this quantity usually tends to 0 quickly as $\tau$ tends to infinity. Intuitively, this is because widely separated events tend to be independent and consequently in that limit $\ave{D_{i}D_{j}}\approx \ave{D_{i}}\ave{D_{j}}=\ave{D_{i}}^2$. Setting $D=1$ for brevity, relation \eqref{100} may be written as,
\begin{align}
\label{xx}
P_c(\tau)=\sum_{k=0}^{\infty}k^2\frac{d^2P(k,\tau)}{d\tau^2},
\end{align} 
where $P_c(\tau)dt d\tau$ denotes the probability density of having an event between 0 and $dt$ and an event between $\tau$ and $\tau+d\tau$ or, equivalently, $P_c(\tau)d\tau$ is the probability density of another event being registered during the time interval $\tau$ and $\tau+ d\tau$, given that an event occurred at $t=0$. In \eqref{xx} $P(k,\tau)$ denotes the probability of $k$ events being registered between $t=0$ and $t=\tau$.

\section{Dark-room picture}\label{dark}

For the sake of illustration let us present a simple picture of regular point processes. The initial point process considered is periodic and consists of events occurring at $t=1,2...$ time units, that is $t_k=k$. Under circumstances to be defined later on (delay times much larger than unity) this process may be viewed as being almost stationary. The density is clearly unity. 

In our picture, one person (representing an electron) enters into a dark room every time unit and wanders randomly in the room until he finds the exit.  This picture may describe regularly-pumped lasers at high power because electrons entering the optical resonator are quickly converted into light quanta. Light quanta wander in the resonator for some time and then get instantly converted into photo-electrons. The point process is written as $t_k=k+\xi_k$, where the $\xi_k$ are independent of one-another and distributed according to the same density $P(\xi_k)\equiv P(\xi)$. An appropriate distribution would be the exponential one.

Let us treat a special case that may be solved almost by inspection, namely the case where  $P(\xi)=1/\tau_r$ if $0≤\xi<\tau_r$ and 0 otherwise, and $\tau,\tau_r$ are large integers. Consider a pair $i,j≠i$ of $k$ values such that $i+\xi_i$ may be in the first time slot $(0,dt)$ and $j+\xi_j$  may be in the second time slot $(\tau, \tau+d\tau)$. Inspection shows that this is possible only if $-\tau_r<i≤0,\tau-\tau_r<j≤\tau$. Ignoring first the restriction $j≠i$, we find that the probability we are looking for is the number of allowed $i,j$ values, that is, the product of the $i,j$ ranges, times $1/\tau_r^2$, namely $\tau_r^2/\tau_r^2=1$. This result is accurate if $\tau≥\tau_r$. But if $\tau<\tau_r$ one must subtract from the numerator of the previous expression the number of $i,j$-values that are equal, namely $\tau_r-\tau$, so that the normalized correlation reads
\begin{align}
\label{corr}
g(\tau)&=1,\quad\qquad &\tau≥\tau_r\nonumber\\
g(\tau)&=\frac{\tau_r^2-\p \tau_r-\tau\q}{\tau_r^2}=1- \frac{\tau_r-\tau}{\tau_r^2},\quad &\tau<\tau_r.
\end{align}
In particular, $g(0)=1-1/\tau_r$, indicating a modest amount of anti-bunching, remembering that  $\tau_r\gg1$. The same result is obtained for the laser model in \eqref{g0} if we set $\tau_r=2\tau_p$ to make the average life-times the same in the two models. 

The reduced photo-events spectrum is obtained from $g(\tau)$ through a Fourier transform according to \eqref{g}
as
\begin{align}
\label{sp}
\N(\Om)&\equiv 2\int_{0}^{1} dx \p x-1\q \cos(\Om\tau_r x)=2\frac{\cos(\Om\tau_r)-1}{\p\Om\tau_r\q^2}\nonumber\\
\spectral(\Om)&=1+\frac{\cos(\Om\tau_r)-1}{(\Om\tau_r)^2/2},
\end{align}
where we have set $x\equiv \tau/\tau_r$, remembering that the density (average rate) $D=1$. We note that $\spectral(0)=0$, as one expects from the fact that the primary process is regular and that no event has been lost or created. The spectral density of the process considered, given in \eqref{sp}, is illustrated in Fig.~\ref{darkroom}. 

\setlength{\figwidth}{0.6\textwidth}
\begin{figure}
\centering
\includegraphics[width=\figwidth]{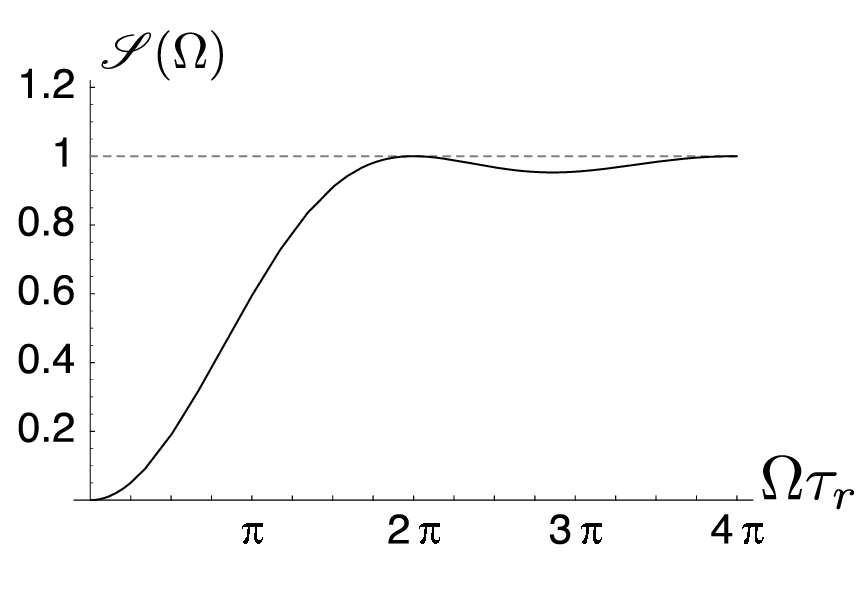}
\caption{Plain line: spectral density corresponding to the dark room picture, see \eqref{sp}. The dotted line corresponds to the shot-noise level.}
\label{darkroom} 
\end{figure}

Using \eqref{v} we obtain in the present model
\begin{align}
\label{nn}
\V(T)&=-1+\frac{\tau_r}{3T},\qquad T≥\tau_r \nonumber\\
\V(T)&=-\frac{T}{\tau_r}+\frac{T^2}{3\tau_r^2},\qquad T<\tau_r. 
\end{align}
It is easy to see that the expression of $g(\tau)$ in \eqref{v'} is verified in that example.

\section{Linewidth}\label {linewidt}

Consider a signal of constant amplitude and phase
\begin{align}
\label{nnn}
\phi(t)=\int_0 ^t  ~dt~\Delta\om(t), 
\end{align}
where $\Delta\om(t)$ is a white, gaussian-distributed, random process of spectral density $\spectral_{\Delta\om}$. Because of linearity, $\phi(t)$ is gaussian as well. It can be shown that the power spectrum of $\exp(\ii \phi(t))$ has a width at half-power
\begin{align}
\label{nnnn}
\de \om\approx \spectral_{\Delta\om}.
\end{align}

\newpage

\chapter{Circuits}\label{circuit}

The basic concepts employed in Circuit Theory will be recalled, and devices useful in radio, microwave, and optical frequency ranges will be described, considering both resonators and transmission lines. The circuit equations provide currents $i(t)$, viewed as (real, linear, causal) responses to specified potential sources $v(t)$. Formally, the circuit theory is based on the Darwin lagrangian that neglects terms of order $1/c^3$, implying that radiation does not occur. Electromagnetic phenomena within a "black box" are determined by the input currents, and therefore they have no independent degrees of freedom. The electromagnetic energy is of the form $\frac{1}{2}(\rho\phi+j.A)$, where $\rho,j$ are respectively the electrical charge density and the current density, and $\phi,A$ are respectively the scalar and vector potentials created by the other charges. This energy is localized at the electrical charges, but it may often be expressed in terms of the free-space field. 

\begin{figure}
\setlength{\figwidth}{0.18\textwidth}
\centering
\begin{tabular}{cccc}
\includegraphics[width=\figwidth]{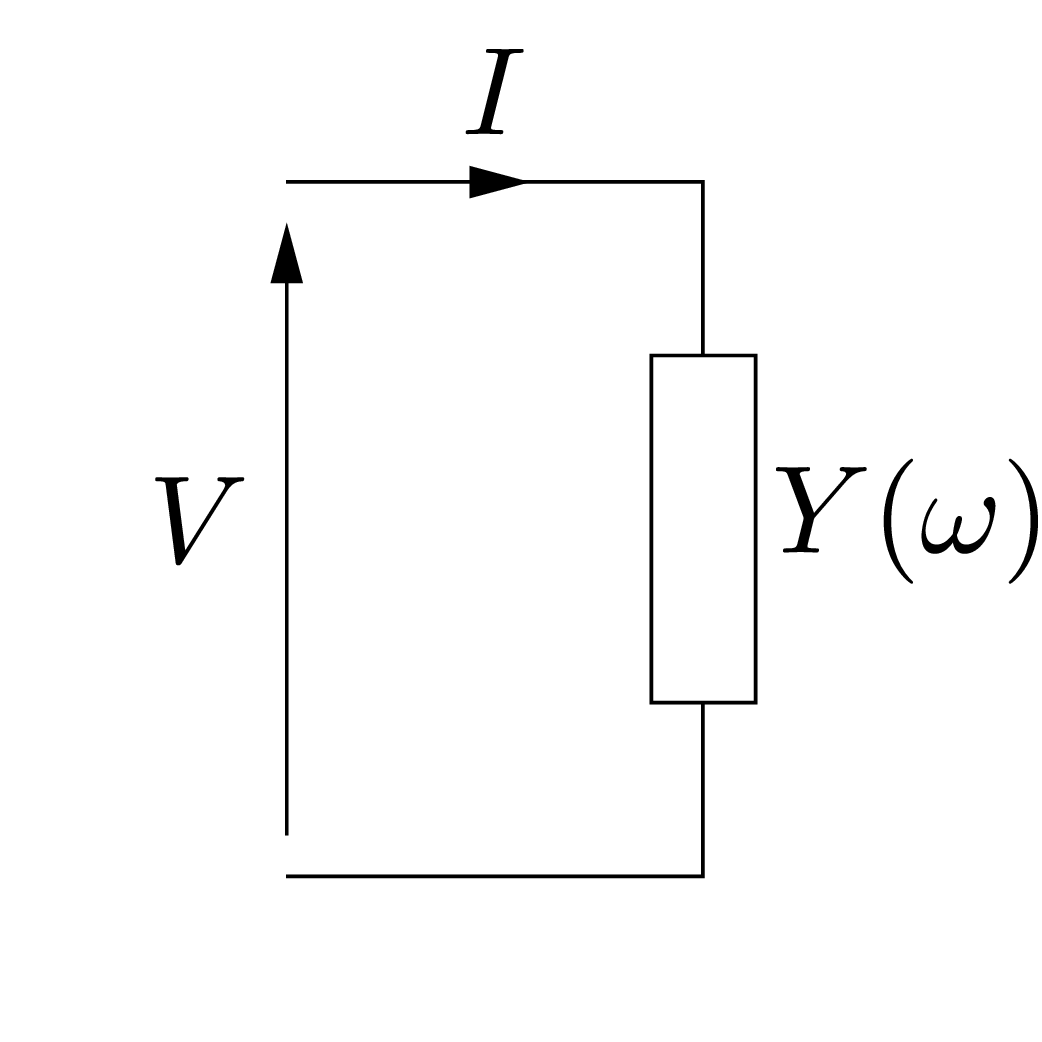} & \includegraphics[width=\figwidth]{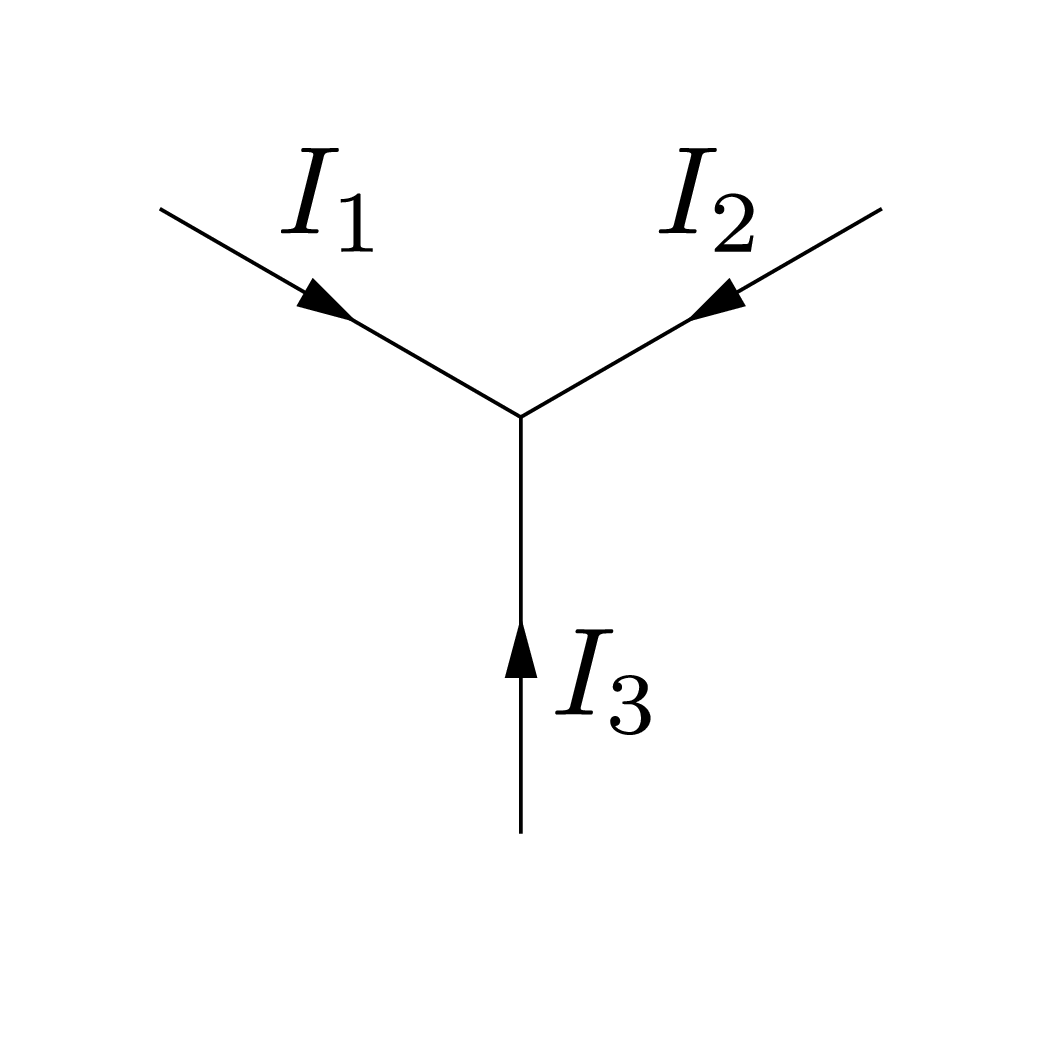} &  \includegraphics[width=\figwidth]{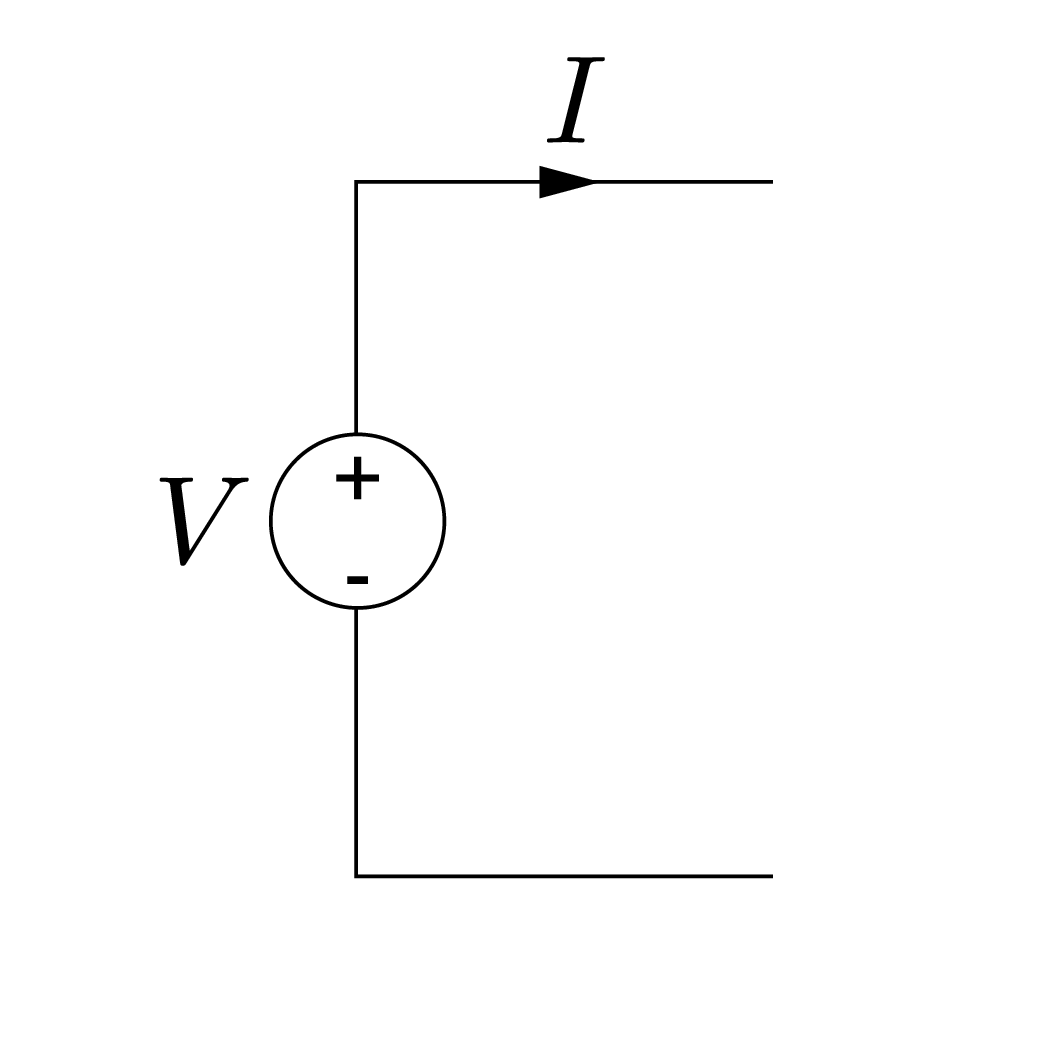} & \includegraphics[width=\figwidth]{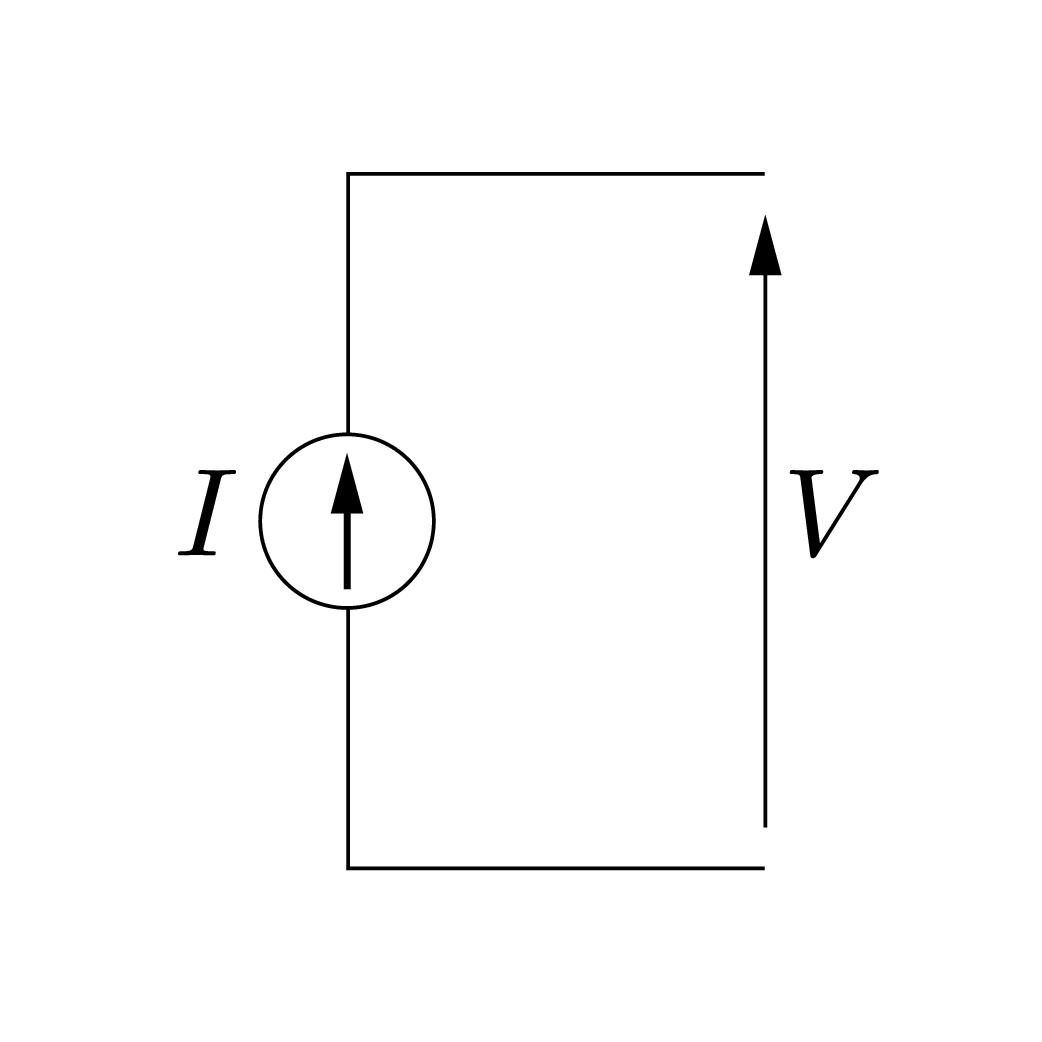}\\
(a) & (b) & (c) & (d) \\
\includegraphics[width=\figwidth]{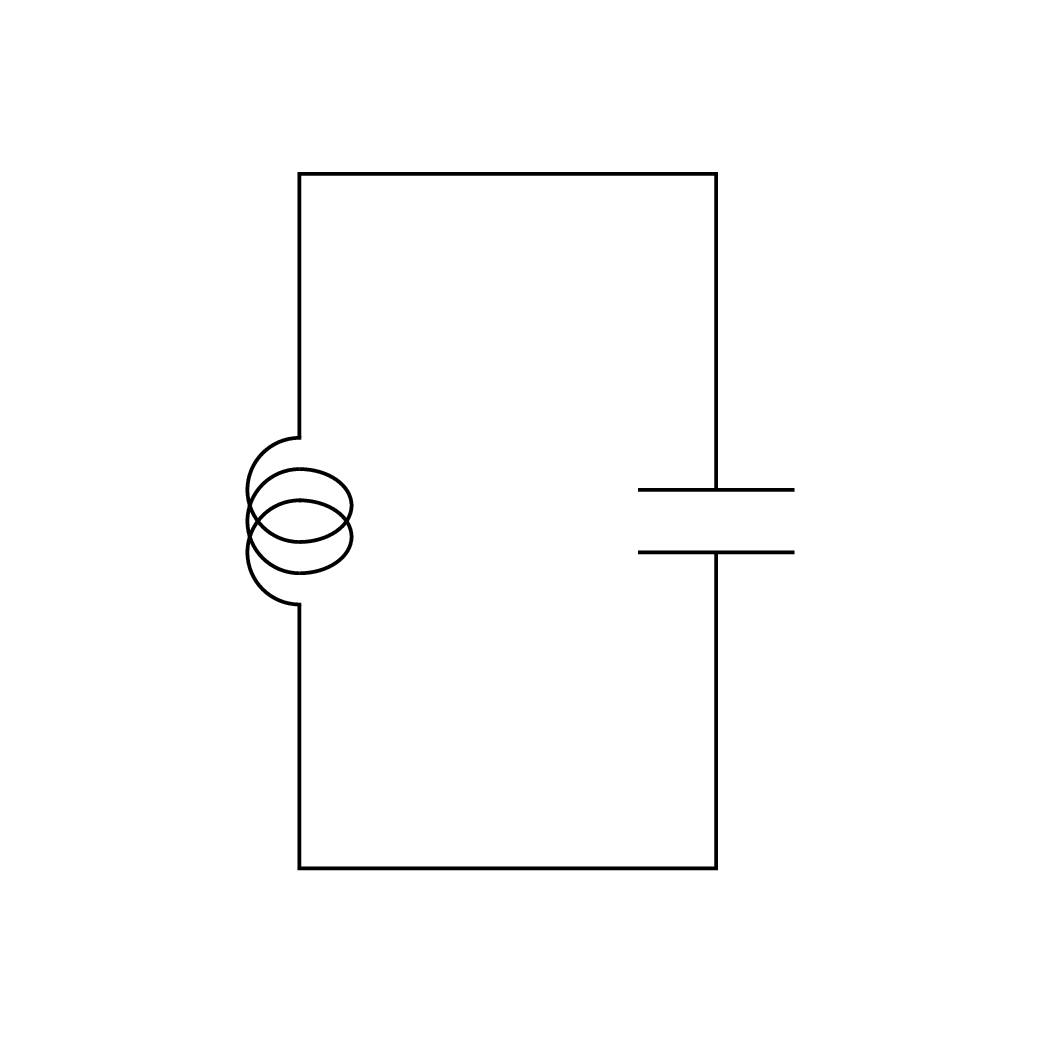} & \includegraphics[width=\figwidth]{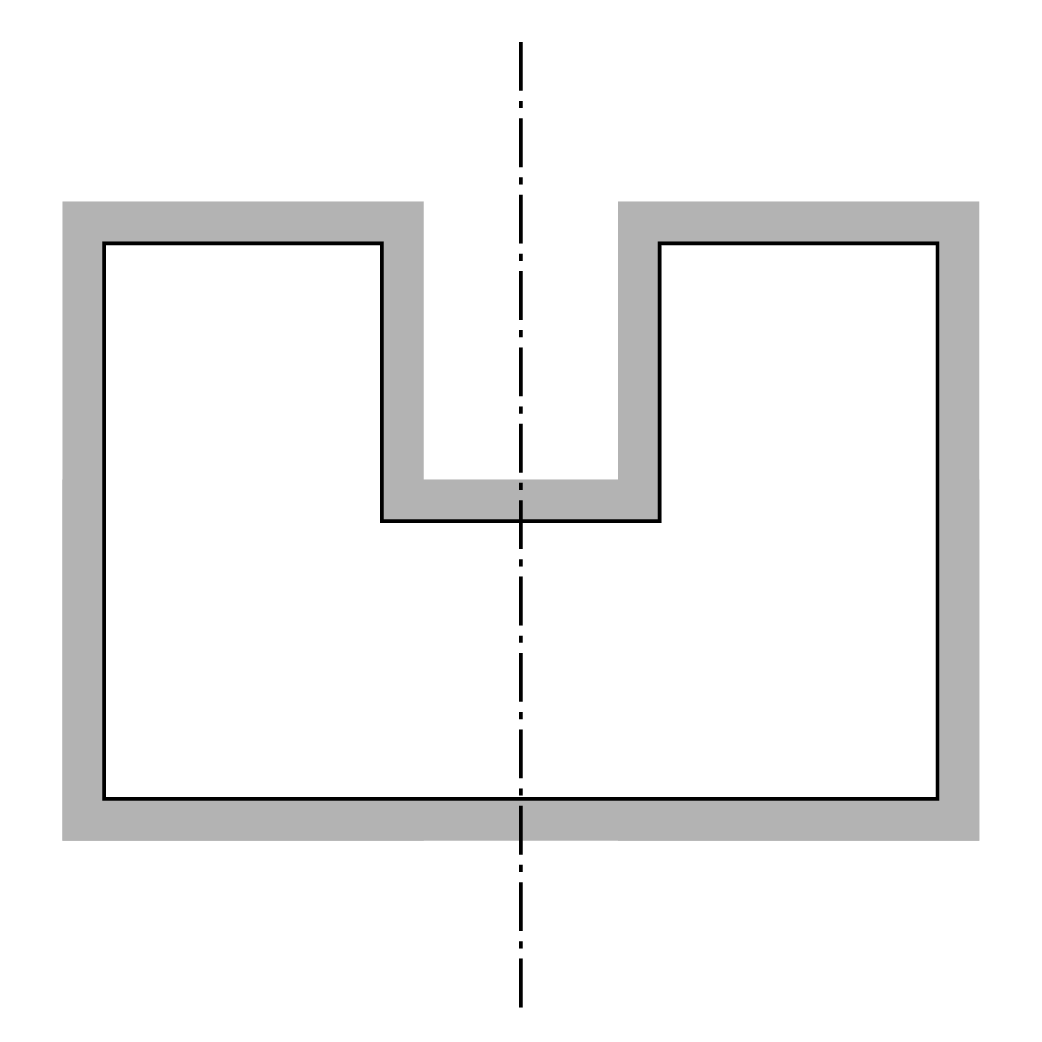} &  \includegraphics[width=\figwidth]{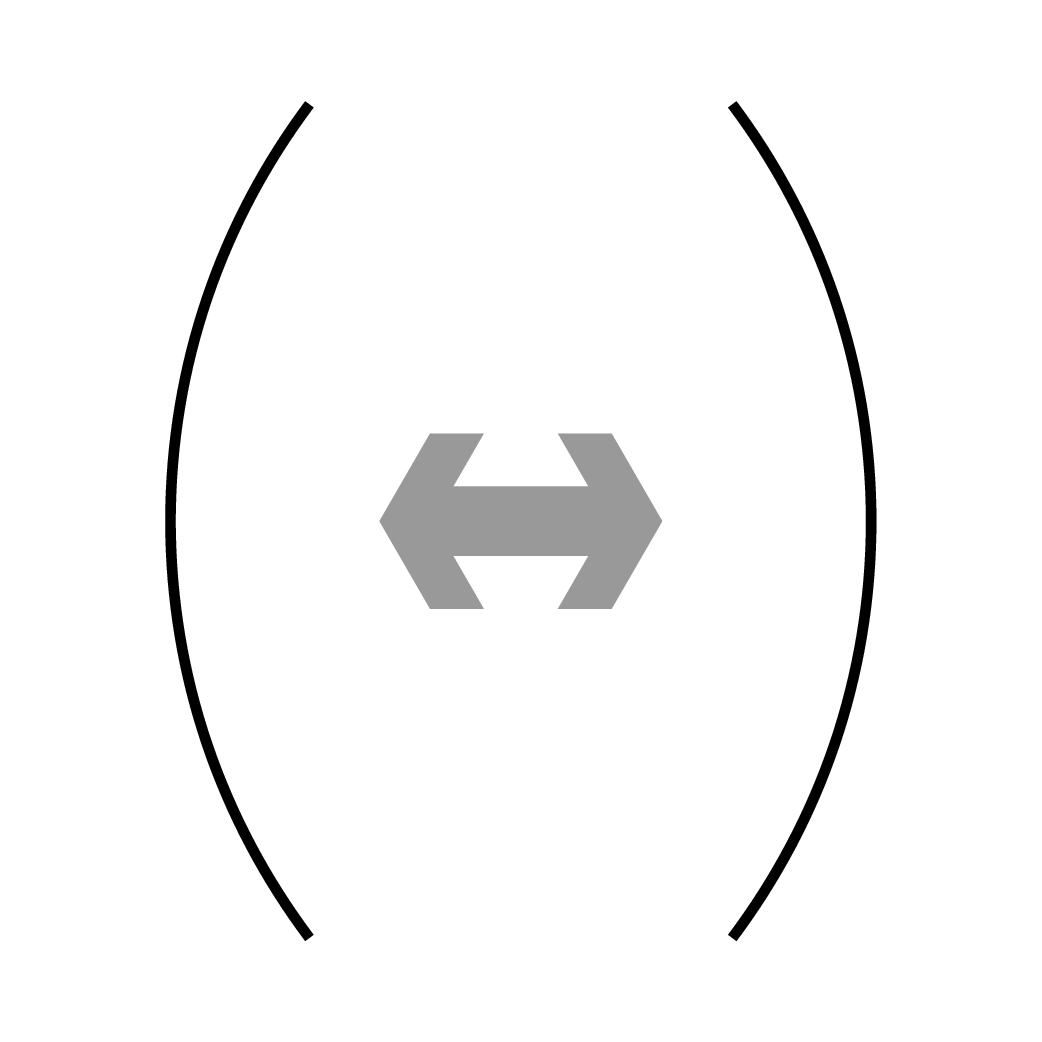} & \includegraphics[width=\figwidth]{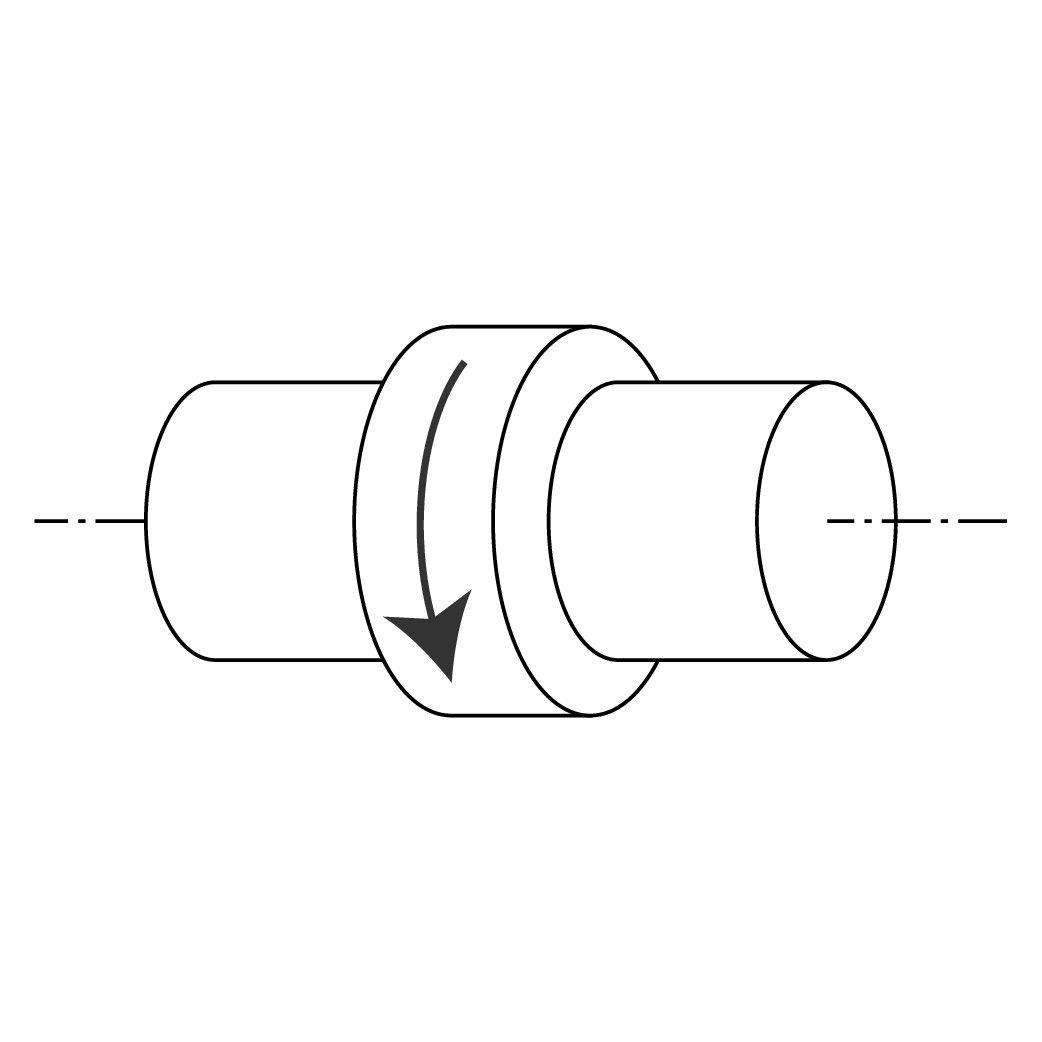}\\
(e) & (f) & (g) & (h) \\
\includegraphics[width=\figwidth]{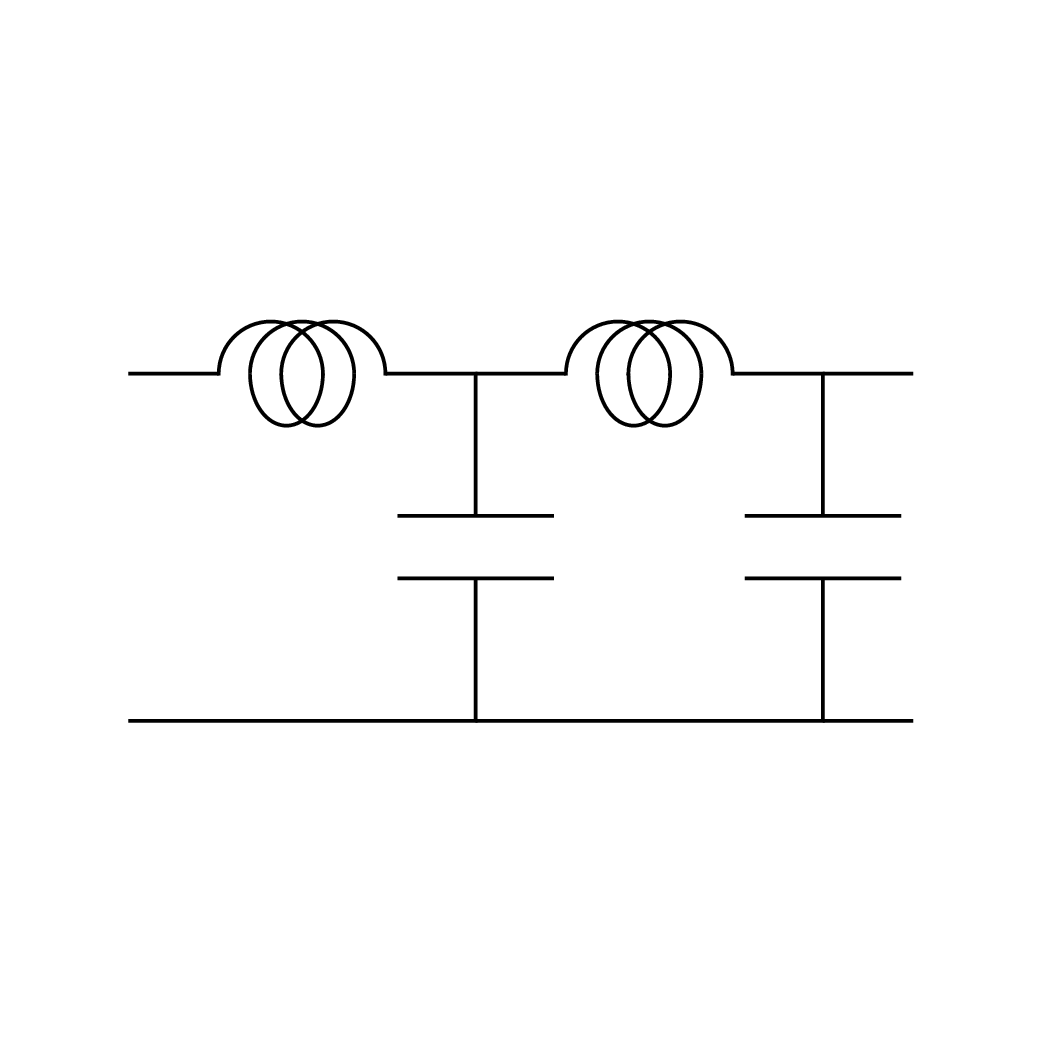} & \includegraphics[width=\figwidth]{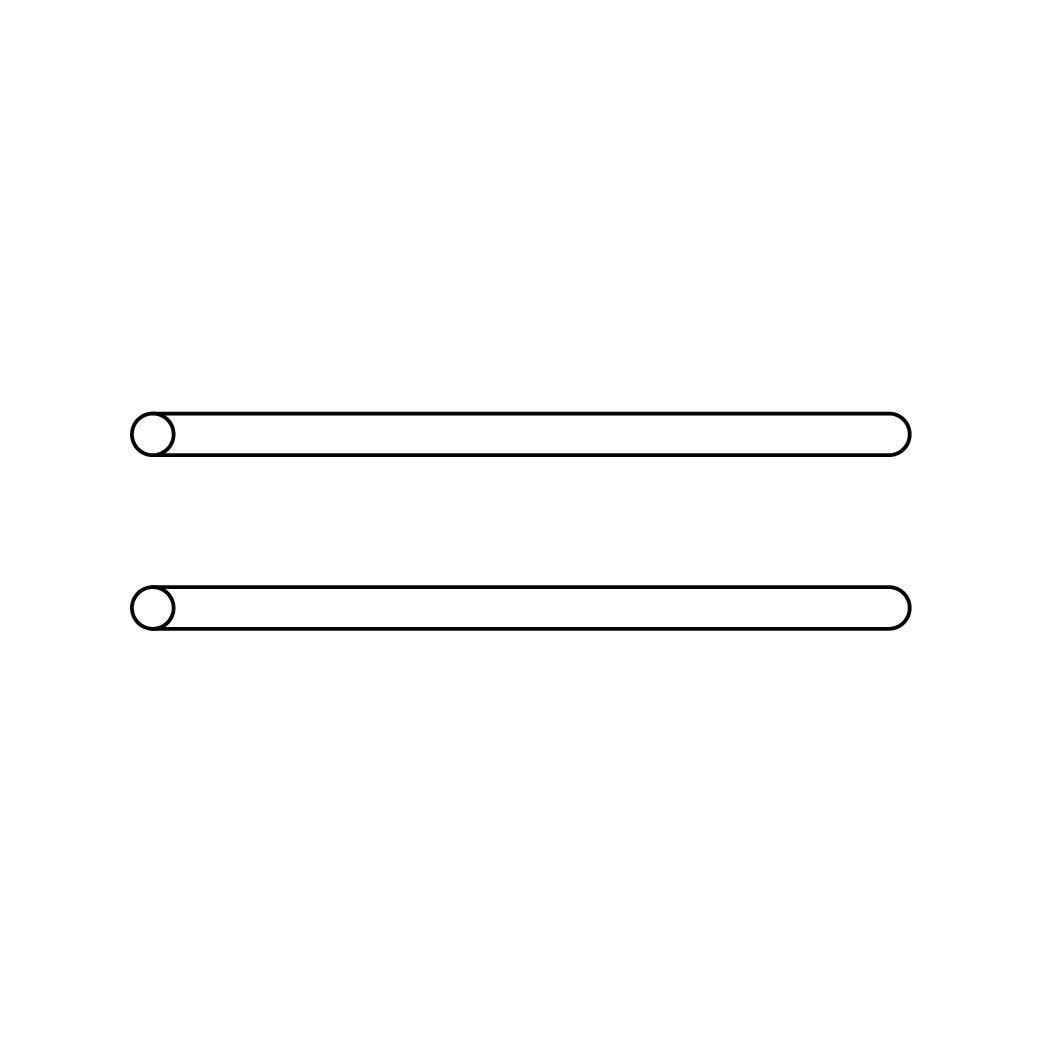} &  \includegraphics[width=\figwidth]{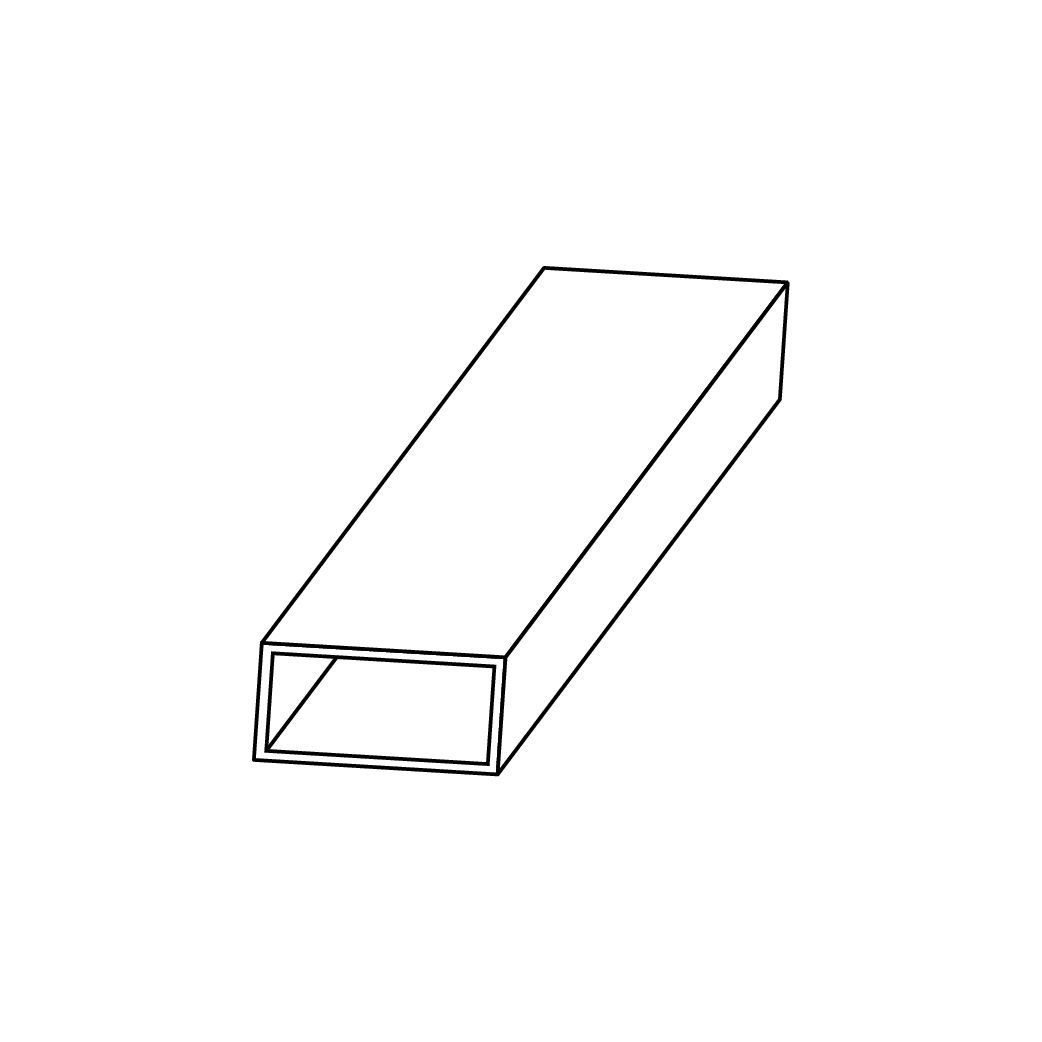} & \includegraphics[width=\figwidth]{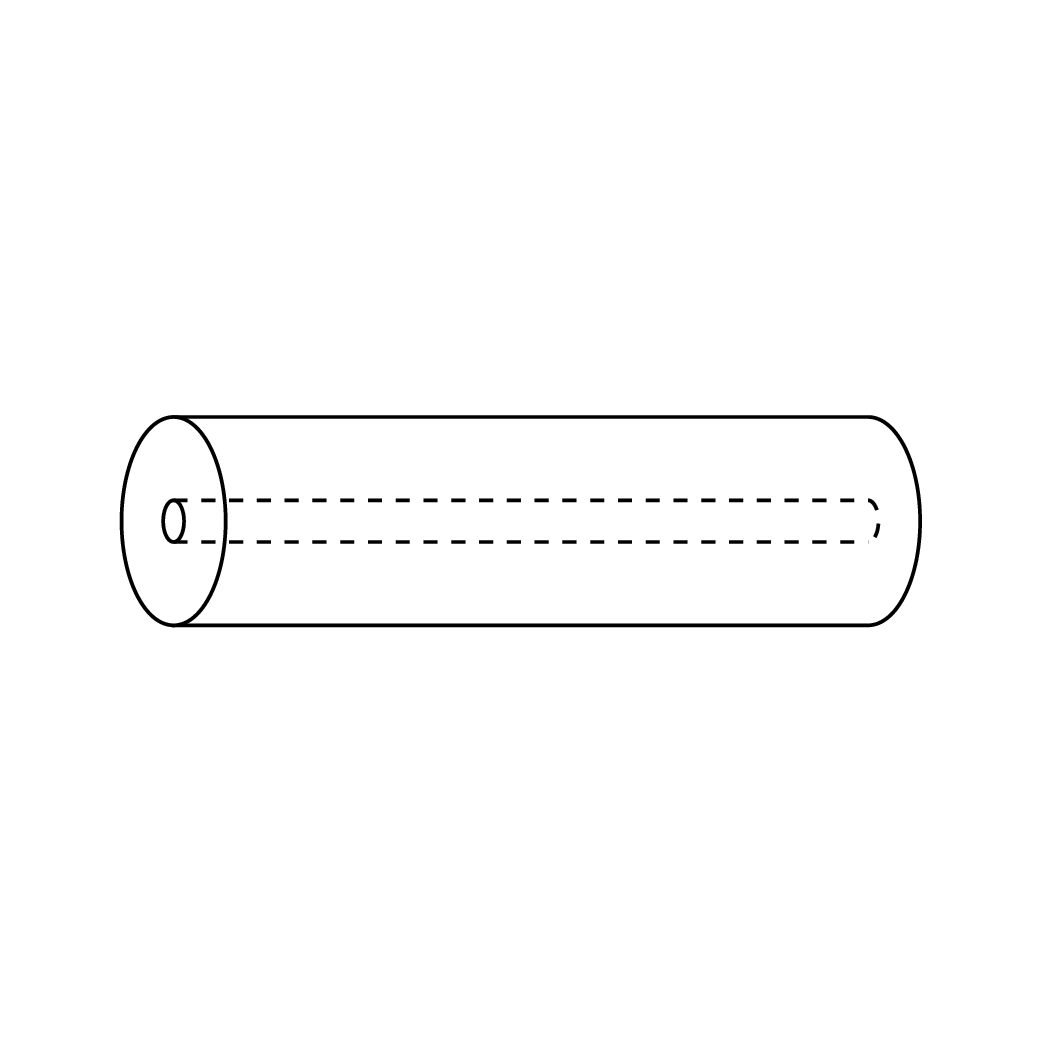}\\
(i) & (j) & (k) & (l) \\
\includegraphics[width=\figwidth]{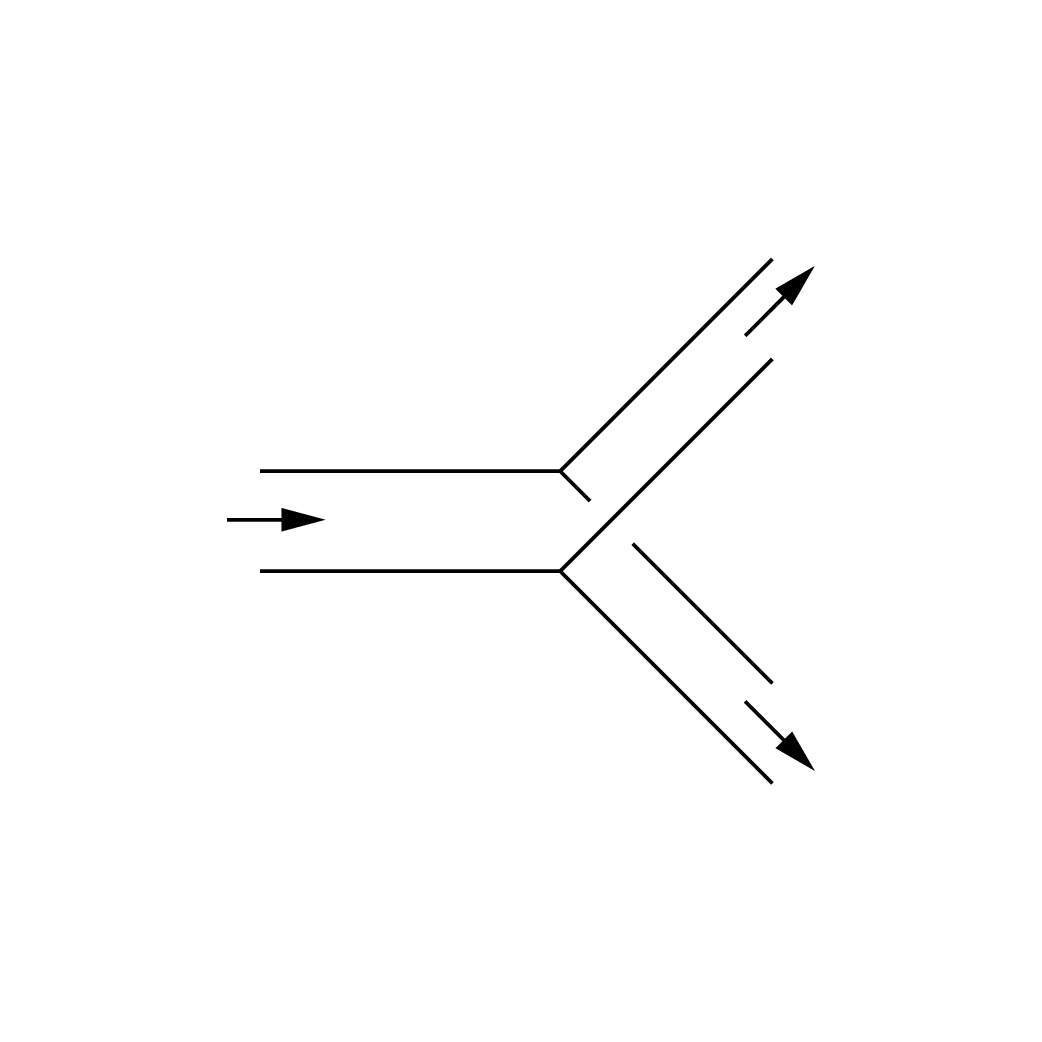} & \includegraphics[width=\figwidth]{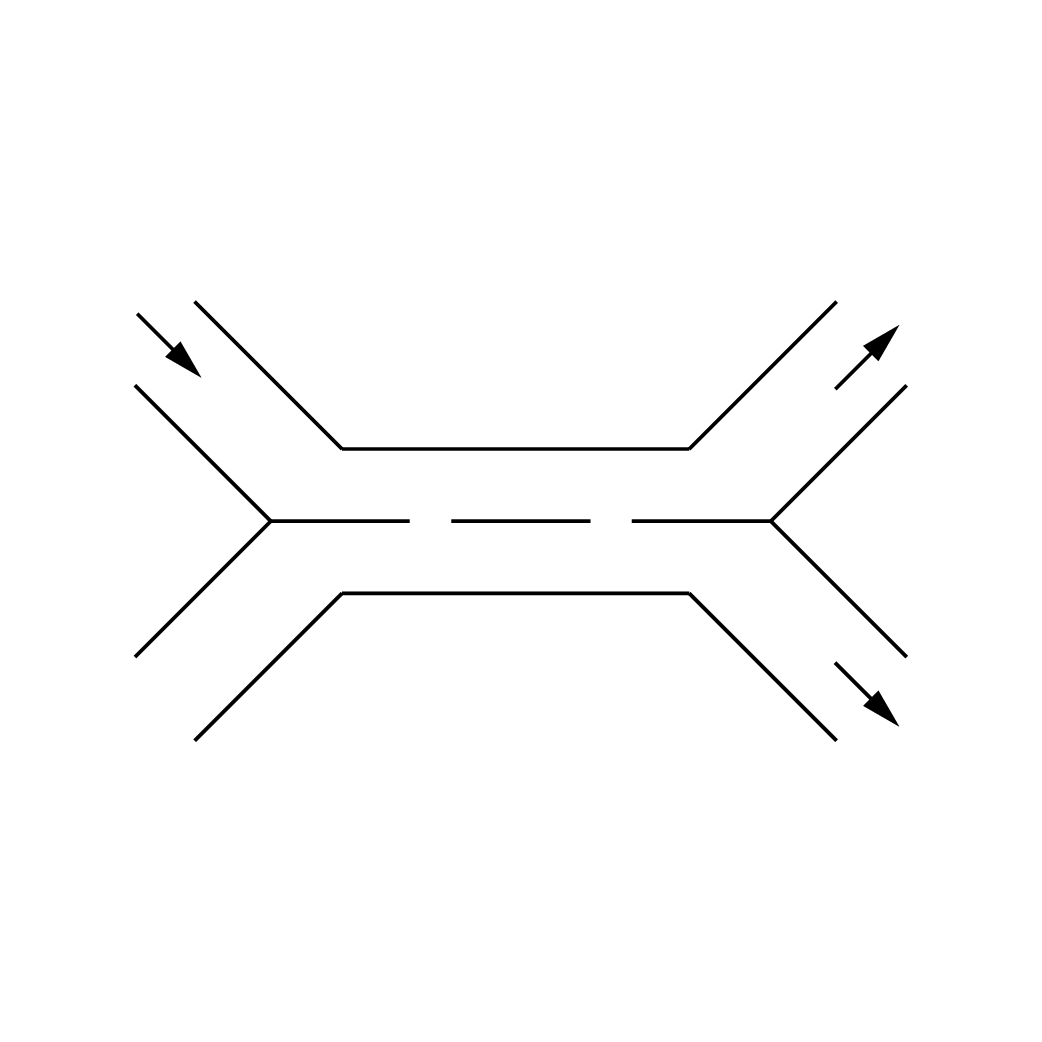} &  \includegraphics[width=\figwidth]{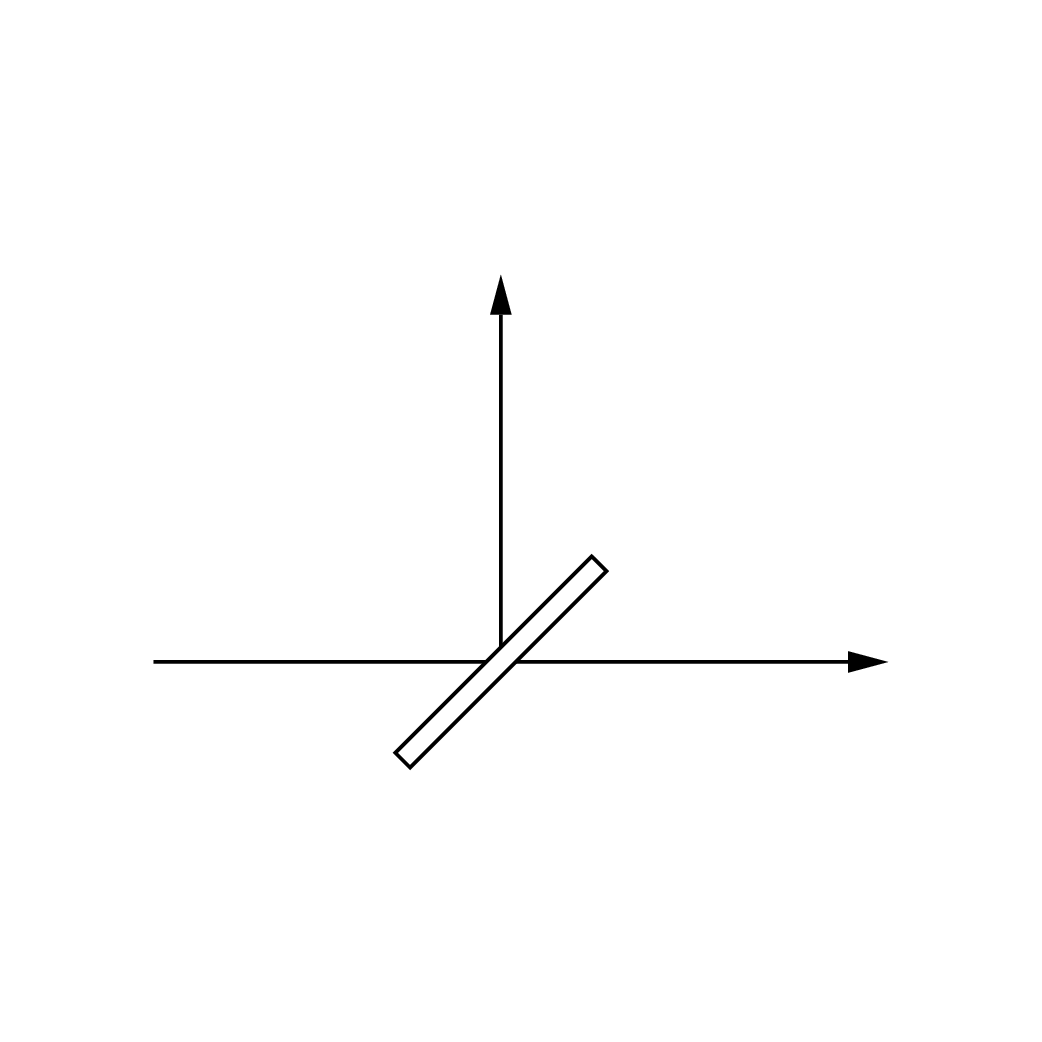} & \includegraphics[width=\figwidth]{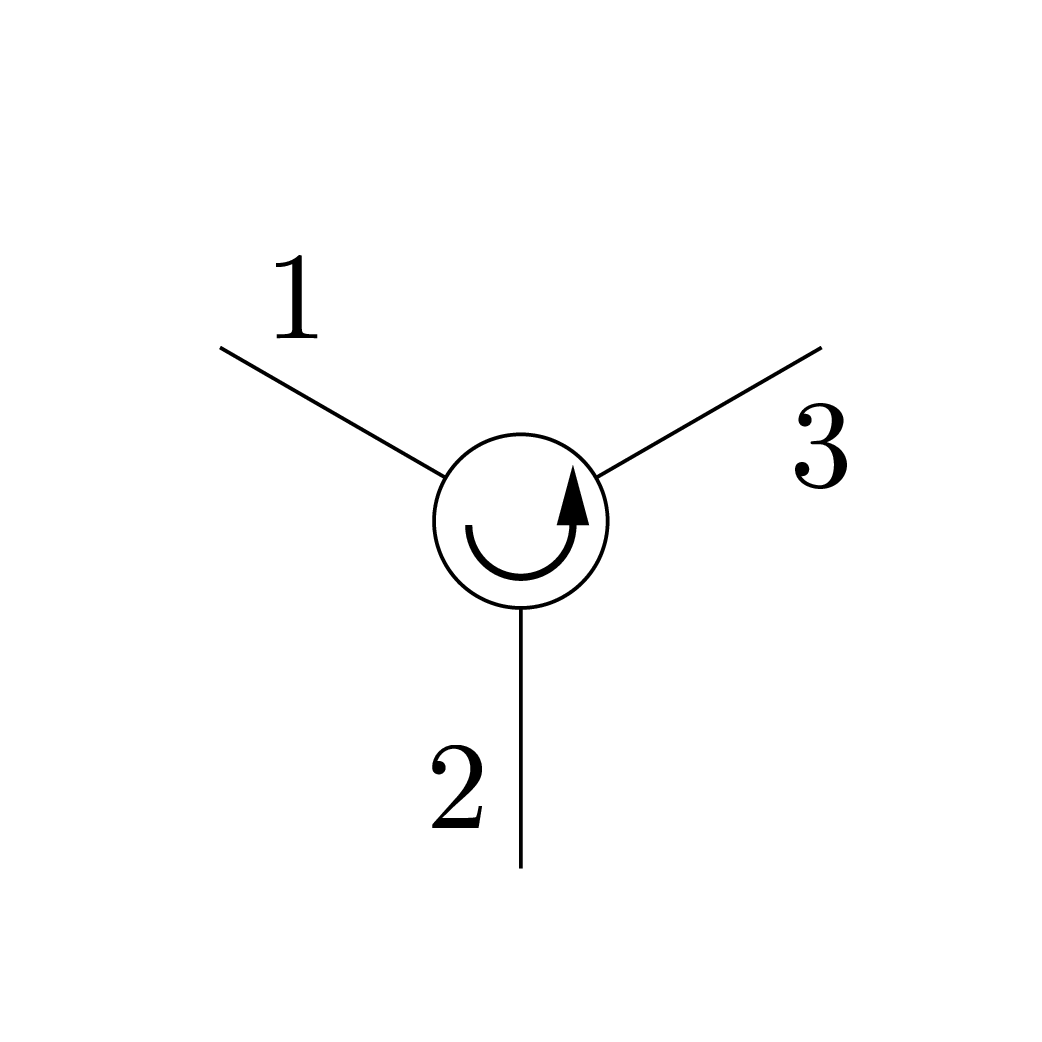}\\
(m) & (n) & (o) & (p) \\
\end{tabular}
\caption{a) There is a linear relation between the potential $V$ and the current $I$ at some frequency $\om$ (generalized Ohm law $I=Y(\om)V$).  b) Illustrates the Kirchhoff law: $I_1+I_2+I_3=0$. c) Represents a potential source with $V$ independent of $I$. d) Represents a current source with $I$ independent of $V$. e) Inductance-capacitance resonating circuit. f) Cavity employed, e.g., in reflex klystrons. g) Fabry-Pérot-type optical resonator with two curved mirrors facing each other. h) Whispering-gallery mode resonator. i) Low-pass filter, j) Parallel conductors, k) Waveguide, l) Optical fiber. Waves may be split in various ways: m) A transmission line is connected to two transmission lines whose characteristic conductances sum up to the original line characteristic conductance. n) Directional coupler. The two holes are spaced a quarter of a wavelength apart. o) The beam splitter is an optical equivalent of the directional coupler. p) The circulator is ideally a loss-less non-reciprocal device.}
\label{circuitfig}
\end{figure}

A resonator at frequency $\omega$ may be viewed as a cavity carved into a good conductor such as copper or some super-conducting material, as shown in Figure \ref{circuitfig} in f). A resonator is, of course, a physical object (rather than a mathematical abstraction such as a "quantization box") consisting of billions of heavy nuclei and billions of electrons that make the system electrically neutral. These electrons act collectively, and should best be viewed as a classical charged fluid\footnote{The discrete character of electrical charges appears only when the charges get accelerated in free space as is the case in vacuum devices.}. Such cavities may contain free atoms, entering or exiting through holes that are too small for the electromagnetic radiation to escape.

We first consider conservative elements (that is, elements that conserve energy) such as capacitances, $C$, and inductances, $L$, and circulators. Circulators require non-reciprocal material described by a relation of the form $B=\mu ~H$, where $\mu$ denotes an Hermitian non-symmetrical matrix. We next consider non-conservative elements such as positive or negative conductances. In most practical cases the conductances are obtained from separate measurements. We postpone to Section \ref{electronmotion} a microscopic conductance model, consisting of a single electron located between two parallel conducting plates. A static potential $U$, and an optical potential $v(t)$ oscillating at an optical frequency $\om$, are applied to the plates. We suppose that there is an exact resonance between $v(t)$ an the electron natural oscillatory motion.  The Quantum-Mechanical averaged induced current $i(t)$ is proportional to the average electron momentum, and, under circumstances to be discussed later on, the ratio $i(t)/v(t)$ may be a real constant $G$. When $eU$ is slightly smaller than $\hbar\om$, the conductance is positive and the optical potential delivers energy to the static potential. On the other hand, when $eU$ slightly exceeds $\hbar\om$ the conductance is negative and the optical field receives energy from the static potential. When $eU$ is precisely equal to $\hbar\om$ the conductance vanishes but fluctuations remain, as is the case when two conductances of opposite signs are connected in parallel. To summarize, the complete system including static and optical potentials conserve energy, aside from an irreversible loss of energy $\abs{eU-\hbar\om}$ that can be made in principle as small as one wishes.

We mainly consider sources (and responses) that vary sinusoidally in the course of time at frequency $\om$. For isolated linear systems the circuit equations have solutions only for discrete complex values $\om_n$ of $\om$. We will be particularly interested in isolated circuits that have only one nearly-real frequency, the other ones having large negative imaginary parts corresponding to strongly damped modes.

\section{Classical devices}\label{circuitbis}

Let us recall some basic results. The complex notation often employed for describing quantities that vary sinusoidally in time was recalled in Section \ref {notation}. For strictly sinusoidal potentials and currents represented by the complex numbers $V$ and $I$, respectively, and linear circuits, we have the generalized Ohm law $I=Y(\om)V$, where the complex constant of proportionality $Y(\om)$, called the admittance, and its inverse the impedance $Z(\om)$, in general depend on the frequency $\om$, which may vary from minus to plus infinity. We set $Y(\om)=G(\om)+\ii B(\om)$ or $Z(\om)=R(\om)+\ii X(\om)$.

If a potential $v(t)$ is applied to a constant conductance $G$ (a real number), we have by definition $i(t)=Gv(t)$, or, using the complex notation, $I=GV$, where $V$, and thus $I$, are in general complex numbers. We consider in the major part of this paper \emph{ideal} conductances defined as follows: They are supposed to be independent of the driving potential $V$ and to be independent of frequency. Furthermore, they are supposed to have a fixed energy content that may be set equal to zero since only energy differences are relevant. A physical model for ideal conductances is a piece of metal having a large number of inelastic scattering centers. Electrons accelerated by the applied field quickly loose their energy, which is converted into heat. Under such circumstances the electron kinetic energy remains negligible, and thus the total energy is fixed. In contradistinction, the input conductance of a loss-less transmission line of characteristic conductance $G_c$ terminated by an ideal conductance $G=G_c$ (matched load) is equal to $G$ at any frequency. But there is in that case a stored energy equal to $G\abs{V}^2\tau$, where $\tau$ denotes the transit time of a pulse along the transmission line (this is power divided by the group velocity times the line length). Thus, a matched transmission line does \emph{not} constitute an ideal conductance in the sense defined above, even though the input conductance $G$ is a real constant. 

\section{Capacitances}\label{capacitance}

A capacitance $C=\epsilon_o A/d_C$ may consist of two parallel perfectly-conducting plates of area $A$ separated by a distance $d_C\ll \sqrt A$. The constant $\epsilon_o$ in this formula is called the free-space permittivity. Its exact numerical value is given at the end of Section \ref{sec_num}. There are two wires connected respectively to the upper and lower plates of a capacitance, so that electrical charges may be introduced or removed.  If an electrical charge $q$ is displaced from the (say, lower) plate to the upper plate a potential $v$ appears between the two plates given by $v=q/C$. The energy stored in the capacitance is $E_C=Cv^2/2=q^2/(2C)$, a result obtained by considering elementary charges $dq$ being displaced from the lower to the upper plate of the initially-uncharged capacitance until a final charge $q$ is reached. If $q$ is a function of time and $C$ is kept constant, we have $v(t)=q(t)/C$. We may set $q=-Ne$, where the number $N$ of electrons is supposed to be so large that $q$ varies almost continuously. As before, $e$ denotes the absolute value of the electron charge. 

Let now $v(t)$ be of the sinusoidal form given above. Because the current $i(t)$ represents a flow of electrical charges into one plate or flowing out from the other plate, $i(t)$ is the time-derivative of the electrical charge: $i(t)=dq(t)/dt$. The relation between the complex current $I$ and the complex potential $V$, as defined above, thus reads $I=-\ii C \om V$. The admittance is in the present case $Y(\om)=-\ii C\om$. If we set $Y=G+\ii B$, we have therefore for an ideal capacitance $G=0$ and $B=- C\om$. The stored energy averaged over a period $2\pi/\om$, reads $\ave{E_C}=C\abs{V}^2/2$. We may consider a light-emitting device driven by a very large capacitance (instead of, say, a battery) with a very large initial charge $q$ such that the potential $U=q/C$ across the capacitance has the desired value, for example 1 volt. If the light-emitter operation duration is denoted by $\mathcal{T}$, the capacitance supplies a current $i$ during that time, and thus loose a charge $\De q=i\mathcal{T}$. Because $q$ is very large we have $\De q\ll q$ provided the experiment does not last too long. As a consequence the potential $U$ across the capacitance does not vary appreciably. We realize in that manner a constant-potential source, that is a source whose potential does not depend appreciably on the delivered current.

A constant-potential source at optical frequency $\om$ may be realized in a similar manner. Again the capacitance $C$ and the initial charge $q$ are supposed to be arbitrarily large, but we now allow the spacing $d_C$ between the capacitance plates to oscillate\footnote{Practically-minded readers may object that mechanical motion may not be feasible at high frequencies. Let us recall here that the numerical value of $\om$ is arbitrary. What we call "optical" frequency $\om/2\pi$ may be as low as 1Hz provided that the other frequencies considered be much lower, e.g., 1 mHz.} at the optical frequency $\om$. This spacing variation entails a fluctuation of the capacitance, and thus of the potential across the capacitance since the charge is nearly constant as was discussed above. The potential across the capacitance may be written as $U+v(t)$. The important point is that the optical potential $v(t)$ as well as the static potential $U$ are independent of the current delivered. That is, if atoms are present between the two capacitance plates, processes occurring in the atomic collection have no influence on the field. We have just described an essential component of our circuit-theory schematic. In contradistinction, the potential across a resonating inductance-capacitance circuit modeling a single-mode cavity does depend on atomic processes. For that resonator configuration the assumption that the optical field is nearly constant holds only in some limit. 

\section{Inductances}\label{inductance}

An inductance $L$ may be constructed from a conducting cylinder of area $A$ and height $d_L\gg \sqrt A$, split along its hight, so that an electrical current may flow along the cylinder perimeter. In that case $L=\mu_o A/d_L$, where $\mu_o$ denotes the free-space permeability. For an inductance $L$, the magnetic flux (or magnetic charge) $\phi(t)=Li(t)$, and the potential across the inductance is $v(t)=d\phi/dt$. It follows that for a constant $L$, $v(t)=Ldi(t)/dt$, or, using the complex notation $V=-\ii L\om I$. Thus $Y(\om)\equiv I/V=\ii/  \p L\om  \q$. The energy stored in an inductance with a current $i$ flowing through it is $E_L=Li^2/2$. For a sinusoidal current represented by the complex number $I$, the time-averaged energy is $\ave{E_L}=L\abs{I}^2/2$. If a large inductance supports a large magnetic flux, the current flowing through the inductance is nearly independent of the potential across the inductance. In that manner, we may realize constant-current sources, either static or oscillating at optical frequencies through a change of $L$ (e.g., by changing the coil length if a coil instead of a simple cylinder is employed). 

The linear relationships outlined in previous paragraphs are sometimes referred to as the "generalized Ohm laws". Let us recall that there is a well-known duality between potentials and currents and between  electrical charges (expressed in coulombs) and magnetic fluxes (expressed in webers), so that expressions obtained for capacitances may be translated into expressions relating to inductances. 

\section{Energy and power}\label{energy}

As an application of the above energy formulas, let us consider a circuit consisting of an inductance $L$ and a capacitance $C$ connected in parallel. Since the system is isolated and loss-less the total admittance must vanish and we obtain the resonance formula $LC\om^2=1$, where $\om$ denotes the resonant frequency. Since the system is isolated, the sum $E$ of the energy $E_L(t)$ located in the inductance and the energy $E_C(t)$ located in the capacitance does not vary in the course of time. This is twice the time-average energy stored in the capacitance (or inductance). Using above formulas we find that the rms (root-mean-square) field across the capacitance is 
\begin{align}
\label{field}
\mathcal{E}=\sqrt{\frac{E}{\eps_o \mathcal{V}}},
\end{align}
where $\mathcal{V}\equiv Ad_C $ denotes the capacitance volume. We later show that when a resonator such as the one presently considered is in a cold environment it eventually reaches a state corresponding to an average energy $\hbar\om/2$, where $\hbar$ denotes the Planck constant (divided by $2\pi$). According to the above formula, the so-called "vacuum (rms) field" reads $\mathcal{E}_{vacuum}=\sqrt{\frac{\hbar\om/2}{\eps_o \mathcal{V}}}$. The two oppositely-charged capacitance plates attract one another with an average force $F=d\p \hbar\om/2\q /d(d_c)=\hbar\om/\p 4d_c\q$.

If two sub-systems are connected to one another by two perfectly conducting wires with a potential $v(t)$ across them and a current $i(t)$ flowing into one of them (the current $-i(t)$ flowing in the other one), the power flowing from one sub-system to the other at some instant $t$ is equal to $v(t)i(t)$. For sinusoidal time-variations, the power averaged over an oscillation period reads $P=\Re\{V^\star I\}$. 

Finally, let us recall that at a node, that is, at the junction between perfectly conducting wires, the sum of the currents entering into the node vanishes as a consequence of the fact that the electric charge is a conserved quantity. For three wires traversed by currents $i_1(t),~i_2(t),~i_3(t)$, respectively, we have at any instant $i_1(t)+i_2(t)+i_3(t)=0$. It follows that the complex currents sum up to zero, $I_1+I_2+I_3=0$, that is, both the real and the imaginary parts of that sum vanish. Such relations are sometimes called "generalized Kirchhoff laws". The above discussion suffices to treat circuits consisting of conductances, capacitances and inductances arbitrarily connected to one another. Some circuits require a more complicated description, for example (non-reciprocal) circulators. These
components are useful to separate reflected and incident waves. 

\section{The tuned circuit}\label{tuned}

For the sake of illustration and later use, let us generalize the resonator previously considered by introducing in parallel with the capacitance $C$ and the inductance $L$ a conductance $G$. The relation between a complex current source $\C$ at frequency $\om$, supposed to be independent of frequency, and the potential $V$ across the circuit reads
\begin{align}
\label{lossreson}
V(\om )=\frac{\C}{Y(\om )}=\frac{\C}{G-\ii \p C\om-1/L\om\q}.
\end{align}

The power dissipated in the conductance $G$ at frequency $\om$ reads
\begin{align}
\label{power'}
P(\om)=G\abs{V(\om )}^2\approx \frac{G \abs{\C}^2}{G^2+4C^2\p \om-\om_o\q^2}
\end{align}
in the small-loss approximation, where $\omega_o\equiv 1/\sqrt{LC}$. Thus $P(\om)$ drops by a factor of 2 from its peak value when $2C\p \om_±-\om_o\q=±G$. The full-width at half power (FWHP) $\de\om$ of the resonance that is, the difference of (angular) frequencies at which the dissipated power drops by a factor of two, is therefore
\begin{align}
\label{width}
\de \om= \om_+-\om_- =\frac{G}{C}\equiv \frac{1}{\tau_p},
\end{align}
where $\tau_p=C/G$ is called the "resonator lifetime". If the resonator is left alone in a cold environment ($T$=0K), at the classical level its energy decays according to an $\exp(-t/\tau_p)$ law. 

For a Fabry-Pérot resonator with mirrors of small power transmissions $T_1,~T_2$, respectively, and spacing $L$, we have
\begin{align}
\label{fp}
\frac{1}{\tau_p}=\frac{T_1+T_2}{2L/v},
\end{align}
where $v$ denotes the group velocity and $2L/v$ is the round-trip time.

The energy contained in the resonator is twice the average energy contained in the capacitance whose expression was given earlier. We then obtain in the small-loss approximation
\begin{align}
\label{en'}
E(\om)=C\abs{V(\om )}^2=  \frac{C\abs{\C}^2}{G^2+4C^2\p \om-\om_o\q^2}\approx \frac{\tau_p \abs{\C}^2/G}{1+x^2},
\end{align}
where $x\equiv2\tau_p \p \om-\om_o\q$. 

\section{Derivative of admittance with respect to frequency}\label{derivative}

For late use, note the expression of the derivative with respect to $\om$ of the admittance of a linear circuit, submitted to a voltage $V$
\begin{align}
\label{dispers}
\ii V^2\frac{dY(\om)}{d\om}=-\ii I^2\frac{dZ(\om)}{d\om}=\sum_{k}{C_k V_k^2-L_k I_k^2},
\end{align}
where the sum is over all the circuit capacitances and inductances. $V_k$ denotes the (complex) voltage across the capacitance $C_k$ and $I_k$ the (complex) current flowing through the inductance $L_k$. The circuit resistances or conductances do not enter in the sum. This relation is readily verified for an inductance in series with a resistance and a capacitance in parallel with a conductance. Thus the relation holds for any combination of elements, connected in series and in parallel.

\section{Integral relation}\label{summ}

Let $G(\om)$ denote the conductance of a linear circuit. We have the 
\begin{align}
\label{intrel}
\int_0^\infty d\om R(\om)=\frac{\pi}{2 ~C_{HF}},
\end{align}
where $C_{HF}$ denotes the high-frequency capacitance.

\section{Matrix formulation}\label{matrix}

For an arbitrary circuit, the task is to "extract", figuratively speaking, the (positive or negative) conductances from the given circuit, each conductance being connected to the conservative circuit that remains after extraction of the conductances. If $N$ (positive or negative) conductances are involved, the circuit becomes an $N$-port conservative device. For an $N$-port circuit, we define the vectors $\boldsymbol{V}\equiv [V_1, V_2,...V_N]^t$ and $\boldsymbol{I}\equiv [I_1, I_2,...I_N]^t$ where the upper $t$ denotes transposition. The linear relation is written in matrix form $\boldsymbol{I}=\boldsymbol{Y}(\om)\boldsymbol{V}$, where $\boldsymbol{Y}(\om)$ is called the circuit admittance matrix. For a conservative circuit the total entering power $\Re\{ \boldsymbol{V}^{t\star} \boldsymbol{I} \}=0$. Since this relation must hold for any source this implies that $\boldsymbol{Y}^{t\star}+\boldsymbol{Y}=\boldsymbol{0}$.

It is convenient to view the connections between the conservative circuit and the conductances as ideal transmission lines of small length and characteristic conductances $G_c$. Supposing that $G_c=1$, the potential $V$ across one of the transmission lines and the current $I$ flowing through the (say, upper) wire, are combined into an ingoing wave whose amplitude is defined as $a=V+I$ and an outgoing wave defined as $b=V-I$. Since under our assumptions the circuit elements are linear, there is a linear relationship between the $a$-waves and the $b$-waves. The relation between $\boldsymbol{b}$ and $\boldsymbol{a}$, defined like $\boldsymbol{I}$ and $\boldsymbol{V}$ above, may be written in matrix form as $\boldsymbol{b}=\boldsymbol{Sa}$, where the $\boldsymbol{S}$ matrix is called the circuit "scattering matrix". Because the circuit is conservative, the outgoing power equals the ingoing power. It follows that the $\boldsymbol{S}$-matrix is unitary, i.e., $\boldsymbol{S}^{t\star} \boldsymbol{S}=\boldsymbol{1}$. We need not assume that the circuit is reciprocal, however, that is, the $\boldsymbol{S}$-matrix needs not be symmetrical.

\section{Various circuits}\label{sec_various}

We have represented a number of important conservative (loss-less, gain-less) components in either their circuit form, their microwave form, or their optical form in Fig.~\ref{circuitfig}. The origin of the differences is that, as one goes to shorter wavelength (higher frequencies) some circuit elements become too small to be fabricated. It should also be noted that metals, such as copper, that are excellent electrical conductors up to microwave wavelengths, do not behave as electrical conductors any more at optical wavelengths because of electron inertia. On the other hand, while it is difficult to find very low-loss dielectrics at microwave frequencies, extremely low-loss glasses exist at optical frequencies. Fig.~\ref{circuitfig} in c-h) represents four resonating circuits, that one may call "0-dimensional" devices. Namely, the inductance-capacitance circuit employed up to about 100 MHz, the cavity employed in reflex klystrons and masers for example, the Fabry-Perot resonator consisting of two mirrors facing each other, and the whispering-gallery-mode dielectric resonator, first demonstrated in the microwave range and now-a-days employed in the optical range. Resonators are primarily characterized by their resonant frequency $\om_o$. Small losses may be characterized by the so-called resonator lifetime $\tau_p$, defined earlier. When the resonator size is large compared with wavelength many resonating modes may be present. In most applications it is desirable that only one of them be loss-less, or nearly so.

Figure~\ref{circuitfig} in i-l) represents four one-dimensional devices called "transmission lines". The circuit form is a periodic sequence of series inductances and parallel capacitances. The microwave form consists of two parallel conductors characterized by a characteristic conductance $G_c$, with waves propagating at the speed of light. Above 1GHz one would rather use waveguides. The optical form is the now-a-day well-known \emph{optical fiber}. A glass fiber (core) in vacuum may guide optical waves by the mechanism of total reflexion. In order to increase the core size without having spurious modes propagating, the core is usually immersed into a lower-refractive-index glass.

Other useful devices are shown in Fig.~\ref{circuitfig} m-p). The power carried by a transmission line of characteristic conductance $G_c$ may be split into two parts simply by connecting it to two transmission lines whose characteristic conductances sum up to $G_c$. The connecting circuit is a three-port reciprocal conservative device. Alternatively, when two transmission lines are put side by side and coupled at two locations separated by a quarter of a wavelength, some of the power incident on a transmission line is transmitted into the other one. This device is called a directional coupler, see Fig.~\ref{circuitfig} n). This 4-port device may be reduced to a 3-port device by putting a matched load at the end of one of the transmission lines. The optical form of a directional coupler is called a beam-splitter, which may simply consist of a flat piece of glass, see Fig.~\ref{circuitfig} o). An important non-reciprocal 3-port device is the \emph{circulator}, which exists in microwave and optical versions. It is intrinsically loss-less: a wave entering into port 1 entirely exits from port 2, a wave entering into port 2 entirely exits from port 3, and a wave entering into port 3 entirely exits from port 1, see Fig.~\ref{circuitfig} p). Such a device is convenient to separate reflected waves from incident waves without introducing losses.

\section{Thin slab with gain}\label{thin}

We consider isotropic materials described by bulk propagation constant $k\equiv (\om/c) n$, and set $n\equiv n'+\ii n''$, where $n'$ represents the real part of the refractive index and $n''$ the imaginary part.  Since the field varies according to $\E(z)=\E(0)\exp(\ii k z)\equiv \E(0)\exp[\ii (k'+\ii k'') z]\equiv \exp[(\om/c)(\ii n'z-n'' z)]$, a medium with loss corresponds to $n''>0$ and a medium with gain corresponds to $n''<0$.

\paragraph{Dielectric waveguide.}

A dielectric waveguide is a cylindrical dielectric embedded into a lower refractive index dielectric. We denotes by $z$ the cylindrical axis. Neglecting dielectric losses and irregularities, a number of modes may be found with real propagation constant $k_{zi}, ~i=0,1,2...$. The power carried by these modes, that one may call "trapped modes", is finite. The field $\E(y)$, where $y$ denotes the transverse coordinate, may be taken as a real function of $y$. (There exist also "radiation modes" that are required to make the mode expansion complete. These will not be considered further here). A particularly interesting wave-guide is the thin slab configuration, analogous to the $\delta(x)$ potentials considered in quantum mechanics. This slab is made up of a thin sheet of loss-less dielectric with $n>1$, immersed into free space. Such a slab supports a single trapped mode, and the field is of the form $\E(y)\propto \exp(-\abs{y})$. 

\paragraph{"Inverted" waveguide.}

In the case of "inverted" waveguides, the inner slab has a refractive index lower, rather than higher, than that of the surrounding medium. Only "leaky modes" then may propagate. Such modes get attenuated along the $z$ propagation axis and the field grows to arbitrary large values in the transverse direction $y$. Accordingly, these modes carry, in principle, infinite power, yet make sense physically near a source.

\paragraph{Waveguide with gain.}

If we now allow the inner slab of an inverted waveguide (inner $n'$ less than outer $n'$, outer $n''=0$) to have gain, that is to have inner $n''<0$, the guide may exhibit gain above some threshold value. This is of course necessarily the case for laser oscillators and amplifiers. Such devices are called "gain-guided amplifiers or oscillators". By adding gain to the inner region, the leaky mode with loss has been converted into an amplified wave. It has not always been recognized that the guided wave then exhibits a nearly usual behavior. That is, the field decays exponentially as one moves away from the inner region, and the mode carries a finite power. But the power grows along the $z$ coordinate, and the wavefront is curved and diverging. If the field is denoted as above by $\E(y)$, where $y$ denotes the transverse coordinate, $\E(y)$ may no longer may be taken as a real function of $y$. Note that this conclusion (complex $\E$-field) applies as well, strictly speaking, to the case where the inner $n'$ \emph{exceeds} the outer $n'$. In that case the inner slab gain may, however, be viewed as a small perturbation of the normally guiding waveguide. From a fundamental view-point, as soon as gain entails that the wave is growing in the $z$ direction, there an no basic difference between inverted and normal waveguides. 

An important parameter is
\begin{align}
\label{gainguided}
K \equiv \abs{ \frac{\int_{-\infty}^{\infty}\abs{\E^2(y)}dy}{\int_{-\infty}^{\infty}\E^2(y)dy}}^2≥1.
\end{align}
In the linear regime (but not above threshold), laser linewidths are enhanced by the $K$-factor. There are more general expression for $K$ applicable to arbitrary media with in-homogeneous gain and loss.

There is a simpler picture involving thin slabs that clarifies the gain-guidance issue. One may transform the configuration described above through a series of steps that do not spoil the basic properties, but enable us to proceed to above-threshold lasers. First, the laser end-faces at $z=0$ and $z=L$ are supposed to be made fully reflecting. The generated power is now dissipated into the outer medium, whose $n''$ is made slightly positive. (Admittedly, such a laser is not a very useful device since the generated power is dissipated internally, instead of propagating outside. But practical considerations are not our main concern at the moment). We have now reduced the initial two-dimensional ($z,y$) configuration into a one-dimensional ($y$) configuration. Indeed, we may simply require that there is an integral number of half wavelengths from $z=0$ to $z=L$, and then ignore the $z$ coordinate. Furthermore, we may reduce the inner slab to a thin slab with gain, immersed into an outer medium with small losses. We may then define an optical potential equal to the product of $\E(y)$ and the width of the device in the $y$ direction, which we take as unity for simplicity. The thin slab is then represented by an admittance $Y_e(\om)$ with negative conductance part, say $-G_e$. The outer medium is a transmission line with loss, whose input admittance consists of a positive frequency-dependent conductance $G(\om)$ and a susceptance $B(\om)$ . Finally, the gain-guided configuration has been reduced to a simple (zero-dimensional) circuit. Analysis of the response of that circuit to random current sources $c(t)$ leads to a linewidth-enhancement factor which coincides with the $K$-factor whose expression has been given above. It thus appears that the original gain-guided configuration is equivalent to a circuit with the negative conductance separated from the positive conductance by a reactive element.

Once this has been done, we may go on to the above-threshold regime, something that could hardly be done by direct analysis of the initial configuration. Above threshold the linewidth enhancement is not given by $K$ anymore, but by a more complicated expression given in Part II, which involves the phase-amplitude coupling factor $\al$. This factor describes how the active material refractive index varies when the material gain varies.

\newpage 

\chapter{Statistical Mechanics}\label{various}

Aside from the law of conservation of energy, the fact that electrons have the same (conserved) electrical charge $e$, the Schroedinger equation and the Pauli principle (if the detailed electronic properties of semiconductors are thought for), the theory presented in this book rests on the introduction of random current sources or random rates. To any conductance $G$ one associates a complex random current $C(t)\equiv C'(t)+iC''(t)$, where $C'(t)$ and $C''(t)$ are uncorrelated and have spectral density $\hbar\omega \abs{G}$, where $\omega$ denotes the operating (angular) frequency. Alternatively one may consider emitted or absorbed rates whose fluctuation spectral densities are equal to the absolute values of the average rates. Sources of different locations or, e.g., collections of atoms in the absorbing state and collections of atoms in the emitting state (which may be represented respectively by positive conductances $G_a$ and negative conductances $-G_e$ connected in parallel) are uncorrelated. These conclusions rest fundamentally on Statistical Mechanics, whose historical origin is the Carnot laws of Thermodynamics. We therefore first present in this chapter a simple derivation of those laws. 

We begin with an historical introduction concerning heat. At the Carnot time, the working agent, that is the material being displaced from a heat reservoir to another at a different temperature while some parameter is being varied, was generally air, contained in a cylinder of variable length. The Carnot finding is that heat-engine efficiencies reach a maximum value depending only on the two bath temperatures when the engine is reversible, and that, for slow processes, the only source of irreversibility is the contacting of bodies at different temperatures. 

We first show that (following Boltzmann) the Carnot discovery concerning heat engine efficiencies and the work produced per cycle, may be viewed as elementary consequences of the concept of probability, no empirical result being needed, using the model of two reservoirs containing identical balls in the earth gravitational field. We obtain straightforwardly the Carnot expressions for average work and efficiency, and the work variance. Note however that the result is directly applicable to the Otto cycle employing as working agent a two-level system. This cycle may stop delivering work even when the two bath temperatures are different.

We next consider more general quantized working agents, involving more than two levels. The Carnot result remains applicable provided a well-defined temperature may be ascribed to such quantized systems. Clearly, this is not the case in general. If a quantized system is in contact with a heat bath at temperature $T$, and a single electron is considered, the probability that a particular level of energy $E_k$ be occupied is proportional to $\exp(-\beta E_k)$ where $\beta$ denotes the temperature reciprocal $1/T$, according to the Boltzmann law. When this quantized system is separated from the bath and allowed to evolve slowly (adiabatic transformation), the level occupations do not vary, and therefore they are unlikely to correspond to any temperature. We will restrict ourselves to quantized systems where this difficulty does not occur. Particularly to the case of evenly-spaced energy levels $E_k=k\epsilon, ~k=...,-1,0,1,2...$, where $\epsilon$ may be allowed to vary, and some states may be forbidden. 

The simplest case is that of a single electron with two levels, akin to the reservoir model mentioned above. This situation is often referred to as a two-level atom gas. We briefly consider a system having $B$ levels, e.g., a bismuth nucleus in a magnetic field. The case where $E_k=(\frac{1}{2}+k)\epsilon, ~k=0,1,2...$ corresponds to the quantized harmonic oscillator. A summation gives the average energy at temperature $T$. Next we consider evenly-spaced levels $E_k=k\epsilon, ~-\infty<k<\infty$ and many electrons (sometimes referred to as a "Fermi sea". For a given spin state, there may be only one electron at each level, according to the Pauli principle. It is supposed that at $T=0$K, electrons fill up the lower states up to some level called the Fermi level. If an energy $r$ is added, some of the electrons must move upward on the energy scale, so that the sum of their upper displacements be $r$ setting $\epsilon=1$. The entropy $S(r)$ is the logarithm of the number of partitions $W(r)$ of $r$. The temperature reciprocal $\beta(r)\approx S(r+1)-S(r) $. Averaging over energy and over the electron number give the canonical and grand-canonical distributions, respectively. The grand-canonical Fermi-Dirac distribution follows from these considerations. Semiconductors, in which a range of energies is forbidden (band gap) may be treated similarly. When the material is "pumped", that is, when $n$ electrons are being transferred from the valence band to the conduction band, the system is no longer in a state of thermal equilibrium. Each band, however, is in equilibrium and (quasi) Fermi levels may be defined for the two bands. Equilibrium within each band may be the result of electrons moving up or down within each band, ending up to two possibly different temperatures. Thermal contact of the electron gas with the crystal and of the crystal with a heat sink may ensure that both temperatures are the same as that of the heat sink when the energy exchanges do not occur too quickly.

The laws of Statistical Mechanics apply primarily to isolated systems. An important question is: how does the isolated system evolve in the course of time, starting from some initial conditions that do not correspond to the equilibrium situation? The answer was given by Einstein in terms of atomic transition probabilities per unit time. Likewise, we consider atoms resonant with a (single mode) optical resonator at frequency $\om$, i.e., atoms whose level energy difference equals $\hbar\om$. This energy is set as unity for convenience, by a proper choice of units. We call $m$ the integer part of the optical resonator energy $E$ that is, the largest integer less than $E$. Whenever an atom jumps from the emitting to the absorbing state, the resonator energy is incremented by one and conversely whenever an atom jumps from the absorbing to the emitting state, the resonator energy is decremented by one. If follows from the Boltzmann distribution relating to the atoms that the probability that the optical resonator energy be comprised between $E$ and $dE$ is proportional to $\exp(-\beta m)dE$. As far as the average value of the resonator energy is concerned, this distribution gives the expression obtained by the Quantum Optics methods that treat the optical resonator as a quantized harmonic oscillator. But the variance of the energy derived from the expression $p(E)\propto\exp(-\beta m)$ gives the classical expression in the high-temperature limit, while the variance evaluated from the Quantum Optics expression does not. Note that $m$ is an integer from its very definition, not as a result of field quantization. This being made clear, we find it convenient to say that, in the present case the resonator contains $m$ "light quanta". The probabilities per unit time that upward or downward atomic transition occur are found to be both proportional to the arithmetic mean of the number of light quanta before and after the jump. 

Atoms perform jumps because they can be observed (through, e.g., methods of selective ionization) only  in one state or another. The Schrödinger equation describes electrons submitted to a static potential and a term representing the optical field supposed to be prescribed, that is, to be independent of the electron motion, a condition fulfilled when $m$ is a large number. We then find the probability that the electron be found in a particular state as a function of time (Rabi equations). In order to describe jumps, however, one must introduce a phenomenological parameter denoted here $\gamma$. One may then evaluate the probability that, given that a downward event occurred at $t=0$, an upward event occurs at time $t$. This is the waiting time, evaluated in a subsequent chapter, with methods similar to those employed to treat resonance fluorescence.

The system of major interest in this book, namely a laser with a pump rate $J$ and a detection rate $D$, is solved with the help of rate equations. These equations involve the transition probabilities per unit time whose expressions follow, as we just said, from the properties of isolated systems.

\section{History of the Carnot discovery}\label{Chistory}

Some notions concerning heat were probably acquired by humans at a very early time. Particularly the fact that heat flows from hot bodies to cold bodies but that the converse never occurs spontaneously. This is a fundamentally non-reversible process. Eventually the two bodies approach a state of thermal equilibrium, that is, reach some intermediate temperature. The first heat engine was probably invented around 100 AD by Hero of Alexendria, in the form of a sphere partially filled with water and with an oblique outlet letting the water vapor escape. Daniel Bernoulli offered in 1738 the following picture: "Air pressure increases not only by compression but also by heat supplied to it, and since it is admitted that heat may be considered as an increasing internal motion of the particles, it follows that this indicates a more intense motion in the air particles". Rumford noted that gun drilling with degraded tools increases the metal temperature without affecting it otherwise. The speed of sound in air calculated by Newton under the assumption of a constant temperature was in error by some 10 per cent. A good agreement with observation was subsequently obtained by Laplace, who supposed instead that the air compression process is adiabatic, that is assuming that there is no heat flow between adjacent air layers. A number of practical heat engines were fabricated in the 18\textsuperscript{th} century and employed, particularly in mining, but they had poor efficiencies. As far as the history of the discoveries preceeding the Carnot work is concerned one should mention the Boyle-Mariotte empirical law $PV=T$ (omitting constants), where $V$ denotes the volume of a container, $P$ the pressure exerted on the wall, and $T$ the absolute temperature. If $V$ changes while the temperature remains constant (isothermal regime) it is straightforward to evaluate the work $W$ performed as the integral of $P ~dV=T~dV/V$, $W=T\log(V_f/V_i)$, where $i,f$ refer to initial and final values, respectively. Finally, it was found (in a different form) that an helium atom has an energy $U=3\frac{T}{2}$ at a temperature $T$, where the number 3 corresponds to the atom translational degrees of freedom. (It is only at very high temperatures that the electrons degrees of freedom should be considered). If the vessel is isolated (adiabatic process) the elementary work is opposite to the change in energy $dW=P~dV=-dU=-\frac{3}{2}dT=-\frac{3}{2}(P~dV+V~dP)$, where the previous relation $PV=T$ has been used. Integration gives $PV^{5/3}$=constant. We may therefore evaluate the work performed in an adiabatic change of $V$, and the final temperature of the gas. More generally, if $f$ denotes the number of degrees of freedom in some temperature range, we find the relation $PV^{\gamma }$=constant, where $\gamma\equiv 1+\frac{2}{f}=\frac{C_P}{C_V}$, where $C_V=f/2$ and $C_P=C_V+1$ denote respectively the specific heats (energy increments per unit temperature increment) at constant volume and constant pressure, respectively. 

Carnot then undertook to understand the principles ruling the operation of heat engines and discovered around 1824 the first and second laws of Thermodynamics. Heat engines operate by retrieving heat from a hot bath and delivering a lesser amount of heat to a cold bath. Their efficiency is defined as the ratio of the work performed and the higher temperature reservoir heat consumption, both being expressed in the same energy units, e.g., in joules. The first law asserts that heat is a form of energy and that isolated systems energies do not vary in the course of time. Let us cite Carnot\,\cite{Kastler1976}: ``Heat is nothing but motive power, or rather another form of motion.  Wherever motive power is destroyed, heat is generated in precise proportion to the quantity of motive power destroyed; conversely, wherever heat is destroyed, motive power is generated''. Carnot calculated that 1 calorie of heat is equivalent to 3.27 J, instead of the modern value: 4.18 J. He proved that engine efficiencies reach their maximum value when they are reversible, from the consideration that energy cannot be obtained for free. He noted that non-reversibility originates only from the contacting of two bodies at different temperatures, acknowledging that a small temperature difference had nevertheless to be tolerated. He discovered a cycle in which temperature differences between contacting bodies could be made arbitrarily small, ending up with the celebrated ``Carnot cycle''. He established that the maximum efficiency of heat engines operating between baths at absolute temperatures $T_l$ and $T_h$, respectively, is $\eta=1-T_l/T_h$, and that this efficiency may be reached under idealized conditions, irrespectively of the working agent employed (Carnot pointed out that, however, for practical reasons, it would be a very poor choice to employ the thermal expansion of solids or liquids to fabricate heat engines). He found that the work performed is $W=(T_h-T_l)S$, where $S$ denotes the entropy ("calorique" in Carnot terminology) transferred from the hot bath to the cold bath. To define entropy in practical terms, note that when an amount of heat $\Delta Q$ is added to a constant-volume body at temperature $T$, the body-entropy increment is $\Delta S=\Delta Q/T$, the variation of $T$ being considered insignificant. If the body constant-volume heat capacity $C_V(T)\equiv \Delta Q/\Delta T$ is known from measurements over a broad range of temperatures, $S(T)$ may be obtained by integration, to within an arbitrary additional constant. It has been recognized later on by Nernst that, in general, $S(0)=0$.

Even now the importance of the Carnot contribution is not fully appreciated.  It is often said that Carnot discovered the second law of Thermodynamics while ignoring the first one, and furthermore, that he confused entropy and heat. One reason for the misunderstanding is that part of the Carnot contribution appeared in print only decades after his early death. A second one is that his work was popularized by Clapeyron in a partly erroneous manner. A third one is the unfortunate use by Carnot of the word ``calorique'' to designate what Clausius later on called ``entropy''. The word ``calorique'' had been formerly employed by Lavoisier to designate some hypothetical heat substance. Let us quote Zemansky and Dittman:\,\cite{PRE9} ``Carnot used \emph{chaleur} when referring to heat in general, but when referring to the motive power of fire that is brought about when heat enters an engine at high temperature and leaves at low temperature, he uses the expression \emph{chute de calorique}, never \emph{chute de chaleur}.  Carnot had in the back of his mind the concept of entropy, for which he reserved the term \emph{calorique}''. 

We are aware of approximately ten papers, written from 1919 up to this time, concerning the early history of Thermodynamics, that attempted to correct this situation. A. Kastler in a colloquium\,\cite{Kastler1976} and the russian scientist Brodiansky in a recent book "Sadi Carnot"\,\cite{PRE10}, maintain that Carnot discovered both the first law and the second law of Thermodynamics. Another paper is by La~Mer (1954) who expressed himself forth-fully as follows: ``Unless the view-point that the Carnot theory is accurate is adopted, one is placed in the position of maintaining that Carnot succeeded in demonstrating some of the most fundamental and profound principles of physical science by the most masterly display of scientific double-talk that has ever been perpetrated upon the scientific world. This view is untenable''. The historian of science R. Fox says: ``Until recently there were very few studies concerning [the physics of Carnot reflexions]. Thanks to the work of Hoyer\,\cite{hoyer}, we now have papers on the logical implications of the Carnot theory, and its analogy with modern thermodynamics. It is not at all obvious to understand how Carnot discovered the mechanical equivalent of heat. Hoyer articles provide complete references to earlier attempts. He explains the exactness of Carnot calculation (which is even more striking if one uses modern values for the specific heats) by noticing that the Carnot theory is entirely accurate''.

The next giant step concerning heat processes was made near the end of the 19\textsuperscript{th} century by Boltzmann, who introduced the concept of probability. Previous work in that direction had been made by Bernoulli and Maxwell.

\begin{figure}
\centering
\includegraphics[scale=1]{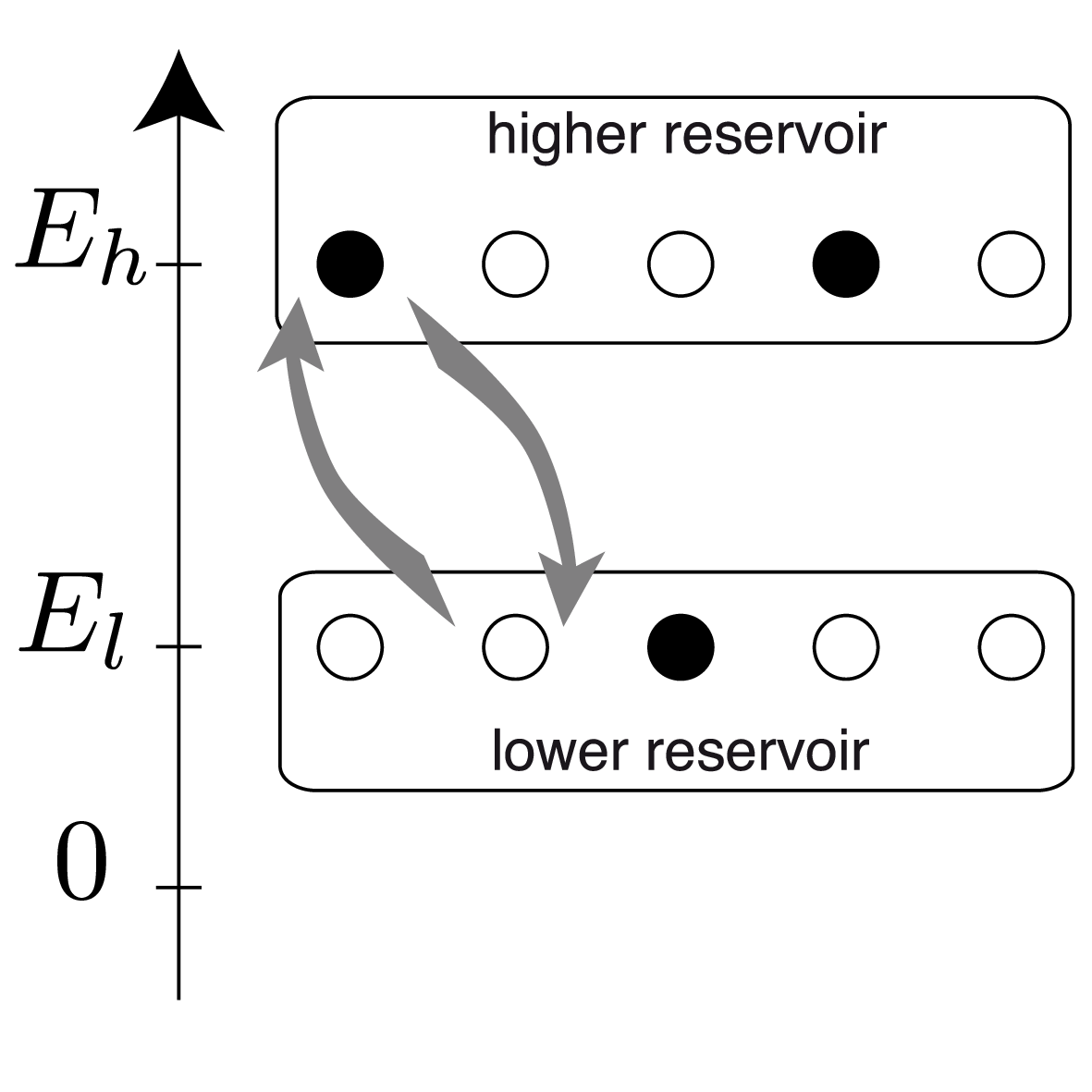}
\caption{Schematic representation of an engine that converts potential energy into mechanical work. The figure represents two reservoirs located at low and high altitudes, $E_l,~E_h$ respectively, with $N_l,~N_h$ possible ball locations ($N_l=N_h=5$ in the figure). The number of weight-1 balls (black circles) is $n_l$ in the lower reservoir, and $n_h$ in the higher reservoir (with $n_l=1,~n_h=2$ in the figure). For each reservoir, every ball configuration is equally likely considering that the energies are the same. A cycle consists of exchanging two randomly picked up balls from one reservoir to the other. Alternatively, one may view the dark circles as electrons in a static field. This figure may then be viewed as representing five atoms numbered from 1 to 5 from left to right. 1 and 4 are in the emitting state, 3 is in the absorbing state, and 2 and 5 are inactive. A Rabi oscillation may convert atom 1 from one state to the other, thereby incrementing or decrementing the resonator energy.}
\label{fig}
\end{figure}

\section{Exchange of balls between two reservoirs}\label{exchange}

Consider two reservoirs at altitudes $E_{l}$ and $E_{h}$ respectively, with $E_{h}>E_{l}$ by convention, as shown in Fig. \ref{fig}. There are $N_l$ possible ball locations in the lower reservoir and $n_l < N_l/2$ weight-one balls. Likewise, there are $N_h$ possible ball locations in the higher reservoir and $n_h < N_h/2$ weight-one balls\footnote{We require that $n < N/2$ so that temperatures are positive.}. A cycle consists of exchanging randomly-picked balls between the two reservoirs\footnote{More formally, a cycle consists in picking in each reservoir one location, uniformly at random and independently, then exchanging the content of these locations. 
If $n \le N$ denotes the number of balls in a given reservoir, it is clear that the probability of picking up a ball inside is $n/N$. Another setting consists in picking a \emph{location} in each reservoir as above, removing the balls if any, and picking a \emph{hole} at random in the other reservoir to receive each removed ball. It is not difficult to see that the two settings are strictly equivalent. In the sequel, we simply say that two random balls are exchanged between the two reservoirs.}.

The energy of  a reservoir at altitude $E$ containing $n$ weight-1 balls is obviously $Q=nE$ (kinetic energy being not considered, the total energy coincides with the potential energy). The letter $Q$ is employed anticipating a correspondence with heat. When a weight-1 ball is added to a reservoir at altitude $E$ the reservoir energy is incremented by $E$. On the other hand, if a ball is randomly picked up from the $N$ locations of a reservoir containing $n$ weight-1 balls and subsequently carried to a reservoir at altitude $E$, the latter reservoir average energy is incremented by $\Delta Q=E~ n/N$. The word ``average'' is omitted for brevity when confusion is unlikely to arise.

Consider now two such reservoirs. One at altitude $E_l$ (lower reservoir) and containing $n_l$ weight-1 balls. The other at altitude $E_h$ (higher reservoir) containing $n_h$ weight-1 balls. A cycle consists of exchanging two random balls between the two reservoirs. From what has just been said and if we set $l\equiv n_l/N_l$, $h\equiv n_h/N_h$, the energies added to the lower and higher reservoirs read respectively 
\begin{align}
\label{a} 
\Delta Q_{l}&=E_{l}(h-l), \nonumber \\
\Delta Q_{h}&=-E_{h}(h-l).
\end{align}
The work performed follows from the law of conservation of energy
\begin{align}\label{b}
 W=-\Delta Q_{l}-\Delta Q_{h}=(E_{h}-E_{l})(h-l).
\end{align}
The engine efficiency, defined as the ratio of the work performed $W$ and the energy $-\Delta Q_{h}$ lost by the higher reservoir, is therefore 
\begin{align}
\label{c} 
\eta\equiv \frac{W}{-\Delta Q_{h}}=1-\frac{E_{l}}{E_{h}}.
\end{align}
The purpose of the next section is to relate the engine described above to heat engines. We will show that when $h\approx l$ the efficiency given in \eqref{c} coincides with the Carnot efficiency and the work given in \eqref{b} coincides with the expression given by Carnot.

\section{Heat engines}\label{heateng}

We consider again reservoirs containing $N$ locations and $n$ weight-1 balls. To relate this device to \emph{heat} engines, let us first recall that the number of ball configurations in a reservoir is $N!/n!(N-n)!$. For example, if $N=3$ and $n=1$, there are 3!/1!2!=3 configurations, namely
$(\bullet\circ\circ)$, $(\circ\bullet\circ)$ and $(\circ\circ\bullet)$. Next, we define the entropy as the logarithm of the number of configurations, the Boltzmann constant being set equal to unity, that is, for a reservoir, 
\begin{align}
\label{d} 
S(n)=\ln \left(\frac{N!}{n!(N-n)!}\right).
\end{align}
Note that 
\begin{align}
S(n+1)-S(n)&=\ln \left(\frac{N!}{(n+1)!(N-n-1)!}\right)-\ln \left(\frac{N!}{n!(N-n)!} \right)\nonumber \\
&=\ln \left(\frac{N-n}{n+1} \right)\approx \ln\left(\frac{N}{n}-1\right),\label{e}
\end{align}
for large $n$.

The absolute temperature of a reservoir is then defined as
\begin{align}
\label{f} 
T(n)&=\frac{Q(n+1)-Q(n)}{S(n+1)-S(n)}\approx \frac{E}{\ln(\frac{N}{n}-1)}.
\end{align}
Temperature is an \emph{intensive} quantity. For example, the temperature of two identical bodies at temperature $T$, considered together, is again $T$. Because heat has the nature of an energy and is an extensive quantity, it is required that $S$ be also an extensive quantity. Since the number of configurations in two separate bodies is the \emph{product} of the configurations (for each configuration of one body one must consider all the configurations of the other body) and the logarithmic function has the property that $\ln(ab)=\ln(a)+\ln(b)$, the above definitions do ensure that $T$ be an intensive quantity. Note that we have chosen a temperature unit such that the Boltzmann constant $k_B$ be unity. By doing so the distinction between extensive and intensive quantities drops out of sight. For example, the energy of a single-mode oscillator $E=k_B T$ reads in our notation $E=T$. However, the distinction may be restored, while keeping $k_B=1$, by writing $E=T\times$ the number of modes. The number of modes depends on volume, while $T$ does not.

The cycle efficiency given in \eqref{c} may now be written in terms of temperatures as
\begin{align}
\eta=1-\frac{E_{l}}{E_{h}}
=1-\frac{T_l}{T_h}\frac{\ln(\frac{1}{l}-1)}{\ln(\frac{1}{h}-1)}\label{g} .
\end{align}
Thus, when $l\approx h$, the last fraction in the above equation drops out and the Carnot efficiency is indeed obtained.  In the limit $l\approx h$ the work $W$ produced per cycle is very small. However, one may always add up the work contributions of any number of similar devices having the same reservoir temperatures (but possibly different values of $E, ~ n$), and achieve any specified work at the Carnot efficiency.

The ball exchange discussed above may increment the reservoir entropies. The number of balls in a reservoir may indeed be incremented by one, remain the same, or be decremented by one. From what was said before, the probability that a weight-1 ball be transferred from the high reservoir to the lower one is $h\equiv n_h/N_h$, and the probability that a weight-1 ball be transferred from the low reservoir to the higher one is $l\equiv n_l/N_l$. Since these events are independent, the lower reservoir entropy increment reads
\begin{align}\label{h}
\Delta S_l=h(1-l)[S(n_l+1)-S(n_l)]+l(1-h)[S(n_l-1)-S(n_l)].
\end{align}
Using \eqref{e} we obtain
\begin{align}\label{i}
\Delta S_l=(h-l)\ln\left(\frac{1}{l}-1\right) .
\end{align}
The increment of the higher reservoir entropy is obtained by exchanging the $h$ and $l$ labels in the above expression, that is
\begin{align}\label{j}
\Delta S_h=-(h-l)\ln\left(\frac{1}{h}-1\right) .
\end{align}

We thus find that, in the limit $n_l/N_l\approx n_h/N_h$ (or $l\approx h$), 
$\Delta S_l\approx -\Delta S_h$ so that there is no net entropy produced. 
Entropy is just carried from the higher reservoir to the lower one.
The Carnot expression for the work recalled in Section 2 may thus be written as
\begin{align}
W=(T_h-T_l)\Delta S_l=-(T_h-T_l)\Delta S_h&\approx \left( \frac{E_h}{\ln\left(\frac{1}{h}-1\right)}- \frac{E_l}{\ln\left(\frac{1}{l}-1\right)} \right)\Delta S_l  \nonumber \\
&\approx (E_h-E_l)(h-l) \label{k},
\end{align}
so that the Carnot general formula for $W$ recalled in Section 2 indeed coincides with the expression for the work performed per cycle evaluated for our model from simple reasoning. 
More precisely, the ratio of the total entropy produced to the work produced tends to zero as $h\to l$, as we explain in the next section.

The final expression tells us that the engine, according to our model, delivers work only if the terms $E_h-E_l$ and $h-l$ are both positive or both negative. But since $E_h>E_l$ by convention, this implies that we must have $h>l$. Going back to the expression of the temperature in \eqref{f}, and remembering that the $\ln(.)$ function is a monotonically increasing function of its argument, we find that work may be produced only if $T_h>T_l$. Whenever $T_h>T_l$ there exist heat engines that may deliver work. But this is not so for every heat engine. In particular, we will later on discuss the properties of an Otto heat engine that stops delivering work when a condition similar to our $l=h$ condition holds, even though $T_h>T_l$.

Conventional heat engines operate with two large baths, or reservoirs, one hot and one cold. Because these baths are not infinite in size, cycle after cycle, the hot bath cools down and the cold bath warms up. Eventually no work is being produced. The same situation occurs in our model. After a very large number of cycles the values of $h$ and the value of $l$ tend to coincide and no work is being produced any more. Because the reservoir temperatures do not equalize, however, one may say that the system has then reached a state of equilibrium, but not a state of \emph{thermal} equilibrium. This is not a peculiarity of our model, but a general property of some heat engines.

\section{Fluctuations}

Laser noise is related to fluctuations. We are thus interested in the fluctuations (precisely the variances) of the quantities considered above, namely the work produced and the total entropy produced. 

Recall that in our model a cycle consists of exchanging simultaneously a ball from the higher reservoir (at altitude $E_h$ and containing $n_h$ balls) and a ball from the lower reservoir (at altitude $E_l$ and containing $n_l$ balls). The probability that a ball be picked up from the higher reservoir is $h\equiv n_h/N_h$ (and therefore the probability that no ball be picked up is $1-h$), The probability that a ball be picked up from the lower reservoir is $l\equiv n_l/N_l$ (and therefore the probability that no ball be picked up is $1-l$). The two events are independent. Recall that $N_l,N_h$ are the numbers of possible ball locations in the reservoirs.

Setting $E\equiv E_h-E_l$, we have seen in the main text that the average work produced per cycle is $\ave{W}=E(h-l)$. We now evaluate $\ave{W^2}$. The probability that a ball falls and none is raised is $h(1-l)$. If this event occurs, the work performed squared is equal to $E^2$. Conversely, the probability that a ball is raised and none falls is $l(1-h)$. If this event occurs, the work performed squared is again equal to $E^2$. 
Because the two other cases (exchange of weight-1 balls or exchange of nothing) produce no work, it follows that $\ave{W^2}=E^2[h(1-l)+l(1-h)] $. Therefore, the variance of the work produced reads
\begin{align}\label{l}
\var(W)\equiv \ave{W^2}-\ave{W}^2=E^2[h(1-h)+l(1-l)] .
\end{align}
In the limit $h\approx l$ considered in the main text, we have 
\begin{align}\label{lbis}
\var(W)\approx2E^2 l(1-l).
\end{align}

Let us now consider the total entropy produced $\Delta S\equiv \Delta S_l+\Delta S_h$. When a ball is being transferred from the high reservoir to the lower one and none from the low reservoir to the higher one, an event that occurs with probability $h(1-l)$, the increment of $S_l$ is, according to \eqref{e}, $\Delta S(n_l+1)-\Delta S(n_l)=\ln(\frac{1}{l}-1)$, and the increment of $S_h$ is $\Delta S(n_h-1)-\Delta S(n_h)=-\ln(\frac{1}{h}-1)$. It follows that the increment in total entropy is $\ln(\frac{\frac{1}{l}-1}{\frac{1}{h}-1})$ with probability $h(1-l)$. When a ball is being transferred from the low reservoir to the higher one and none from the high reservoir to the lower one, an event that occurs with probability $l(1-h)$, the increment of $S_l$ is, according to \eqref{e}, $\Delta S(n_l-1)-\Delta S(n_l)=-\ln(\frac{1}{l}-1)$, and the increment of $S_h$ is $\Delta S(n_h+1)-\Delta S(n_h)=\ln(\frac{1}{h}-1)$. It follows that the increment in total entropy is $\ln(\frac{\frac{1}{h}-1}{\frac{1}{l}-1})$ with probability $l(1-h)$.

The average increment in total entropy is therefore
\begin{align}
\ave{\Delta S}&=h(1-l)\ln \left( \frac{\frac{1}{l}-1}{\frac{1}{h}-1} \right) +l(1-h)\ln \left(\frac{\frac{1}{h}-1}{\frac{1}{l}-1} \right)\nonumber\\
&=(h-l) \ln \left(\frac{\frac{1}{l}-1}{\frac{1}{h}-1}\right) \geqslant 0.\label{m}
\end{align}
As we said in the main text, when $h\approx l$, $\Delta S\approx 0$ and the system tends to be reversible and to achieve the highest efficiency. Note that the entropy increment is non-negative for both a heat engine ($h>l$) and a heat pump $l>h$). More precisely, noting that to first order in $\delta \equiv h-l$ we have $\ln[(1/l-1)/(1/h-1)]\approx \delta/[l(1-l)]$, 
\begin{align}
\ave{\Delta S}\approx\frac{\delta^2}{l(1-l)}.\label{mbis}
\end{align}
The whole model presented makes sense because the generated entropy is proportional to $\delta^2$ while the work produced is proportional to $\delta$, so that, for small $\delta$, reversibility does not imply zero work.

Finally, we evaluate the variance of the total entropy increment. From the above expressions, it follows that 
\begin{align}
\ave{(\Delta S)^2}&=h(1-l)\left[\ln \left(\frac{\frac{1}{l}-1}{\frac{1}{h}-1}\right)\right]^2+l(1-h)\left[\ln \left(\frac{\frac{1}{h}-1}{\frac{1}{l}-1}\right)\right]^2\nonumber\\
&=(h+l-2lh)\left[ \ln \left(\frac{\frac{1}{l}-1}{\frac{1}{h}-1}\right) \right]^2,\label{n}
\end{align}
and the variance reads
\begin{align}
\var(\Delta S)&\equiv \ave{(\Delta S)^2}-\ave{\Delta S}^2\nonumber\\
&=(h+l-2lh-(h-l)^2)\left[ \ln \left(\frac{\frac{1}{l}-1}{\frac{1}{h}-1}\right)\right]^2\nonumber\\
&=\left[(h(1-h)+l(1-l)\right]\left[ \ln\left(\frac{\frac{1}{l}-1}{\frac{1}{h}-1}\right)\right]^2,\label{o}
\end{align}
which vanishes, as well as the average entropy produced, when $h\approx l$. To first order in $\delta\equiv h-l$, we have
\begin{align}
\var(\Delta S)\approx 2\ave{\Delta S},\label{p}
\end{align}
a remarkably simple result.
 
\section{Relation to practical heat engines}

Real heat engines retrieve energy from two arbitrarily large baths, one at temperature $T_l$ and the other at temperature $T_h>T_l$. A working agent such as a gas-filled cylinder of length $L$ is put in contact alternately with the hot bath and the cold bath. There are four steps. A parameter such as the length $L$ of a cylinder or a magnetic field (see below) may vary during these four steps.Net work may be obtained in that manner. We consider below in particular the so-called Otto cycle, which describes in an idealized form the gasoline engine,discovered by Beau de Rochas in 1862 and Otto in 1876. In that cycle, the parameter does not vary when the working agent is in contact with either bath. The parameter varies only during the adiabatic transitions from one bath to the other. Furthermore, we suppose that the working agent is a collection of electrons (whose magnetic moments are $\pm \mu)$ submitted to a magnetic field $B_l$ in the lower (cold) bath and $B_h$ in the higher (hot) bath. We denote by $E_l$ and $E_h$ the corresponding differences in electronic level energies. Otto cycle efficiencies approach the Carnot efficiency when the work produced per cycle tends to zero. This is exactly what our ball model describes.

To see more clearly their relationship, consider the special case where, in the Otto cycle, $E_l/T_l=E_h/T_h$. Then, according to the Boltzmann statistics, the two electron level populations are the same in either bath. Indeed, calling $n$ the upper state population and $N-n$ the lower-state population, we have $(N-n)/n=\exp(E/T)$, a relation to which subscripts $l$ and $h$ may be appended. It follows that $n_l/N_l=n_h/N_h$. It is also known that during the (slow) adiabatic transformations the level populations do not vary. We conclude that the populations are the same at all time. The work produced when the electrons are carried from the hot bath to the cold bath is therefore opposite to the work performed when the electrons are carried from the cold bath to the hot bath, so that the net work $W$  delivered in the special circumstance presently considered vanishes. But the relation $n_l/N_l=n_h/N_h$ coincides with the relation $l=h$ introduced in our ball model for the case where $W=0$, since we have defined $l\equiv n_l/N_l, h\equiv n_h/N_h$. In both our model and the Otto cycle, whenever $h$ slightly exceeds $ l$ there is some work produced almost at the Carnot efficiency, and whenever $l$ slightly exceeds $ h$ there is some work absorbed almost at the Carnot efficiency (ideal heat pump). Thus we conclude that the ball model realistically describes \emph{some} common heat engines. For an exact description of Carnot heat engines see the generalization of the ball model given below in \ref{carnotforballs}.
   
To summarize, the only concepts involved in the present section are those of potential energy and of uniform probability. We have been able to prove that the efficiency and work coincide with the Carnot expressions in some limit. Fluctuations are obtained straightforwardly.

\section{Relation to lasers}

As in the previous section we view the figure dark circles as electrons in a static field. Figure \ref{fig} may then be viewed as representing five atoms numbered from 1 to 5 from left to right. 1 and 4 are in the emitting state, 3 is in the absorbing state, while 2 and 5 are inactive. A Rabi oscillation resulting from the field at frequency $\om$ with $\hbar\om=E_h-E_l$ of a resonator, may convert the atom 1 from one state to the other, thereby incrementing or decrementing the resonator energy. We suppose that atom 1 only is concerned. The Rabi period, inversely proportional to the square root of the resonator energy, corresponds to the cycle duration. The probability that atom 1 be in the emitting state is $h(1-l)$ and the probability that atom 1 be in the absorbing state is $l(1-h)$, using the previous $l,h$ notation. Once the Rabi exchange has been performed, the resonator has received, on the average, an energy $W=(E_h-E_l)\p h(1-l)-l(1-h)\q=(E_h-E_l)(h-l)$. This is the same result as obtained before in \eqref{b} for the thermodynamic cycle. Note that the population inversion may be used to define a negative temperature $T$, as usual, according to the Boltzmann law: $\frac{l(1-h)}{h(1-l)} \equiv \exp{  \frac{E_h-E_l} {T}  }$.

The laser schematic could be completed by letting the resonator interact with an atomic collection similar to the one shown in \ref{fig}, but absorbing energy on the average, that is, with $l>h$ and corresponding to a positive temperature. Efficiency and fluctuations may be obtained for some laser systems on the basis of such a model. The efficiency of a laser is defined as the ratio of the optical power produced to the pump power. In the present model the output energy is $W$ per cycle. In order to sustain the oscillation one must add steadily emitting-state atoms and remove absorbing-state atoms.

\section{Full Carnot cycle}\label{carnotforballs}

In the two-reservoir model previously described the Carnot efficiency was reached only in the limit of small work produced per cycle. That model suffers from some restrictions since, obviously, with only two reservoirs reversing the cycle is immaterial. In the present section the model is generalized, and the exact properties of Carnot cycles are obtained. The higher reservoir is being replaced by a very large number of higher reservoirs, and likewise the lower reservoir is being replaced by a very large number of lower reservoirs. A ball randomly picked in one reservoir is carried to the next in a cyclic manner. If the cycle is being reversed the heat engine gets converted into a heat pump. We call "forces" the ratios $n/N$, and denote altitudes by $E$ as before.

There are $m$ low-level reservoirs (referred to as sub-reservoirs) at
altitudes $E_{l,1},E_{l,2},\dots E_{l,m}$, with forces
$f_{l,1},f_{l,2},\dots f_{l,m}$, respectively. Likewise, we suppose that there are $m$ high-level
sub-reservoirs at altitudes $E_{h,1},E_{h,2},\dots E_{h,m}$, with forces
$f_{h,1},f_{h,2},\dots f_{h,m}$, respectively. We obtain
\begin{align}
\label{multilev}
Q_{l}&=E_{l1}f_{hm}+\sum_{i=1}^{m-1}f_{li}(E_{l,i+1}-E_{l,i})
-E_{lm}f_{lm} \nonumber \\
Q_{h}&=E_{h1}f_{lm}+\sum_{i=1}^{m-1}f_{hi}(E_{h,i+1}-E_{h,i})
-E_{hm}f_{hm}.
\end{align}
 
We have considered above arbitrary sub-reservoir forces. We now restrict the generality by supposing
that the forces depend on the reservoir altitudes according to laws of the form

\begin{eqnarray}
\label{forces} f_{i}=f(\beta E_{i}).
\end{eqnarray}
where $f(.)$ denotes some function to be specified later, and $E_{i}$, with $i=1,2 \dots m$
denotes as before the altitude of the (lower or higher) sub-reservoirs. The parameter $\beta$ takes
the values $\beta_{l}$ for the lower sub-reservoirs and the value $\beta_{h}$ for the higher one.
Under that restriction, the above expressions for the reservoir energies read

\begin{align}
\label{gen}
\beta_{l}Q_{l}&=L_{1}f(H_{m})+\sum_{i=1}^{m-1}f(L_{i})(L_{i+1}-L_{i})-L_{m}f(L_{m}) \nonumber \\
\beta_{h}Q_{h}&=H_{1}f(L_{m})+\sum_{i=1}^{m-1}f(H_{i})(H_{i+1}-H_{i})-H_{m}f(H_{m}).
\end{align} 
where we have defined $L_{i}=\beta_{l}E_{l,i}$ and $H_{i}=\beta_{h}E_{h,i}$,
$i=1, 2 \dots m$.

If the $E$-values do not vary much from one low sub-reservoir to the next, and likewise, do
not vary much from one high sub-reservoir to the next, sums may be replaced by integrals and we have

\begin{align}
\label{smooth}
\beta_{l}Q_{l}&=s(L_{1},H_{m})-s(L_{m}) \nonumber \\
\beta_{h}Q_{h}&=s(H_{1},L_{m})-s(H_{m}),
 \end{align} 
where

\begin{align}
\label{sss}
 s(x,y) &\equiv x f(y) - \int_{}^{x}f(x') dx' , & s(x) &\equiv s(x,x) ,
\end{align}
the lower limit of the integral being unimportant.

When $L_{1}=H_{m}\equiv L$ and $H_{1}=L_{m}\equiv H$ the maximum efficiency is reached. Indeed, the
quantity $\beta_{l}Q_{l}+\beta_{h}Q_{h}$ then vanishes. If we set $\beta_{l}\equiv 1/T_{l}$,
$\beta_{h}\equiv 1/T_{h}$, the efficiency reads

\begin{eqnarray}
\label{ss}
\eta=1-\frac{T_{l}}{T_{h}}.
\end{eqnarray}
The work performed is then the product of the change of $T$ and the change of $s$, namely
\begin{eqnarray}
\label{work'} W= \left( T_{h}-T_{l} \right)(s(H)-s(L)). 
\end{eqnarray}
We have thus recovered the properties of Carnot cycles. Fluctuations could be obtained as was done in the previous section.

Even though the above relations were obtained for a rather peculiar kind of heat engine, the results are general. Different working agents differ only by the expression of the $f(x)$ function introduced above, and therefore by the $s(x,y)$ function. It is important to notice that these functions are unaffected by the addition of a constant to $f$. As a consequence, if the working agent employed is a single-mode resonator, the term $\hbar\omega/2$ in the expression of the average energy $\ave{E}$ of the resonator, that is the additional term $\hbar/2$ in the expression of the oscillator action $f\equiv \ave{E}/\omega$, affects neither the work produced nor the efficiency.

\section{Two-states agent}

The ball-reservoirs considered in this paper model may reach equilibrium with baths located at the same altitude $\epsilon$ provided they have the same $h$ or $l$ values. This is achieved by allowing balls to be displaced between the ball-reservoir and the bath. Usually, working agents and baths exchange energy, but not particles. If we insist on the condition that only energy be exchanged, our model needs be modified. Instead of balls one should consider electrons immersed in a magnetic field. An electron has two energy levels depending on its spin state. The lower one may be set at 0 (corresponding to empty locations) and the other at 1 (corresponding to a weight-1 ball). What we previously called the ball-reservoir ``altitude'' corresponds here to the magnetic field in which the electrons are being immersed.

There is an alternative model close to the one shown in the figure. Instead of weight-1 balls in the earth gravitational field, one may have electrons in a constant electrical field $V/d$ created by two parallel conducting plates spaced a distance $d$ apart, with a potential difference $V$. The potential between the plates is a constant if they are connected to an arbitrarily large capacitance. There may be a collection of $n_l<N_l$ electrons at altitude $E_l$ and a collection of $n_h<N_h$ electrons at altitude $E_h$. $N_l,~N_h$ then represent the number of states in the lower and higher bands. When an electron drops from altitude $E_h$ to altitude $E_l$, it delivers an energy $eV(E_h-E_l)/d$. Conversely, when an electron at altitude $E_l$ is raised to $E_h$ it absorbs an energy $eV(E_h-E_l)/d$. These energies increment or decrement the large capacitance charge by $q$, such that $qV$ equals the energy delivered by the electron. In that model, the work $W$ evaluated above appears in the form of an electrical energy corresponding to a current $i=q/\tau$ lasting during the cycle duration $\tau$. The work is positive if there is population inversion $n_h/N_h>n_l/N_l$.

\section{Quantized working agents}\label{working}

We have treated above a rather peculiar type of heat engine consisting of balls at different altitudes. We would like to consider now more conventional heat engines involving a "working agent" depending on a varying parameter that we denotes $\omega$ (in view of the case where the working agent would be an oscillator at frequency $\omega$). This working agent is displaced from a heat-temperature bath to a low-temperature bath and back. In the historical part of this chapter the thermal properties of ideal classical gas have been recalled, that enabled Carnot to propose a specific ideal cycle. From now on we consider quantized working agents, when the quantized energy levels factorize according to $\epsilon_k(\omega)=h(k)g(\omega)$. If this were not the case, the concept of temperature could not be defined accurately, and therefore the Carnot requirement that two bodies should be contacted only if they have the same temperature would loose part of its meaning. This factorization condition is fulfilled in particular for a particle moving in a power-law potential of the form $V(x)\propto x^{2\kappa}$. The case $\kappa\to\infty$ corresponds to a particle moving back and forth between two walls, and the case $\kappa=1$ corresponds to harmonic oscillators. We may consider also the case of $B$ evenly spaced levels, where $B=2$ corresponds to the two-level case (e.g., an electron in a magnetic field), $B=10$ is applicable to bismuth nuclei in a magnetic field, and $B\to\infty$ corresponds again to harmonic oscillators. The model applies also to a large collection of single-spin non-interacting electrons occupying evenly spaced levels (one electron per level). Let us begin by recalling the properties of oscillators.

\section{Harmonic oscillators}\label{harmonic}

In the present section we first recall the classical properties of oscillators and particularly the theorem that says that if the resonant frequency $\omega$ is changed slowly, the resonator average energy $U$ is proportional to $\omega$, so that the resonator action $f\equiv U/\om$ is a constant\footnote{Such considerations are relevant to black-body radiation, which, historically, is at the origin of Einstein theory of light quanta (later on called "photons" by Lewis), in particular in relation of the Wien displacement law. Note that a black-body may be considered as a large collection of resonators that are not directly coupled to one another. Rather than considering the black body as a whole, it is simpler to consider first the properties of a single resonator. At the time of Planck the word "resonator" was referring to atoms or molecules. In the present text "resonator" always refers to electrical (or optical) resonating circuits. As far as material objects are concerned we employ the term "oscillators". Even though optical resonators and (harmonic) oscillators have close formal resemblance, they are distinguished in this book.}.

\paragraph{Adiabatic transformation}

A transformation is called "adiabatic" (literally "does not go through") when there is no heat exchange with the outside, and the process is slow. We treat here the adiabatic transformation of classical harmonic oscillators. We evaluate the work produced when the resonant frequency $\om$ vary slowly. Next, we observe that the result is valid for any energy distribution, and applies therefore also to semi-classical and quantized harmonic resonators.
 
For concreteness, let us consider a \emph{classical} inductance–capacitance $L-C$ circuit resonating at angular
frequency $\omega$. To begin with, note that the oscillator considered involves a huge number of electrons that act collectively, and that, accordingly, it would be meaningless to quantize these electrons (this, of course, is \emph{not} what is meant by "quantizing the resonator" in Quantum Optics). The electrical charges on the capacitor plates oscillate sinusoidally in the
course of time. The resonator is initially neutral, so that the electric charge appearing on
one plate is opposite to the electrical charge appearing on the other plate, and the plates
always attract each other. It follows from the Coulomb law that the average force $F = U/2a$,
where $a$ denotes the plate separation, and $U$ the oscillator energy. If $a$ is incremented by
$da$ slowly so that the oscillation remains almost sinusoidal, the elementary work performed
by the oscillator is $dW = -F da = U da/2a$. On the other hand, it follows from the
well known resonance condition $LC\omega^2 = 1$ and the fact that $C \propto 1/a$, where $\propto$ denotes
proportionality, that $2 d\omega/\omega = da/a$. The elementary work may therefore be written as
$dW=-(U/\omega) d\omega \equiv-f d\omega$, where we have introduced the generalized force, or "action", $f = U/\omega$. It may be that $U$ varies slightly when $a$ is incremented by $da$, but this variation does not affect
$dW$ to first order. Accordingly, the above discussion holds both for isolated oscillators, as presently considered, and for
oscillators in contact with a bath, even though U varies differently in these two situations. It follows that if the initial frequency $\omega_1$ of an oscillator of energy $U_1$ is slowly changed to $\omega_2$, the energy $U_2$ is given by $U_2=U_1(\omega_2/\omega_1)$. Because the oscillator presently considered is isolated (adiabatic transformation), the work performed is $W=U_1-U_2=U_1(1-\omega_2/\omega_1)$. 

One may apply the above classical result to more general systems involving a distribution $P(U)$ of energy. If $U_i$ is not known precisely, but the probability density $P_1(U_1)$ is known, the probability density of $U_2$ is $(\omega_1/\omega_2) P_1\p U_2(\omega_1/\omega_2)\q$. It follows that one may evaluate the cumulants of the oscillator energy after an adiabatic transformation. In particular, the average energies are related as
\begin{align}
\label{ent}
\ave{U_2}=\ave{U_1}(\omega_2/\omega_1).
\end{align} 
The average work performed $\ave{W}=\ave{U_1}(1-\omega_2/\omega_1)$ is given formally by the same relation as for sure energies. 

\paragraph{Oscillator in contact with a heat bath}\label{oscillatorbath}

It remains to evaluate the average energy $\ave{U}\equiv \ave{E}$ of an oscillator in contact with a bath at absolute temperature $T$. The average energy of a classical oscillator at temperature $T$ equals $T$ because an oscillator possesses two degrees of freedom. For a quantized harmonic oscillator with energy levels $E_k=(\frac{1}{2}+k)\epsilon$, $\epsilon\equiv \hbar\om$, we have, using the Boltzmann distribution and after summation
\begin{align}
\label{enis}
\ave{E}=\frac{\hbar \om}{2}\frac{\exp(\beta \hbar \om)+1}{\exp(\beta\hbar \om)-1},
\end{align}
where $\beta\equiv1/T$. 
We therefore have now all the information required to study thermal cycles employing oscillators as working agents. In particular, we can evaluate the work delivered in a Carnot cycle employing as working agent a collection of independent oscillators. The above expressions apply both to mechanical quantized oscillators and semi-classical optical resonators, since they have the same average energy at some temperature.

In a cycle, one may start from a frequency $\om_1$ when the resonator is in contact with the low-temperature bath $T_l$. It then possesses an average energy given by \eqref{enis} with $T=T_l$. Then, the resonator is carried to the high-temperature bath at temperature $T_h$ and the frequency is made to evolve slowly from $\om_1$ to a value $\om_2$ such that the average energy given in \eqref{ent} coincides with the value given in \eqref{enis} with $T=T_h,~\om=\om_2$. If this is the case the resonator average energy is unaffected when contacting the hot bath. The second step consists of changing $\om_2$ to $\om_3$ while the resonator remains at temperature $T_h$. The third and fourth steps are similar. We are therefore in position to evaluate the work performed by the engine and its efficiency. The point of this discussion is to emphasize that the working-agent entropy is irrelevant. Accordingly, one may consider that the resonator field does not possess any entropy without contradicting experimental facts.

\section{Many electrons with evenly-spaced levels}

For the analysis of semi-conductors, it is important to know the Fermi-Dirac distribution. This distribution is based on the Pauli principle that forbids two electrons to occupy the same state (only one spin state is considered). We assume for simplicity evenly-spaced energy levels, with energy separation $\epsilon$. This assumption would be valid if the electrons were moving in a square-law potential. But in a two-dimensional semiconductor the density of state is  constant, so that the assumption of evenly-spaced levels entails little errors\footnote{In fact, a similar observation was made by Boltzmann, who introduced an energy spacing $\epsilon$ for the purpose of evaluating the number of configurations, and found this assumption accurate for a two-dimensional box. However, Boltzmann eventually let $\epsilon$ tend to zero. In contradistinction, Planck and Einstein kept non-zero $\epsilon$ values.}. Our model is approximately applicable to any semiconductor because the density of state may often be taken as a constant near the operating point. For simplicity, we further take $\epsilon$ as the energy unit (e.g., 1 meV). In the figure it is assumed that, at $T=0$K, all the states are occupied from level $k=0 $ to $k=6$. Because the added energy $r$ that we shall consider is less than 6, we may as well suppose that the negative $k$ levels are all occupied).  The Fermi level $\mu$ employed below corresponds to the highest occupied level plus 1/2.
%%%%%%%%%%%%
\begin{figure}
\setlength{\figwidth}{1.2\textwidth}
\centering
{\small
\begin{tabular}{|c|c|ccccccccccc|c|c|c|c|}
%\hline
%\multicolumn{2}{|c|}{}&\multicolumn{15}{c|}{$r=6$}\\
\hline
$k$&$\kappa$&\multicolumn{11}{c|}{microstate}&$m_{k}(6)$&$\ave{N_\kappa}$&$\ave{N_\kappa}_\mathrm{C}$&$\ave{N_\kappa}_\mathrm{FD}$\\
\hline
7&6.5&&&&&&&&&&&&0&0.000&0.032&0.032\\
6&5.5&$\bullet$&&&&&&&&&&&1&0.091&0.052&0.053\\
5&4.5&&$\bullet$&&&&&&&&&&1&0.091&0.084&0.087\\
4&3.5&&&$\bullet$&$\bullet$&&&&&&&&2&0.182&0.131&0.138\\
3&2.5&&&&&$\bullet$&$\bullet$&$\bullet$&&&&&3&0.273&0.202&0.213\\
2&1.5&&&&&$\bullet$&&&$\bullet$&$\bullet$&$\bullet$&&4&0.364&0.300&0.313\\
1&0.5&&&$\bullet$&&&$\bullet$&&$\bullet$& $\bullet$&&$\bullet$&5&0.455&0.430&0.435\\
\hline
0&-0.5&&$\bullet$&&$\bullet$&&&$\bullet$&$\bullet$&&$\bullet$&$\bullet$&6&0.545&0.570&0.565\\ 
$-1$&-1.5&$\bullet$&&&$\bullet$&&$\bullet$&$\bullet$&&$\bullet$&$\bullet$&$\bullet$&7&0.636&0.700&0.687\\
$-2$&-2.5&$\bullet$&$\bullet$&$\bullet$&&$\bullet$&&$\bullet$&&$\bullet$&$\bullet$&$\bullet$&8&0.727&0.798&0.787\\
$-3$&-3.5&$\bullet$&$\bullet$&$\bullet$&$\bullet$&$\bullet$&$\bullet$&&$\bullet$&& $\bullet$&$\bullet$&9&0.818&0.869&0.862\\
$-4$&-4.5&$\bullet$&$\bullet$&$\bullet$&$\bullet$&$\bullet$&$\bullet$&$\bullet$&$\bullet$&$\bullet$&&$\bullet$&10&0.909&0.916&0.913\\
$-5$&-5.5&$\bullet$&$\bullet$&$\bullet$&$\bullet$&$\bullet$&$\bullet$&$\bullet$&$\bullet$&$\bullet$&$\bullet$&&10&0.909&0.948&0.947\\
$-6$&-6.5&$\bullet$&$\bullet$&$\bullet$&$\bullet$&$\bullet$&$\bullet$&$\bullet$&$\bullet$&$\bullet$&$\bullet$&$\bullet$&11&1.000&0.968&0.968\\
\hline
\end{tabular}
}
\caption{System containing seven single-spin electrons allocated to evenly-spaced levels. An energy $r=6$ is added to the system. The microstate column exhibits the $W(6)=11$ ways of incrementing the energy by that amount. The level occupations in the micro-canonical (isolated), canonical (thermal exchange with a heat bath) and grand canonical (exchange with a  heat and electron bath) situations are given in the columns $\ave{N_\kappa}$, $\ave{N_\kappa}_\mathrm{C}$ and $\ave{N_\kappa}_\mathrm{FD}$, respectively. In the so-called "Thermodynamics limit", all these occupations tend to coincide.}
\label{fermi}
\end{figure}

%%%%%%%%%%%%%
The probability that a level $k$ be occupied at some temperature $T>0$ is given by the Fermi-Dirac distribution
\begin{align}\label{FD}
p(k)=\frac{1}{q^{\mu-k}+1}\qquad q\equiv \exp(-\beta)\equiv \exp(-1/T).
\end{align}
To understand the significance of this formula it is useful to consider first the isolated system (micro-canonical ensemble) represented in Fig. \ref{fermi} on the left. At $T=0$K, only states labeled by $\kappa<0$ are occupied. Note that $\mu=6+1/2$ in the case of the figure, and $\kappa\equiv k-\mu=-1/2$. 
If an energy $r$ is added to the system ($r$=6 in the case of the figure), electrons are moving upward until the total energy acquired be equal to $r$. This may be accomplished in $W(r)$ different manners, where $W(r)$ denotes the number of partitions of $r$ (see the mathematical section, where $W(r)$ is denoted $p(r)$). We obtain the entropy $S$ as the logarithm of $W(r)$, and the temperature by derivation. Referring to \eqref{asymptotic}, we have, for large $r$ values, $S=\log(W(r))\approx \pi\sqrt{2r/3}$. It follows that  $r=3S^2/2\pi^2,~T=dr/dS=\sqrt{6r/\pi}$. The (constant-$\epsilon$) electronic heat capacity is therefore $C_V\equiv dr/dT=\pi^2 T/3$.

Next we introduce the so-called canonical distribution by allowing $r$ to vary according to the Boltzmann distribution. Finally, we allow the number of electrons to fluctuate. The occupations in the micro-canonical $\ave{N_{\kappa}}$, canonical $\ave{N_{\kappa}}_c$ and grand canonical $\ave{N_{\kappa}}_{FD}$ ensembles are shown on Fig. \ref{fermi} on the right. Following this three-steps procedure, the celebrated Fermi-Dirac distribution quoted above is recovered. We also obtain three distinct expressions for the constant-$\epsilon$ electronic heat capacity. These expressions tend in the so-called "Thermodynamics" (high energy) limit to the value given above, namely $C_V=\pi^2~T/3$. This heat capacity vanishes at $T=0$, as it must, if the entropy is to be finite when the integration begins at $T=0$K. Generalization of these considerations to account for the two electron spin states has been done.

\newpage

\chapter{Spectral density of random sources}\label{spect}

The present chapter describes four different (partly heuristic) ways of establishing the expression of the spectral density of the random (current or rate) sources. The basis of these methods and the approximations made are listed below.

\begin{itemize}

\item  Section \ref{ratebis} 

Any of the media described in chapter \ref{various} could in principle be used to evaluate the probability $P(m)$ of having $m$ light quanta in a resonator, from which we may obtain the variance of $m$, where $m$ denotes the integer part of the optical energy divided by $\hbar\om$. For reasons of simplicity, this is done only for $N$ two-level atoms (and one electron per atom). The theory assumes that we are dealing with a Markov process, and it is valid when the number $n$ of atoms in the emitting state and the number $m$ of light quanta are large compared with unity so that the time-derivatives of $n~,m$ make sense. We evaluate the random rates $r(t)$ that ensure that the calculated variance of $m$ is obtained. 

\item Section \ref{heuristic}
 
Here, we look directly for a generalization of the classical expression for the average resonator energy, and do not consider random sources.
We employ the classical expression $\ave{E}=T$ of the average resonator energy, and observe that the resonator action $f=\ave{E}/\om$ obeys a Ricatti equation. By adding a constant $(C/2)^2$ on the right-hand-side of this equation, the well-known divergence of the classical black-body radiation is removed. Next we argue that when cold resonating atoms interact with a high-energy resonator, the energy in the resonator gets eventually uniformly distributed between 0 and $\hbar\om$, so that the average energy is $\hbar\om/2$. This condition at $T=0$ entails that the constant $C$ introduced above is equal to $\hbar$. The Planck law is then obtained. This method requires that the resonator interacts with a large number of atoms, but any value of $m$ is allowed.

\item Section \ref{ener} 

We calculate the energy of a (single-mode) resonator driven by two current sources, one associated with atoms in the excited state and the other associated with atoms in the absorbing state. The population ratio of these two populations is given by the Boltzmann law. The requirement that the average resonator energy be $T$ for large $T$-values gives the spectral density of the random current sources. A symmetry condition between emission and absorption is also employed.

\item Section \ref{sec_wait} 

We first draw a comparison between reflex klystrons (classical microwave-generating devices) and lasers. Next we consider a single electron interacting with a single-mode resonator. The optical field is prescribed, which implies that the number $m$ of light quanta in the resonator is large compared with unity. (Otherwise one should account for the back action of the evolving electron on the amplitude and phase of the optical field, a random process. This will not be done here). The Rabi equations, modified by a phenomenological decay constant $\gamma$, are valid as long as no transition event has occurred. If $\gamma$ is large, the random rate spectral density is shown to be equal to the average rate, that is, the event rate is Poisson distributed. The same result holds for any $\gamma$-value if a large number of independent electrons are present (i.e., $n\gg 1$).

\end{itemize}

\section{Isolated resonator with $N$ two-level atoms}\label{ratebis}

In 1916 Bohr proposed that atomic electrons may occupy only discrete energy levels and that these electrons may perform jumps from one state to another, thereby explaining the appearance of absorption lines and related dispersion effects. The probabilities of occupation of these levels obey the classical Boltzmann statistics. In previous sections we have taken for granted the Boltzmann distribution for the integer part $m$ of the resonator energy divided by $\hbar\om$. In the present section we consider a closed system that contains $N$ two-level atoms and a single-mode resonator. Let us recall that from the semi-classical view-point the optical field is a reservoir of energy having no independent degree of freedom and therefore no entropy. We first consider the average number of excited-state atoms, the average value of $m$ (conveniently called the number of light quanta), and subsequently the variance of these quantities. We establish the expression of the random rates and the expression of the random currents.

The basic properties of thermodynamical cycles have been recalled in the previous sections. We have shown how the work fluctuation may be evaluated. As we indicated there, our ball-reservoir model is applicable to electrons immersed in a magnetic field. Such electrons have two spin states corresponding to two energy levels, which we set at 0 and $\epsilon$ by convention, where $\epsilon$ is proportional to the magnetic field. In the present section the magnetic field is kept constant and for simplicity we assume that the level energies are 0 and 1. Equivalently, we may consider two-level atoms of level energies 0 and 1, respectively, the direct coupling between atoms being non-zero so that an equilibrium may be reached, but is vanishingly small. 

The number of atoms that are in the upper state is denoted by $n$, and the
number of atoms in the lower level is therefore $N-n$.  Within our conventions the atomic energy equals $n$. There is population inversion when $n>N/2$. The statistical weight $W(n)$ of the atomic collection is the number of distinguishable configurations corresponding to some total energy
$n$. For $N$ identical
atoms, the statistical weight (number of ways of picking up $n$ atoms
out of $N$) is  
\begin{align}
    W(n)=\frac{N!}{n!(N-n)!}\qquad  Z\equiv \sum_{n=0}^{N}W(n)=2^{N}   .
    \label{Wofn}
\end{align}

Suppose next that these atoms are coupled to a single-mode resonator. The integer part of the resonator energy is denoted $m$. Following Bohr, we suppose that the atoms reside in either one of the two levels, but may possibly perform jumps from one level to the other in the course of time. By conservation of energy, whenever an atom drops from the level of energy 1 to the level of energy 0, $m$ is incremented by 1 ($m\to m+1$). Conversely, whenever an atom jumps from the level of energy 0 to the level of energy 1, the oscillator energy is decremented by 1 ($m\to m-1$). The basic principle of Statistical Mechanics asserts that in isolated systems all states of equal energy are equally likely to occur.  Accordingly, the probability $p(m)$ that some $m$ value occurs at equilibrium is proportional to $W(N-m)$, where $W(n)$ is the statistical weight of the atomic system.  As an example, consider two (distinguishable) atoms ($N$=2).  A microstate of the isolated (matter$+$field) system
is specified by telling whether the first and second atoms are in
their upper (1) or lower (0) states and the value of $m$.  Since the
total energy is $N=2$, the complete collection of microstates (first
atom state, second atom state, field energy), is: (1,1,0), (1,0,1),
(0,1,1) and (0,0,2).  Since these four microstates are equally likely,
the probabilities that $m=0,1,2$ are proportional to 1,2,1 respectively.  This is in agreement with the fact stated earlier
that $p(m)$ is proportional to $W(n)\equiv W(N-m)$.  After normalization, we obtain for example that p(0)=1/4.
 
The normalized probability reads in general
\begin{align}\label{P}
p(m)=\frac{W(N-m)}{Z}=\frac{N!}{2^{N}m!(N-m)!}
\end{align}
The moments of $m$ are defined as usual as
\begin{align}\label{mr}
 \ave{m^{r}}\equiv \sum_{m=0}^{N}m^{r} p(m)
\end{align}
where brakets denote averagings.  It is easily shown that $\ave{m}=N/2$ and $\mathrm{var}(m) \equiv
{\ave{m^{2}}-\ave{m}^{2}} = N/4$:  The statistics of $m$ is
sub-Poisson, with a variance less than the mean\footnote{For a single atom with $B$ evenly-spaced levels and $n≤B$ electrons, two electrons may not occupy the same level according to the Pauli principle (only one spin state is considered). The variance of $m$ is found to be $(B+1)/6$ times the average of $m$. This result is obtained by considering the shifting of electrons downward beginning with the lowest one, until the specified energy $m$ is subtracted. For some subtracted $m$ value the statistical weight is the number $p(n,m)$ of partitions of $m$ into at most $n$ parts, none of which exceed $B-n$. According to \eqref{hfun}, $\ave{m}=\frac{1}{2}n(B-n)$ and variance($m$)=$\frac{1}{12}n(B-n)(B+1)$. The same result holds for any number $N$ of $B$-level atoms. This more general result coincides with the one given above if we set $B=2$.}.

The expression of $p(m)$ just obtained has physical and
practical implications.  Suppose indeed that the equilibrium cavity
field is allowed to escape into free space at some time, thereby generating an
optical pulse containing $m$ light quanta.  It may happen, however, that no
pulse is emitted when one is expected, causing a counting error.  From
the expression in (\ref{P}) and the fact that $\ave{m}=N/2$, the
probability that no light quantum be emitted is seen to be
$p(0)=4^{-\ave{m}}$.  For example, if the average number of light
quanta $\ave{m}$ is equal to $20$, the communication system
suffers from one counting error (no pulse received when one is
expected) on the average over approximately $10^{12}$ pulses.  Light
pulses of equal energy with Poissonian statistics are inferior to the
light presently considered in that one counting error is recorded on
the average over $\exp(\ave{m})=\exp(20)\approx 0.5~10^9$ pulses.

Let us generalize the previous result, supposing that initially all the atoms are in the lower state ($n$=0) and that $m$ also vanishes. Let an energy $U$ with integer part $u$ be introduced in the system. At equilibrium, we have by conservation of energy $u=n+m$. The probability to have $m$ light quanta reads, replacing in the previous expression $n$ by $u-m$
\begin{align}\label{Q}
p(m)\propto\frac{N!}{(u-m)!(N-u+m)!}
\end{align}
From this expression, given $u$ and $N$, we may evaluate $ \ave{m}$ and then $\ave{n}=u-\ave{m}$. Under the assumption that $u\ll N$, we may define a temperature $T$ from the average atomic populations according to the Boltzmann distribution
\begin{align}\label{Qb}
\exp(1/T)=\frac{N-\ave{n}}{\ave{n}}
\end{align}
We compare below $p(m)$ as given by \eqref{Q} to the Boltzmann law $\propto \exp(-m/T)$, where $T$ is obtained from the above expression, for $N=100$ and $u=20$.
\begin{center}\begin{tabular}{r|cccc}
$m$   &0 & 1 & 2  &3 \\ \hline                         
$p(m)$    &0.7578 &0.187  & 0.043  & 0.009  \\
Boltzmann &0.754 & 0.185 &0.045   & 0.01 
\end {tabular}\end{center}
The comparison shows that the Boltzmann distribution is applicable. A formal proof could be obtained from the Stirling approximation.

Because no entropy is ascribed to the resonator, the system total entropy reads
\begin{align}\label{Z}
S=\ln\p\sum_{n=0}^N W(n)\q=\ln(Z)=N\ln(2)
\end{align}
In an artificial manner this total entropy may be split, however, into a matter entropy and a field entropy. 

\paragraph{Stimulated transitions}\label{stim}

Historically, Einstein explained the law of black-body radiation by considering a highly multi-moded cavity and assuming that atoms have a probability $BE$ of performing a transition from one state to the other, per unit time, where $E$ is a measure of the energy in the resonator (stimulated emission and absorption). In addition, atoms in the upper state have a probability $A$ per unit time of decaying spontaneously. Under those assumptions, the Planck law of black-body radiation is indeed recovered\footnote{Einstein went one step further by establishing under what conditions the Maxwell atomic distribution is recovered at equilibrium. In this book, however, atoms are supposed to have extremely large nuclei masses, and their motion is not of concern to us.}. We consider here only a single-mode resonator. In that case spontaneous emission is represented by the number 1 in the expression $m+1$ of decay probability, while the atom promotion probability is proportional to $m$. Note also that the Rabi theory does not predict transition rates proportional to time, as postulated by Einstein. This result requires the introduction of a phenomenological parameter $\gamma$, as is discussed later on.

We thus consider below processes by which an atom in the upper state decays to the lower state under the influence of the resonator field (stimulated emission) or is promoted to the upper level (stimulated absorption), the equilibrium situation being known from the principles of Statistical Mechanics. The rate of absorption is found equal to $m$, while the rate of emission is found equal to $m+1$, to within a common constant. (We will indicate at the end of the present section how the symmetry between absorption and emission may be restored. In brief, this amounts to considering $m$ at jump time, defined as the arithmetic average of $m$ just before and just after a transition. From that view-point, an observer cannot tell whether a movie reporting the transition events is being run forward or backward).

As said earlier  electrons perform jumps from one state to
another in response to the optical field, so that the number of atoms
in the upper state is some function $n(t)$ of time.  If $m(t)$ denote the
number of light quanta at time $t$ (defined earlier as the integer part of the average energy), the sum $n(t)+m(t)$ is a conserved
quantity. This is the total atom+field energy to within an additive constant comprised between 0 and 1. If $N$ atoms in their
upper state are introduced at $t=0$ and $m=0$,
part of the atomic energy gets converted into field energy as a result
of the atom-field coupling and eventually the equilibrium situation is
reached.  

Let us now evaluate the probability $p(m,t)$ of having $m$ light quanta at time $t$.  Note that here $m$ and $t$ represent two independent variables.  A particular realization of the process is sometimes denoted $m(t)$.  It is hoped that this simplified notation will not cause confusion. Let $R_e(m)dt$ denote the probability that, given $m$ at time $t$, this number jumps to $m+1$ during the infinitesimal time interval [$t, t+dt$], and let $R_a(m)dt$ denote the
probability that $m$ jumps to $m-1$ during that same time interval
(the letters "$e$" and "$a$" stand respectively for "emission"
and "absorption"). The probability $p(m,t)$ obeys the relation
\begin{align}\label{FP}
    p(m,t+dt) =  p(m+1,t)R_a(m+1)dt+p(m-1,t)R_e(m-1)dt \nonumber\\+p(m,t)[1-R_a(m)dt-R_e(m)dt].
 \end{align}
Indeed, the probability of having $m$ light quanta at time $t+dt$ is the sum
of the probabilities that this occurs via states $m+1$, $m-1$ or $m$
at time $t$.  All other possible states are two or more jumps away
from $m$ and thus contribute negligibly in the small $dt$ limit.  After a sufficiently long time, one expects
$p(m,t)$ to be independent of time, that is $p(m,t+dt)=p(m,t)\equiv
{p(m)}$.  It is easy to see that the "detailed balancing" relation 
\begin{align}\label{ball}
    p(m+1)R_a(m+1)=p(m)R_e(m)
\end{align}
holds true because $m$ cannot go
negative. When the expression of
$p(m)$ obtained in \eqref{P} is introduced in \eqref{ball}, one finds that
\begin{align}\label{EA}
    \frac{R_e(m)}{R_a(m+1)}=\frac{p(m+1)}{p(m)}=\frac{N-m}{m+1}.
\end{align}
$R_e$ must be proportional to the number $n=N-m$ atoms in the upper state while $R_a$ must be proportional to the number $N-n=m$ of atoms in the lower state. We therefore set $R_e(m)=\p N-m\q f(m), R_a(m)=mg(m)$, where $f(m)$ and $g(m)$ are two functions to be determined. Substituting in \eqref{EA} we find that
\begin{align}\label{Ei}
    f(m)=g(m+1).
\end{align}
Because we assume that atoms emit or absorb a single light quantum at a time, the two functions $f(m)$ and $g(m)$ must be of the linear form $f(m)=am+b$ and $g(m)=cm+d$, where $a,b,c,d$ are constants. But $R_a$ is required to vanish for $m=0$ since, otherwise, $m$ could go negative, and thus $d=0$. Substituting into \eqref{Ei}, we find the relation $am+b=c(m+1)$ which must hold for any $m$-value. Therefore, $a=b=c$. Setting for brevity $a=b=c=1$ amounts to fixing up a time scale. Then $R_e(n,m)=n\p m+1\q, R_a(n,m)=\p N-n\q m$. We note here a lack of symmetry between the rate of stimulated emission (proportional to $m+1$) and the rate of stimulated absorption (proportional to $m$). Since according to the Schrödinger equation the two processes should be similar, we are led to define the resonator average energy as $\ave{E}=m+1/2$, and to express $R_e$ and $R_a$ in terms of the integer resonator energy at jump time, defined as the arithmetic average of $m$ just before and just after the jump. If we do so, we finally obtain, setting $n\equiv n_e$ and $N-n\equiv n_a$
\begin{align}\label{Ei'}
R_e(n_e,m)&=n_e\ave{E}\\
R_a(n_a,m)&=n_a \ave{E},
\end{align}    
and the symmetry is indeed restored. Thus, the rates per atom are both equal to the average resonator energy at jump time. In the practical applications to be later considered, $m$ is a large number and consequently the above detailed considerations are of little consequence. 

\paragraph{Time evolution}

We now show that the average transition rates $R_e,~R_a$ must be supplemented by random rates $e(t),~a(t)$ if the variance of $m$ derived from Statistical mechanics is to be recovered.
We restrict our attention to large $N,~m~n$
values.  Since $m$ is large, it may be
viewed as a continuous function of time with a well-defined
time-derivative.  Because the standard deviation $\sqrt{N/4}$ of $m$
is much smaller than the average value, the so-called "weak-noise
approximation" is permissible. Within that
approximation, the average value of any smooth function $f(n,m)$ may
be taken as approximately equal to $f(\ave{n},\ave{m})$.
  
The evolution in time of a particular realization $m(t)$ of the
process obeys the classical Langevin equation
\begin{align}\label{L1}
    \frac{dm}{dt}= \mathfrak{E}-\mathfrak {A},
\end{align}
where  
\begin{align}\label{aux}
   {\mathfrak{E}} \equiv R_e(m)+e(t) \qquad {\mathfrak{A}} \equiv R_a(m)+a(t).
\end{align}    
In these expressions, $e(t)$ and $a(t)$ represent uncorrelated
random rates whose spectral densities are set equal to $ \alpha R_e(\ave{m})$ and $\alpha R_a(\ave{m})$, respectively, where $\alpha$ is a constant to be determined. 

Let us show that the variance of $m$ obtained from the above Langevin equation coincides with the result obtained directly from
Statistical Mechanics only if $\alpha=1$.  Without the noise sources, the evolution of
$m$ in (\ref{L1}) would be deterministic, with a time-derivative
equals to the drift term $R_e(m)-R_a(m)$.  If the expressions (\ref{Ei})
are employed, the Langevin equation (\ref{L1}) reads
\begin{align}
    \frac{dm}{dt} & =  Nm-2m^2+e-a \nonumber\\
	\spectral_{e-a} & =  \alpha \p R_e+R_a\q=\alpha N\ave{m}=\alpha N^{2}/2,
	\label{L3}
\end{align}
where the approximation $N\gg1$ has been made.
  
Let $m(t)$ be expressed as the sum of its average value $\ave{m}$ plus
a small deviation $\Delta m(t)$, and $Nm-2m^{2}$ in (\ref{L3}) be
expanded to first order.  A Fourier transformation of $\Delta m(t)$
with respect to time amounts to replacing $d/dt$ by $j\Omega$.
The Langevin equation now reads
\begin{align}\label{L4}
    j\Omega \Delta m= -N \Delta m+e-a\qquad \spectral_{e-a}=\alpha N^{2}/2,
\end{align}
where $m$ has been replaced by its average value $N/2$. 

Since the spectral density of some random function of time $z(t)=ax(t)$, where $a\equiv{a'+ja''}$
is a complex number and $x(t)$ a stationary process, reads
$\spectral_{z}(\Omega)=|a|^2 \spectral_{x}(\Omega)$, see Section \ref{random}, one finds from (\ref{L4}) that the spectral density of the $\Delta m(t)$ process is
\begin{align}\label{spb}
    \spectral_{\Delta m}(\Omega)=\frac{\alpha N^{2}/2}{N^{2}+\Omega ^{2}}.
\end{align} 
The variance of $m$ is the integral of $\spectral_{\Delta m}(\Omega)$ over
frequency ($\Omega/2\pi$) from minus to plus infinity, that is
var$(m)=\alpha N/4$. There is agreement with the previous result derived from the basic Statistical Mechanics rule only if $\alpha=1$. It follows that the spectral density of fluctuation rates such as $r(t)$ must be equal to the average rates, say $R$.

As seen earlier, the probability density of a resonator energy is
\begin{align}
\label{pd}
P(E)=(1-q)q^{m}\qquad m\equiv int(E),
\end{align}
where $int(x)$ denotes the largest integer not exceeding $x$. We have set for brevity $\hbar\omega=1$, and $q\equiv\exp(-1/T)$. Consider  in particular the $T=0$ limit. Initially, we know nothing about the resonator energy $E$, and we may suppose that the distribution of $E$ is uniform up to very large values. Let this resonator interact with a large collection of cold resonating atoms, that is atoms in the ground state. Each atom interacting with the oscillator has a non-zero probability of exiting in the upper state and thereby removing from it a unity amount of energy, as long as the resonator energy remains positive. It follows that after some time the oscillator energy is uniformly distributed between 0 and 1. This distribution leads to an average energy 1/2. This zero-temperature average energy is the same as the one predicted by Quantum Optics methods, but the variance obtained for $E$ is different. In the large $T$-limit, the above probability law in \eqref{pd} gives cumulants that coincide with the ones obtained from the Boltzmann distribution $P(E)\propto\exp(-E/T)$. This is not the case for the even cumulants (particularly the second cumulant, i.e., the variance) obtained from the Quantum Optics distribution. The difference is -1/12.

\section{Comparison with the classical expression}\label{heuristic}

In the classical regime, single-mode resonators at frequency $\om$ have a probability $\exp(-\beta E)$ of having an energy $E$, according to Classical Statistical Mechanics. Thus the average energy reads
\begin{align}\label{n4}
\ave{E}=\frac{ \int_0^\infty E\exp(-\beta E) dE } { \int_0^\infty \exp(-\beta E) dE }=\frac{1}{\beta}=T.
\end{align}
The action $f$ of an oscillator is defined as the ratio $\ave{E}/\omega$ of its average energy and frequency, and accordingly $f(x)=1/x$, setting $x\equiv\beta \ave{E}= \om/T$. Thus $f(x)$ obeys classically the Riccati equation
\begin{align}
\label{ric}
\frac{df(x)}{dx}+f(x)^2=0.
\end{align}

However, as was noted at the end of the 19\textsuperscript{th} century, the expression $\ave{E}=T$ leads to an infinite cavity energy since the number of electromagnetic modes is infinite in a cavity with perfectly reflecting walls. It was apparently not observed at the time that this difficulty is resolved simply by adding a constant $\p \hbar/2\q^2$ on the right-hand side of \eqref{ric}, that is, supposing that
\begin{align}
\label{ric'}
\frac{df(x)}{dx}+f(x)^2=\p\frac{\hbar}{2}\q^2,
\end{align}
where $\hbar$ is a universal constant with the dimension of action. The solution of this modified equation reads
\begin{align}
\label{solter}
x=\int \frac{df}{\p\hbar/2\q^2-f^2}=\frac{1}{\hbar}\log(\frac{2f+1}{2f-1})+x_o,
\end{align}
where $x_o$ denotes an arbitrary constant. For $x_o$=0 we obtain 
\begin{align}
\label{sol}
f(x)=\frac{\hbar}{2}\frac{\exp(\hbar x)+1}{\exp(\hbar x)-1}
\end{align}
which coincides with the Planck formula. The arbitrary constant $x_o$ on the right-hand-side of \eqref{solter} must vanish to obtain agreement with the classical result\footnote{ It was noted by Einstein and Stern in 1913, that $f(x)-1/x\to0$ if $x\to 0$, that is, the expansion of $f(x)$ is of the form
\begin{align}
\label{solbis}
f(x)=\frac{1}{x}+ax+...,
\end{align}
where $a$ is a constant, without an $x$-independent term.}. The finite black-body argument would still hold if the right-hand-side of \eqref{ric'} were multiplied by a number different from one. But then the requirement that at $T=0$ we have $\ave{E}=\hbar\omega/2$ would not hold.

To conclude, the classical expression of the average resonator energy generalizes to 
\begin{align}
\label{entotbis}
\ave{E}=\frac{\hbar \om}{2}\frac{\exp(\beta \hbar \om)+1}{\exp(\beta\hbar \om)-1}.
\end{align}

If we next consider a cavity with perfectly-conducting walls, solutions of the Maxwell equation exist only for a series of real resonating frequencies $ \om_1, ~\om_2, \cdots $. Each of these modes of resonance is ascribed an average energy given by the above expression with $\om$ replaced by $ \om_1, ~\om_2,\cdots$. For a d-dimensional cavity the mode density $\rho(\om)$, where $\rho(\om)d\om$ denotes the number of modes whose frequencies are between $\om$ and $\om+d\om$, grows in proportion of $\om^{d-1}$. It follows that the total energy is apparently infinite. This is perhaps why, in his original work, Planck subtracted the vacuum energy $\hbar \om/2$ from the expression given in \eqref{entotbis}. The total energy in a cavity of volume $\V$ in thermal equilibrium at absolute temperature $T$ is then found to be finite and proportional to the fourth power of $T$. If the cavity is pierced with a small hole that does not perturb much the state of thermal equilibrium, the measured output-power spectral density is supposed to be proportional to the internal energy spectrum, that is, to the product of $\ave{E(\om)}-\hbar\om/2$ and the mode density $\rho(\om)$. This leads to the famous black-body spectrum measured around 1899, and obtained by Planck in 1900. This formula agrees very well with measurements. In our formalism we consider that the power radiated from the small hole pierced in the cavity is in fact collected by an absorber at $T=0$K having its own noise source. A finite result is then obtained from the expression of $\ave{E(\om)}$ as given in \eqref{entotbis}, without having to subtract the $T=0$ energy term.

\section{Energy of random sources driven oscillators.}\label{ener}

Having established that the average energy of a cold resonator is $\ave{E(\om)}=\hbar\om/2$, let us show that this result is consistent with the expressions given earlier for the spectral densities of the random currents $C'(t), ~C''(t)$.

A single-mode resonator may be in a state of thermal equilibrium with a small negative conductance and a small larger positive conductance (in absolute values). We  consider an inductance-capacitance ($L-C$) oscillator resonating at frequency $\om_o$ ($LC\omega_o^2=1$) with a small negative conductance $-G_e$  (subscript "$e$" for "emitting") and a small positive  conductance $G_a$ (subscript "$a$" for "absorbing") in parallel.  $G_a$ exceeds $G_e$ so that there is positive absorption: the resonator is slightly damped. We set $G_a-G_e\equiv G>0$. The conductance $G_a$ is supposed to represent the absorption by two-level atoms in the lower state, while $G_e$ represents the emission from atoms in the upper state. A near-resonance condition at frequency $\omega_o$ holds. According to the Schrödinger equation (see the section \ref{electronmotion}), the conductances are proportional to the corresponding numbers of atoms, $n_a$ and $n_e$, respectively. 
\begin{figure}
\setlength{\figwidth}{0.45\textwidth}
\centering
\begin{tabular}{cc}
\includegraphics[width=\figwidth]{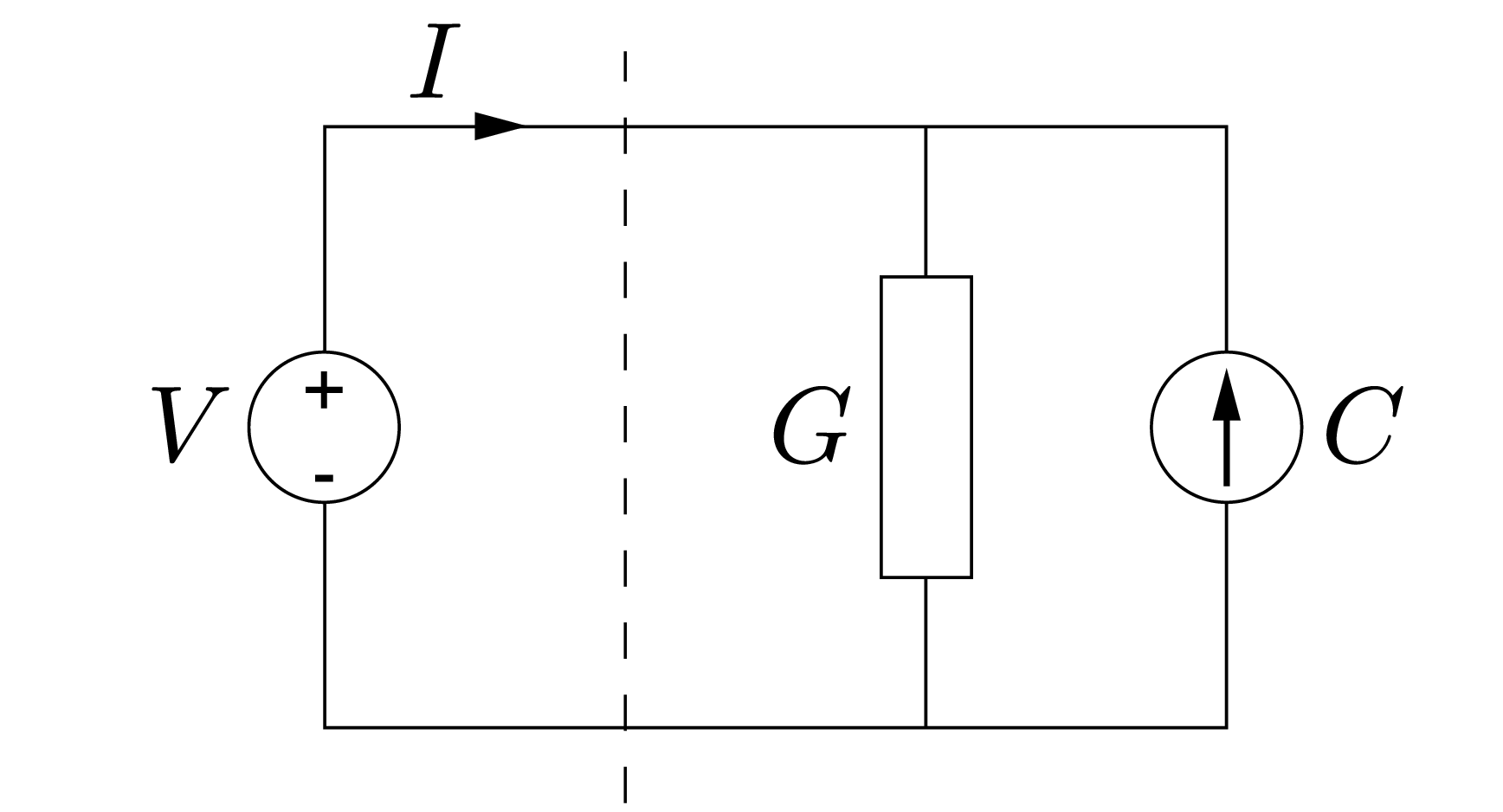} & \includegraphics[width=\figwidth]{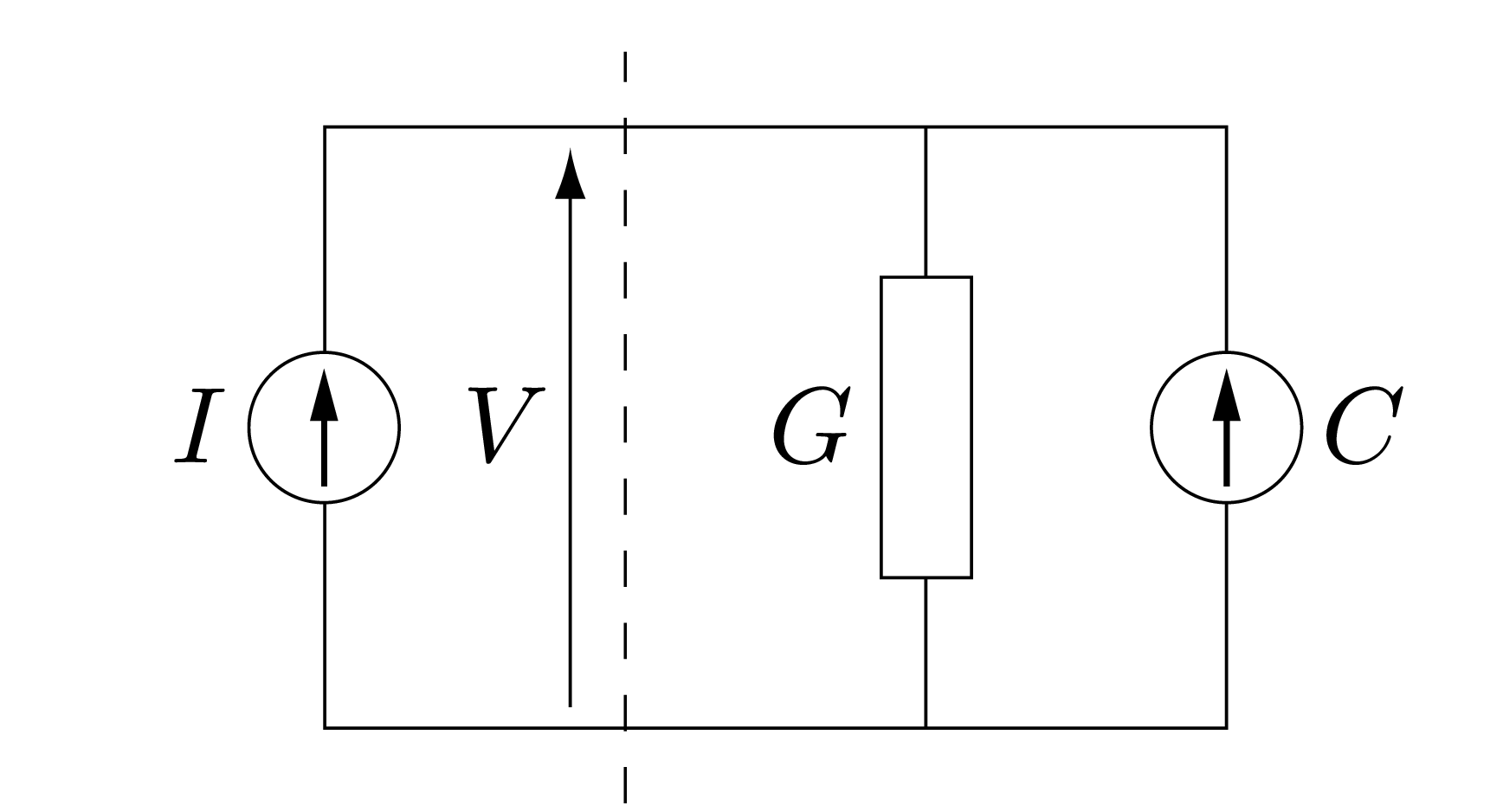} \\
(a) & (b) \\
\includegraphics[width=\figwidth]{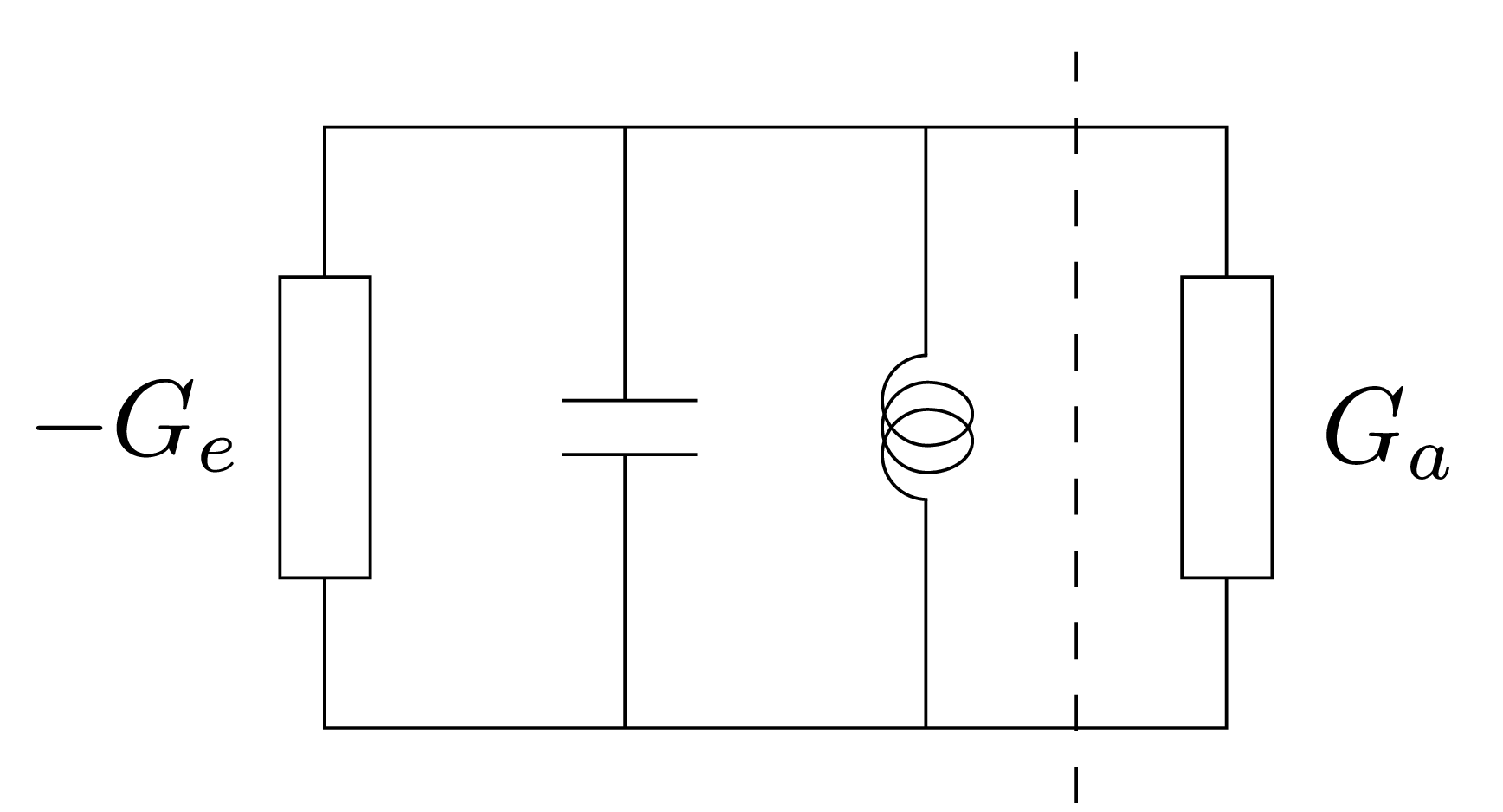} & \includegraphics[width=\figwidth]{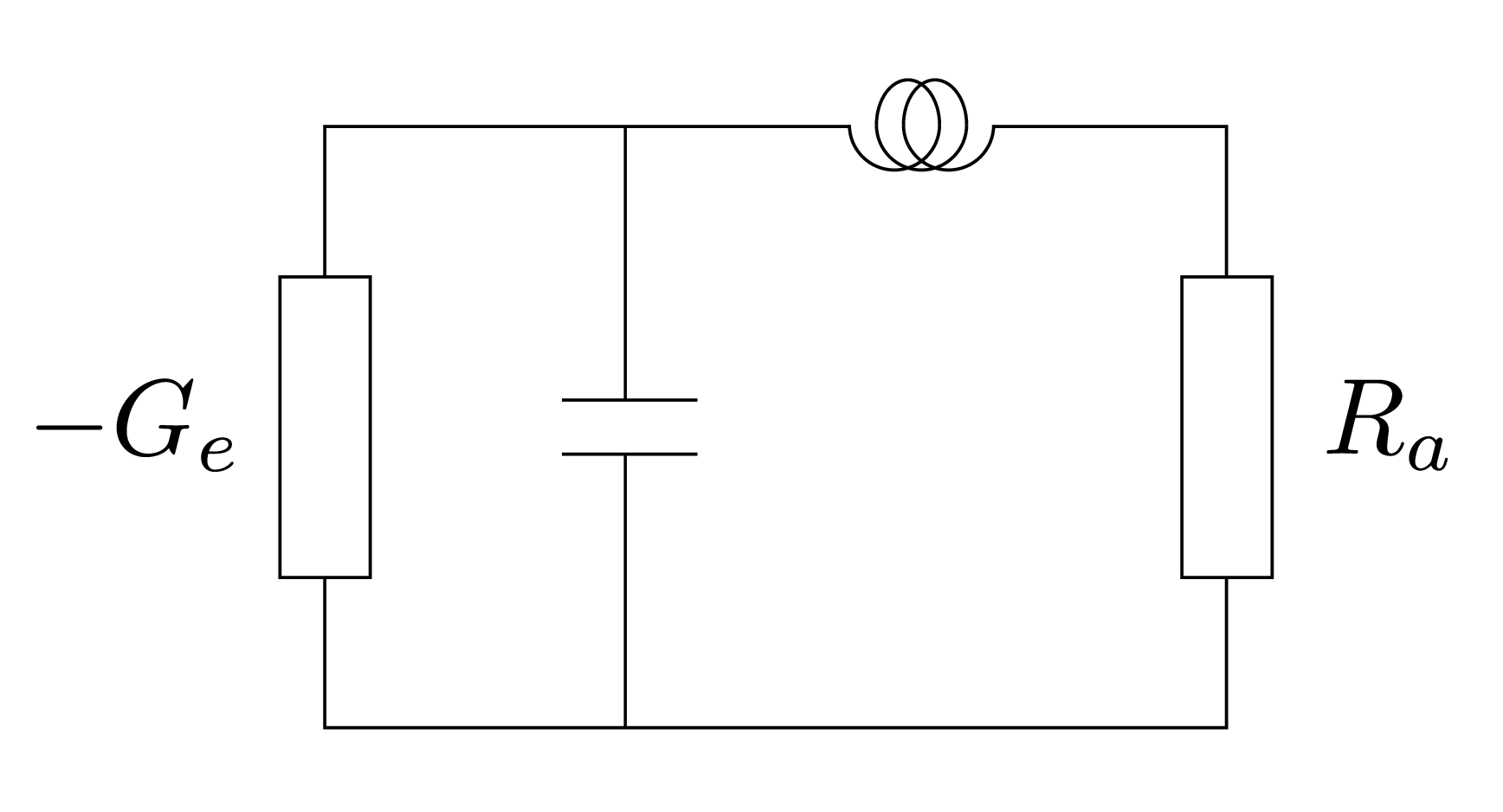} \\
(c) & (d) \\
\end{tabular}
\caption{a) represents a potential source $V$ applied to a conductance $G$. $C$ represents the random current source associated with $G$. We are mostly concerned with the power flowing from left to right through the dashed line, b) represents a conductance driven by a current source $I$, c) is a laser model with a negative conductance $-G_e$, a resonating circuit, and a positive conductance $G_a$, representing the detector of radiation, connected in parallel, d) represents a laser model that exhibits a linewidth-enhancement factor.}
\label{circuitlaser}
\end{figure}

Because the conductances considered are constant the circuit is linear. The potential $V$ across the circuit is therefore equal to $\C/Y(\om)$, where $\C$ denotes the driving current, assumed to be independent of frequency, and the resonating circuit admittance $Y(\om)=G+\ii B(\om)$ where $B(\om)$ represents the sum of the $L$ and $C$ susceptances. Referring to \eqref{en'} the oscillator energy is
\begin{align}
\label{ennmk}
E= \int_{-\infty}^{+\infty}\frac{d\om}{2\pi}\frac{C\abs{\C}^2 }{G^2+4C^2\p \om-\om_o\q^2}
= \frac{\abs{\C}^2 }{4G} \frac{1}{\pi} \int_{-\infty}^{+\infty}\frac{dx}{1+x^2}=\frac{\abs{\C}^2 }{4G}.
\end{align}

Let this oscillator be in thermal contact with a bath at absolute temperature $T$. We replace the deterministic current $\C$ by a complex random function of time\footnote{Note that in the present linear regime the regulation mechanism at work in above-threshold lasers does not occur, and the fluctuations of $V$ are comparable to average values. Supposing that the current source is gaussian distributed, this is also the case for the optical potential $V$ and optical current $I$. Power should in general be evaluated as the real part of $V^\star ( I+ C(t))$, but in the linear regime presently considered the term $C(t)$, much smaller than the fluctuations of $I$, may be neglected. The power $\Re\{V^\star  I\}$ is Rayleigh-distributed.}. Because the processes are stationary we expect that the statistical properties of the random source $C(t)\equiv C'(t)+\ii C''(t)$ are unaffected by an arbitrary phase change, that is, we require that $C(t) \exp(\ii \phi)$ has the same statistical property as $C(t)$ for any phase $\phi$. This entails that $C'(t)$ and $C''(t)$ are uncorrelated and have the same statistical density. We thus set $\spectral_C'=\spectral_C''\equiv \spectral$. We have seen that for $n$ independent atoms in some state both $G$ and $\spectral$ are proportional to $n$. We therefore expect that $\spectral=\al G$, where $\al$ is a constant to be determined. 

Because of the symmetry between stimulated absorption and stimulated emission implied by the Schrödinger equation the spectral densities have the same form for positive conductances $G_a$ and negative conductances $-G_e$, namely $\spectral_a=\al G_a$ and $\spectral_e=\al G_e$, with the same constant of proportionality $\al$. Note that $C'$ and $C''$ contribute equally and that double-sided spectral densities for $C'$, $C''$ in the Fourier $\Om$-domain are employed. If the conductances $G_a$ and $-G_e$ are connected in parallel the total conductance is $G=G_a-G_e$, as said above, and the total spectral density is $2\al \p G_a+G_e\q$. The average resonator energy follows from \eqref{ennmk}
\begin{align}
\label{emk}
\ave{E}=\frac{2\al (G_a+G_e) }{4G}=\frac{\al}{2}\frac{G_a/G_e +1}{G_a/G_e -1}.
\end{align}

Classical Statistical Mechanics tells us that at thermal equilibrium $\frac{n_a}{n_e}$ is equal to $\exp(\frac{E_e-E_a}{T_m})$. It follows that
\begin{align}
\label{ennm}
\frac{G_a}{G_e}=\frac{n_a}{n_e}=\exp\p\frac{E_e-E_a}{T}\q=\exp\p\frac{\hbar \om_o}{T}\q.
\end{align}
Classical Statistical Mechanics also tells us that in the classical limit the average energy equals $ T/2$ per degree of freedom, and thus $\ave{E}= T$ for the oscillator considered when $T\gg \hbar\om_o$. According to \eqref{emk} this is the case if and only if
\begin{align}
\label{em}
\al=\hbar \om_o.
\end{align}

We have thus obtained expressions of the spectral densities of the random current sources associated with a conductance that lead to the average energy of a resonator in contact with a bath at temperature $T$.

\section{Waiting time}\label{sec_wait}

The mechanism behind the probability per unit time that an electron jump from one state to another under the influence of the optical field was at the Einstein time, and even later, rather mysterious. In the present section we consider an optical field which is prescribed, that is, the field is not significantly affected by the electron motion. This is a permissible approximation when the number of light quantas $m$ is much larger than unity. Under that approximation, the optical field may be treated as a time-varying potential added to the static potential in the Schrödinger equation. It is found that the probability that an electron, being in the lower state at $t=0$ (say), be found in the upper state upon a measurement at time $t$ is $\sin^2(\frac{\Om_R}{2}t)$, where the Rabi frequency $\Om_R$ is proportional to the optical field $\E$. This result in itself does not lead to a transition probability per unit time as required by the Einstein concept. It is necessary to introduce a phenomenological parameter $\gamma$. When this parameter is large enough we do indeed obtain a waiting time which is Poisson distributed. This result, derived in the present section, leads to the desired probability concept. It also enables us to recover the result obtained in the previous chapter, namely that a Poisson random rate must be ascribed to any absorbed or emitted rate. If $m$ is not large compared with unity, the present treatment is inapplicable. For example, if $m=0$ and the electron is in the lower state at $t$=0, conservation of energy requires that that electron remains in the lower state at any time. Since the optical field $\E$ may not be strictly equal to zero, one must presume that, instead of being prescribed as assumed above, it evolves in the course of time under the influence of the electron motion in such a way that the conclusion just stated holds.  

\section{Induced current}\label{electronmotion}

We consider in this chapter a single electron submitted to an optical field, considering first the classical treatment applicable, for example, to reflex klystrons. We then undertake the quantum treatment. Provided a phenomenological constant $\gamma$ is introduced in the well-known Rabi equations, we obtain the average emitted rate $\ave{ R}$ and the fluctuation $\Delta R$ of that rate. If $\gamma$ is large, the emitted rate is Poisson. If $\gamma$ is not large, the emitted rate is sub-Poisson, as is the case in resonance fluorescence. However, the superposition of a large number of point distributions being known to be Poisson, for a large number of independent electrons the emitted rate is Poisson in any event. This is the conclusion that we have reached before by various means based on Statistical Mechanics, which is relevant because the non-zero constant $\gamma$ term makes the one-electron system part of a larger system. The pure-state density matrix that would describe an electron submitted to a prescribed field becomes a mixed-state density matrix.

Electron motion is described by a solution of the Schrödinger equation in which enters the sum of a static potential that determines the electron level energies, and a potential of the form $x\E(t)$, where $x$ is a spatial coordinate and $\E(t)$ denotes a classical electric field varying sinusoidally in time at frequency $\omega$. A resonance occurs when $\hbar\omega$ is equal to the energy difference between two atomic levels, according to the Schrödinger equation, the other levels being ignored. When this is the case, the probability that the electron be found in the lower (or upper) state varies sinusoidally as a function of time (Rabi oscillation at frequency $\Omega_R\ll\omega)$. In other words, there is a continuous back and forth exchange of energy between the electron and the device that generates the alternating field. It follows that the average conductance "seen" by the field source vanishes. At that point one must introduce a mechanism allowing Rabi oscillations to be somehow interrupted. In the case of resonance fluorescence, this interruption is caused by spontaneous emission. One supposes instead here that the electron is not fully confined by the static potential but may escape through tunneling. In any case, a phenomenological parameter $2\gamma$ need be introduced. As a result, the field source "sees" a well-defined conductance $G$, with events that interrupt the Rabi process. As said above, if $n$ non-directly coupled electrons are present instead of a single one, the average conductance is multiplied by $n$, and the event-times statistics  tend to be Poisson-distributed for large $n$ values.

We will point out a close similarity between the configuration just described and (idealized) reflex klystrons. These vacuum tubes are microwave oscillators discovered by the Varian brothers before the second world war. The electron motion there occurs between a cathode and a reflector, approximately at the same potential. The electron moving back and forth between the cathode and the reflector, interacts resonantly with a resonator made of two parallel grids (permeable to electrons but ensuring a uniform alternating field) and an inductance. This resonator, located between the cathode and the reflector, is positively biased with respect to the cathode at potential $U$. 

The motion of successive electrons is perturbed by the field in such a way that these electrons get bunched, that is, grouped together in the form of a periodic sequence of electronic packets. In return, these bunches induce a current back to the field source. However a definite (negative) conductance appears only if the electrons are allowed to be captured by the anode. The probability that this capture occurs is somewhat analogous to the $\gamma$ parameter mentioned above. In fine, the constant potential $U$ delivers a power which, ideally, would be entirely converted into electromagnetic energy. The converse process may also occur.

There are therefore strong similarities between the lasers considered and reflex klystrons. The differences relate mainly on the approximations that are allowed in the treatment of klystrons, but not in the case of lasers. To wit, in klystrons 1) the electron motion under the influences of the static and alternating fields is treated according to the laws of classical mechanics, 2) the current induced in the resonator is essentially the electron momentum, and 3) a probability that the electron be captured by the anode when it has lost most of its energy is given. In the case of lasers, the electron is described by the Schrödinger equation, 2) the induced current is evaluated by the methods of quantum mechanics and 3) instead of the capture probability mentioned above, a phenomenological parameter $\gamma$ is introduced that induces a decay in the Rabi equations. Note that space-charge effects (Coulomb interaction between electrons) have not been considered in the above discussion, even though this is an important factor in klystrons\footnote{Space charge renders the electron arrival events sub-Poisson, which is a favorable effect. On the other hand space charge prevents electrons to be fully bunched. The latter effect reduces the device efficiency.}. We are now going to review the above steps one at a time.

Consider first a single electron located between two parallel conducting plates. A static fixed potential source $U$ and an alternating potential source $v(t)$ at the optical frequency $\om$ are applied to the plates. The potential $v(t)$ is presently supposed to be independent of the current that the electron motion may induce. The sign convention is the one given in Fig. \ref{circuitfig} in (c). In both the Classical and Quantum Theories, the induced current $i(t)$ is, to within a constant, equal to the electron momentum $p(t)$, although the interpretations of $i(t)$ and $p(t)$ differ. The power $v(t)i(t)$ supplied by the optical potential, once averaged over an optical period, is denoted by $P(t)$. 

The over-all effect of the electron motion is to transfer energy from the static source to the alternating source (stimulated emission) or the converse (stimulated absorption). Spontaneous emission in the usual sense does not occur because the optical cavity is closed and coupling to free-space modes does not take place. What may occur is that the electron is emitted by the cathode and captured by the anode (resonator), or remains in the interaction region for a limited time. In the Classical treatment, one first evaluate 1) the electron motion under the static field, 2) the perturbation caused by alternating field, and 3) the electron momentum and the induced current. The same steps are taken in the Quantum treatment. Namely, we consider the stationary states of the electron submitted to the static field, the perturbation of those states due to the alternating field, and finally evaluate the induced current from the electron momentum. 

\section{Classical Equations of Electron Motion}\label{classicalelectronmotion}

The equations of motion of an electron of charge $-e$ and mass $m$ are first established for the case of a static (time-independent) potential. As an example consider an anode at zero potential and an electron emitted from a cathode at potential $-U$ in vacuum, and look for the electron motion and the induced current. If $i(t)$ denotes the current delivered by the potential source, the power $Ui(t)$ must be equal at any instant to the power delivered to the electron, which is the product of the velocity $p(t)/m$, where $p$ denotes the electron momentum, and the force $eU/d$ exerted upon it, where $d$ denotes the electrode spacing. Since the potential $U$ drops out from this equation, the current is
\begin{align}
\label{currentb}
i(t)=\frac{e}{md}~p(t).
\end{align}

Solving the equations of motion, we find that the diode current $i(t)=\frac{e^2U}{md^2} t$ increases linearly with time and drops to zero when the electron reaches the anode. Thus, each electron freed from the cathode entails a triangularly-shaped current pulse. If $i(t)$ is integrated over time from $t=0$ to $t=\tau$ we obtain of course the electron charge $e$. We will neglect the pulse duration (or transit time) $\tau$, so that triangularly-shaped current pulses are approximated by $e\de(t)$-functions. The above theory is applicable only when few electrons are emitted so that the initial electron velocities and space-charge effects are neglected. 

\begin{figure}
\centering
\begin{tabular}{cc}
\includegraphics[scale=0.35]{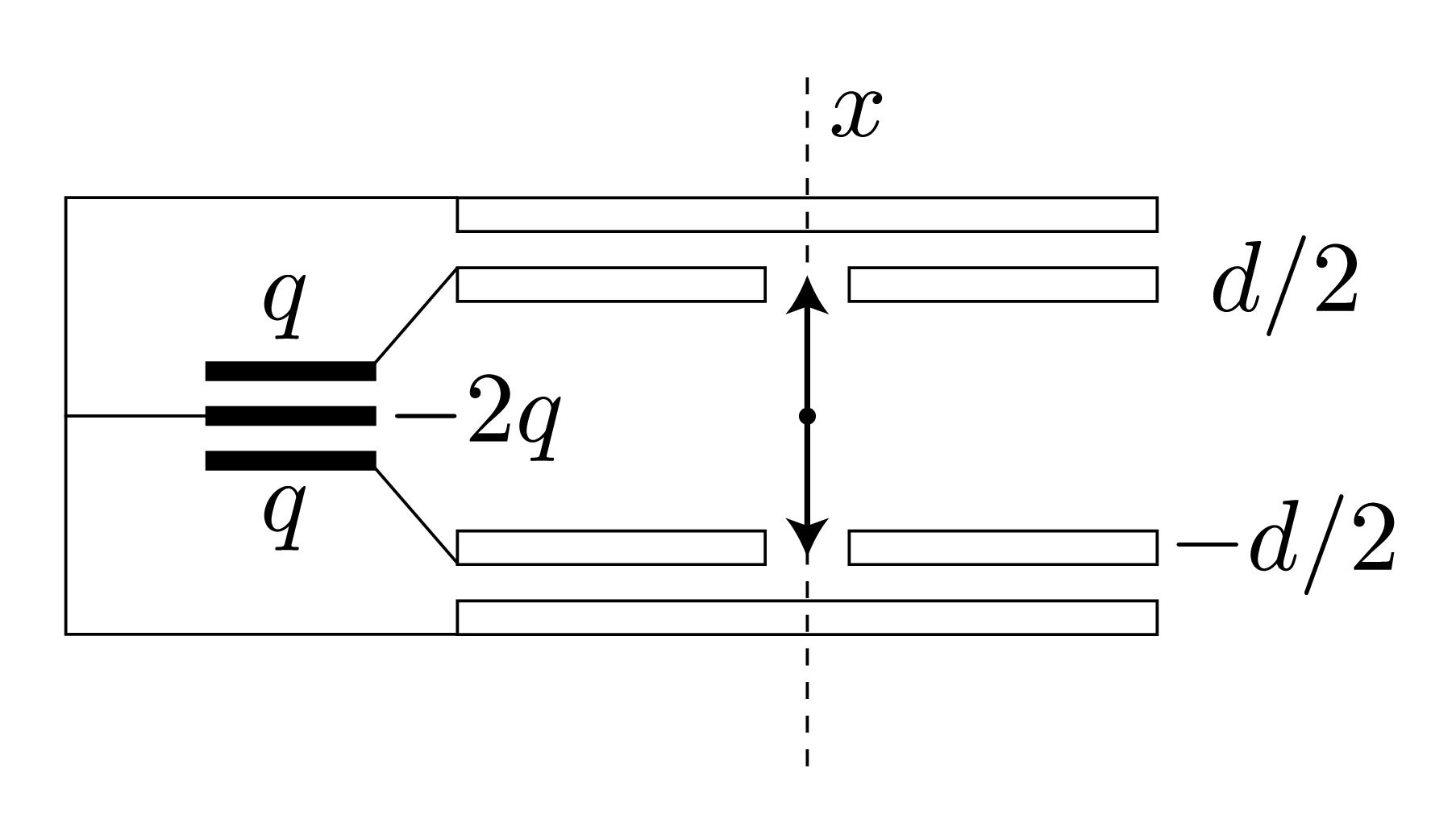}   &  \includegraphics[scale=0.35]{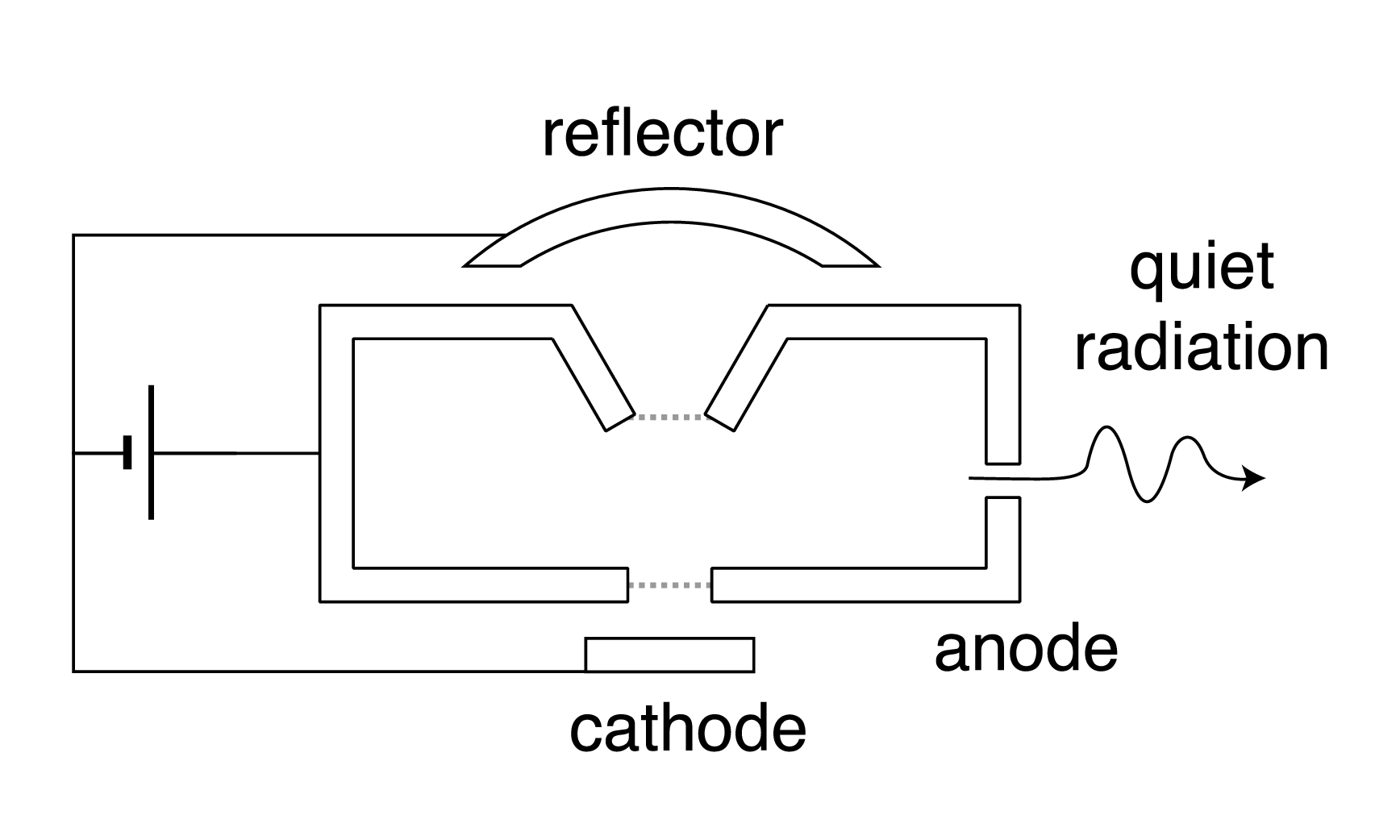}  \\
(a) & (b) \\
\includegraphics[scale=0.35]{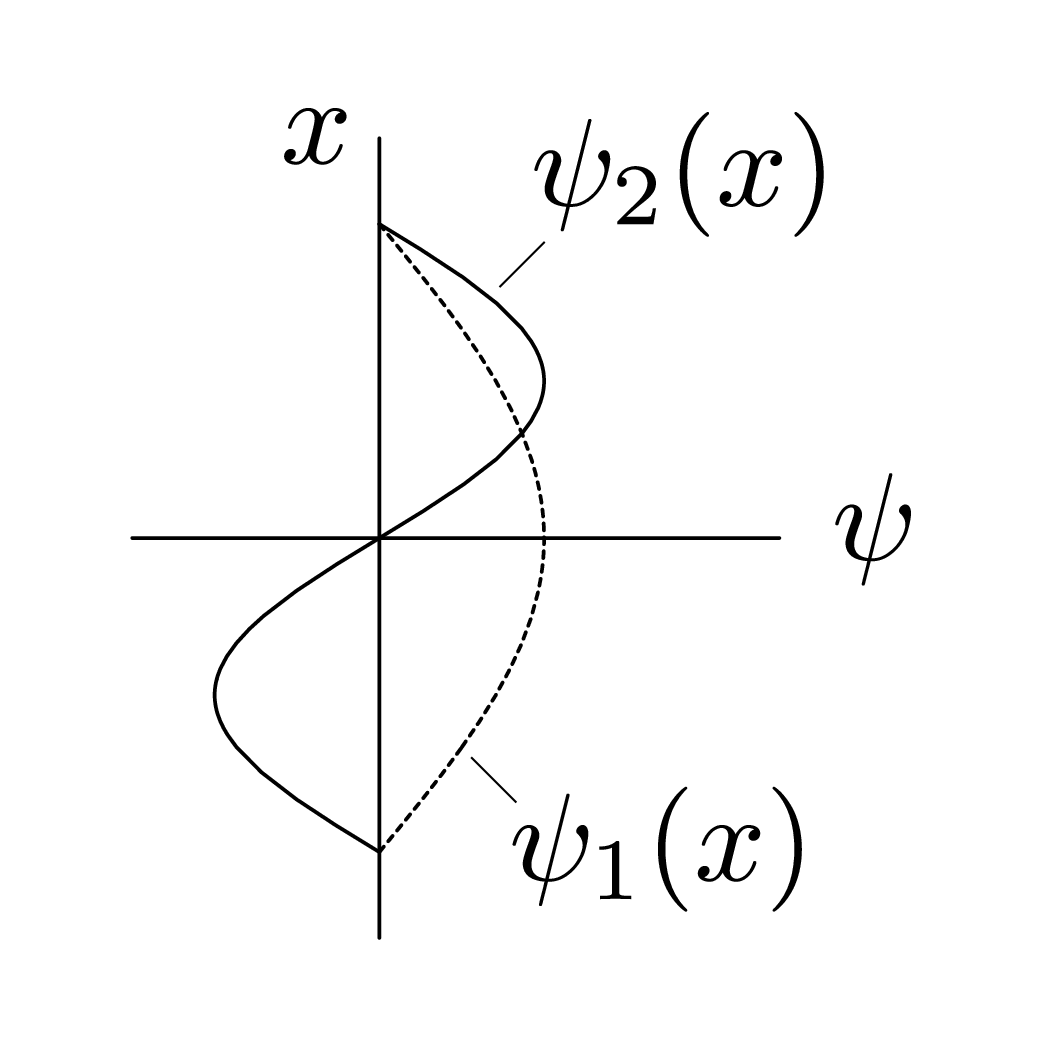}   &  \includegraphics[scale=0.35]{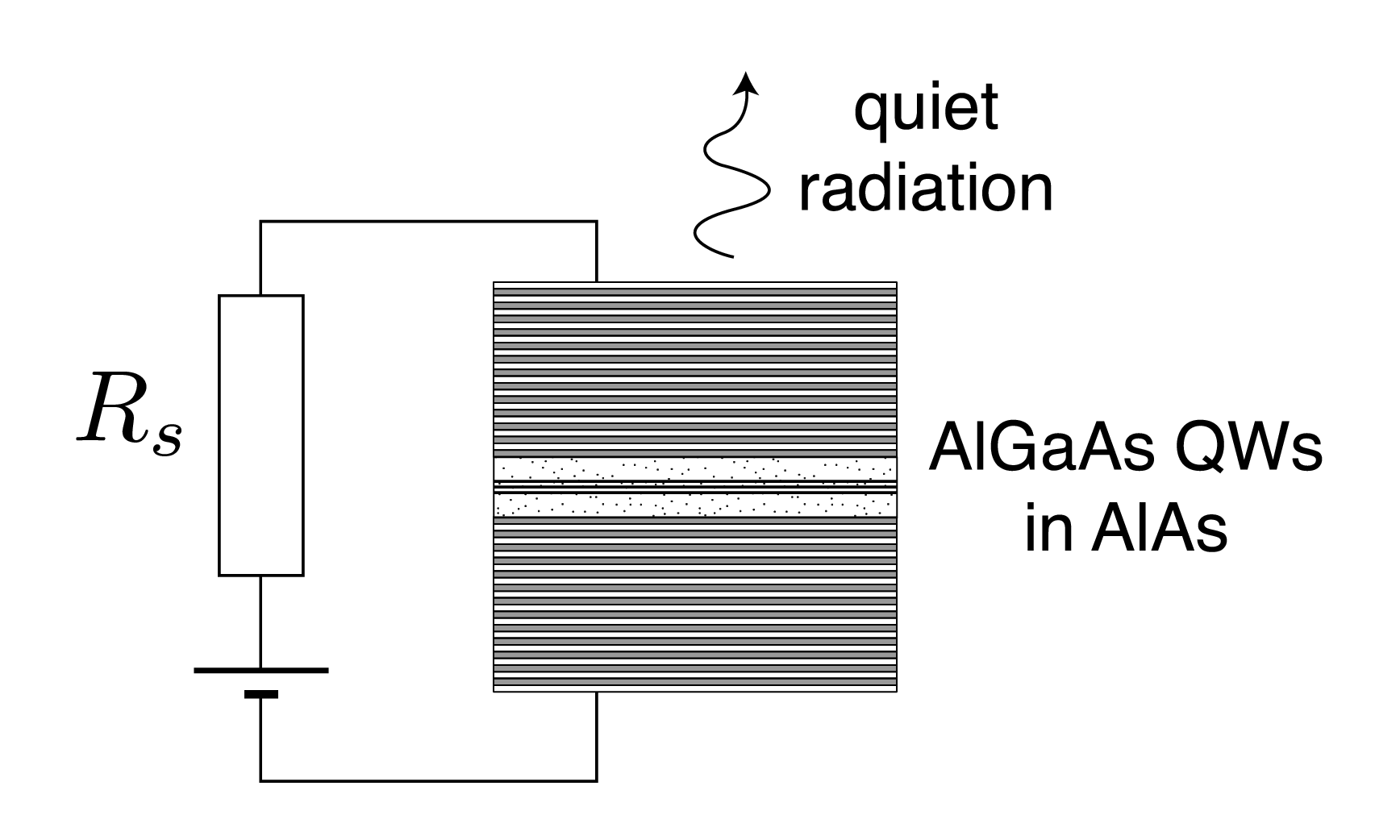}  \\
(c) & (d)
\end{tabular}
\caption{a) Illustrates the potential created by two anodes (inner electrodes) and two cathodes located just outside the anodes. The potential is generated by a large, charged, capacitance, shown on the left. According to the Classical Picture the space-time electron trajectory is almost  a zig-zag path, with slight incursions of the electron between the anodes and the cathodes. b) Represents a reflex klystron, which is similar to the previous schematic, but with a resonator added to it. The current is regulated by a space-charge limited cathode. c) Pictures the wave-functions of the ground state and first excited state of a square potential well. d) Represents a "surface-emitting" laser diode. The current is regulated by a large cold resistance $R_s$.}
\label{klystron}
\end{figure}

As a second example, consider a one-dimensional square-well, whose potential is equal to 0 for $\abs{x}<d/2$ and infinite (or nearly so) beyond. This potential may be generated by parallel anodes at potential 0 and cathodes at potential $-U$, as shown in Fig.~\ref{klystron}. The electron space-time trajectories $x=x(t)$ consist of straight lines with slopes $dx(t)/dt=±p/m$, where $p^2/2m=E$ is the electron energy, which may be selected arbitrarily from 0 to $eU$ so that the electron is not captured by the cathodes. The electron is prevented from being captured by the anodes by a strong magnetic field in the $x$-direction. The quick electron incursions between anodes and cathodes are here neglected. We may consider in particular a lower electron energy $E_1$ and a higher electron energy $E_2$, corresponding to small and large slopes in the $x=x(t)$ diagram, respectively.  

The electron motion induces an electrical current $i(t)$ in the potential source proportional to the electron momentum $p(t)$, as said above. In the case of a static potential source the induced current does not correspond to any power delivered or received by the source on the average, so that the electron motion may go on, in principle, for ever.

If now the static potential $U$ is supplemented by a sinusoidal potential $v(t)$ of small amplitude, whose frequency is resonant with the electron motion described above, the \emph{unperturbed} electron momentum $p(t)$ does cause the alternating potential source to receive or generate power, depending of the electron state. However, if we consider a large collection of unperturbed electrons, the power averages out again to zero. It follows that a net energy transfer may be obtained only if we take into account the fact that the alternating potential \emph{perturbs} the electronic motion. In the present classical picture this amounts to bringing all the electrons with the appropriate phase with the alternating potential, an effect called "bunching". This name originates from the fact that electrons initially spread out uniformly on the time scale eventually are forced by the alternating field to form periodic "bunches". This desirable bunching effect is limited by the velocity spread of electrons originating from the hot cathode and the fact that electrons tend to repel each others (space-charge effect).

This is not however the end of the story. The electron, initially in the lower energy state, may gain enough energy to be captured by a cathode. Once in a cathode, the electron flows through the potential source to the anode, delivering an energy $eU$ to that source, and is emitted by the anode back into the lower energy state. The net effect of these processes is that some power is being transferred from the alternating potential source to the static potential source, or the converse, the electrons playing an intermediate role.

In more general situations, the Classical Equations of Motion of electrons of charge $-e$, mass $m$, and potential energy $-eu(x,t)$ are best based on the Hamiltonian formulation. The particle total energy $E(t)$ is expressed as a function of position $x$, momentum $p$, and time $t$ according to the relation
\begin{align}
\label{ham}
H(x,p,t)-E(t)\equiv \frac{p^2}{2m}-eu(x,t)-E(t)=0,
\end{align}
where $p^2/(2m)$ represents the kinetic energy. The Hamiltonian equations read
\begin{align}
\label{eqmot}
\frac{dx(t)}{dt}&=\frac{\partial H(x,p,t)}{\partial p }=\frac{p(t)}{m}\\
\frac{dp(t)}{dt}&=-\frac{\partial H(x,p,t)}{\partial x}=e\frac{\partial u(x,t)}{\partial x}.
\end{align}
The first equation says that the particle momentum $p(t)=m dx(t)/dt$, and the second equation may be written, with the help of the first equation, in the Newtonian form $m~d^2x(t)/dt^2=e~\left(\partial u(x,t)/\partial x\right)_{x=x(t)}$. Going back to the first example in this section, let us consider two parallel plates located at $x=0$ and $x=d$, and at potentials $0$ and $-u(t)$, respectively. We have $u(x,t)=-u(t)x/d$, and thus the equation of motion is $md^2x(t)/dt^2=-eu(t)/d$. The electron decelerates if $u(t)>0$. The electron is repelled by the negatively-charged cathode.

As far as static conditions are concerned, an electron submitted to a static potential source $U$ is analogous to an electron submitted to the Coulomb potential created by positively-charged nuclei. The potentials, on the order of 1 volt (corresponding to potential energies of 1.6 10$^{-19}$ joules) are comparable in the two situations. In the case of atoms, however, the Bohr radius, which is roughly equivalent to our distance $d$, is on the order of 0.05 nanometers while, in the case of two conducting plates, the distance can hardly be less than 100 nanometers for practical reasons. As a consequence there exist in the two-plate model many states whose energy is comprised between the lower-state energy $E_1\approx 0$ and the upper-state energy $E_2\approx eU$. This is why a classical mechanical treatment is appropriate in the case of microwave vacuum tubes. In both the microwave and the optical cases the conductance (ratio of the induced current to the applied potential) is initially equal to zero and grows in time linearly until the electron is somehow absorbed or leaves the interaction region. 

\section{Quantum Equations of Electron Motion}\label{quantum}

The configuration that we have in mind is again the one shown in Fig. \ref{klystron} in (a), but with the electron motion quantized as in (c). In this picture, as was discussed earlier, the electron is submitted to a static potential source generated by a charged capacitance of arbitrarily large value. This potential is applied between inner electrodes and outer electrodes. The electron is constrained to move along the $x$-axis with the help of a magnetic field (not shown on the figure). Classically, the electron performs a zig-zag $x(t)$ path. From the time-independent Schrödinger-equation view point, the electron may reside only in a lower state 1 and an upper state 2. In real masers or lasers the potential is generated by the static potential of fixed, positively-charged point-like nuclei (because of their large mass, plates or nuclei recoils may be neglected). However, from our view-point, the two configurations differ only in the form of the wave-functions and the value of the transition element later on denoted by $x_{12}$. 

The static potential is supplemented by an alternating potential source, at a frequency on the order of 10 GHz for klystrons and 300 THz for lasers, generated in the picture of Fig. \ref{klystron} in (a) on the left by a sinusoidal motion of the inner capacitance plate. In a real klystron, the alternating potential is generated by an inductance-capacitance circuit, as shown in Fig. \ref{klystron} in (b), In general, this oscillator, resonant with the electron alternating motion, cannot be considered as a \emph{source}, because the potential depends on the induced current (remember that according to our terminology "sources" supply potentials or currents that are independent of the load). It is only in the limit where the tuned-circuit capacitance would be extremely large and the inductance extremely small (the resonating frequency remaining the same), that this tuned circuit could be considered as an alternating potential \emph{source}. Indeed, in the limit considered, for a given alternating field, the tuned circuit energy is extremely large and little affected by the electron motion\footnote{In quantum optics, this limiting situation is described by saying that "the number of photons in the cavity is supposed to be extremely large, so that a classical treatment of the field is adequate".}. Let us emphasize that the configuration treated in the present section is only one idealized component of a complete laser device. We need to characterize this component accurately (in terms of conductances and event time statistics) before going on.

What is missing in the schematics of Fig. \ref{klystron} in (a) is the absorber of radiation. In that schematics this absorber could be realized by adding on the right a triple-plate capacitance, as already shown on the left. Similarly, in Fig. \ref{klystron} in (b) the wavy line, symbolizing the escape of radiation, could be replaced by a potential configuration similar to the one shown on the left, but with a slightly different static potential, so that power flows from the potential source on the left to the potential source on the right, rather than the opposite. The escaping radiation is collected by a detector, and eventually converted back to an electrical current. The electron motion and the alternating field may be viewed as playing intermediate roles. If this is the case, one may wonder why complicated devices are needed to merely transfer energy from one capacitance (or battery) to another. The answer of course is that either in the microwave or optical forms energy may be carried over large distances with little absorption or diffraction loss. High-frequency electromagnetic waves also serve as sensors, e.g., in the radar.

The Quantum Equations of Motion of an electron of charge $-e$ and mass $m$ are first established for a static (i.e., time-independent) potential source. As an example we consider a one-dimensional square-well, whose potential is equal to 0 for $\abs{x}<d/2$ and infinite (or nearly so) beyond. This potential may be generated by parallel anodes at potential 0 and cathodes at potential $-U$, as shown in Fig.~\ref{klystron}. We solve the time-independent Schrödinger equation and obtain in particular a state 1 with lower energy $E_1$ and a state 2 with higher energy $E_2$. As we shall see, these two states correspond to wave-functions $\psi_1(x)=\cos(x)$ and $\psi_2(x)=\sin(2x)$, respectively, leaving aside constants. In the case of a static potential there is no energy exchange between the potential source and the electron when the electron is initially in a stationary state, so that the electron remains in stationary states, in principle, for ever. There are no energy exchange either if we perform a time averaging when the electron is in a superposition of stationary states. This situation may be compared to the one discussed classically above. 

Let now the static potential source $U$ be supplemented by a sinusoidal potential source $v(t)$ of small amplitude, whose frequency is (in some sense to be defined later) resonant with the electron motion described above. A net energy transfer may be obtained only if we take into account the fact that the alternating potential \emph{perturbs} the electronic motion. In the Classical picture this amounts to bringing all the electrons with the appropriate phase, an effect called "bunching", as said previously. In the Quantum picture (time-dependent Schrödinger equation), the electron wave function $\psi(x,t)$ is the weighted sum of the unperturbed states defined above, with time-dependent weights. The theory leads to (Rabi) oscillations between the two states. Initially, the induced current is equal to zero and grows in proportion to time, but the conductance vanishes on the average. A non-zero positive conductance is obtained if the electron initially in the lower state does not remain permanently in the interaction region.  We may then evaluate the average conductance "seen" by the optical potential source.

The quantum treatment is based on the Schrödinger equation
\begin{align}
\label{sch}
[H(x,p,t)-E]\psi(x,t)=0,\quad E=\ii \hbar\partial/ \partial t,\quad p=-\ii \hbar\partial/ \partial x,
\end{align}
where the sign "$\partial$" denotes partial derivation.  $\psi(x,t)$ is called the wave-function, whose initial value $\psi(x,0)$ is supposed to be known, and $H(x,p,t)=\frac{p^2}{2m}-eu(x,t)$ as in the Classical Equations of Motion, but $p$ and $E$ are now operators of derivation. It is easily shown that, provided $\psi(x,t)$ decreases sufficiently fast as $x\to±\infty$, the integral over all space of $\abs{\psi(x,t)}^2$ does not depend on time. It therefore remains equal to 1 if the initial value is 1, a result consistent with the Born interpretation of the wave function. Because of linearity the sum of two solutions of the Schrödinger equation is a solution of the Schrödinger equation (superposition state). The wave-functions add up, but not in general the probabilities.

Let us state the first Ehrenfest equation
\begin{align}
\label{eh}
\frac{d\ave{x(t)}}{dt}=\frac{\ave{p(t)}}{m}.
\end{align}
Thus the classical relation $p=m\frac{dx}{dt}$ still holds provided $x$ and $p$ be replaced by their QM-averaged values.

\section{Static potentials}\label{static}

Let us suppose that $u(x,t)\equiv u(x)$ does not depend on time. In that case solutions of the above equation of the form $\psi(x,t)=\psi_n(x)\exp(-\ii\om_nt)$ may be found, where $n=1,2...$. The $\psi_n(x)$ are real functions of $x$ and $E_n\equiv \hbar\om_n$ that form a complete orthogonal set of functions. For $n=1,2$ the wave functions obey the differential equations
\begin{align}
\label{diffwave}
\frac{\hbar^2}{2m}\frac{d^2\psi_1(x)}{dx^2}+eu(x)\psi_1(x)+E_1\psi_1(x)=0\nonumber\\
\frac{\hbar^2}{2m}\frac{d^2\psi_2(x)}{dx^2}+eu(x)\psi_2(x)+E_2\psi_2(x)=0.
\end{align}
with the appropriate boundary conditions. They may be ortho-normalized such that
\begin{align}
\label{norm}
\int_{-\infty}^{+\infty}dx~ \psi_m(x) \psi_n(x)=\de_{mn},
\end{align}
where $\de_{mn}=1$ if $m=n$ and 0 otherwise, and they form a complete set. 

\section{Potential well}\label{free}

As an example consider an electron of mass $m$ moving along the $x$ axis be reflected by boundaries at $x=-d/2$ and $x=d/2$ where the wave-function is required to vanish, that is, $\psi(±d/2)=0$. The lowest-energy state $n=1$ and the first excited state $n=2$ are
\begin{align}
\label{wf}
\psi_{1}(x,t)&=\sqrt{2/d} \cos(\pi x/d)\exp(-\ii \om_1 t)\\
\psi_{2}(x,t)&=\sqrt{2/d} \sin (2\pi x/d)\exp(-\ii \om_2 t).
\end{align}
 Notice that $\psi_{1}(x)$ is even in $x$, while $\psi_{2}(x)$ is odd in $x$. Substituting these expressions in the Schrödinger equation \eqref{sch} with $u(x,t)=0$, we obtain that
\begin{align}
\label{news}
\frac{\hbar^2}{2m}\frac{d^2\psi_n(x)}{dx^2}+\hbar \om_n \psi_n(x)=0
\end{align}
provided
 \begin{align}\label{En}
E_{n}\equiv \hbar \om_n=\frac{\pi^2 \hbar^2}{2md^2}n^2\qquad n=1,2.
\end{align} 
We will see later on that optical fields at frequency $\om_o=\om_2-\om_1=\p 3\pi^2\hbar\q/\p2md^2\q$ may cause the system to evolve from state 1 to state 2 and back. Numerically, $\hbar\om_o\approx 1.12$ electron-volt if $d=1$ nano-meter.

For later use let us evaluate
 \begin{align}\label{x12}
x_{12}&\equiv \int_{-d/2}^{d/2}dx~x~\psi_{1}(x)\psi_{2}(x)\nonumber\\
&=\frac{2}{d}\int_{-d/2}^{d/2}dx~x~\cos(\pi x/d)\sin (2\pi x/d) \nonumber\\ &=\frac{16d}{9\pi^2},
\end{align} 
where we have used the mathematical relation
 \begin{align}\label{int}
\int_{-\pi/2}^{\pi/2}t\cos(t) \sin(2t)dt=\frac{8}{9}.
\end{align}
The parameter $x_{12}$ determines the strength of the 
atom-field coupling. It is convenient to define a dimensionless 
\emph{oscillator strength} 
\begin{align}\label{fparam}
f\equiv \frac{2m\omega_{o}}{\hbar }x_{12}^2=\frac{256}{27\pi^2}\approx 
0.96.
\end{align} 
The maximum value of $f$ is 1.

\section{Perturbed motion}

We next suppose that a potential source $v(t)=\sqrt2V\cos(\om_o t)$ is applied between the two anodes in Fig.~\ref{klystron}. Since the potential varies linearly with $x$ the electron is submitted to a space-independent optical field $\E(t)=\E_o\cos(\om_o t), \E_o=\sqrt2V/d$, where $\om_o\equiv \om_2-\om_1$ is the 1-2 transition frequency defined in the previous section. In that case \eqref{sch} reads 
\begin{align}
\label{solve}
H\psi\equiv\left(\frac{p^2}{2m}-e\E_o\cos(\om_o t)x-E\right)\psi(x,t)=0,\quad E=\ii \hbar\partial/ \partial t,\quad p=-\ii \hbar\partial/ \partial x,
\end{align}
remembering that for stationary states $\psi_n(x)$
\begin{align}
\label{solvebis}
\left(\frac{p^2}{2m}-\hbar\om_n\right)\psi_n(x)=0.
\end{align}

The wave function may be expressed as an infinite sum of $\psi_n(x)\exp(-\ii\om_n t)$ functions with slowly time-varying coefficients $C_n(t)$, that is
\begin{align}
\label{solveter}
\psi(x,t)=\sum_{n=1}^\infty C_n(t)\exp(-\ii \om_n t)\psi_n(x).
\end{align}
We first evaluate 
\begin{align}
\label{soveter}
\left(\frac{p^2}{2m}-e\E_o\cos(\om_o t)x\right)\psi(x,t)&=\sum_{n=1}^\infty C_n(t)\exp(-\ii \om_n t)\left(\hbar\om_n -e\E_o\cos(\om_o t)x\right)\psi_n(x)\nonumber\\
E\psi(x,t)&=\sum_{n=1}^\infty\exp(-\ii \om_n t)\left(\hbar\om_n  C_n(t) +\ii\hbar \frac{dC_n(t)}{dt}\right)\psi_n(x).
\end{align}
If we subtract the first expression from the second and substitute this expression into the Schrödinger equation, taking \eqref{solvebis} into account, we obtain 
\begin{align}
\label{solebis5}
0=\sum_{n=1}^\infty\exp(-\ii \om_n t)\left(\ii\hbar\frac{dC_n(t)}{dt}+e\E_o\cos(\om_o t)~x~C_n(t)\right)\psi_n(x).
\end{align}
If we multiply \eqref{solebis5} throughout by $\psi_m(x)$, integrate with respect to $x$, and take into account the ortho-normality of the $\psi_m(x)$ functions, we obtain an infinite number of exact ordinary differential equations that can be solved numerically. 

Considering only states 1 and 2, we set
\begin{align}
\label{solvter}
\psi(x,t)= C_1(t)\exp(-\ii \om_1 t)\psi_1(x)+C_2(t)\exp(-\ii \om_2 t)\psi_2(x).
\end{align}
Introducing the resonance condition $\om_o=\om_2-\om_1$, we obtain from \eqref{solebis5}
\begin{align}
\label{lveter}
0=\ii\hbar\frac{dC_1(t)}{dt}+\exp(-\ii \om_o t)\cos(\om_o t)\E_oex_{12}C_2(t),\nonumber\\
0=\ii\hbar\frac{dC_2(t)}{dt}+\exp(-\ii \om_o t)\cos(\om_o t)\E_oex_{12}C_1(t),
\end{align}
where $x_{12}$ is given in \eqref{x12}. Because the wave-functions $\psi_1(x),\psi_2(x)$ are real $x_{12}$ is real, and because of the symmetry of the wave-functions $x_{11}=x_{22}=0$.

The rotating-wave approximation consists of keeping only the slowly-varying terms, that is, replacing $\exp(-\ii \om_o t)\cos(\om_o t)$ by 1/2. Thus, the complex coefficients $C_1(t),C_2(t)$ obey the differential equations
\begin{align}
\label{formbis}
 \frac{dC_1(t)}{dt}=\ii\frac{\Om_R}{2}C_2(t)\qquad
 \frac{dC_2(t)}{dt}=\ii\frac{\Om_R}{2}C_1(t)\qquad C_1(t)C_1^\star(t)+C_2(t)C_2^\star(t)=1, 
\end{align}
where $\Om_R\ll\om_o$ is the Rabi frequency given by
\begin{align}
\label{rabi}
\hbar\Om_R=\E_oex_{12}.
\end{align}
For the potential considered and the value obtained in \eqref{fparam}, the above relation reads
\begin{align}
\label{om}
\hbar\Om_R=\frac{16}{ 9\pi^2}e\sqrt 2V\approx 0.17 ~e\sqrt 2V.
\end{align}
Normalization requires that $\abs{C_1(t)}^2+\abs{C_2(t)}^2=1$. The pair of first-order differential equations in \eqref{formbis} is easily solved. Assuming that the electron is initially ($t=0$) in the absorbing state, we have the initial condition $C_2(0)=0$. The wave function thus reads
\begin{align}
\label{form}
\psi(x,t)&= C_1(t)\exp(-\ii \om_1 t)\psi_1(x)+C_2(t)\exp(-\ii \om_2 t)\psi_2(x)\nonumber\\
&=\cos(\frac{\Om_R}{2} t)\psi_{1}(x)\exp(-\ii\om_1t)+\ii \sin(\frac{\Om_R}{2} t)\psi_{2}(x)\exp(-\ii\om_2t).
\end{align}

\section{Waiting time evaluation}\label{damped}

The equations in \eqref{formbis} generalize to
\begin{align}
\label{formx}
 \frac{dC_1(t)}{dt}=\ii\frac{\Om_R}{2}C_2(t)-\gamma C_1(t)\qquad
 \frac{dC_2(t)}{dt}=\ii\frac{\Om_R}{2}C_1(t).
\end{align}
The same final expression may be obtained from the density-matrix method. Starting from a pure state as above, constraints appear in the time evolution of $\boldsymbol{\rho}$, so that once a $2\gamma$ parameter has been selected for the decay of $\rho_{11}$, a corresponding decay of $\rho_{12}$ is required. We shall follow here the simpler route of using the $C_1(t),C_2(t)$ functions. However, it should be noted that the equations in \eqref{formx} are valid only as long as no event has occurred. They do not describe the electron wave-function.

The equation obeyed by $C_1(t)$ is obtained by deriving the first equation with respect to time and employing the second equation. We obtain
\begin{align}
\label{formx2}
 \frac{d^2C_1(t)}{dt^2}+\gamma \frac{dC_1(t)}{dt}+ (\frac{\Om_R}{2})^2 C_1(t)=0.
\end{align}
When the electron is initially in the upper-energy state 2, the initial condition is $C_2(0)=1,~C_1(0)=0$, and we find from \eqref{formx2}
\begin{align}
\label{formy}
C_1(t)=\frac{\ii\Om_R}{2\al}\left(\exp(\frac{-\gamma+\al}{2}t)-\exp(\frac{-\gamma-\al}{2}t)       \right)\qquad \al\equiv\sqrt{\gamma^2-\Om_R^2}.
\end{align}
The quantity $C_1(t)C_1^\star(t)$ represents the probability that the electron resides in the lower state \emph{as long as no jump occurs}. 

We obtain directly from \eqref{formy} the waiting-time probability density
\begin{align}
\label{wait}
w(t)=2\gamma C_1(t)C_1^\star(t)=\frac{\gamma\Om_R^2}{2\al^2}\{ \exp (-(\gamma-\al) t)+ \exp(-(\gamma+\al) t) -2\exp(-\gamma t)\} .
\end{align}
The quantity $w(t)dt$ is the probability that, given that the electron is in the upper state at $t=0$, it performs a jump from state 1 to state 2 \emph{for the first time} between $t$ and $t+dt$. When such a jump occurs, the same process starts again. Thus the jumps form an ordinary renewal process. The average inter-event time
\begin{align}
\label{avt}
\ave{\tau}=\frac{1}{R}\equiv\int_0^\infty dt~t~w(t)=\frac{1+2\gamma^2/\Om_R^2}{\gamma}\equiv\frac{1+a}{\gamma}\qquad a\equiv2\gamma^2/\Om_R^2,
\end{align}
where $R$ denotes the average jump rate. It is straightforward to go from the waiting time probability density $w(t)$ evaluated above to the event probability density $G(t)$. The concept is that the probability density of an event occurring at $t$ is the sum of the probabilities that this occurs through 1 jump, 2 jumps,...It follows that $G(t)=w(t)+w(t)*w(t)+w(t)*w(t)*w(t)+...$, where the middle stars denote convolution products. 

The Laplace transform $w(p)$ of $w(t)$ in \eqref{wait} reads
\begin{align}
\label{rhomp}
w(p)\equiv\int_0^\infty dt ~\exp(-pt)w(t)=\frac{\gamma\Om_R^2}{p^3+3\gamma p^2+(2\gamma^2+\Om_R^2)p+\gamma\Om_R^2}.
\end{align}
Thus the Laplace transform $G(p)$ of $G(t)$ is the sum of an infinite geometric series, which may be written in terms of the Laplace transform $\tilde{w}(p)$ of $w(t)$ as        
\begin{align}
\label{rhoml}
G(p)=\frac{w(p)}{1-w(p)}.
\end{align}

The jump rate may be written in general as $R(t)=R+r(t)$ where $r(t)$ represents a small fluctuation. The quantity we are interested in is the (double-sided) spectral density $\spectral_{R(t)}(\Om)$ of the jumps at Fourier (angular) frequency $\Om$. According to the Wiener-Khintchine theorem, the spectral density is the Fourier transform of the event correlation. It may thus be obtained directly from $\tilde{G}(p)$ after some rearranging as
\begin{align}
\label{oml}
\frac{\spectral_{R(t)}(\Om)}{R}&=1+G(\epsilon+\jj\Om)+G(\epsilon-\jj\Om)\qquad \epsilon\to 0\nonumber\\
&=2\pi R\de(\Om)+1-\frac{3a}{(1+a)^2+a(5a/4-1)(\Om/\gamma)^2+(a^2/4)(\Om/\gamma)^4}\nonumber\\
&\equiv2\pi R\de(\Om)+\frac{\spectral_r(\Om)}{R}.
\end{align}
The first term $2\pi R\de(\Om)$ which simply expresses that the average rate equals $R$ is henceforth omitted. The subsequent terms may be obtained by subtracting $R/p$ from $G(p)$, setting $p=\jj\Om$, and rearranging. We consider particularly the $\Om\to 0$ limit of $\spectral_r(\Om)$
\begin{align}
\label{ol}
\frac{\spectral_r(0)}{R}=1-\frac{3a}{(1+a)^2}\qquad a\equiv2\gamma^2/\Om_R^2.
\end{align}
In the large-$\gamma$ limit the shot-noise level $\spectral_r(0)=R$ is recovered. Otherwise, the event sttistics is sub-Poisson, as is the case in resonance fluorescence.

In the case of stimulated absorption, the role of the upper and lower states should be interchanged. Stimulated absorption occurs in optical detectors. Then the jumps previously considered correspond to photo-electron emission events. Unless the optical power is very large detectors are linear. This weak-field condition corresponds in previous expressions to the case where $\Om_R\ll \gamma$. The fluctuations are then seen to be at the shot-noise level. Alternatively, one may employ in that limiting situation random current sources and conductances.

\section{Semiconductor electrical properties}

We will not enter into much details here, but recall the basic facts. A resonant atom may be represented by a series $l,c,r$ circuit, possibly connected in parallel with the optical resonator described by an $L,C$ circuit, as previously described. The conductance of the $l,c,r$ circuit reads 
\begin{align}
\label{atom}
g(\om)=\frac{r}{r^2+\p  l\om-1/(c\om)     \q^2},
\end{align}
which exhibits a Lorentzian (bell-like) shape, with a maximum value $g_{max}=1/r$ when $lc\om^2=1$. This circuit corresponds to an atom in the absorbing state when $g>0$ and an atom in the emitting state when $g<0$. For semi-conductors, it is in fact unnecessary to know the value of $g$. Indeed, the different semi-conductor states (analogous to detuned atoms) are so closely spaced (spacing $\epsilon$) that is suffices to perform an integration of $g(\om)$ over $\om$ from 0 to $\infty$. The population inversion factor is derived from the fermi-Dirac distribution at the semiconductor temperature $T$. Next we employ the causality relation to obtain the susceptance and derive the $\alpha$ factor. Alternatively, we may sum up the susceptance part of each transition. The parameters $h$ or $K\equiv 1+h^2$ depend on the complete circuitry. They cannot be considered as a property of the semi-conductor alone. Finally, we may consider a "gain compression" parameter $\kappa$. The origin of this parameter is that conductances (gain or loss) $G(n,R)$ depend explicitly on the emitted rate $R$. This implies a departure from the Fermi-Dirac distribution, which, on the other hand, tends to be restored by Auger-like or thermal effects. 

\newpage

\chapter{Linear systems}\label{C}

First we consider C-states. These are defined as light beams that generate in detectors photo-electrons having Poisson statistics, irrespectively of the carrier phase. We show that current or potential sources radiate light in the C-state. We will examine what happens when the incident light is split into parts, perhaps by using linear conductances connected in parallel, or an array of beam splitters. As an application, balanced photo-detection is discussed. The noise added by linear loss or linear gain is considered.

\section{$C$-states}

Let us first give an example of Poisson-distributed electrons. The current emitted by a cathode whose emission is limited by the cathode temperature (instead of space charge) consists of independently emitted electrons\footnote{The current flowing in a conducting wire consists of a large number of slowly-moving electrons. One should consider the electron flow as a continuous incompressible fluid. But once emitted in vacuum, accelerated and possibly dispersed, the electrons, all of them with the same charge $-e$, should be considered individually. Neglecting the space charge amounts to treating each electron as a solution of the one-electron Schrödinger equation, with some static potential.}. Let the emitted current be denoted by $e\mathcal{J}\equiv eJ+e\De J(t)$, where $J$ denotes the time-averaged emitted electron rate, where $e$ denotes the absolute value of the electron electrical charge. The spectral density of the rate fluctuation $\De J(t)$ is given by the well-known formula
\begin{align}\label{7}
\spectral_{\De J}&=J 
\end{align}
where, as in the rest of this book, double-sided spectral densities are employed. Relation \eqref{7} says that the average power dissipated in a 1$\Om$ resistance following a 1 Hz band-pass filter centered at any low frequency (white noise) is given by the shot-noise formula $2eJ$.

Light beams may be received from free space or from transmission lines made up of two parallel ideally conducting wires with characteristic conductance $G_c$. Let this transmission line be terminated by a cold detector of conductance $G=G_c$ in which case the light beam is fully absorbed.  From the photo-detection events record one may define an average event rate $D$ and a fluctuation $\De D(t)$ that may be Fourier analyzed to provide a spectrum $\spectral_{\De D}(\Om)$. Such incident waves will be said to be in the C-state\footnote{C-state beams resemble the so-called "coherent" states of light employed in Quantum Optics. However, C-states are fundamentally states of propagating light, defined by the outcome of a detection process, while coherent states are primarily states of the field in optical resonators. In the context of Quantum Optics, Glauber has shown in 1963 that a classical prescribed current (which we would call a current source) radiates light in the so-called "coherent state". When coherent states are incident on a photo-detector the statistics of the photo-electrons is Poisson. The results therefore are similar.} if they generate Poisson-distributed photo-electrons events irrespectively of the carrier phase\footnote{Concretely, the carrier phase may be changed by inserting on the optical beam, before detection, a second-order all-pass filter. This is a conservative (i.e., lossless, gainless) device that changes the carrier phase without changing the amplitude. We suppose that this circuit bandwidth is very small compared with the frequencies of the fluctuations of interest. In that case, the carrier phase may be changed arbitrarily from 0 to 2$\pi$ through a slight filter-frequency change. The fluctuations, on the other hand, are essentially unaffected by that filter. }. We will show that potential or current sources radiate light waves in the C-state. 

Thus, if $\mathcal{D}(t)\equiv \ave{D}+\Delta D$ denotes the detection rate, the spectral density of $\Delta D$ must be equal to $\ave{D}$, that is $\spectral_{\Delta D}=D$, the averaging sign being omitted. If we express $\mathcal{D}$ in terms of the wave amplitude $b$, we have to first order $\mathcal{D}=\abs{b+\Delta b}^2\approx b^2+2b\Delta b$. Note that there is a random source associated with the matched load of conductance $G$. However, to first order, the detected rate depends only on the incident wave $b$ and its fluctuation. Thus $\Delta D=2b\Delta b\equiv bx$, where we have defined for brevity $x\equiv 2\Delta b$. We have $\spectral_{\Delta D}=b^2\spectral_{x}=D X$, if $X$ denotes the spectral density of $x$. It follows that for a C-state we have $X=1$. Considering now the phase, we set $b\equiv b'+\ii b''$, define $x'\equiv 2\Delta b',~x''\equiv 2\Delta b''$. $X$ denotes the spectral density of $x'$ and $Y$ denotes the spectral density of $x''$. Proceeding as above we conclude that a light beam is in the C-state if and only if $X=Y=1$.

The studied configurations consist of conservative elements such as inductances, capacitances, non-reciprocal devices such as circulators, and positive and negative conductances $G$ to which one associates random current sources $c(t)$ of spectral density $\hbar\omega_o \abs{G}$, where $\hbar$ denotes the Planck constant (divided by $2\pi$). Because we are interested only in narrow bands around the optical frequency $\om_o$, this noise source is considered to be frequency filtered by a narrow-band filter centered at $\om_o$ and whose spectral width is much larger than the relevant Fourier-frequency range, but much smaller than $\omega_o$. $c(t)$ may then be written as $c(t)=C'(t)\cos(\om_o t)+C''(t) \sin (\om_o t)$, where $C'(t), C''(t)$ are uncorrelated and have spectral densities $\hbar\omega_o \abs{G}$. The optical field is treated as a classical function of time $\approx \cos(\omega_o t)$, with some amplitude and phase fluctuations. Potentials and currents varying at, or near, the optical frequency are called "optical potentials", $V$, and "optical currents", $I$, respectively, to distinguish them from the static potentials $U$ and static currents $eJ$. The simplest configurations, treated in the present chapter, are of course those in which the conductances are constant. The response of linear systems to specified sources is well known. The linear regime is applicable to lasers below the so-called "threshold" driving current and usually to attenuators and amplifiers. We shall introduce the $h$ (or $K\equiv 1+h^2$) factor, viewed as resulting from the fact that negative and positive conductances are separated from one another by conservative elements. 

\section{Radiation from potential and current sources}\label{rad}

Suppose first that $G$ is positive constant. We are referring to it as a "cold load" (i.e., at T=0K) or, if this conductance describes a detector, as a "cold detector". A cold detector is a collection of atoms that are in the ground state most of the time and quickly revert to the ground state non-radiatively (through a static potential $U$) whenever they get promoted to the excited state under the influence of the incident optical field. Let us see how the above formalism fits with previously-quoted shot-noise formulas. We consider first a potential source and subsequently a current source. We find that in both cases the emitted light is in the C-state, that is, the photo-electron events generated in a (cold, ideal) photo-detector are Poisson distributed, irrespectively of the light carrier phase. 

For a potential source, the proof follows straightforwardly from previous considerations. Let the potential source be denoted $V\equiv V'+\ii V''$. The positive conductance $G$ with its associated noise source 
$C(t)\equiv C'(t)+\ii C''(t)$ represents the photo-detector, see Fig.\ref{circuitlaser} in $a$. Since the conductance $G$ is submitted to the potential $V$, the current flowing through $G$ is $GV$. Taking into consideration the random source $C(t)$ (for the sign, see the figure), the total current delivered by the
potential source is $GV-C(t)$. The power entering into the photo-detector is the power flowing from left to right through the dotted line in the figure. This power reads $Re\{V^\star \p GV-C(t)\q\}$. It consists of a steady-state value $P=GV^\star V$, and a fluctuating power $\Delta P(t)=-Re\{\p V'-\ii V''\q \p C'(t)+\ii C''(t)\q\}=-V'C'(t)-V''C''(t)$. Let us recall that the spectral density of $z(t)\equiv ax(t)+by(t)$, where $a,b$ are constants and $x(t),y(t)$ uncorrelated random functions of time of spectral densities $\spectral_x$, $\spectral_y$, respectively, is $\spectral_z=\abs{a}^2 \spectral_x+\abs{b}^2 \spectral_y$. The spectral density of the fluctuating power is therefore, setting in this formula $a=-V', b=-V''$ and $\spectral_x=\spectral_y= \hbar \om_o G$, $\spectral_{\Delta P}=\p V'^2 +V''^2\q \hbar \om_o G=\hbar \om_o P$. The spectral  density of the rate fluctuations $\Delta D=\Delta P/\hbar \om_o $, is therefore $\spectral_{\Delta D}=\spectral_{\Delta P}/(\hbar\om_o)^2=\hbar \om_o P/ (\hbar\om_o)^2=D$. This shows that the detector event rate is at the shot-noise level irrespectively of the phase of $V$ (see the above discussion on the $C$- light beams). We have not considered dispersive elements and therefore the fluctuations are "white", that is, have a uniform spectrum (i.e., a constant spectral density).

\paragraph{Current source.}

Consider next a non-fluctuating complex current $I\equiv I'+\ii I''$ that flows through a cold conductance $G$ as shown in Fig.\ref{circuitlaser} in $b$, endowed with a current noise source $C\equiv C'+\ii C''$ as above. Using Ohm's law, the power entering the detector through the dashed line is
\begin{align}
\label{14}
\mathcal{P}\equiv P+\De P=Re\{V^\star I\}=Re\{\frac{ I^\star+C^\star }{G}I \}= \frac{ I^\star I}{G}+Re\{\frac{(C'-\ii C'')(I'+\ii I'')}{G}\}.
\end{align}
The power fluctuation is therefore
\begin{align}
\label{15}
\De P=\frac{I'C'+I''C''}{G}.
\end{align}
Since $C'$ and $C''$ are uncorrelated, the spectral density of $\De P$ is the weighted sum of the spectral densities $\hbar \om_o G$ of $C'$ and $C''$ and thus
\begin{align}
\label{16}
\spectral_{\De P}=\frac{I'^2\spectral_{C'}+I''^2 \spectral_{C''}}{G^2}= I^\star I\frac{\hbar \om_o}{G}=P \hbar \om_o, 
\end{align}
It follows that
\begin{align}
\label{17}
\spectral_{\De D}=\spectral_{\De P/\hbar \om_o}=\frac{P}{\hbar\om_o} =D.
\end{align}
in agreement with the shot-noise formula. We have thus proved that current sources radiate light in the C-state. According to the expression in \eqref{16}, $\spectral_{\De D}$ does not depend on the phase of the source $I$ as one expects on physical grounds.

\section{Current source with a random modulation.}\label{withindep}

At that point, one may wonder how a quiet detector output could possibly be obtained from a semi-classical theory, as was asserted from the very beginning of this book. Indeed, we have just seen that a prescribed current source $I$ gives a detector output at the shot-noise level. If $I$ fluctuates randomly and independently of the previous random currents, the detector-output fluctuation may only be \emph{above} the shot-noise level since the random current source $C$ is uncorrelated with $I$. This is in fact the case for the \emph{current} source presently considered as shown below.

We thus generalize the result of the previous section, ascribing to the current source $I$, shown in Fig.\ref{circuitlaser} in b), the form $\De I(t)\equiv \De I'(t)+\ii \De I''(t) $ specified from the outside and therefore independent of $C(t)$. This may be achieved by an amplitude modulator driven by a random potential. The new term in the expression of the detector-power  fluctuation reads, neglecting cross products of small quantities,
\begin{align}
\label{18}
\De P= 2 \frac{I'\De I'+I'' \De I''}{G}.
\end{align}

Because a fixed average frequency $\om_o$ is considered throughout this book, we may take without loss of generality $\hbar \om_o$ as the energy unit and suppress it. Thus, from now on, $D\equiv P$. Supposing further $I$ real for simplicity, the full expression for the spectral density of $\De D=\De P$ reads
\begin{align}
\label{19}
\spectral_{\De D}=4D^2 \spectral_{\De I'/I}+D.
\end{align}
It follows from this relation that the relative noise $\N$ defined by
\begin{align}
\label{19'}
D\N\equiv \frac{\spectral_{\De D}}{D}-1=\spectral_{2\De I'/I}.
\end{align}
From the above expression and the fact that the spectral density of any measurable quantity is non-negative, it follows that $\spectral_{\De D}>D$, that is, the detector rate fluctuation always \emph{exceeds} the shot-noise level. 

Sub-Poisson photo-electron statistics may nevertheless be obtained from \emph{correlated} current and potential sources. Because these two sources refer to the same conductance they may be correlated. There are circumstances where the detected-rate fluctuation actually vanishes (ideal quiet lasers), as is shown in Section \ref{noise}. 

\section{Balanced detection}\label{dett}

The relative noise $\N$ could in principle be measured with a single (ideal) detector.  If the average
photo-electron rate is measured to be $D$, a measurement of the rate spectral density at some Fourier frequency $\Om$ gives $\N$ from its definition in \eqref{19'}\footnote{Relative noises should be expressed in seconds. Let us recall that the photo-current $i(t)=e\D(t)$, so that expressions in terms of rates may alternatively be written in terms of electrical currents}. 

Unfortunately, detectors suffer from various defects, optical loss at the entrance window and non-unity
quantum efficiency, for example.  A better scheme is therefore to employ a calibrating light beam having the same average power level as the light beam under test, and known to exhibit Poisson statistics.  Such a calibrating light source may be obtained from a light-emitting-diode provided its spectral width be much larger than the detector response time reciprocal. If $\spectral$ denotes the spectral density of the light beam under test  and $\spectral_{o}$ the spectral density of the reference source, the ratio $(\spectral/\spectral_{o}-1)/D$ provides a measure of $\N$. However, any strongly attenuated light tends to exhibit a Poisson statistics, so that in many circumstances $\N$, whether positive or negative, is small in absolute value.  The balance-detection scheme discussed below allows one to measure small values of $\N$ more accurately than it would be possible with a single detector. It is important, though, that the two detectors employed in that scheme be almost identical.

We denote by $Q$ the event rate that \emph{would} be measured by a detector located just after the laser. The measuring apparatus now involves a lossless beam splitter. A fraction $\R$ of the incident light power  is reflected toward detector 1, while a fraction $\T=1-\R$ is transmitted toward detector 2. Under ideal conditions, the average output rates from detectors 1 and 2 are therefore
 \begin{align}\label{bal1}
D_{1}=\R Q \qquad D_{2}=(1-\R)Q 
\end{align}
The spectral densities of the detected rates are, by straightforward application of previous formulas 
  \begin{align}\label{bal2}
\spectral_{\De D_{1}} &= D_{1}+D_{1}^2 \N\\
 \spectral_{\De D_{2}} &= D_{2}+D_{2}^2 \N\\
\spectral_{\De D_{1}\Delta D_{2}} &= D_{1}D_{2} \N.
\end{align}
It follows that the relative noise $\N$ equals $\spectral_{\De D_{1}\De D_{2}}$ normalized by $D_{1}D_{2}$. $\spectral_{\De D_{1}\De D_{2}}$ is measured by performing by electronic means the product of the two photo-currents, and time averaging (recall that the processes considered are stationary and ergodic).

\section{Splitting light into many beams}\label{split}

We have considered in the previous section the case where an incident light beam is split into two parts with the help of a beam splitter. We consider presently the more general situation where an incident beam is split into $N$ secondary beams. The light beam perhaps expands as a result of diffraction and is collected by a mosaic of detectors. The purpose of this section is to  show that the relative noise is unaffected\footnote{It is worthwhile recalling the photon picture of this effect. Light at frequency $\om_o$ is supposed to consist of particles called photons, each of them carrying an energy $\hbar \om_o$. Attenuation of light is supposed to imply that photons are randomly deleted. For the case of an ideal detector each photon generates a photo-electron. It follows that the photo-electron statistics is "thinned" (we employ here the language of point-processes, see Chapter \ref{math}). It is known that thinning does not affect the normalized correlation function, or the relative noise, of a point process. In subsequent calculations no such considerations enter. We employ only the concept of random current sources, and do not consider meaningful the statistical properties of light itself. We later consider the cross correlation between two detectors. Here again the configuration is often modeled by a series of beam splitters. It is supposed that the incident light beam consists of a stream of tiny particles called "photons" incident on the beam splitters. These photons are independent and are ascribed some probabilities of being transmitted or reflected. It can be shown that the statistics of the output light beams obtained in that manner are accurate. The photon (viewed as a particle) picture fails, however, in more general situations.}.

Another difference with the previous section is that the light source is viewed as a resonator containing $m$ light quanta\footnote{$m$ is defined as the integer part of the ratio of the energy, $E$, and $\hbar\om_o$. The expression "light quantum" does not refer to any particle property of light. }. When the output light is absorbed by a single element (e.g., a detector) at rate $\Q(t)\equiv 
Q+\Delta Q$, where $Q$ denotes the average rate 
and $\Delta Q(t)$ the fluctuation, the rate fluctuation is of the form 
 \begin{align}\label{deQ}
\De Q=Q \frac{\De m}{m}+q,
\end{align}
where $\De m$ denotes the fluctuation of $m$, and $q$ the noise term associated with the detector, whose spectral density is equal to the average rate $Q$\footnote{Note that if the term $q$ in the above equation were omitted the detector fluctuations would simply reflect the $m$-number fluctuations. This assumption, made in many optical-engineering papers, is incorrect, except at very high noise levels. }. The relative fluctuation of $m$ given in \eqref{deltamt} may be written in the form
 \begin{align}\label{deltam}
\frac {\De m}{m}=x+aq,
\end{align}
where $x$ is uncorrelated with $q$, and $a$ denotes some complex 
constant. This expression accounts for the correlation that may exist between $\De m$ and $q$ if $a≠0$. However, we do not intend to solve here the complete laser equations. We only need the general form in \eqref{deltam}.

Let us now suppose that the total rate $\Q$ is collected by 
$N$ detectors whose individual rates are denoted $\D_{k}$, with $k=1,2\ldots,N$, that is
\begin{align}\label{deltaD}
\Q=\sum _{k=1}^{N}\D_{k}\qquad 
\De Q=\sum _{k=1}^{N}\De  D_{k}.
\end{align} 
The form of each $\De D_{k}$ is the same as in (\ref{deQ}), namely
\begin{align}\label{deltad}
\De D_{k}=D_{k} \frac{\De m}{m}+d_{k},
\end{align}
where the $d_{k}$ are independent noise sources whose spectral densities 
are equal to the corresponding average rates $D_{k}$. The noise term $q$ is the sum 
over $k$ of 
the $d_{k}$. Since the $d_{k}$ are independent and have spectral 
densities 
$D_{k}$, the spectral density of $q$ is equal to $Q$.

If the expression of $\Delta m/m$ in \eqref{deltam} is substituted into (\ref{deltad}), we obtain
 \begin{align}\label{newdeltaD}
\De D_{k}=D_{k}x+(aD_{k}+1)d_{k}+D_{k}a\sum_{j\neq k}d_{j}.
\end{align}
The spectral density of $\Delta D_{k}$ is therefore
 \begin{align}\label{newdeltaD2}
\spectral_{\De D_{k}}=D_{k}^2 S_{x}+\vert aD_{k}+1\vert ^2 D_{k}+D_{k}^2 \vert 
a\vert ^2(Q-D_{k}).
\end{align}
It follows from \eqref{newdeltaD} that the relative noise
 \begin{align}\label{deltamb}
\N\equiv \spectral_{\De D_{k}/D_{k}}-\frac{1}{D_{k}}=\spectral_{x}+a+a^\star+a^\star aQ
\end{align}
is the same for all the absorbers.

The cross-spectral density between the rate at absorber (detector) $k$ and the rate at absorber $l$ is obtained similarly. We use the expression in (\ref{newdeltaD}) twice, once with subscripts $k$ and once with subscript $l\neq k$. In the product, only terms with the same subscripts are retained since $d_{k}$ and $d_{l}$ are uncorrelated. We obtain
\begin{align}\label{cross}
\spectral_{\De D_{k}\De D_{l}}&=D_{k} D_{l}\spectral_{x}+(D_{k}a+1)^\star aD_{k}D_{l}\nonumber\\
&+(D_{l}a+1)^\star a^\star D_{k}D_{l}+D_{k}a^\star a(Q-D_{k}-D_{l}).
\end{align}
After simplification, and recalling the result previously derived for the case where $k=l$, we obtain
\begin{align}\label{cross'}
\spectral_{\De D_{k}\De D_{l}}=\de _{kl}D_{k}+D_{k}D_{l} \N,
\end{align}
where $\de _{kl}$ equal 1 if $k=l$ and 0 otherwise. It follows from this expression that if the incident beam is Poissonian ($\N=0$) the secondary beams are uncorrelated.

\section{Linear attenuators}
\label{att}

Previous results relating to the noise properties of cold linear attenuators are now expressed in terms of propagating waves. Let us recall that the Ohm law reads $I=Y(\om)V$, where $I$ denotes the electrical current, $V$ the electrical potential at optical frequency $\om$, and $Y(\om)$ the circuit admittance. It is sometimes convenient to describe circuits in terms of forward-propagating waves of complex amplitude $a$ and backward-propagating waves of amplitude $b$ instead of voltages $V$ and currents $I$, see Fig. \ref{gainb} in a). Taking the transmission line characteristic conductance as unity for simplicity these quantities are related by
\begin{align}
a&=\frac{V+I}{2}\qquad b=\frac{V-I}{2},\nonumber\\
\label{32}
V&=a+b\qquad I=a-b.
\end{align}
In this formalism, $\abs{a}^2$ and $\abs{b}^2$ represent respectively the forward and backward propagating rates, setting for convenience $\hbar \om=1$. The difference $\abs{a}^2-\abs{b}^2=(V^\star I+I^\star V)/2=Re\{I^\star V\}$ represents the power dissipated in the load admittance $Y$. 

We suppose that forward-propagating and backward-propagating waves are separated physically from one another with the help of the (ideal) circulator shown in Fig. \ref{gainb}. It is then appropriate to call the $a$-wave the input wave and the $b$-wave the output wave. $Q\equiv\abs{a}^2$ is the rate that a detector placed just behind the light source would measure. Because there is no reflexion back to the source, $Q$ may be viewed as representing the input rate. We suppose that the load at the second port of the circulator is a cold positive conductance, that is, $Y=G$ with $0≤G<\infty$, and we take $\ave{a}, ~\ave{b}$ real for simplicity, see \ref{gainb}. The input rate fluctuation is thus the fluctuating part of $\ave{Q}+\Delta Q=\abs{\ave{a}+\Delta a}^2=\ave{Q}+2\ave{a}\Delta a'$, to first order, where $\Delta a'$ denotes the real part of $\Delta a$. If the output wave is detected by an ideal cold detector $D\equiv\abs{b}^2$ represents the output rate. The fluctuating part of $\ave{D}+\Delta D=\abs{\ave{b}+\Delta b}^2=\ave{D}+2\ave{b}\Delta b'$, to first order, where $\Delta b'$ denotes the real part of $\Delta b$.

\begin{figure}
\begin{center}
\includegraphics[scale=0.5]{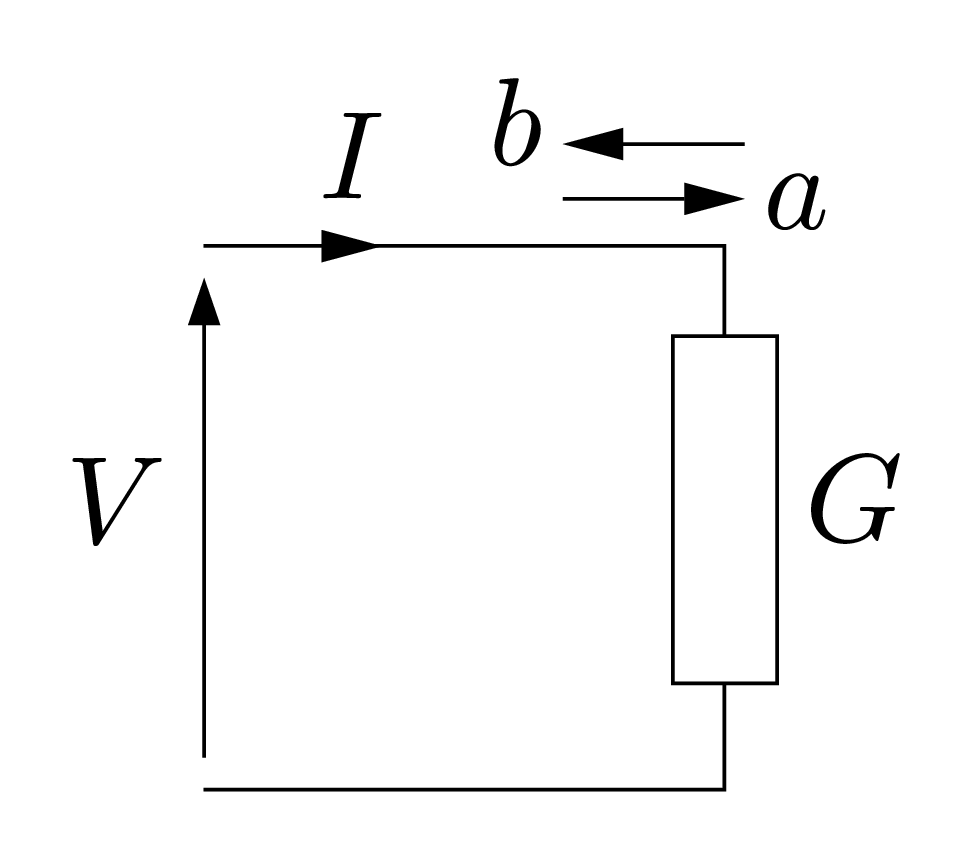}
\caption{Transmission line with forward a-wave and backward b-wave. The noise source is not shown. }
\label{gaina}
\end{center}
\end{figure}
\setlength{\figwidth}{0.6\textwidth}
\begin{figure}
\centering
\includegraphics[width=\figwidth]{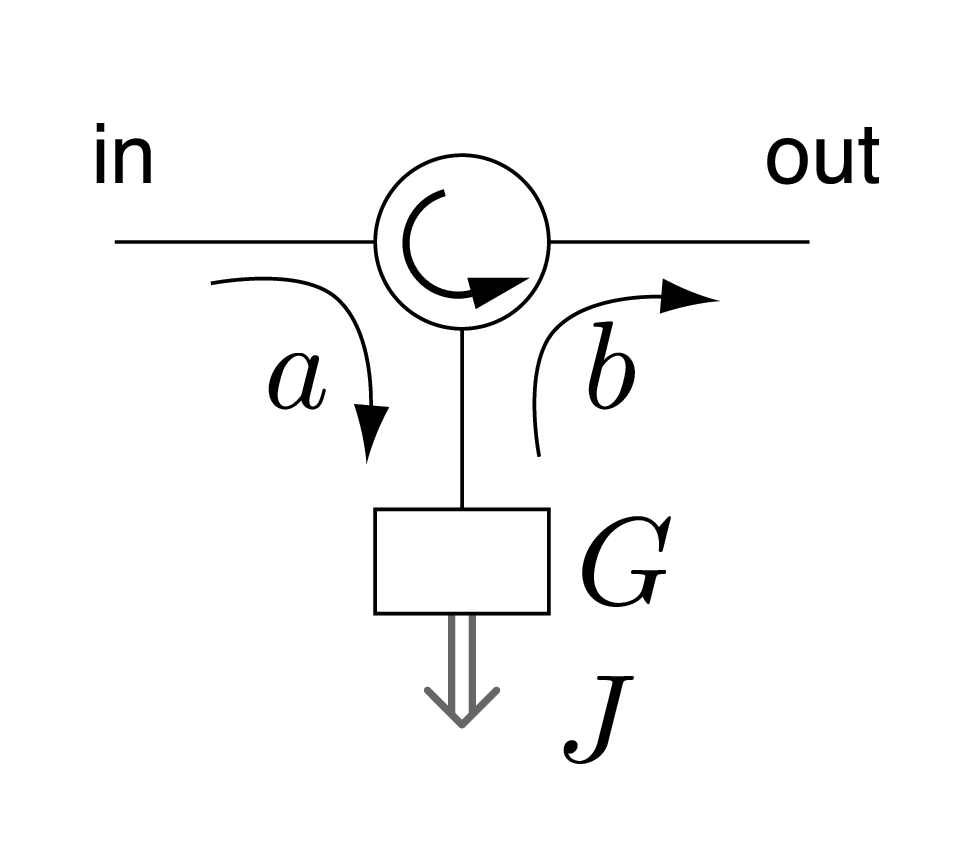}
\caption{ The a and b- waves may be separated from one another with the help of a lossless non-reciprocal device called a circulator.}
\label{gainb} 
\end{figure}

The noise-source $C$ spectral density equals $G$. For simplicity, all the quantities introduced in the present section are supposed to be real. According to the Ohm law we have $I=GV-C$. Replacing $I$ and $V$ by their expressions in \eqref{32} we obtain that $a-b=G(a+b)-C$. Solving this relation for $b$ 
\begin{align}
\label{33}
b=\frac{1-G}{1+G}a+\frac{C}{1+G}.
\end{align}
The relation between the $a$ and $b$-waves average values is obtained from the above equation by suppressing the $C$-noise term, that is
\begin{align}
\label{::}
\ave{b}=\frac{1-G}{1+G} \ave{a}\Longrightarrow \gf\equiv \frac{\ave{b}}{\ave{a}}=\frac{1-G}{1+G}.
\end{align}
The ratio of the average output to input rates is therefore (the letter \G~ stands for ``power gain'' even though in the present situation this quantity is less than unity and in fact expresses a loss)
\begin{align}
\label{35}
\G\equiv \gf^2=(\frac{1-G}{1+G})^2≤1.
\end{align}

Consider now a small variation $\Delta a$ of $a$ from its average value $\ave{a}$, corresponding to an input rate fluctuation $\Delta Q=2a \Delta a$ (when no confusion may arise we omit the brackets representing average values).
The output $b$-wave fluctuation reads according to \eqref{33}
\begin{align}
\label{34}
\Delta b=\frac{1-G}{1+G}\Delta a+\frac{C}{1+G}.
\end{align}
If, as supposed above, the reflected b-wave is incident on a cold detector delivering an electron rate $\mathcal{D}\equiv D+\Delta D=(\ave{b}+\De b)^2\approx b^2+2b \De b$. It follows from \eqref{::} and \eqref{34} that the detected-rate fluctuation is
\begin{align}
\Delta D&=2b \Delta b=2\frac{1-G}{1+G}a\left(\frac{1-G}{1+G}\Delta a+\frac{C}{1+G}\right)\nonumber\\
\label{37}
&=\G\Delta Q +2 \frac{1-G}{(1+G)^2}aC.
\end{align}
since $\Delta Q=2a\Delta a$.

Because the internal noise source $C$ is independent of the input wave fluctuation $\Delta Q$ and its spectral density equals $G$, the spectral-density of the detected rate $\Delta D$ reads
\begin{align}
\spectral_{\Delta D}&=\G^2\spectral_{\Delta Q}+4\frac{G(1-G)^2}{(1+G)^4}Q\
\label{39}
\end{align}
It follows after (much) rearranging that
\begin{align}
\mathcal{N}_{out}&\equiv S_{\frac{\Delta D}{D}}-\frac{1}{D}= S_\frac{\Delta Q}{Q}-\frac{1}{Q}\equiv \mathcal{N}_{in}. 
\end{align}
where $\N_{out}$ denotes the output relative noise. We thus observe once more that cold linear attenuators preserve the relative noise, that is, $\N_{out}=\N_{in}$. In particular, if the input light fluctuates at the shot-noise level, that is, if $S_{\Delta Q}=Q$, the output light also fluctuates at the shot-noise level, $S_{\Delta D}=D$.

\paragraph{Complex form.}

Generalizing \eqref{34} to the complex form we obtain
\begin{align}
\Delta b'&=\frac{1-G}{1+G}\Delta a'+\frac{C'}{1+G},\nonumber\\
\Delta b''&=\frac{1-G}{1+G}\Delta a''+\frac{C''}{1+G},\nonumber\\
\label{42}
S_{C'C''}&=0\qquad S_{C'}=S_{C''}=G.
\end{align}

Setting $2\Delta a \equiv x'_{in}+\ii x''_{in},~ 2\Delta b\equiv x'_{out}+\ii x''_{out}$ and $2C/(1+G)\equiv u'+\ii u''$, we may write \eqref{42} as
\begin{align}
x'_{out}&=\gf x'_{in}+u',\nonumber\\
x''_{out}&=\gf x''_{in}+u'',\nonumber\\
\label{46}
S_{u'u''}&=0\qquad S_{u'}=S_{u''}=1-\mathcal{G}\ge 0, 
\end{align}
Indeed, the spectral densities of $u'\equiv 2C'/(1+G)$ and $u''\equiv 2C''/(1+G)$  are $4G/(1+G)^2=1-\p(1-G)/(1+G)\q^2\equiv 1-\mathcal{G}$.

Let further $X,Y$ denote the spectral densities of the $x',x''$ noise terms respectively, with subscripts "in" and "out" appended where needed. Since the internal $u$ noise and the input noise are independent of one another, the input and output noise spectral densities are related, according to \eqref{46}, by
\begin{align}
X_{out}&=\mathcal{G}X_{in}+1-\mathcal{G},\nonumber\\
\label{48}
Y_{out}&=\mathcal{G}Y_{in}+1-\mathcal{G}.
\end{align}
We say that a light beam is in the C-state whenever $X=Y=1$. It follows that when a C-state beam ($X_{in}=Y_{in}=1$) suffers from any cold attenuation it remains in the C-state  (i.e., $X_{out}=Y_{out}=1$), as we have seen before in various ways.

The conductance $G$ receives a net rate $J=Re\{I^\star V\}=a^\star a-b^\star b$. If this conductance describe an ideal cold photo-detector, the photo-detection rate fluctuation denoted $\Delta J$ is, in terms of the $x$-fluctuations
\begin{align}
\label{48bis}
\Delta J&=\Delta(a^\star a-b^\star b)=2a\Delta a'-2b\Delta b' \nonumber\\
&=ax'_{in}-bx'_{out}=a(x'_{in} -\gf x'_{out}) 
\end{align}
where we have supposed again $\ave{a}, \ave{b}$ real for simplicity. The noise source associated with $G$ enters in the expression of $\Delta J$ explicitly when $x'_{out}$ is expressed in terms of $x'_{in }$ as given in \eqref{46}. 

\section{Linear amplifiers}\label{amp}

We considered above a collection of 2-level atoms, all of them residing most of the time in the lower state, which we called a cold absorber (or cold detector) and described it in terms of a positive conductance $G$. We now consider a collection of 2-level atoms that reside most of the time in the upper state, and call it an amplifying medium. Full population inversion is assumed for simplicity, and spontaneous electronic decay from upper to lower states is neglected. Such an atomic collection is characterized by a constant negative conductance denoted $-G$. The conductance $-G$ terminates a transmission-line whose characteristic conductance is unity. We require that $0<G<1$, the restriction $G<1$ being needed to avoid a singularity.

 As in the previous sub-section, the input $a$-wave and the amplified $b$-wave are supposed to be separated from one another with the help of a circulator, as shown in Fig.\ref {gainb}. Because in-phase as well as quadrature fluctuations are equally amplified in the present set-up the system is called a "phase-insensitive" amplifier.
The amplifier field gain $\gf$ and power gain $\G$ are given by the same formula as in \eqref{35} but with $G$ changed to $-G$, that is
\begin{align}
\label{48bisb}
\gf&\equiv \frac{\ave{b}}{\ave{a}}=\frac{1+G}{1-G}≥1\\
\G&=\gf^2=\left(\frac{1+G}{1-G}\right)^2\qquad 1≤\G<\infty.
\end{align}

We denote by $Q+\Delta Q$ the input rate (which could be detected, but is not in the present configuration) and by $D+\Delta D$ the output (detected) rate. As far as the average rates are concerned, we have $D=\G Q$. 
Proceeding as in the previous section, the spectral density of the photo-current rate fluctuation $\Delta D$ is found to be
\begin{align}
\label{49}
S_{\Delta D}=S_{\mathcal{G}\Delta Q}+(\mathcal{G}-1)D=\G^2 S_{\Delta Q}+(\mathcal{G}-1)D.
\end{align}
The first term in the above expression results from amplified input fluctuations while the second term has been introduced by the amplification process. 

In the special case where the input fluctuations are at the shot-noise level, that is when $S_{\Delta Q}=Q$, the above equation tells us that
\begin{align}
\label{49bis}
S_{\Delta D}=(2\mathcal{G}-1)D=2(\mathcal{G}-1)D+D
\end{align}
The second expression of $S_{\Delta D}$ above, namely $2(\mathcal{G}-1)D+D$, has been split into two terms. The first one, $2(\mathcal{G}-1)D$, is sometimes interpreted as resulting from the beat between the signal and the spontaneously emitted noise, while the second one is viewed as the shot noise associated with the optical power incident on the detector. These interpretations however are only of historical interest.

Using the notation introduced above, namely $x"_{in}+\ii x''_{in}\equiv 2\Delta a,~x'_{out}+\ii x''_{out}\equiv 2\Delta b$, where $\Delta a,~\Delta b$ denote the input and output wave fluctuations, we have
\begin{align}
x_{out}&=\gf x_{in}+u',\nonumber\\
\label{51}
y_{out}&=\gf y_{in}+u'',\\
\label{52}
S_{u'u''}&=0\qquad S_{u'}=S_{u''}=\mathcal{G}-1\ge 0. 
\end{align}
Denoting as before by $X,Y$ the spectral densities of the $x',x''$-noise terms, respectively, with appropriate subscripts "in" and "out", and remembering that the internal $u$-noise and the input noise are independent, the above expressions show that the input and output noise spectral densities are related by
\begin{align}
X_{out}&=\mathcal{G}X_{in}+\mathcal{G}-1,\nonumber\\
\label{55}
Y_{out}&=\mathcal{G}Y_{in}+\mathcal{G}-1.
\end{align}
For a C-state input beam in particular ($X_{in}=Y_{in}=1$) the spectral densities of the in-phase and quadrature fluctuations are both equal to $X_{out}=Y_{out}=2\G-1$. 

If two amplifiers of gains $\G_{1}$ and $\G_{2}$ respectively are placed in sequence, it is straightforward to show on the basis of \eqref{55} that the output noise is the same as for a single amplifier of gain $\G=\G_{1}\G_{2}$ as one expects, a result that can be generalized to any sequence of linear amplifiers. The case where the active medium inversion is incomplete may be treated by the same method.

An attenuator receives a greater rate at the input than it delivers at the output. As a consequence, by conservation of energy an attenuator delivers an electron rate $J=a^\star a-b^\star b$ as we have seen earlier. Conversely, an amplifier delivers a greater rate at the output than it receives at the input. Accordingly, it delivers an electron rate $J=a^\star a-b^\star b$ which is negative. It is then appropriate to call $-J>0$ the amplifier pumping rate. The fluctuation $\Delta J$ of $J$ may be employed as a modulating signal. 

Let us summarize the main results given above relating to linear attenuators and amplifiers. The circuit-theory formalism consists of associating a complex noise source $C'+\ii C''$ with  positive conductances $G$ or negative conductances $-G$. In both cases $G≥0$. Full population inversion is assumed. Here $C',C''$ are uncorrelated random functions of time whose spectral densities are both equal to $\abs{G}$ (setting $\hbar \om_o =1$). We considered a transmission line of characteristic conductance unity and called $a$ the amplitude of the forward-propagating wave (input wave of average rate $Q=\ave{a}^2$) and $b$ the amplitude of the backward-propagating wave (output wave of average rate $D=\ave{b}^2$). The latter is supposed to be detected, that is, converted into a photo-electron flow. We have $\G\equiv D/Q=\p(1-G)/(1+G)\q^2≤1$ for the loss case and $\G\equiv D/Q=\p(1+G)/(1-G)\q^2≥1$ for the gain case. We define the input relative noise $\N_{in}\equiv \spectral _{\Delta Q/Q}-1/Q$ and the output relative noise $\N_{out}\equiv \spectral _{\De D/D}-1/D$. Our main result is that cold linear losses do not affect the relative noise, that is $\N_{out}=\N_{in}$. For the case of loss and gain we have respectively
\begin{align}
\label{220}
\N_{out}&=\spectral_{2\frac{\De Q}{Q}}-\frac{\G}{D} \qquad\G≤ 1,\\
\label{221}
\N_{out}&=\spectral_{2\frac{\De Q}{Q}}+\frac{\G-2}{D}\qquad \G≥ 1.
\end{align}
Of course the two formulas coincide when $\G= 1$.

\section{Linear oscillators with incomplete population inversion}\label{linearosc}

We have represented in Fig.\ref{circuitlaser} in c) a simple laser model consisting of a frequency-independent negative conductance $-G_e$ and a positive frequency-independent conductance $G_a>G_e$, in parallel with an $L-C$-circuit resonating at frequency $\om_o$. The noise sources associated with these conductances lead us to the celebrated Shawlow-Townes (ST) linewith formula. The spectrum is Lorentzian with full-width at half-power (FWHP)
\begin{align}
\label{Schbis}
\de\om_{ST} ~Q ~\tau_p^2=1,
\end{align}
where $\tau_p=C/G_a$ is the resonator lifetime. This is the average time that a light pulse would spent in the circuit if no laser action were taken place (i.e., if $G_e=0$), and $Q$ the output rate, that is, the power dissipated in the load $G_a$ divided by $\hbar\omega_o$. Note that $G_e$ and $G_a$ may rather large, so that the circuit is strongly damped when $G_e$ is being suppressed. In that case the resonator lifetime is ill-defined. It is then preferable to give the expression of the product $\de\om Q$ in terms of the various elements that constitute the device, rather than in reference to the ST formula.

In the above expression, \eqref{Schbis}, it was assumed that the conductance $G_a$ is at temperature $T=0$K. It was further assumed that the population inversion in the laser material is complete. We now relax the latter assumption. If the population inversion is incomplete the laser material contains electrons in the lower state and a greater number of electrons in the upper state. If the laser material is represented by a negative conductance $-G_e$ in parallel with a positive conductance $G_a$ we define the conductance-inversion factor
\begin{align}
\label{eta}
\frac{G_e+G_a}{G_e-G_a}\approx\frac{n_e+n_a}{n_e-n_a}>1.
\end{align}
The second expression in \eqref{eta} holds for isolated atoms, the conductances being in that case proportional to the number of atoms. It could alternatively be expressed in terms of a negative temperature. In parallel with the laser conductances there is a load conductance that we denoted $G_o$, and we have $G_o\approx G_e-G_a$ that is $G_a+G_o\approx G_e$, so that the circuit is again highly resonant. The linewidth for the circuit presently considered is the same as before if we replace $G_a$ in the previous expression by $G_a+G_o$. However, what we now call radiated power $P$ is the power dissipated in $G_o$ \emph{alone}, not in $G_a+G_o$. Furthermore the photon lifetime $\tau_p$ obtained by suppressing the laser material, that is both $G_a $ and $G_e$ being set equal to 0, is now defined as $\tau_p=C/G_o$. The end result of the calculation is that the ST linewidth in \eqref{Schbis} should be multiplied by a population-inversion factor
\begin{align}
\label{np2}
n_p=\frac{G_e}{G_e-G_a}.
\end{align}

\section{Dispersive linear oscillators.}\label{dispersive}

We consider again linear oscillators. The optical potentials and currents are simply responses of a linear circuit to the current sources associated with conductances or resistances. For simplicity, we maintain that the active laser material is represented by a frequency-independent conductance $-G_e$ and assume complete population inversion. We consider the case where the circuit consists of a capacitance $C$ and an inductance $L$ in \emph{series} with a cold positive resistance $R_a$, as represented in Fig. \ref{circuitlaser} in d). 

Let $Y(\om)$ represent the admittance of the circuit. We have
\begin{align}
\label{hx}
Y(\om)&\equiv G(\om)+\ii B(\om)=-G_e-\ii C\om+\frac{1}{R_a-\ii L\om}\nonumber\\
&=\frac{R_a}{R_a^2+L^2\om^2}-G_e+\ii\p  \frac{ L\om}{R_a^2+L^2\om^2}-C\om      \q    .         
\end{align}
The resonant frequency $\om_o$ corresponds to a vanishing susceptance $B(\om_o)=0$, that is $\om_o=\sqrt{1/\p LC\q-\p  R_a/L\q^2}$, which implies that $R_a<\sqrt{L/C}$. We also assume that the circuit is only very slightly damped, which implies that $G_e$ is almost equal to $R_a C/L$. 
At the resonant frequency, the positive conductance is frequency-dependent. Derivating the conductances and susceptances given above with respect to $\om$ or using \eqref{dispers} we may evaluate the $h$ parameter defined as 
\begin{align}
\label{hparam}
h\equiv\p \frac{dG/d\om}{dB/d\om}\q _{\om=\om_o}.
\end{align}
The non-zero value of the $h$ factor may be viewed as a consequence of the fact that gain and loss do not occur at the same location. That is, if the gain and loss are represented by two conductances in parallel, the $h$-factor vanishes, and there is no linewidth-enhancement factor. But in the present situation the conductance $-G_e$ and the resistance $R_o$ are separated from one another in the circuit by the inductance $L$, and the $h$-factor is non-zero.

When the negative conductance is suppressed, the resonance frequency acquires a negative imaginary part that may be employed to define the resonator lifetime
\begin{align}
\label{hhh}
\tau_p\equiv \frac{dB/d\om}{2G_e}.
\end{align}
In the case of a classical $L-C$-circuit, we have $dB/d\om=2C$, the inductance and capacitance contributing equally, and the previous formula $\tau_p\equiv C/G_e$ is recovered from \eqref{hhh}.

Detailed calculations show that the $h$ factor defined above results in a linewidth-enhancement factor $K=1+h^2$ with respect to the ST result. For the $C,L,R_a$ circuit presently considered we obtain
\begin{align}
\label{K}
K=\frac{1}{1-R_a^2C/L}.
\end{align}
According to \eqref{K}, the $K$-factor is unity when $R_a\ll\sqrt{L/C}$ but tends to infinity when $R_a$ approaches $\sqrt{L/C}$.

To conclude, simple oscillators are subjected to linewidth enhancement. Similar conclusions concerning various linear circuits have been reached. A related linewidth-enhancement factor was discovered by Petermann in relation to the so-called "gain-guiding" lasers. Using an appropriately simplified schematics one can show that the above linewidth-enhancement effect is the same as the one given by Petermann. Since linewidth enhancement occurs for single-mode resonators, as we have just seen, this effect does not seem to be fundamentally related to mode non-orthogonality as other treatments suggested.

\section{Cavity linear oscillators}\label{cavity}

The linewidth may be obtained for an arbitrary cavity containing (linear, time-independent) dispersive and space dependent dielectrics $\epsilon(x,\om)$ (or even more generally bi-anisotropic media) in terms of the resonating fields. Here $x$ represents the three spatial coordinates. The detector is considered a part of the cavity, rather than being an external device, so that the system is closed.

Let us consider a cavity with perfectly conducting walls containing a medium with permittivity $\epsilon(x,\om)\equiv \epsilon'(x,\om)+\ii \p\epsilon''_a(x,\om)- \epsilon''_e(x,\om)\q,~\epsilon''_a(x,\om), \epsilon''_e(x,\om)≥0$. The imaginary part of the permittivity consists of two (possibly overlapping) spatial distributions. One, with subscript $a$ is absorbing, and may correspond physically to a photo-electron generating medium, while the other, with subscript $e$ is emitting and receives some pump power. For positive $\epsilon''$ all the electrons are in the lower state (zero temperature) and for negative $\epsilon''$ all the electrons are in the upper state. The spatial distributions are such that there is only one mode in the cavity whose frequency is slightly damped, the other modes being strongly damped may be ignored. Under those conditions the power generated by the medium with gain is nearly equal to the power absorbed by the medium with loss, but their spatial distributions may be entirely different. Furthermore, the resonating mode frequency $\om_o$ is nearly real in the limit considered. For simplicity we let the medium permeability be a real constant $\mu$. The first step consists of evaluating the linear response of the resonator to some given electrical distribution. Next, this current distribution is taken to be the random current density associated with both $\epsilon''_a$ and $\epsilon''_e$. The power absorbed by $\epsilon''_a$ slightly exceeds the power generated by $\epsilon''_e$, so that the response is stable but sharply peaked. We will only give the result of the calculation.

The product of the (full-width at half power) linewith $\de\om$ and power $P$ dissipated in the $\epsilon''_a$-part of the medium (or generated by the $\epsilon''_e$-part of the medium), reads
\begin{align}
\label{hhhh}
\de\om~P=\frac{4\p  \int dx~ \om\epsilon''_e(x,\om)\abs{\E}^2   \q^2}{\abs{ \int dx~ [\frac{d(\om\epsilon(x,\om))}{d\om}\E^2-\mu \mathcal{H}^2 ]}^2}.
\end{align}
where $\E,\mathcal{H}$ denotes the (complex) resonating electrical and magnetic fields. All the quantities are evaluated at the real frequency $\om_o$. Here $dx$ stands for $dx~dy~dz$ and the integrals are over the full cavity volume. This fairly general formula may be specialized to the one dimensional or zero-dimensional formulas given earlier.

Usually, the power delivered by the active medium is not dissipated internally as we supposed above. Instead, part of the cavity wall transmits power to some external detector. Formally, we may in that case apply the previous expression to a large cavity enclosing both the part of the medium considered as being the laser and the part of the medium considered as the detector. One expects intuitively that the linewidth measured by some cold reflexion-less detector does not depend on the distance between the laser and the detector, the laser being separated from the detector by a loss-less, dispersion-less transmission line. Let us show that this is indeed the case. Changing the laser-detector distance clearly does not affect the numerator of \eqref{hhhh} since $\epsilon$ in the intermediate region is real. The denominator is not affected either because $\epsilon$ does not depend on $\om$ in the intermediate region, and thus $d(\om\epsilon(\om))/d\om=\epsilon$. For a matched transmission line we have $\epsilon \E^2+\mu \mathcal{H}^2=0$. It follows that the additional term in the denominator of \eqref{hhhh} corresponding to the transmission-line volume, that is, the space comprised between the laser and the detector, does not contribute to the above expression.

\section{Propagating wave oscillators with gain.}\label{sec_gainguided}

In the present section we consider linear propagating-wave oscillators. That is, we consider either a dielectric wave-guide exhibiting gain and terminated at planes $z=0$ and $z=L$ by partially-transmitting mirrors, or a dielectric wave-guide with no net gain or loss, terminated at planes $z=0$ and $z=L$ by perfectly-transmitting mirrors. In the latter case, the generated power is dissipated internally. This configuration is clearly of little practical interest, but it helps comprehend the gain-guidance fundamentals.

Let us first recall simple properties of dielectric waveguides. We consider a medium uniform in the propagation direction $z$, with permittivity $\epsilon(x)$, and a source at some fixed frequency $\om$. The medium may be characterized instead by the "free wave-number" $k(x)$ with $k(x)^2=\epsilon(\om)\mu\om^2$. Because $\om$ is a constant this argument is omitted. For simplicity we take the permeability $\mu$ as being a real constant. 

Recall that if the medium is $z$-invariant we may define "transverse modes" as solutions of the wave equation of the form $\psi(x)\exp(\ii k_zz)$, where $\psi(x)$ may represent the electric field within the scalar or weakly-guiding approximation, and the propagation constant $k_z$ may be complex valued. We suppose that the real part of $k_z$ is positive and that accordingly the wave propagates in the positive $z$-direction. If the imaginary part of $k_z$ is positive the wave amplitude decays as it propagates, and conversely the wave amplitude grows if the imaginary part of $k_z$ is negative. The function $\psi(x)$ is some complex function of $x$.

If the permittivity is real (that is, the medium is loss-less, gain-less) and decreases as $\abs{x}$ increases, there exist a number of solutions (trapped modes) corresponding to real $k_z$-values and real $\psi(x)$-functions decaying exponentially in the outer medium. In that case the wave is said to be "index-guided". Some higher-order modes, however, may be "leaky", in which case the wave amplitude decays along the $z$-axis while $\psi(x)$\emph{grows} exponentially as a function of $\abs{x}$ and is no longer real\footnote{To avoid a confusion let us note that by introducing fictitious planes far-away from the guiding structure, where the field is required to vanish, we obtain the so-called "radiation modes". These modes are needed from a mathematical stand-point to expend the actual field into a complete set of functions.}. As a matter of fact, when the permittivity \emph{increases} as a function of $\abs{x}$, only leaky-modes can be found.

The question we are driving at is the following. Let us consider a leaky mode and suppose that the medium has gain for $\abs{x}<d/2$, but is loss-less, gain-less, for $\abs{x}>d/2$. Intuitively, we feel that the inner-medium gain is, figuratively speaking, fighting against the leaky-mode loss, and that, provided the gain is large enough, the wave amplitude may grow along the $z$-axis, instead of decaying. There is indeed a threshold gain when this does occur. What is not so obvious is that, above that threshold, the wave ceases to be leaky, that is, $\psi(x)$ \emph{decays} exponentially at large $\abs{x}$ values.

To prove this statement, let us suppose that the loss-less gain-less outer-medium is homogenous with a real propagation constant $k$, that is, for $\abs{x}>d/2$, we have $k_x^2+k_z^2=k^2$ real. If there is a net gain the imaginary part  $k_z$ is negative while the real part is positive. On the other hand, for a growing wave propagating away laterally ($x$ axis) we must have similarly that the imaginary part of $k_x$ negative while the real part is positive. But since $k$ is real these conditions are inconsistent. Indeed the condition $0=Im\{k^2\}=2k'_xk''_x+2k'_zk''_z$ cannot be fulfilled since the right-hand-side of this expression is the sum of two negative quantities. This argument shows that there exist no basic difference between index-guided and gain-guided configurations. In both cases, as long as there is a net gain along the $z$-axis, the field decays exponentially in the outer medium and the modes may be viewed as being "trapped". The wave may be pefectly matched to an incident beam at plane $z=0$, and perfectly matched to an outgoing beam at the output plane $z=L$. In such mode-matching conditions the power delivered by the active medium is simply the difference between the outgoing and ingoing powers. However the $\psi(x)$ function is complex in such waveguides with gain. This corresponds to a diverging wavefront. As a consequence, there is a $K$-factor larger than unity in both cases. It must be recognized, however, that the $K$-factor is likely to be much greater in the case of an inverted real-index profile than in the case of the usual real-index profile.
 
As a specific example we may consider a uniform dielectric slab of thickness $d$ in the $x$ direction, with free wace-number $k$ for $\abs{x}<d/2$, and wave-number squared $k_o^2=k_x^2+k_z^2$ in the outer medium, $\abs{x}≥d/2$. In the (thin-slab or reactive-surface) limit where $d\to 0$, the ($y$-directed) electrical field obeys the following boundary condition $d\psi/dx+\p k^2d/2\q \psi=0$ for $x\to 0^+$. It follows that $\ii k_x+k^2d/2=0$ or $k_x=\ii k^2d/2$. In the usual index-guiding conditions, $k_o$ is real (loss-less, gain-less outer medium) and $k$ is real and larger than $k_o$. In that case the wave propagate along the $z$-axis with a constant amplitude, and the wave amplitude decays exponentially away from the slab. When the medium has gain, that is when $k''<0$, the wave grows along the $z$-axis with some power gain per unit length. The field $\psi(x)$ is complex-valued. from its expression we may calculate a $K$-factor greater than unity. 

Alternatively, we may consider in place of a propagating wave a standing-wave by introducing perfectly-reflecting mirrors at $z=0$ and $z=L$. However, since in the present model no loss occur along the $z$-axis, the loss is introduced by ascribing slight losses to the outer medium. In such a resonator with inner gain and outer loss the field varies sinusoidally as a function of $z$. We may represent such a system by a simple circuit, with the slab with gain being represented by a negative conductance $-G_e$ and the lossy outer medium by a dispersive absorbing medium. We are thus led to the circuit considered earlier, and the same $K$-factor is obtained.

In order to treat a configuration similar to the gain-guiding configuration initially introduced in connection with laser diodes\footnote{The double-hetero-junction described in Section \ref{el} guides optical waves in the $x$-direction perpendicular to the layers, but is anti-guiding along the transverse $y$-direction because current injection tends to make the effective refractive index to decrease. The problem of gain-guidance then occurs along the $y$, rather than $x$, direction, as we suppose in the present discussion.}, one must envisioned two parallel anti-guiding slabs.

\chapter{Nonlinear operation}\label{rate-eq}
In the linearized regime generally applicable to above threshold lasers, we first solve the circuit or rate equations ignoring the random-current sources. We then approximate the conductances by introducing terms of the form $dG(n)/dn$, evaluated at the average value of $n$, where $n$ denotes the number of electrons. That number itself is considered to be much larger than unity. It obeys a rate equation of the form $dn/dt=J-R(t)$, where $J$ denotes the constant injected rate and $R(t)$ the output rate. Line-widths are evaluated from the spectral density of the "instantaneous" frequency deviation $\Delta\om(t)$. The latter are obtained by stating that the total current (including the random currents) vanish at any time.

\section{Laser rate equations}\label{noise}

Lasers are open systems with a source of energy called the pump, and a sink of energy presently viewed as an ideal optical detector.  It is
natural to suppose that the probabilities of atomic decay or atomic promotion that were found consistent with the laws of statistical mechanics still hold when there is a supply of atoms in the emitting state (the pump), and an absorber of light power (the detector).

We evaluate the relative noise of lasers driven by a non-fluctuating pump of rate $J$. Let us recall that we associate to any absorbed or emitted rate of average value $R$ a random rate $r(t)$ whose spectral density equals $R$. Rates of different origins are uncorrelated. This is also the case when a conductance both emits and absorbs. The random rates, say $r_a$ and $r_e$, are independent and, accordingly, their spectral densities add up. They are supposed to be negligibly affected by the small fluctuations of the system elements.

We consider an ideal single-mode cavity resonating at frequency $\om$ containing resonating atoms whose electrons may be either in the upper state of energy $\hbar\om\equiv1$ or the lower state of zero energy. $n\gg 1$ denotes the number of electrons in the emitting state, and $m\gg1$ denotes the number of light quanta: the integer part of the field energy. If the cavity were an isolated system we would have $m+n=$ constant. This is not presently the case because of pumping and detection.

\section{Time evolutions of $m$ and $n$.}

The evolution equation for $m$ is obtained by subtracting the detection rate $\mathcal{Q}$ from the net
light quanta generation rate $\mathfrak{E}-\mathfrak{A}$, see section \eqref{aux}.  Since the system is not isolated, $n+m$ may now fluctuate and the expressions of $R_e(n,m)$ and $R_a(n,m)$
given in (\ref{Ei'}) must be employed.  A second equation describing
the evolution of the number $n$ of atoms in the upper state is needed,
which involves the prescribed pump rate $J$. Thus, the evolution equations for $m$ and $n$ are
\begin{align}\label{Laserm}
    \frac{dn}{dt}&=J-\mathfrak{E}+\mathfrak{A},\nonumber\\
    \frac{dm}{dt}&=\mathfrak{E}-\mathfrak{A}- \mathcal{Q}= J- \frac{dn}{dt}- \mathcal{Q}, 
\end{align}
where
\begin{align}\label {E(t)}
	{\mathfrak{E}}&\equiv R_e(n,m)+e(t), \qquad {\mathfrak{A}}\equiv
	R_a(n,m)+a(t),\\
\label {deltaq} 
	{\mathcal{Q}}&=\frac{m}{\tau_p}+q(t),\\
\label{E(n,m)}
	&R_e(n,m)=nm,\qquad R_a(n,m)=(N-n)m,\\
\label{noiseb}
    &\spectral_{e}=R_e,\qquad \spectral_{a}=R_a,\qquad \spectral_{q}=Q.
\end{align}

In the following, averaging signs are omitted when no confusion is likely to arise. The average light output rate $Q$ (or detection rate) is the ratio of $m$ and the resonator lifetime $\tau_p$. In the steady-state, the right-hand-sides of \ref{Laserm}) vanish, and we have: $J=R_e-R_a=Q$, that is
\begin {align} \label{nseb}
   J=Q=(2n-N)m=\frac{m}{\tau_p }
\end{align}
The latter relation expresses the fact that the stimulated emission gain coefficient
$n$ minus the stimulated absorption loss coefficient
$N-n$ equals in the steady state the linear loss coefficient $1/\tau_p$.

When the above equations are
linearized and $\De{m}, \De{n}$ are Fourier transformed, one
obtains
\begin{align}\label{Lm}
    j\Om \De m=-j\Om \De n-\frac{ \De m}{\tau_p}-q
\end{align} 

\begin{align}\label{Ln}
	j\Om \De n=-2m \De n -\frac{ \De m}{\tau_p} -e+a.
\end{align}

Let us recall that $e$, $a$ and $q$ are uncorrelated processes whose spectral densities are equal to the corresponding average rates (i.e., the corresponding capital letters).  After elimination of $\Delta n$ from the above two
equations, $\Delta m$ may be expressed in terms
of uncorrelated noise sources as
\begin{align}\label{deltamt}
	\De m=\frac{j\Om(e-a)-(2m+j\Om)q} {j\Om
	2m+ 2m/\tau_p-\Om^2}.
\end{align}

Evaluating the spectral density of $\De m$ and integrating over frequency to obtain the
variance of $m$ we obtain
\begin{align} \label{varm}
	\frac{\mathrm{var}(m)}{m} =
	\frac{N+1/\tau_p}{4m}+\frac{1}{2}.
\end{align}
In the limit that $N\tau_p\gg1$ and $m=N/2$, the
right-hand-side of (\ref{varm}) is 1 while the corresponding result for
the isolated cavity is 1/2.  This is due to the
singular behavior of the spectral density of $\De m$ at $\Om=0$
in the limit considered.  Physically, this means that small losses
allow $m(t)$ to drift slowly.

We are however mostly interested in the detection rate fluctuation $\De
Q=\De m/\tau_p + q$.  Notice that $m$ and $q$ \emph{are}
correlated.  Proceeding as in the previous section, we obtain
\begin{align}\label{X}
	Q\N\equiv \frac {\spectral_{\De Q}(\Om)}{Q}-1=\frac{[(N\tau_p+1)/4
	\ave{m}^{2}]\Om
	^2-1}{(\Om\tau_p)^2+[1-\Om^2\tau_p/2 \ave{m}]^2}.
\end{align}
In the limit that $N\tau_p\gg1$, $\ave{m}=N/2$, the above result
reduces to the one given earlier at high frequencies $\Om\tau_p \gg
\sqrt {N\tau_p}$.
As expected, the spectral densities of the
photodetection process go to zero at zero Fourier frequencies.  For the parameter values $N=100$, $\tau_p=0.05$ and $m=N/2$ for example, a small
relaxation oscillation peak appears. In the large optical power limit
($m \gg 1$), the above expression reduces to
\begin{align}\label{Xbis}
	\frac{\spectral_{\Delta Q}(\Omega)}{Q}=1-\frac{1}{(\Omega\tau_p)^2+1}.
\end{align}
In that limiting form it is even clearer that the spectral density of the photo-current vanishes at zero Fourier frequency, $\Om=0$. An erroneous result would be obtained if the correlation between $q$ and $\De m$ were ignored.

\section{Arbitrary dependence of the gain on the electron number}

In general the gain is some function of the number of atoms in the emitting state, or of the number of electrons in the conduction band in a semi-conductor, again denoted $n$. For example, for a semi-conductor at $T=0$ we have at the optical frequency at which the gain is maximum $G_{max}\propto n^{1/3}$. Furthermore, there are in general no simple relation between the population inversion factor $n_p$ and $n$. We need therefore perform a calculation similar to the one just performed, but in a more general setting.

We set
\begin{align}\label{ref1}
\frac{dn}{dt}&= J-\mathfrak{R},   \qquad  \frac{dm}{dt}= \mathfrak{R}- \mathcal{Q}=J-\frac{dn}{dt}-\mathcal{Q},\nonumber\\
    \mathfrak{R}&\equiv \mathfrak{E}- \mathfrak{A}=G(n)m+r,\qquad  \spectral_{r}=(2n_p-1)R,\nonumber\\
   \mathcal{Q}&= \frac{m}{\tau_p}+q(t),\qquad \spectral_{q}=Q,\nonumber\\
   J&=R=Q=\frac{m}{\tau_p},
    \end{align} 
where the last expression relates to average values, and we have defined the population inversion factor $n_p\equiv R_e/(R_e-R_a)$. 

To evaluate the spectral density of $\Delta Q$ we consider first-order variations and set $d/dt=\jj \Om$. The last equation on the first line of \eqref{ref1}
\begin{align} \label{ref2}
\Delta \mathfrak{R}&=\frac{gm}{\tau_p n}\Delta n+\frac{\De m}{\tau_p}+r\qquad\spectral_{r}=(2n_p-1)Q
\nonumber\\
\Delta \Q&=\frac{\De m}{\tau_p}+q \qquad \spectral_{q}=Q
\end{align}
where $g\equiv(n/G(n))(dG(n)/dn)$. 

Therefore, replacing $dm/dt$ by $\jj\Om \De m$ and $dn/dt$ by $\jj\Om \De n$, we obtain, setting $b\equiv gm/n$ and $\Om^{\circ}\equiv \Om \tau_p$ for brevity,
\begin{align} \label{ref5}
\p1+\jj \Om^{\circ}\q \frac{\De m}{\tau_p}+\jj \Om^{\circ} \frac{\De n}{\tau_p}+q&=0\nonumber\\
\p1+\frac{b}{\jj \Om^{\circ}}\q \jj \Om^{\circ}  \frac{\De n}{\tau_p} +\frac{\De m}{\tau_p}+r&=0\qquad \Om^{\circ}\equiv \Om \tau_p
\end{align}
Substituting $\jj\Om\De n$ from the first above equation to the second and replacing $\De m/\tau_p$ by $\De Q-q$, we obtain
\begin{align} \label{nppp}
\De Q-q+r=\p  1+\frac{b}{\jj\Om^{\circ}}     \q\p (1+\jj \Om^{\circ})(\De Q-q)+q  \q 
\end{align}
that is
\begin{align} \label{nnpp}
\De Q=\frac{r+(\jj\Om^{\circ}+b)q}{\jj\Om^{\circ}+b+b/\jj\Om^{\circ}} 
\end{align}
Since the spectral density of $z=ax+by$ is $\spectral_z=\abs{a}^2\spectral_x+\abs{b}^2\spectral_y$ when $x,y$ are uncorrelated, and $\spectral_q=Q,~\spectral_r=(2n_p-1)Q$, we obtain
\begin{align} \label{npmp}
Q \N \equiv  \frac{\spectral_{\De Q}}{Q}-1&=\frac{{\Om^{\circ}}^2+b^2+2n_p-1}{b^2+(\Om^{\circ}-b/\Om^{\circ})^2}-1\nonumber\\
&=\frac{(2n_p-1+2b){\Om^{\circ}}^2/b^2-1}{{\Om^{\circ}}^2+(1-{\Om^{\circ}}^2/b)^2}
\end{align}
We note that at high power $b\to \infty$ we obtain again
\begin{align} \label{npmp2}
Q \N =-\frac{1}{1+{\Om^{\circ}}^2}
\end{align}
which is equal to -1 when $\Om^{\circ}\to 0$, that is, the detected fluctuation vanishes.

In the cavity model treated earlier, the number of atoms in the emitting state is $n\equiv n_e$ and the number of atoms in the absorbing state is $N-n\equiv n_a$. It follows that the net rate $R(n,m)\equiv G(n)m=R_e(n,m)-R_a(n,m)=\p G_e(n)-G_a(n)      \q m$ we have $G_e(n)= n, ~G_a(n)=N-n$ and thus $G(n)= n-(N-n)=2n-N$, which means that the net gain is positive, that is, there is population inversion (negative temperature), when $n>N/2$\footnote{$G(n)$ should not be confused with the constant resonator load $G$. Oscillation occurs when $G_e\approx G+G_a$. Note further that with an appropriate normalization we need not distinguish the gain and the conductance, both being denoted by the letter $G$. }. The dimensionless differential gain reads
\begin{align} \label{diff}
g\equiv \frac{n}{G} \frac{dG}{dn}= \frac{1}{1-N/2n}.
\end{align}
The spectral density of the net random rate is $\spectral_{e-a}=\spectral_{e}+\spectral_{a}=R_e+R_a$, and we define a population inversion factor
\begin{align} \label{np}
n_p\equiv \frac{n_e}{n_e-n_a}=\frac{n}{2n-N}=\frac{1/2}{1-N/2n}=\frac{g}{2}.
\end{align}
Note that in the present model this population-inversion factor approaches unity (ideal case) when $n$ approaches $N/2$. 

The general expression in \eqref{npmp} gives again \eqref{X} when the substitution in \eqref{diff} are made, namely $g=2n_p=1/(1-N/2n)=2n\tau_p$ implying that $b\equiv gm/n=2\tau_p m$.

\section{Laser diodes with evenly-spaced levels}\label{lowp}

The method of obtaining above-threshold-laser relative noise given in the previous section is applicable to laser diodes. Stimulated transitions occur between the valence band and the conduction band of a semi-conductor. For simplicity, both bands are described by $B$ levels separated in energy by $\epsilon$, with $\epsilon\ll T\ll B\epsilon$, where $T$ denotes the temperature. The two bands are separated from one another by a band-gap, which, though essential to determine the oscillation frequency, does not enter explicitly in our calculations. We give in the present section the result for the case where the electron distributions in both bands is given by quasi Fermi-Dirac distributions corresponding to the same lattice temperature $T$. Spectral-hole burning is neglected. The diode driving current corresponds to a constant rate $J$. The general procedure is the same as the one discussed in the previous section.

We define $\J\equiv J\epsilon/T$, and $\tau_p$ denotes as before the resonator lifetime (absorption is assumed to be due entirely to the detector), $\Omega_r^2=\frac{(\tau_p^2-1)\J}{2\tau_p^2}$ is the square of the (angular) relaxation frequency, and $F\equiv \p \frac{\Om}{\Om_r}\q^2$.
The relative noise at frequency $\Om$ is found to be given by the expression
 \begin{align}\label{loxp}
J\N(\Om)=\frac{\frac{\tau_p+1 }{\tau_p-1}\frac{F}{\J}-1}{\frac{\tau_p^2-1}{2 \tau_p^2}\J F+(1-F)^2}.
 \end{align}

Thus, the relative noise can be evaluated as a function of the Fourier frequency $\Om$ by an explicit expression, the parameters being the injected rate $J$, the resonator lifetime $\tau_p$, the temperature $T$, and the spacing $\epsilon$ between adjacent states. For typical values ($\tau_p=2$, $\J=0.58$) we obtain a relaxation frequency of 74 MHz. The expression given in \eqref{loxp} is in excellent agreement with the relative noise evaluated through a purely numerical method, described in the next section, when the power is not too high. Otherwise, spectral-hole burning should be taken into account, see Fig. \ref{fig3}.

\section{Numerical simulation of high-power laser diodes}\label{nu}

This section reports results obtained from a numerical simulation that keeps track of the number (0 or 1) of electrons in each state of a semiconductor and of the number of light quanta $m$ in the resonator. A constant electron injection rate $J$ is assumed as before, and spontaneous inter-band recombination  is neglected, as in previous sections. As the laser diode power increases, significant departures from the Fermi-Dirac distributions occur. To first order, such departures could be taken into account by introducing the concept of gain compression but such a formulation has not been considered yet in detail. At low optical power a very good agreement between the numerical calculations and the theoretical formulas given in the previous section that employs the Fermi-Dirac distribution is noted. At high powers, a "spectral hole" is found to occurs in the electronic distribution in both bands, as expected. The relaxation oscillation is damped and the relative noise peak gets displaced toward lower Fourier frequencies. 

The Monte Carlo simulation keeps track of each level occupancy in the conduction and valence bands and of the integer resonator energy $m$. At high power levels, new effects occur (spectral-hole burning, temperature fluctuations, statistical fluctuations of the optical gain) that are difficult to handle analytically. The final outcome of the numerical procedure is the relative noise as a function of the Fourier frequency. The curves show that the frequency domain over which the photo-current spectral density is below the shot-noise level becomes narrower as the optical power increases. Exhaustive Monte Carlo simulations of lasers have apparently not been made in the past, probably because of the high computer time required in the case of bulk semiconductor lasers.  But dramatic size reductions have been obtained with micro-cavity quantum dot lasers and two-dimensional photonic-band\-gap quantum-dots lasers. Because the active layers involve few optically active quantum dots in the gain region an individual account of each level occupancy is manageable.  Such devices are likely to be the next generation semiconductor lasers. Note that in optical computing applications only small optical output powers are needed and small driving currents are desired. For conventional applications, e.g., optical communications, milliwatts of powers are usually required.  In that case, our simulation results have to be scaled up since it would be impractical to account for every level occupancy in the case of bulk semiconductors numerically.  Scaling laws applicable to the linearized theory need generalization if one wishes to take into account advanced effects such as spectral-hole burning.  The numerical results presented in this section would help ascertain the validity of more advanced theories.

The main processes involved in laser light generation are recalled in subsection \ref{processes}, and the numerical procedure is explained in subsection \ref{birth-death}. Numerical results concerning photo-detection noise are illustrated in subsection \ref{results}.  

\section{Laser diode processes}\label{processes}

In the present section the basic Physics of Semiconductors, and processes relevant to resonators containing semiconductors in contact with a thermal bath, are considered.  Pumping and optical absorption are discussed subsequently.

One-electron energy levels in semiconductors are supposed to be of the form $\epsilon_{k} = k \epsilon $, with $k$ an integer and $\epsilon$ a constant.  The lasers considered may incorporate quantum dots in the gain region.  The evenly-spaced-level assumption may then be justified by the mechanism of level `repulsion' in nanometer-scale irregular particles. The probability that adjacent levels be separated by $\epsilon$ is of the form $\epsilon^4 \exp(-\epsilon^2)$, a sharply peaked function of $\epsilon$.  Note also that quantum wells exhibit levels that are evenly spaced \emph{on the average} within each sub-bands.  That is, the density of states is a constant.

\begin{figure}
\centering
\includegraphics[scale=0.6]{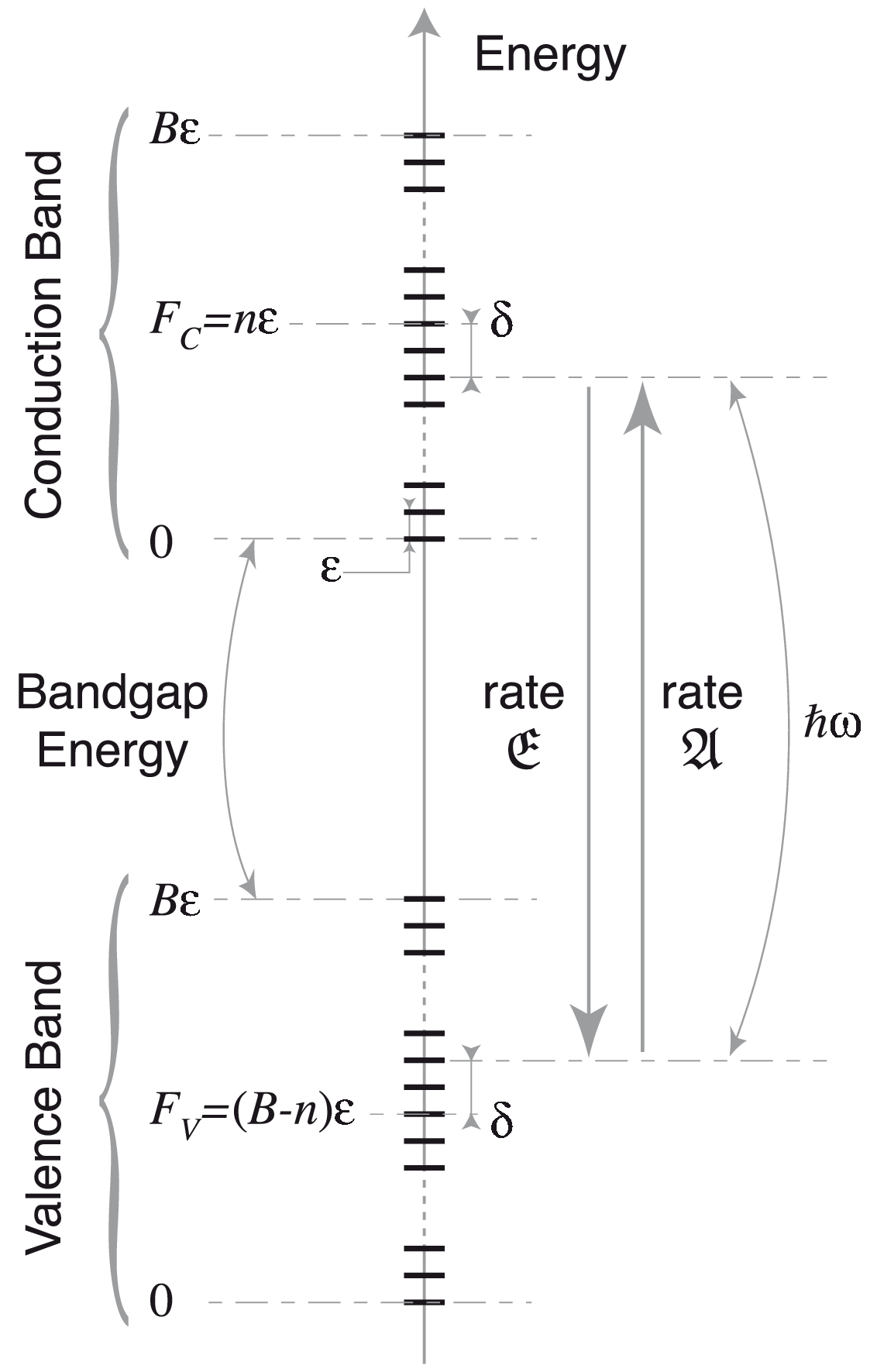}
\caption{Schematic view of the energy level system considered.}
\label{fig1}
\end{figure}

Some levels are allowed while others are forbidden.  Allowed levels may be occupied by at most one electron to comply with the Pauli exclusion principle, the electron spin being ignored. In semiconductors the allowed electronic levels group into two bands, the upper one being called the \emph{conduction} band (CB) and the lower one the \emph{valence} band (VB).  We suppose that both bands involve
the same number $B = 100$ of levels, and are separated by $G_p$ forbidden levels, as shown in Fig.~\ref{fig1}. The band gap energy is instrumental in determining the laser oscillation frequency but it will not enter in our model because of simplifying assumptions to be later discussed.  Only one valence band is considered, but it would be straightforward to take into account the heavy-hole, light-hole and split-off bands found in most semiconductors.  $N = B$ electrons are allocated to the allowed energy levels.  For pure semiconductors at $T = 0$~K, the $N$ electrons fill up the valence band while the conduction band is empty.  The electron-lattice system is electrically neutral.

Without a thermal bath, the electron gas reaches an equilibrium state through Auger-type transitions: an electron gets promoted to upper levels while another electron gets demoted to lower levels in such a way that the total energy remains the same.  The two electrons may belong to the same band or to distinct bands, but only the former situation is presently considered.  Auger transitions ensure that all the system microstates are being explored in the course of time so that electron gases possess well-defined temperatures at any instant in each band.  Nothing, however, prevents these temperatures from fluctuating in the course of time. When $\epsilon \ll T \ll B \epsilon $ the Fermi-Dirac (FD) distribution is recovered with great accuracy.  But $T$ needs not be very large compared with $\epsilon$ in mesoscopic devices, \emph{e.g.}, short quantum wires or quantum dots. Note that laser noise depends in general, not only on active levels \emph{average} occupancies, but also on the fact that, even in the equilibrium (or quasi-equilibrium) state, electrons keep moving in and out these levels, causing the optical gain to fluctuate. Such fluctuations are automatically accounted for in Monte-Carlo simulations. It is only in the linearized theory that statistical gain fluctuations may be ignored. 

Let us now consider the process of thermalization between the electron gas and the lattice.  To enforce thermalization in the numerical model, each electron is ascribed a probability $p$ per unit time of being demoted to the adjacent lower level provided this level is empty, and a probability $p q$, where $q \equiv \exp(-\epsilon / T)$ of being promoted to the adjacent upper level if it is empty.  Strictly speaking, this thermalization model would be applicable to solids with $\hbar \omega_{phonon} = \epsilon$, but the detailed modeling turns out to be rather unimportant. If $p$ is large, thermalization is very efficient.  This implies that
electron-gas temperatures in both bands are equal to the lattice temperature, i.e., are constant in the course of time. The main purpose of this section is to consider the noise spectrum when $p$ is not large, in which case electron gas temperatures are ill defined. 
Near equilibrium, it is immaterial whether Auger or thermalization transitions are dominant since both lead to well-defined temperatures and, in the appropriate limit, to the FD distribution.  But because lasers are out of equilibrium systems, intensity-noise spectra do depend on which one of the Auger or thermalization processes dominates. 

Consider now an isolated system consisting of a single-mode resonator at angular frequency $\omega$. We suppose that the coherent interaction takes place between the middle of the conduction band and the middle of the valence band, that is: $\hbar \omega = (G_{p} + B) \epsilon$. The optical field enters only through the number of light quanta $m$. Stimulated absorption is modelled by assigning a probability $m$ to electrons in the lower working level to be promoted to the upper working level (if that level is empty).  Stimulated emission is modeled by assigning a probability $m + 1$ to electrons in the upper working level to be demoted to the lower working level (if that level is empty)\footnote{In the laser analytic theory the ``1'' of the Einstein expression $m + 1$ may be neglected in the steady-state because $m$ is a large number. The term ``1'' must be kept in the Monte Carlo simulation because the initial value of $m$ considered is 0. Without that term laser start-up would not occur.}.  Setting as unity the factor that multiplies the expressions $m$ or $m + 1$ amounts to selecting a time scale.

Optical pumping would be modeled by assigning some constant probability to electrons in low valence-band levels to be promoted to high conduction-band levels, provided these levels are empty, \emph{and} almost the same probability for the opposite transition.  In that case the pump-rate fluctuations would be close to the shot-noise level. But the electrical current generated by cold high-impedance electrical sources is almost non-fluctuating as a consequence of the Nyquist theorem.  The nature of the detected light depends on the
ratio of this impedance to the intrinsic dynamic resistance of the laser. We restrict ourselves to perfectly regular electrical pumping, \emph{i.e.} to infinite cold
impedances.  Quiet electrical pumping is modeled by promoting low-lying electrons into high-lying levels \emph{periodically} in time\footnote{If the lowest level happens to be unoccupied or if the highest level happens to be occupied, a rather infrequent circumstance, the program searches for the next adjacent levels.}, every $\Delta t =0.2$~ns, corresponding to a pumping rate $J = 5 \text{ns}^{-1}$.  Because the time period considered is very short in comparison with the time scales of interest, this prescription implies that the pumping rate is nearly constant.  This has been verified numerically.

\section{Laser noise from a birth-death process}\label{birth-death}

The method is best explained by considering first time intervals $\delta t$ small enough that the probability that an event of a particular kind occurs within it is small compared with unity, and that the probability of two or more events occurring is negligible,
\emph{e.g.}, $\delta t = 10^{-15}$\,s.  A typical run lasts $\T = 1\;\mu$s, corresponding to $10^9$ elementary time intervals. The total number of events per run is on the order of $10^8$. 
Averaging is made over 20 independent runs.  Instead of the above 
pedestrian approach, the algorithm actually employed accounts more 
rigorously for the birth-death process and minimizes the CPU time.

Stimulated decay of an electron during an elementary time interval $\delta t$ is allowed to occur with probability $(m +1) \delta t$, $m$ being incremented by $1$ if the event does occur.  Likewise, stimulated electron promotion is allowed to occur with probability $m \delta t$, $m$ being reduced by 1 if the event occurs.  The fact that the proportionality constant has been omitted amounts to selecting a time unit, typically, 1 ns, as said earlier.

Thermalization is required for steady-state laser operation.  In the computer model, thermalization is ensured by ascribing to each electron a probability $p \delta t$ to decay to the adjacent lower level if that level is empty, and a probability $q p \delta t$ to be promoted to the adjacent upper level if that level is empty, where $q \equiv \exp(-\epsilon / T)$ denotes the Boltzmann factor.  We select $T = 100$~K, corresponding to $q = 0.891$. Without absorption and pumping ($\tau_p = \infty$, $J = 0$), the program gives level occupancies very close to those predicted by the Fermi-Dirac distribution.

Regular pumping is considered with an electron at the bottom of the VB promoted to the top of the CB every $\Delta t = 0.2$~ns.  This period corresponds to a pumping rate $J = 1/\Delta t = 5$ events per ns, and a pump electrical current in the nA range.  For a 1~$\mu$m--long quantum wire with a 10~nm $\times$ 10~nm cross-section this corresponds to $10^7$~A/cm$^{3}$.

Each light quantum is ascribed a constant probability $ \delta t/\tau_p$ of being absorbed by the photodetector, with $\tau_p = 2$ns.  The average number of light quanta in the cavity follows from the average-rate balancing condition $J= \ave{m}/\tau_p$, that is, $\ave{m} = 10$.

\begin{figure}
\centering
\includegraphics[scale=0.6]{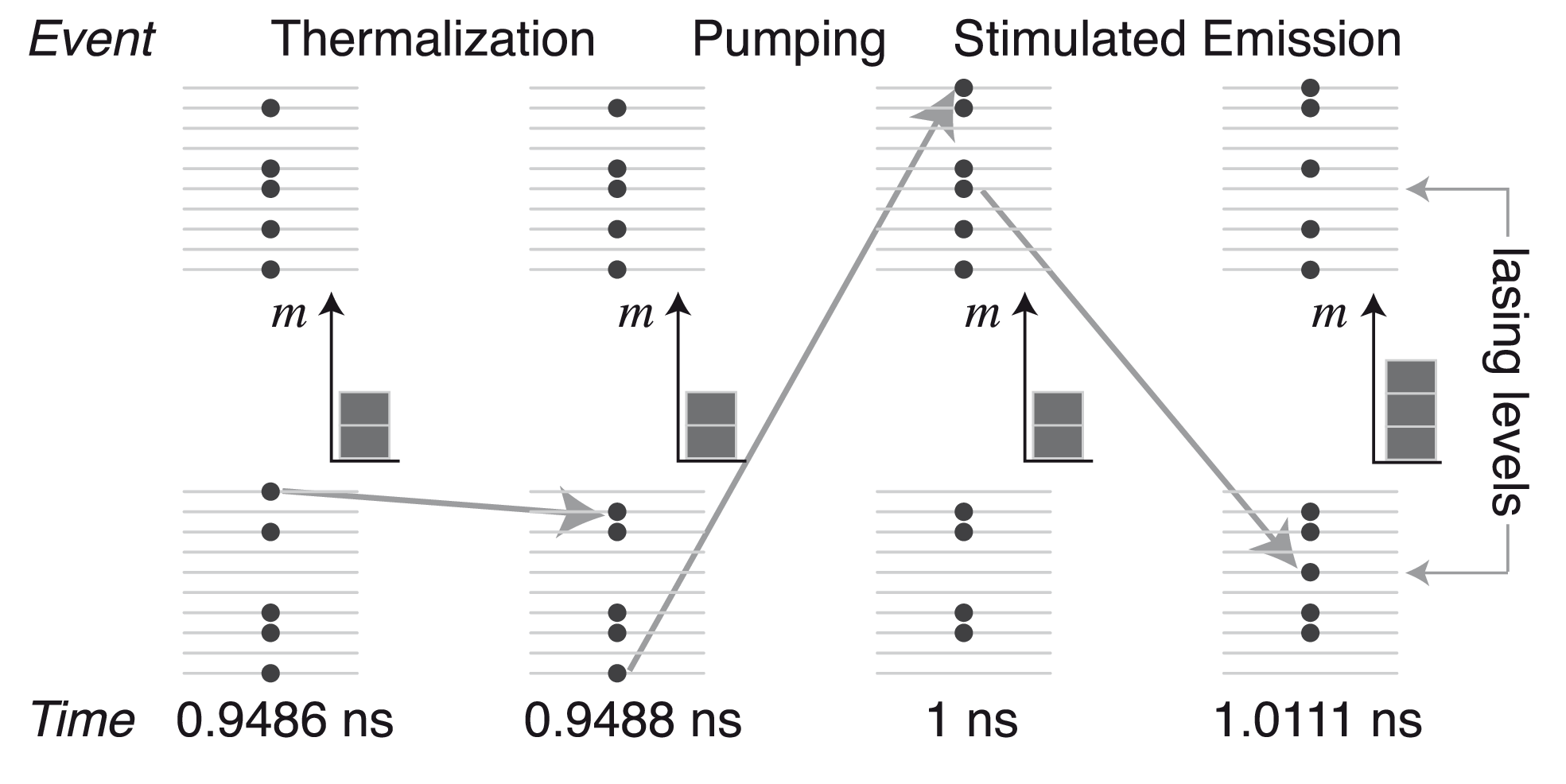}
\caption{Four frames extracted from a Monte Carlo sequence for 10 equally-spaced energy levels in VB and CB. Time in
arbitrary units increases from left to right.  Insets give the number of light quanta $m$ stored in the cavity at some
time. 
Arrows indicate electron displacements from one energy level to another.}
\label{fig2}
\end{figure}

Figure~\ref{fig2} illustrates three of these elementary processes by means of a sequence of four frames extracted from a computer simulation involving 10 levels in each bands.  At the start, the system has already reached a stationary regime.  A sample of the electron distribution is shown on the left.  The corresponding time and the number of light quanta stored in the cavity are respectively $\tau_{0} = 0.9486$~ns and $m = 2$.  The first event at $\tau_{1} = 0.9488$~ns is a VB thermalization.  Its effect is to decrement the system energy by an amount $\epsilon$ since an electron is demoted by one energy step.  The second event at $\tau_{2} = 1$~ns is an electrical pumping event that promotes the VB electron occupying the lowest energy level to the highest energy level of the CB. The third event illustrates stimulated emission between the prescribed lasing levels at $\tau_{3} = 1.0111$~ns.  As a result, the number of light quanta is incremented from $m = 2$ to $m = 3$.

Except for the pumping events, all the processes are governed by a Poisson probability law.  It follows that the whole system aside from pumping events also obeys a Poisson probability law.  Knowledge of the laser microscopic state at time $\tau_{i}$ implies knowledge of the next-jump density function.  An exact Monte Carlo simulation of the laser evolution is then easily obtained by randomly picking up the next event time $\tau_{i+1}$ from a Poisson law and, next, picking the event type from a uniform law weighted by the count of potential events for each type.  This method is more rigorous and more efficient than simulations based on infinitesimal time steps.  The time required to obtain a photo-detection spectrum is on the order of a few hours on desk computers.

The times $t_{k}$ of occurrence of photo-detection events are registered once a steady-state regime has been reached, as is always the case for the kind of lasers considered.  The detection rate $Q(t)$ is the sum over $k$ of $\delta(t-t_{k})$, where $\delta(.)$ denotes the Dirac distribution.  Considering that the photo-detection events are part of a stationary process, the two-sided spectral density of the detection rate fluctuation $\Delta Q(t) \equiv Q(t)-\ave{Q}$ is
\begin{equation}\label{sq} 
\spectral_{\Delta Q}(\Omega)=\frac{1}{\T} \ave{\left|
\sum_{k}\exp(-j\Omega t_{k}) \right|^{2}}
\end{equation}
where the sign $\ave{.}$ stands for averaging and $\Omega \equiv 2 \pi n/\T$ with $n$ a nonzero integer.  For uniformly-distributed independent events, \emph{i.e.} for a Poisson process, Eq.~(\ref{sq})
gives the shot-noise formula $\spectral_{\Delta Q}(\Omega) = \ave{Q}$.

\section{Numerical results}\label{results}

Simulations will be reported for three values of the thermalization parameter, namely $p = 25\,000 \; \text{ns}^{-1}$,
$1\,000 \;
\text{ns}^{-1}$, $250 \; \text{ns}^{-1}$, and $T=100$~K. Large values of $p$ enforce well-defined, constant temperatures
to the electron gas.  Conversely, small values of $p$ correspond to large carrier-lattice thermal resistances.

The total numbers of events during a run lasting 1\,$\mu$s are, respectively, about $6.2\,10^8$, $3.9\,10^7$ and
$1.1\,10^7$.  The number of events of various kinds are listed in Table \ref{evts} for the three values of $p$
considered.  The number of VB (resp., CB) ``cooling'' events is the number of downward electron transitions in the
valence band (resp., conduction band) due to thermalization. 
Likewise, VB (resp., CB) ``heating'' refers to upward electron shifts.

\begin{table}
\begin{center}
\begin{tabular}{ r r r r }
\hline $p$ (ns$^{-1}$) & 25\,000 & 1\,000 & 250\\
\hline Pumping & 5\,000 & 5\,000 & 5\,000 \\
Detection & 5\,004 & 5\,000 & 5\,000 \\
Stimulated abs. & 882 & 865 & 876 \\
Stimulated emi. & 5\,883 & 5\,865 & 5\,876 \\
VB ``cooling'' & 155\,286\,151 & 9\,819\,009 & 2\,848\,060\\
VB ``heating'' & 155\,036\,238 & 9\,569\,413 & 2\,619\,666\\
CB ``cooling'' & 155\,303\,449 & 9\,779\,404 & 2\,824\,115\\
CB ``heating'' & 155\,058\,595 & 9\,534\,806 & 2\,599\,576\\
\hline
\end{tabular}
\end{center}
\caption{Number of events of various kinds during a $1\;\mu$s-run}
\label{evts}
\end{table}

The following observations can be made: 

\begin{itemize}

\item The first two lines of the table show that, over a run, the number of detection events is essentially equal to the
number of pumping events.  Since the number of pumping events does not vary from run to run, it follows that the number
of photo-detection events does not vary either, \emph{i.e.}, the light is ``quiet''.  Non-zero variances of the
photo-count would appear only over much shorter durations.

\item The next two lines show that the difference between the numbers of stimulated emission and absorption events is
nearly equal to the number of photo-detection events.  Because of the band symmetry, the numbers of stimulated events
are almost independent of $p$.

\item The difference between the number of cooling and heating events corresponds to the power delivered by the pump in
excess of the power removed by the detector.  This difference is almost independent of $p$.

\end{itemize}

\begin{figure}
\begin{center}
\includegraphics[scale=0.6]{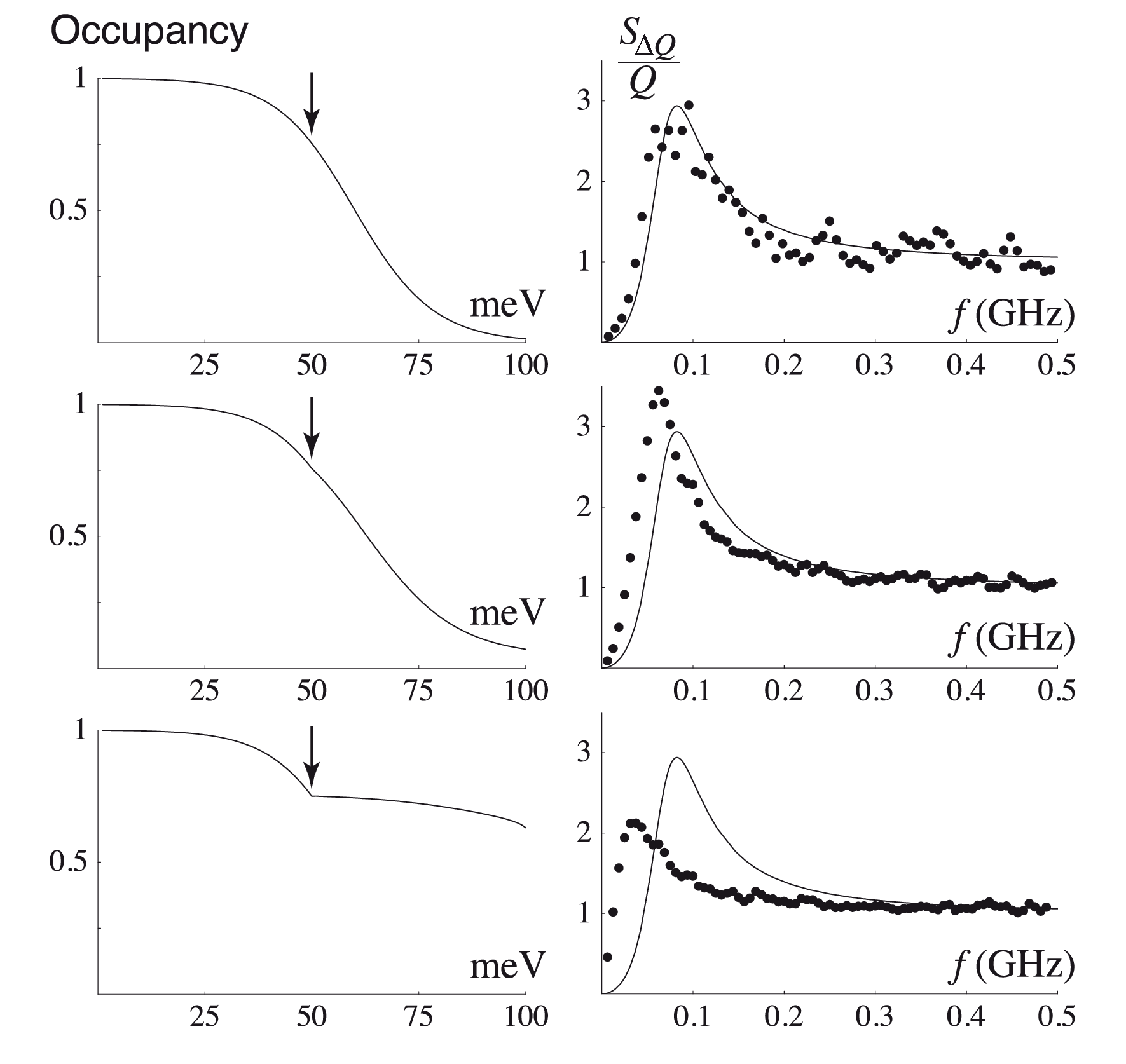}
\caption{On the left, electron occupancy as a function of the energy referred to the bottom of the conduction band. 
Arrows point to the lasing level.  On the right, normalized photo-detection spectra.  Dots are Monte Carlo results and
solid lines are from \eqref{loxp}.  The thermalization rates are $p = 25\,000 \; \text{ns}^{-1}$ (top), $1\,000 \;
\text{ns}^{-1}$ (middle), and $250 \; \text{ns}^{-1}$ (bottom).}
\label{fig3}
\end{center}
\end{figure}

The CB level-occupancies are represented on the left-hand part of Fig.~\ref{fig3}.  The CB electron-occupancies and the
VB \emph{hole} occupancies are symmetrical with respect to the middle of the bandgap. 

\begin{itemize}
    
    \item For $p = 25\,000 \; \text{ns}^{-1}$, electron occupancies
    are very close to the Fermi-Dirac (FD) distribution, except near
    the edges of the band.  A least-square fit shows that the carrier
    temperature is $T_{c} = 105.6$~K for both bands.  The quasi-Fermi
    levels (referred to the bottom of the bands) are respectively
    $\mu_{CB} = 60.2$ and $\mu_{VB} = 40.8$.

    \item For $p = 1\,000 \; \text{ns}^{-1}$, a fit gives $T_{c} =
    132$~K for both bands, $\mu_{CB} = 63.5$ and $\mu_{VB} = 37.5$. 
    There is a dip due to `spectral hole burning' (SHB) at the lasing
    level shown by an arrow.  This dip is difficult to see on the
    figure, but it nevertheless influences importantly the noise
    properties of the laser.

    \item For $p = 250 \; \text{ns}^{-1}$, the dip at the lasing level
    is conspicuous.

\end{itemize}

The right part of Fig.~\ref{fig3} gives a comparison between spectra calculated from Monte Carlo data using Eq.~(\ref{sq}) and the spectrum obtained from the elementary laser-diode noise theory using the same set of parameters, $(\tau_p, T, \epsilon, J )$.  The top part of the figure corresponds to efficient thermalization. The spectral density is below the shot-noise level up to a frequency of 42~MHz.  Notice the strong relaxation oscillation. There is good agreement between the Monte Carlo simulation and the linearized theory. For moderate thermalizations the spectral density is below the shot-noise level up to a frequency of 25~MHz and no more agrees with the linearized theory.  An increase of temperature from 100~K to   132\,K does not suffice to reproduce the observed shift. The change in spectral density may be attributed to spectral-hole burning and carrier heating. The bottom curve corresponds to poor thermalization.  The relaxation oscillation is strongly damped.  The frequency range where the spectral density is below the shot-noise level now extends only up to 8~MHz.

To conclude the present section, a Monte Carlo computer program keeping track of the occupancy of each level in the conduction and valence bands may be applied to regularly-pumped mesoscopic
laser diodes having equally-spaced levels in each band. When the electron-lattice thermal contact is good, theoretical results based on linearization and the Fermi-Dirac distribution are recovered.  In particular, it is verified that sub-Poissonian light may be obtained.  But when the thermal contact is poor, as is the case when lasers are driven to high powers, the simple theory is inaccurate.  Some of the changes observed may be accounted for by temperature increase and gain compression (due to spectral-hole burning).  But unexpected effects are also found. In particular, an increase of the spectral density at low frequencies is noted.

\section {Multilevel atoms.}\label{sec_multi}

Instead of supposing a constant pump rate $J$ as in most previous sections, we consider four-level atomic lasers that are pumped from level 0 to level 3 and from level 3 to level 0
with the same probability $P$. This corresponds to strong optical pumping, e.g., by thermal light. The single-mode resonator is resonant with transitions 1-2. Output power regulation originates from the spontaneous decay from level 3 to level 2 (time constant $\tau_u$) and the spontaneous decay from level 1 to level 0 (time constant $\tau_d$). $u,d$ for "up" and "down", respectively. These spontaneous decays have fluctuating rates at the shot-noise level. They play a role somewhat similar the the resistance employed to generate non-fluctuating drive currents in laser diodes.

The photo-current spectral density are evaluated on the basis of rate equations. 
According to that approach, fluctuations are caused by jumps in emitting \emph{and} detecting atoms. The conditions under which the output light exhibits sub-Poisson statistics are considered in detail. Analytical results, based on linearization, are verified by comparison with Monte-Carlo simulations.  An essentially exhaustive investigation of sub-Poisson light generation by four-level lasers has been made. 

\begin{figure}
\centering
\includegraphics[scale=0.6]{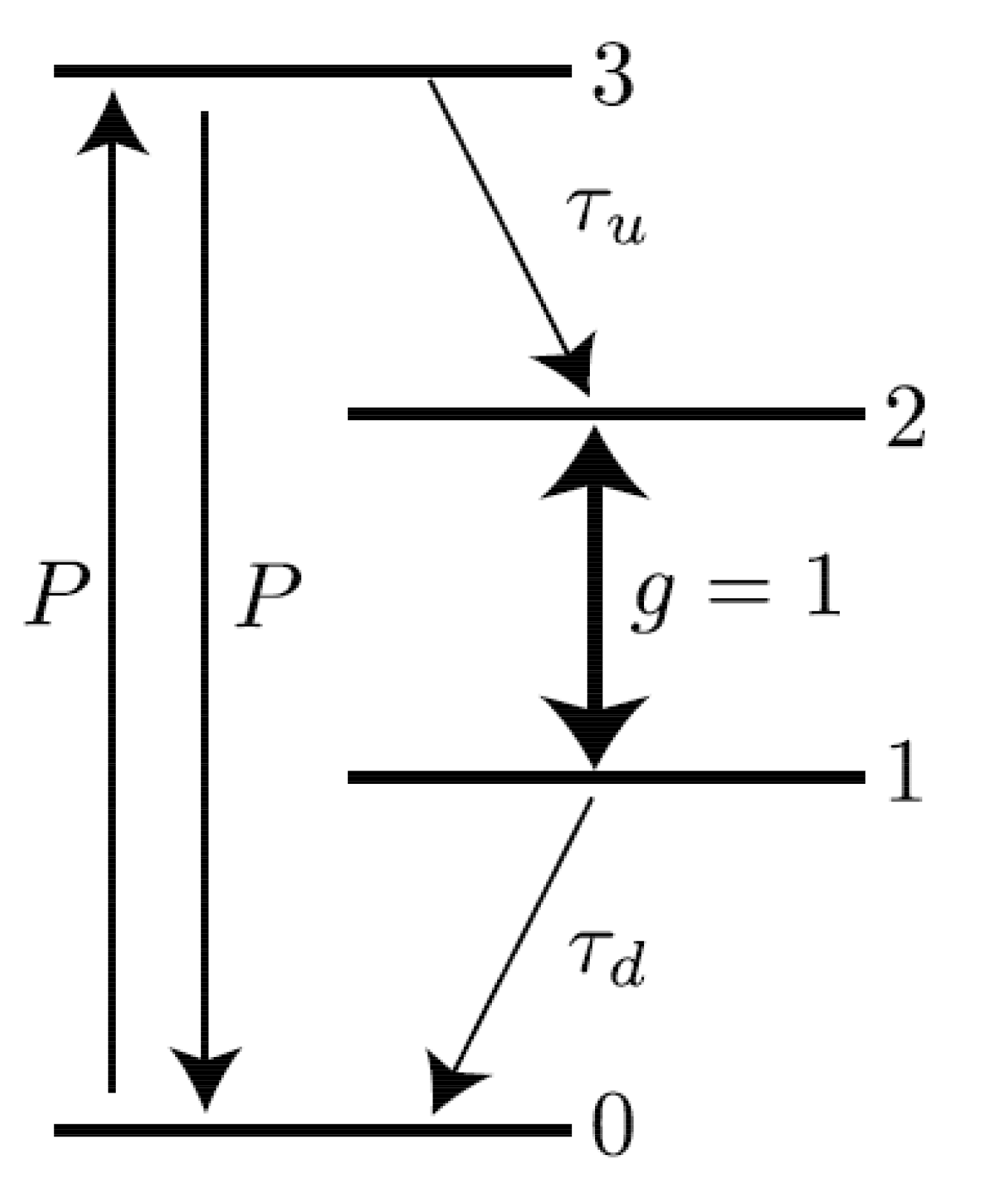}
\caption{Level schemes for a 4-level atomic laser. 
Assuming a gain $g =1$ (this $g$ should not be confused with the normalized differential gain, also denoted $g$ in some sections) normalizes rates to the corresponding time unit. $\tau_u,~\tau_d$ are spontaneous decay times. The working levels are 1,2. Spontaneous decay from 2 to 1 is neglected.}
\label{fig:0}
\end{figure}

Rate equations treat the number $m$ of light quanta in the resonator as well as the numbers of atoms in each state as classical functions of time. The theory rests on the consideration of transition probabilities. Let us emphasize that every absorption event reacts on the number $m$ in the resonator.  A single ideal detector is considered that collects all the generated light. Analytical expressions are obtained from rate equations in a straightforward manner as solutions of a few linear equations.  The analytical expressions are sometimes too lengthy to be exhibited here.  But symbolic calculus enables us to easily determine the optimum conditions of operation, for example the parameter values that minimize the photo-detection noise at some prescribed Fourier frequency. We neglect the rate of spontaneous decay from the upper to the lower working levels (2-1), see Fig. \ref{fig:0}.  We first give details on the rate-equation approach to laser modeling. The weak-noise approximation is compared to the results of Monte-Carlo simulations.  When both the
number of atoms and the pumping level increase, the Monte-Carlo simulation computing time becomes prohibitively large. In that case, analytical results are essential. The photo-current spectral density at zero frequency, the photo-current spectrum and the intra-cavity Fano factor are obtained and shown to be sub-Poisson.

\paragraph{Laser model}\label{sec.Model}

The active medium is a collection of $N$ identical four-level atoms, one of them being represented in Fig.~\ref{fig:0}. 
Level separations are supposed to be large compared with $T$, where $T$ denotes the
system temperature (the Boltzmann constant $\kB=1$), so that
thermally-induced transitions are negligible.  Levels $\ket{1}$ and $\ket{2}$
are resonant with the field of a single-mode optical cavity. 

%Figure 1

The probability per unit time that a transition from level
$\ket{1}$ to level $\ket{2}$ occurs is taken as equal to $m$, and the
probability per unit time that a transition from $\ket{2}$ to $\ket{1}$ occurs is
$m+1$.  Light quanta are absorbed with probability $ m/\tau_p$, where $\tau_p$ denotes the resonator lifetime. It models the transmission of light through mirrors with subsequent absorption by a detector. Levels $\ket{0}$ and $\ket{3}$ need not be sharp.  Instead, they may consist of narrow bands for improved coupling to broad-band pumps. 

%\begin{table}\label{optimum}
%\caption{Minimum value of the zero-frequency photo-current
%spectral density $\spectral_{min}$ and intra-cavity Fano factor
%$\fano$ for three- and four-level atomic lasers.  The conditions on $P$,
%$p_{u}$ and $p_{d}$ are given.  Spontaneous decay from the upper
%working level is neglected and $N \gg \alpha$ is assumed.}
 %   \begin{ruledtabular}
%\begin{tabular}{c|c|c|c}
%Laser & $\spectral_{min}$ & $\fano$ & Conditions \\
%\hline
%$\Lambda$--type three--level\footnotemark[1] & $1/2$ & $3/4$ &
%$p_{d} = 2 P$ \\
%$\Lambda$--type 3--level\footnotemark[2] & $2/3$ & $5/6$ &
%$p_{d} = 3 P$ \\
%$\V$--type 3--level\footnotemark[1] & $1/2$ & $3/4$ &
%$p_{u} = \frac{1}{2} P$ \\
%$\V$--type 3--level\footnotemark[2] & $5/6$ & $11/12$ &
%$p_{u} = \frac{3}{2} P$ \\
%4--level\footnotemark[1] & $1/3$ & $2/3$ & $p_{u} = P$,
%$p_{d} = 2 P$ \\
%4--level\footnotemark[2] & $3/7$ & $5/7$ & $p_{u} = 2 P$,
%$p_{d} = \frac{4}{3} P$ \\
%\end{tabular}
 %   \end{ruledtabular}
%\end{table}

According to the previous model, the laser-detector assembly is treated as a birth-death
Markov process illustrating the evolution of the number $m$ in the resonator, from which the Fano factor
$\fano =\var(m)/\ave{m}$ can be extracted.  Similarly, the instants $t_{k}$ when light quanta
are being absorbed provide us with the spectral density of the photo-current, whose normalized value
$\spectral$ is unity for Poisson processes.  In the following, the normalized spectrum
is denoted $\spectral(\Omega)$, where $\Omega$ denotes the Fourier frequency.

Monte-Carlo results have been improved by averaging over runs and
concatenating neighboring frequencies to produce the final data
with error bars.  There is a fair agreement between Monte-Carlo simulations and analytical formulas. We conclude that first-order variation of rates ruling laser populations is a valid assumption. This holds as least when the number of active atoms exceeds $N=100$. Finally, notice that even with one billion light-quanta absorption events, Monte-Carlo spectra exhibit large error bars. The analytical method is thus preferable.
%The results are summarized in Fig.\ref{MC}

Under the conditions of negligible spontaneous decay and
$N\tau_p\gg 1$, the intra-cavity Fano factor then depends linearly of the
zero-frequency normalized photo-current spectral density
: $\spectral = 2 \fano-1$. This relation does not 
hold in general.

\section{Arbitrary media.}\label{arb}

Arbitrary media could be made discrete with $N$ cells. The larger is the number of cells the more accurate is the system description. The noise properties are obtained through the introduction of random current sources, or rate sources, as explained earlier, and the inversion of an $N\times N$ \emph{bi-complex} scattering matrix. The theory is capable of providing the spectral densities of fluctuations of the photo-detection rate, the phase of the output field (that is, from our view point, the outcome of some appropriate homodyne receiver), and the potential $U$ across the laser diode, as well as the correlations between these three quantities. If the device under study may radiate light into free space, free space should also be described as a circuit. This circuit must be terminated at some large distance by perfect absorbers. It is important to take into account the random currents associated with these absorbers. This concept, it seems, is more realistic than the so-called "box quantization" often employed.

The circuit could consist of any conservative elements such as capacitance, inductance, and circulators, with $N$ ports. For simplicity we have consider only the case of two ports, and a single susceptance $B(\om)$, representing the optical resonator, connected across the input-output wires. This susceptance consists of a capacitance and an inductance connected in parallel. The circuit could be treated by conventional methods. However we found the bi-complex notation useful. Agreement with expressions obtained previously is noted when the corresponding simplifying assumptions are made. The detailed discussion is however omitted here.

\section{Single electron laser.}\label{single}

We consider a laser involving a single electron permanently interacting with the field and driven by a constant-potential battery, and point out a similarity with reflex klystrons. The spectral density of the photo-current spectral density is found to be only 7/8 of the shot-noise level at zero Fourier frequency. It is therefore sub-Poisson. Our calculations are related to resonance-fluorescence treatments but have a different physical interpretations. 

Going back to the configuration represented in Fig. \ref{klystron}, note that the battery represented on the left delivers a measurable average electron rate $J$ that may be increased by increasing the battery potential $U$ slightly above $\hbar\om/e$. The rate $R$ generated by the electron is given in \eqref{avt}. Finally, radiation escaping from the hole shown on the right of the resonator is eventually absorbed by an ideal detector at a rate $D=\mu/\tau_p$, where the lifetime $\tau_p$ depends on the hole size. Evaluating $\tau_p$ is a classical electromagnetic problem that we assume solved. Thus, the steady state condition $J=R=D$ reads explicitly 
\begin{align}
\label{dy}
J=\frac{\gamma}{1+2\gamma^2/\Om_R^2}=\frac{\mu}{\tau_p}.
\end{align}
Accordingly, given the average electron-injection rate $J$ and the resonator lifetime $\tau_p$, we may evaluate the reduced resonator energy $\mu\equiv E/\hbar\om=J\tau_p$. Next, given the capacitance volume $\V$, we may evaluate the square of the Rabi frequency $\Om_R^2$ and the decay constant $2\gamma$ from \eqref{dy}. This value of $\gamma$ corresponds to some value of the static potential $U$ slightly above $\hbar\om/e$.

\paragraph{Noise of single-electron lasers.}\label{lasernoise}

What we call "laser noise" refers to photo-current fluctuations. The result given in \eqref{ol} provides the rate-fluctuation spectral density for an electron submitted to an alternating potential independent of the electron motion. But in lasers the reduced resonator energy $\mu(t)=\mu+\De \mu(t)$ fluctuates. Because this fluctuation is small in above-threshold lasers the fluctuation $r(t)$ previously evaluated is supposed to be unaffected. The rate equation is, remembering that $U$ and therefore $\gamma$ is held constant
\begin{align}
\label{rat2}
\frac{d\mu(t)}{dt}=R(t)-D(t)\qquad R(t)=R+\frac{dR}{d\mu}\De \mu(t)+r(t)\qquad D(t)=\frac{\mu(t)}{\tau_p}+d(t),
\end{align}
where $d\mu(t)/dt$ represents the rate of increase of the integer resonator energy. This is the difference between the in-going rate $R(t)$ and the out-going (or detected) rate $D(t)$. Note that the in-going rate involves a term expressing the fact that $R$, as given in \eqref{avt}, depends on $\mu$ and that $\mu$ is now allowed to fluctuate. The outgoing rate is fully absorbed by an ideal cold detector at an average rate $D$, supplemented by a fluctuating rate $d(t)$, whose spectral density is equal to the average rate $D=R=J$. Because the noise sources $d(t)$ and $r(t)$ have different origins they are independent. 

Considering only fluctuating terms at zero Fourier frequency ($\frac{d}{dt}\to0$), we obtain $\De R(t)=\De D(t)$, that is, explicitly
\begin{align}
\label{rat}
\frac{dR}{d\mu}\De \mu(t)+r(t)=\frac{\De \mu(t)}{\tau_p}+d(t).
\end{align}
Solving this equation first for $\De \mu(t)$, with $\tau_p=\mu/R$, and substituting the result in the expression for $\De D(t)$, we obtain
\begin{align}
\label{rat2a}
\De D(t)\equiv \frac{\De \mu(t)}{\tau_p}+d(t)=\frac{r(t)-Ad(t)}{1-A}\qquad A\equiv \frac{\mu}{R} \frac{dR}{d\mu}=\frac{a}{1+a}\qquad a\equiv \frac{2\gamma^2}{\Om_R^2},
\end{align}
according to \eqref{avt}. Because $r(t)$ and $d(t)$ are independent, the spectral density of the photo-detection rate is, with $\spectral_r/D=1-3a/(1+a)^2$ from \eqref{ol} and $\spectral_d/D=1$,
\begin{align}
\label{rat5}
\spectral_{\De D}=\frac{\spectral_r+A^2~\spectral_d}{(1-A)^2}=(2a^2-a+1)D.
\end{align}
The smallest detector noise, obtained when $a=1/4$, is 7/8 of the shot-noise level. Therefore, a sub-Poissonian laser may be realized with constant static potential sources. As an example, suppose that $\mu=1$ (that is $E=\hbar\om$) we find using \eqref{rat5} that the maser capacitance volume should be $\V=244\tau_p^2$ if minimum noise is to be achieved. With $\tau_p=1\mu$s and $d=0.44\mu$m as in Section \ref{static}, the capacitance size $\sqrt A=23$ mm.

\section{Fourier frequency representation}\label{rep}

As we emphasized in the introduction we attach a physical significance only to electrical currents or potentials at Fourier frequencies, and therefore the Fourier-frequency representation of the system considered is important to describe the end results of the calculations. Such a (low-frequency) circuit is outlined below. The electrical schematic (operating at Fourier frequency $\Om$) consists of a capacitance $C$ representing the resonator, followed by a resistance $1+\kappa$, where $ \kappa$ denotes the gain-compression factor, and a resistance of value -1. Such a negative resistance has been realized with a positive resistance preceded by a so-called "negative-impedance converter". The latter device employs an operational amplifier (high gain, high input impedance). The response measured on this circuit very well agrees with the analytical formulas.

\newpage

\chapter{Linewidth in the linearized approximation}\label{line}

When the linewidth of above-threshold lasers is investigated it is permissible to neglect amplitude fluctuations. We are seeking primarily an expression for the frequency fluctuation $\De \om(t)$ whose (double-sided) spectral density is denoted $\spectral_{\De\om}$. Assuming that this fluctuation is gaussian-distributed the laser linewidth is obtained from conventional signal theory as $\de\om=\spectral_{\De\om}$.

\section{Steady-state.}

The complete system under consideration, namely the pumped semiconductor (with pumping rate $J$), the resonator (a conservative device), and the load (perhaps a detector delivering photo-electrons at rate $D$) is represented by a single admittance (ratio of the complex current $I$ and the complex potential $V$), namely $Y(n,\om)$, where $n$ denotes a real parameter (perhaps the number of electrons in the semiconductor conduction band) and $\om$ the real oscillating frequency. For the time being, random current sources are ignored. Since this admittance is isolated and $V≠0$, the steady state is given by the condition that $Y(n,\om)=0$. Separating the real and imaginary parts of this relation (remember that $Y\equiv G+\ii B$), we obtain two equations. Solving these two equations we may obtain the steady-state values of $n$ and $\om$, denoted $\ave{n}$ and $\om_o$, respectively (the averaging sign is omitted when no confusion is likely to arise. We assume that a solution of the two equations exists and is unique. In the present linearized regime we consider that $V(t)$ is a large potential. If the injected electron rate is $J$ and the detector conductance is $G$, conservation of energy tells us that $G\abs{V}^2=J\hbar \om_o$. Indeed, the current $eJ$ is delivered to the semiconductor at a static potential $U\approx \hbar \om/e$. Because we are not concerned in the present section with potential fluctuations, we consider from now on that $V$ is a fixed known quantity that one may assume to be real for simplicity. 

\section{Deviation from the steady state.}

Next, let us introduce a complex random current source $C(t)$, whose statistical properties will be defined later on, driving the admittance $Y$. The circuit equation reads now $C(t)=Y(n(t),\om(t))V$, where we have written explicitly the dependence of $n$ and $\om$ on time. Expanding to first order, we have
   \begin{align}\label{ab}
C(t)=\frac{\partial Y}{\partial n}\Delta n(t)+\frac{\partial Y}{\partial \om}\Delta \om(t)
  \end{align}
where the partial derivatives are evaluated at the steady-state values. Setting $C(t)\equiv C'(t)+C''(t)$, and separating the real and imaginary parts of the above equation, we obtain
   \begin{align}\label{sep}
C'(t)&=\frac{\partial G}{\partial n}\Delta n(t)+\frac{\partial G}{\partial \om}\Delta \om(t)\nonumber\\
 C''(t)&=\frac{\partial B}{\partial n}\Delta n(t)+\frac{\partial B}{\partial \om}\Delta \om(t).
  \end{align}
  
We may now eliminate $\Delta n(t)$ from these two equations, and obtain $\De \om(t)$ as a linear function of $C'(t)$ and $C''(t)$. Denoting partial derivatives by subscripts, we obtain
   \begin{align}\label{freq}
\De\om=\frac{B_n C'-G_n C''}{B_n G_\om-G_n B_\om}.
  \end{align}
It follows that the laser (full width at half power) linewidth reads
   \begin{align}\label{linew}
\de \om=\spectral_{\De \om}=\frac{B_n^2\spectral_ {C'}+G_n^2\spectral_ {C''}}{(B_n G_\om-G_n B_\om)^2}.
  \end{align}

\section{Complete inversion}

\paragraph{Model.}

We now specialize the previous result by supposing that
\begin{itemize}
\item There is complete population inversion in the active semiconductor (i.e., $G_a=0$). It is therefore represented by an admittance $-G_e(n)-\ii B_e (n)$, where the subscript $e$ means as before "emitting", and we have changed the sign of the conductance so that $G_e$ is positive and expresses the optical gain. The term $B_e$ is a consequence of causality relations. Because the conductance depends on frequency (the gain depends on frequency at the average frequency) there must be some reactive component present. Of course the conductance and the susceptance depend on the same electron number $n$. We denote $B_{en}/G_{en}\equiv \al$. This Lax-Haken-Henry parameter $\al$ is on the order of 2.
\item The load usually considered is a fixed conductance $G$ representing the detector. For generality, we consider a load $Y(\om)\equiv G(\om)+\ii B(\om)$ and define the parameter $h=G_\om/B_\om$. As an example the load may consist of two elements connected in parallel: a capacitance $C$ and an inductance $L$ in series with a resistance $R$. The laser power is the power dissipated in that resistance. Comparison with previous expressions obtained in the linear regime shows that the Gordon-Petermann factor $K=1+h^2$.
\item The random currents $C'(t) $ and $C''(t)$ have the same spectral densities, given in the present model by
 \begin{align}\label{ew}
\spectral_{C'}=\spectral_{C''}= \hbar \om_o (G_e+G(\om_o)).
 \end{align}
\end {itemize}
Remember that the fact that $G$ depends on $\om$ does not affect the spectral density of the noise source because the load is linear.

\paragraph{Results.}

From the above derivations, we readily obtain a linewidth enhancement factor with respect to the linear result, for the same output power 
 \begin{equation}\label{osc}
\de\om=\de\om_{linear}\frac{1+\alpha_A^2}{2}K\qquad\alpha_A\equiv\frac{\De\om'}{\De\om''}=\frac{\alpha+h}{1-\alpha h}
\qquad K=1+h^2.
  \end{equation}
Here $\alpha$ is the Lax-Haken-Henry phase-amplitude coupling factor of the active material employed and $h$ is the dispersion factor introduced earlier. The parameter $\alpha_A$ is defined from the complex frequency deviation of the circuit $\De\om\equiv \De\om'+\ii\De\om''$ corresponding to some departure of $n$ from $n_o$. The formula in \eqref{osc} is the correct way of combining the effects of the $\alpha$ and $K$ factors. These two effects are not independent and they should not be simply multiplied.

\section{Multiple active elements.}\label{multiple}

We consider the linewidth of a laser with multiple elements submitted to the same field. This circumstance holds when the laser end mirrors have a reflectivity close to unity, so that the optical field does not vary much along the length (ignoring the fast variations due to the standing waves.

\paragraph{Model.}

The present model differs from the previous one in that the active admittance $-Y_e$ is replaced by a set of admittances $-Y_k$, $k=1,2...$ connected in parallel. These admittances are driven by possibly different electrical currents $J_k$ and they may posses different phase-amplitude coupling factors $\alpha_k$ and population-inversion factor $n_{pk}$. The dispersion of the load $G_a$is expressed as in previous sections by a factor $h\equiv (dG_a/d\om)/(dB_a/d\om)$. The steady-state oscillation condition requires that the sum of the $G_k$ be equal to $G_a$, and that the total susceptance vanishes.

\paragraph{Result.}

The product of the (FWHP) linewidth $\de\om$ and the photo-electron rate $D$ is given by a formula in the ST form (linear solution) with multiplicative factors. The number $k$ labels the different elements connected in parallel, with the total injected electron rate $J=J_1+J_2+...$. The parameters $\alpha_k$ and $n_{pk}$ denote respectively the phase-amplitude factor and the population-inversion factors for element $k$. $h$ is the dispersion factor which occurs for example when the resistive load is connected in series with the inductance.
 \begin{equation}\label{multi}
\de\om D=\frac {1} {\tau_p^2} \frac{\ave{n_p/(1+\alpha^2\q}}{\p1-\ave{\alpha} h\q^2},
\qquad \tau_p\equiv \frac{2G}{dB/d\om}.
  \end{equation}
where the averaging sign is presently defined for any quantity $a_k$ as
 \begin{equation}\label{multibis}
\ave{a}\equiv \frac{\sum a_k J_k}{\sum J_k}.
  \end{equation}

\section{Detuned inhomogeneously-broadened lasers}\label{sec_inhom}

The present model is similar to the previous one but we consider a collection of detuned atoms, with a Lorentzian distribution of the resonant frequencies. The origin of this de-tuning may be the environment of rare-earth atoms. The total injection electron rate $J$ splits into as many driving rates as they are resonant atoms. The fluctuations of the individual injected rates is evaluated as before.
\paragraph{Model.}

For some mirror reflectivity, there is a minimum number $N_o$ of active atoms required to reach the threshold of oscillation. Our first parameter is the ratio $n\equiv N/N_o$ of the actual number $N$ of atoms and $N_o$. If the atom homogeneous line-width is $1/\tau_o$ and the spectral width of the atom resonant frequency distribution is denoted $1/\tau_i$, we define a second parameter $r\equiv 1+\tau_o/\tau_i$. Finally, the resonant optical cavity may be detuned with respect to the center of the Lorentzian atomic-frequency distribution. The difference normalized to the Lorentzian spectral width is called $\de$. For each element, the pumping rate is supposed to be proportional to the number of atoms in the lower state.

\paragraph{Result.}

Remarkably, the $\de\om~ D$ product, where $\delta\om$ denotes the full width at half power linewidth, and $D=Q$ denotes the photo-electron rate, may be expressed in closed form for any value of the parameters $n\equiv N/N_o$, $r$ and $\de$. In the limit $r\to\infty$ we obtain for example
 \begin{equation}\label{multiter}
\de\om~ D~\tau_p^2= \frac{1/n^2+10++5n^2}{32}.
  \end{equation}
an expression that reduces to previous ones when $n=1$, namely 1/2.

\section{Spatially-varying $\alpha$-factors.}\label{alpha}

We consider in the present section ring-type laser oscillators with the wave propagating in a single direction (with the help of some nonreciprocal element). We are concerned with the laser (full-width at half power) linewidth $\de\om$.

\paragraph{Model.}

We consider a closed path with coordinate $z$, implying that after a round-trip (at $z=L$) the field recovers its value at $z=0$. Of course the linewidth may not depend on the point along the path at which the $z$-axis origin is selected. The propagating wave experiences both a power gain $\gamma(z)$ (complete population inversion is assumed) and a power loss $\ell(z)$ (a temperature $T=0$K is assumed), distributed arbitrarily along the path. A localized loss of the form $\ell(z)\propto \de(z-a)$, where $\delta(.)$ denotes the Dirac distribution, describes a partially-reflecting loss-less mirror located at $z=a$. In addition to the gain and the loss, one must take into account the phase-amplitude coupling factor $\alpha(z)$. As one recalls, this factor is defined as follows:  We consider a piece of the active medium with permittivity $\epsilon$. Under a small change of the electron number, the real and imaginary parts of $\epsilon$ vary and we defined $\alpha\equiv \De \epsilon'/\De \epsilon''$. Frequency dispersion is neglected in the present model. 

\paragraph{Result.}

The product of the (full-width at half-power point) linewidth $\de\om$ and the power $P$ transferred from the gain medium to the loss medium reads

 \begin{equation}\label{varal}
\de\om~ D~\tau^2=\frac{1}{2} \int_{1}^{\Gamma}\frac{d\gamma}{\ell}\int_{1}^{\Gamma}d\gamma\frac{\p 1+\alpha^2 \q \ell}{\gamma^2}.
  \end{equation}

where $\tau$ denotes the round trip transit time and $\Gamma$ the round-trip gain (or loss). In this expression the power gain (varying from 1 to $\Gamma$) is employed as an integration variable instead of the coordinate $z$ (varying from 0 to $L$). This means that $\ell$ and $\alpha$ are considered as functions of $\gamma$ instead of functions of $z$. Even though this is not immediately obvious, the above expression does not depend on the selected origin of the $z$-axis. It reduces to previously known expressions for simpler configurations.

\newpage

\chapter{Compression and feedback}\label{ampfeed}

The two subjects treated in this chapter (gain compression and electrical feedback) are fundamentally independent. However we will see at the end of this chapter that gain compression may be introduced without affecting the conclusions concerning the feedback amplifier.

\section{Gain compression}\label{compr}

Gain compression expresses an explicit dependence of the gain on the emitted light quanta rate $R$, We have to deal now with $G(n,R)$ instead of $G(n)$, leaving a possible dependence on $\om$ aside. The most conspicuous effect of gain compression is to damp the laser relaxation oscillations. In a typical case, a large relaxation oscillation peak (in relative noise or modulation) of a factor 10 is cut by half with only $\kappa\approx 0.05$.    

In the optical-engineering literature it is usually considered that the gain depends on the optical field (or, in our notation, on $V$) rather than on $R$. Of course the dependence of $G$ on $R$ may be re-expressed as a dependence on $V$. But a number of considerations show that if we do so, the random rates are no longer at the shot-noise level. The physical mechanism behind gain compression is most likely spectral-hole burning. That is, at high power levels, the lasing levels get depopulated because the electrons are removed at such a high rate that thermal effects are unable to restore the Fermi-Dirac statistics near the lasing level. This is so even if we consider that the total number $n$ of electrons in the conduction band is hardly affected. It is thus natural to consider that $G$ depends on $R$. A second argument is that one may convert a standard (i.e., constant) conductance into a conductance that depends on the emitted rate (but remains real and frequency-independent) with the help of a beam splitter and two fibers that exhibit Kerr effects of opposite signs. It is found that the random currents are unaffected in that configuration. The details of the calculation will be omitted. Finally, an indirect argument is that, most remarkably, the result described in the next section showing that light beams in the C-state (generating, as one recalls, photo-electron rates at the shot-noise level irrespectively of the carrier phase) is preserved, still holds exactly when the gain-compression effect is introduced. This would not be the case if alternative representations of gain compression were employed.

\section{Standard linear amplifier}\label {feedb}

We have described in section \ref{amp} a linear phase-insensitive amplifier employing an ideal circulator. The wave to be amplified enters in Port 1, a negative conductance $-G, ~G>0$ is in port 2, and the output is in Port 3. Let us recall our notation: $a\equiv \ave{a}+\Delta a$ is the amplitude of the input wave and $b\equiv \ave{b}+\Delta b$ the amplitude of the output wave. Averaging signs are omitted when no confusion is likely to arise. For simplicity, $\ave{a},\ave{b}$ are taken as real. $\G\equiv (b/a)^2=\p(1+G)/(1-G)\q^2$ denotes the power gain. The main result was that
\begin{align}
X_{out}=\mathcal{G}X_{in}+\mathcal{G}-1\qquad Y_{out}=\mathcal{G}Y_{in}+\mathcal{G}-1.
\label{55feed}
\end{align}
where $X$ represents the spectral density of $2\Delta a'$ and $Y$ represents the spectral density of $2\Delta a''$, subscripts being added where needed. In particular if the input is in the C-state ($X_{in}=Y_{in}=1$), the output is $X_{out}=Y_{out}=2\mathcal{G}-1>1$ if $\G>1$. That is, an amplifier of that kind always adds noise to C-state beams.

\section{Feedback amplifier without gain compression}\label {sec_feedb}

Our purpose in the present section is to show that it is, in principle, possible to construct an amplifier whose input-output relations are instead
\begin{align}
\label{177}
X_{out}=\frac{X_{in}+\mu}{\mu X_{in}+1}\qquad Y_{out}=\frac{Y_{in}+\mu}{\mu Y_{in}+1}\qquad \mu\equiv\frac{\mathcal{G}-1}{\mathcal{G}+1}
\end{align}
This result holds exactly (to first order). These expressions show that it is possible to construct an amplifier of power gain $\G>1$ such that, if the input is in the C-state ($X_{in}=Y_{in}=1$), the output is also in the C-state ($X_{out}=Y_{out}=1$). It has been shown in Quantum Optics that a coherent state cannot be amplified linearly without adding noise. The present situation is different. The amplifier is non-linear. Furthermore, coherent states differ somewhat from what we call in this book the C-state. In fact, whenever a light beam in the C-state is absorbed, and a new power-full beam in the C-state is generated, there is formally amplification without noise added. Of course, in such a trivial example, the information (modulation) does not carry through. In the feedback amplifier presently considered there is transmission of information.

We drop in the present section the subscripts "in" for the $a$-wave (Port 1), reserving that subscript for the left-most input. We thus set $x\equiv 2\Delta a\equiv x'+\ii x''$, and $x_{out}\equiv 2\Delta b\equiv x'_{out}+\ii x''_{out}$. We let $X$ denotes the spectral density of $x'$, $Y$ the spectral density of $x''$. At the output, we have as before $x_{out}\equiv x'_{out}+\ii x''_{out}$. $X_{out}$ is the spectral density of $ x'_{out}$, and $Y_{out} $ is the spectral density of $x''_{out}$ (Port 3). The starting equations are therefore the same as the ones given in section \ref{amp} with the "in" subscripts dropped, namely
\begin{align}
\label{79}
x'_{out}&=\gf x'+u',\qquad x''_{out}=\gf x''+u''\qquad S_{u'u''}=0\qquad S_{u'}=S_{u''}=\G-1> 0\nonumber\\
\gf&\equiv b/a = \sqrt{\G}=\frac{1-G}{1+G}\nonumber\\
\De J&=\De (\abs{b}^2-\abs{a}^2)=a(\gf x'_{out}-x')
\end{align}
The output rate fluctuation $\Delta J$ from the conductance is amplified by an electrical amplifier (e.g., a maser) of gain $F$ (electrical feedback factor, properly normalized). The amplified current next modulates the phase of the \emph{input} optical beam. Once modulated, the input beam is transmitted through a phase-shifter that interchanges in-phase and quadrature components, as shown in the figure. The device that exchange in-phase and quadrature components will hence-forth be referred to as an "all-pass filter". The resulting optical signal is the $a$-wave entering into Port 1 of the circulator, see \ref{feedb}
 
\setlength{\figwidth}{0.6\textwidth}
\begin{figure}
\centering
\includegraphics[width=\figwidth]{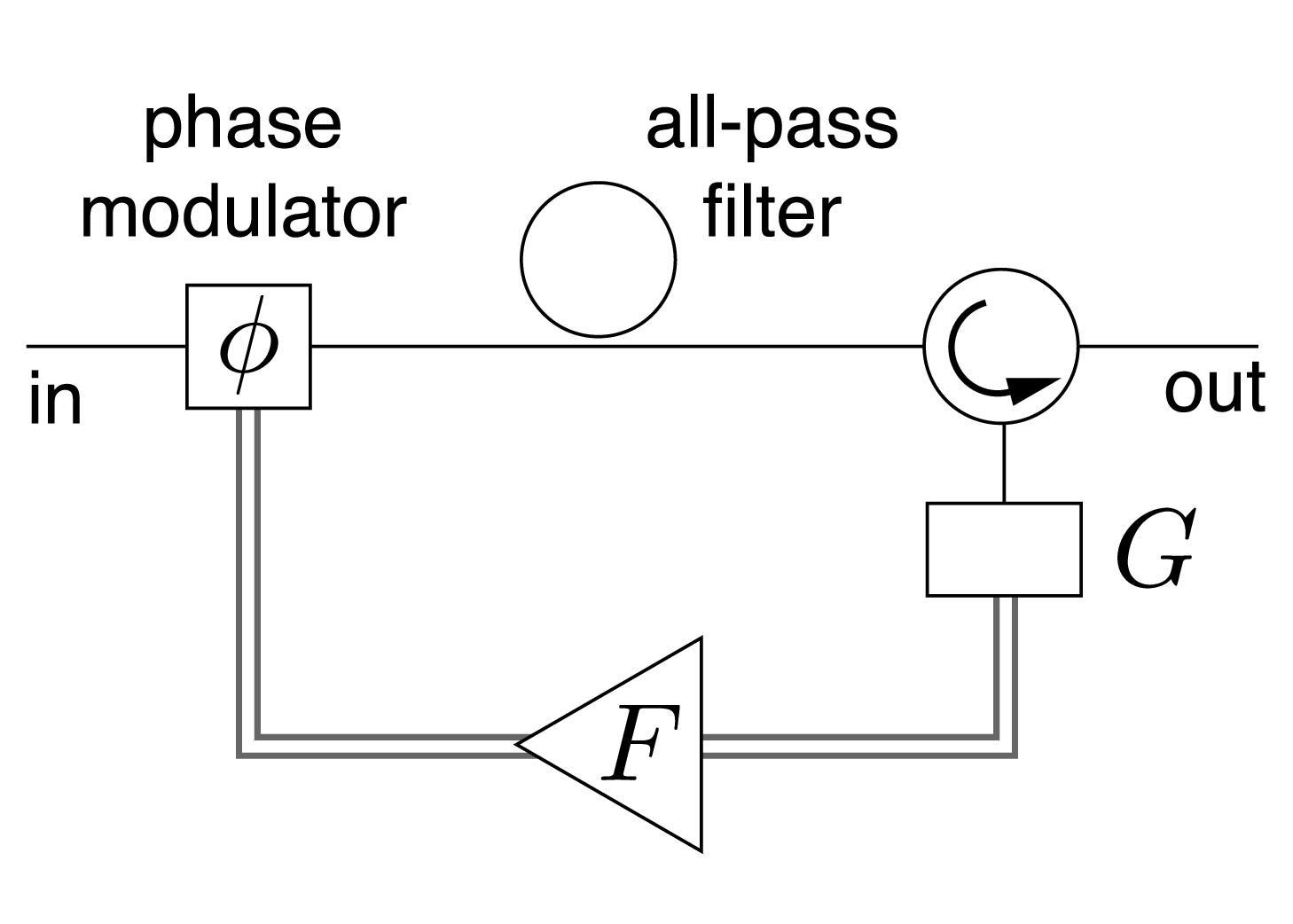}
\caption{Feedback amplifier involving a linear amplifier, a phase modulator, and a quadrature components exchanger (detuned all-pass filter). F is a low-frequency amplifier (e.g., a maser) assumed to be loss-less and linear.}
\label{fig_feedb} 
\end{figure}

The in-phase input component $x'_{in}$ is unaffected by the phase modulator and is converted to $x''=x'_{in}$ by the all-pass filter. Therefore, according to \eqref{79}, $x''_{out}=\gf x'_{in}+u''$. Since $x'_{in}$ and $u''$ are uncorrelated, the spectral densities are related by
\begin{align}
\label{83}
Y_{out}=\G X_{in}+ \G-1
\end{align}

The second relation is more difficult to establish. The electrical fluctuation of the current generated (or absorbed) by the conductance $G$ reads $\De J=\De (\abs{b}^2-\abs{a}^2)=a(\gf x'_{out}-x')$ according to \eqref{79}. This current is multiplied by the electrical gain (or feedback factor) $F$ and modulates the phase of the input beam. The quadrature component, initially $x''_{in}$ gets therefore incremented by $Fa(\gf x'_{out}-x')$    After quadrature interchange, the quantity $x''_{in}+Fa(\gf x'_{out}-x')$ just evaluated becomes the in-phase component $x'$. Thus 
\begin{align}
\label{84}
x'=x''_{in}+Fa(\gf x'_{out}-x').
\end{align} 
Setting for brevity $1+Fa\equiv-f$, the above relation reads
\begin{align}
\label{85}
f x'=(1+f)\gf x'_{out}-x''_{in}
\end{align}
But, since $x'_{out}=\gf x'+u'$, we obtain, multiplying \eqref{85} throughout by $\gf$ and replacing $\gf x'$ by $x'_{out}-u'$
\begin{align}
\label{86}
[f(\G-1)+\G] x'_{out}=\gf x''_{in}-fu'.
\end{align}
Since $u'$ is independent of $x''_{in}$ and has a spectral density equal to $\G-1$, the spectral density $X_{out}$ of $x'_{out}$ reads
\begin{align}
\label{87}
X_{out}=\frac{\G Y_{in}+f^2(\G-1)}{[\G+f(\G-1)]^2}.
\end{align}
For a fixed value of $\G$ and an input quadrature spectrum $Y_{in}$, the quantity $X_{out}$, as given above, reaches its minimum value when $f=Y_{in}$\footnote{$X(f)=(cY+(c-1)f^2)/(c+(c-1)f)^2$, where $c,Y$ denote constants, is stationary when $f=Y$. The reciprocal of $X$ then reads: $1/X=c-1+c/Y$. In the present situation we have: $c=\G, Y=Y_{in}$.}. For that particular value of the normalized feedback factor $f$, we find after rearranging that
\begin{align}
\label{88}
\frac{1}{X_{out}}= \frac{[\G+Y_{in}(\G-1)]^2}  {\G Y_{in}+Y_{in}^2(\G-1)}=\frac{\G}{Y_{in}}+\G-1
\end{align}

To summarize, the input-output relations in the present configuration are
\begin{align}
\label{188}
Y_{out}&=\G X_{in}+ \G-1\nonumber\\
\frac{1}{X_{out}}&=\frac{\G}{Y_{in}}+\G-1
\end{align}

\paragraph{Two feedback amplifiers in chain.}
Let us now suppose that two such amplifiers, with gains $\mathcal{G}_{1}$ and $\mathcal{G}_{2}$ respectively, are assembled one behind the other. If $X,Y$ denote intermediate noise values, the input-output relations read
\begin{align}
\label{73}
Y&=\mathcal{G}_{1}X_{in}+\mathcal{G}_{1}-1\\
\label{74}
1/X&=\mathcal{G}_{1}/Y_{in}+\mathcal{G}_{1}-1
\end{align}
and
\begin{align}
\label{75}
Y_{out}&=\mathcal{G}_{2}X+\mathcal{G}_{2}-1\\
\label{76}
1/X_{out}&=\mathcal{G}_{2}/Y+\mathcal{G}_{2}-1
\end{align}

If we eliminate $X,Y$ and we select $\mathcal{G}_{2}=2-1/\mathcal{G}_{1}$, the total gain $\mathcal{G}=\mathcal{G}_{1}\mathcal{G}_{2}=2\mathcal{G}_{1}-1$. We obtain 
\begin{align}
\label{77}
X_{out}=\frac{X_{in}+\mu}{\mu X_{in}+1}\qquad Y_{out}=\frac{Y_{in}+\mu}{\mu Y_{in}+1}
\end{align}
where
\begin{align}
\label{78}
\mu\equiv\frac{\mathcal{G}-1}{\mathcal{G}+1}
\end{align}
It follows from these expressions that if the input beam is in the C-state ($X_{in}=Y_{in}=1$), the output beam is also in the C-state ($X_{out}=Y_{out}=1$), but is amplified. We call this device a C-amplifier.

\section{Feedback amplifier with gain compression. }

We now suppose that the gain (or conductance) depends explicitely, not only on the number $n$ of electrons in the conduction band, but also on the emitted light quanta rate $R$, that is, we have $G(n,R)$. We define the parameter $\kappa\equiv-(R/G)(\partial G/\partial R)$. 
The previous result may be generalized (exactly) to the case where the conductance suffers from gain compression, that is, depends on the emitted rate, $\kappa>0$. The relations become 
\begin{align}
\label{89}
(1+\kappa \gf)x_{out}&=(\gf+\kappa) x_{in}+u',\\
\label{90}
\gf&\equiv \sqrt{\G}\qquad\qquad 
S_{u'}=1-\G. 
\end{align}
The expression of $X_{out}$ as a function of the normalized feedback factor $f$ reads now
\begin{align}
\label{91}
X_{out}=\frac{(\gf+\kappa)^2Y_{in}+f^2(\G-1)}{(f(\G-1)+\G+\kappa \gf)^2}.
\end{align}
If we employ the general result of the previous section we find that the reciprocal of the minimum value of $X_{out}$ reads 
\begin{align}
\label{92}
\frac{1}{X_{out}}=\frac{\G}{Y_{in}}+\G-1,
\end{align}
which coincides with the expression obtained in the previous section, applicable to the special case where the gain compression may be neglected, i.e., $\kappa=0$.

\newpage

\paragraph{Conclusion}\label{conclusion}

We have shown that the photo-electron spectrum originating from a detector submitted to non-fluctuating-pump laser light may be understood in semi-classical terms. Rates are written in the form $R+\De R(t)+r(t)$, where $R$ denotes the average rate, $\De R(t)$ is proportional to the fundamental noise sources in the linearized regime, and the spectral density of the fundamental noise source $r(t)$ is equal to $R$. This conclusion has been reached from different approaches, essentially requiring agreement with Classical Statistical Mechanics formulas. We have treated for the sake of illustration, linear amplifiers and attenuators, and quiet lasers in the linearized regime. We considered power-rate fluctuations and oscillation linewidth.

\bibliographystyle{ieeetr}
\bibliography{book}

\end{document}